\documentclass[11pt,twoside,openright,a4paper]{book}
\usepackage{amssymb}\usepackage{amsfonts}
\usepackage{amsmath}
\usepackage{graphicx}
\numberwithin{equation}{chapter}{}
\def\gsim{ \lower .75ex \hbox{$\sim$} \llap{\raise .27ex \hbox{$>$}} }
\def\lsim{ \lower .75ex\hbox{$\sim$} \llap{\raise .27ex \hbox{$<$}} }

\title{\LARGE\bf RADIATIVE PROCESSES IN HIGH ENERGY ASTROPHYSICS}

\author{\Large Gabriele Ghisellini\\
\\
{\it INAF -- Osservatorio Astronomico di Brera}} 
{}

\begin{document}

\maketitle {}{} 
\cleardoublepage

\tableofcontents 


\chapter{Some Fundamental definitions}

\section{Luminosity}

By luminosity we mean the quantity of energy irradiated per second [erg s$^{-1}$].
The luminosity {\it is not} defined per unit of solid angle.
The {\it monochromatic} luminosity $L(\nu)$ is the luminosity per unit
of frequency $\nu$ (i.e. per Hz).
The {\it bolometric} luminosity is integrated over fequency:
\begin{equation}
L\, =\, \int_0^\infty L(\nu)\, d\nu
\end{equation}
Often we can define a luminosity integrated in a given energy (or frequency)
range, as, for instance, the 2--10 keV luminosity. 
In general we have:
\begin{equation}
L_{[\nu_1-\nu_2]} \, =\, \int_{\nu_1}^{\nu_2} L(\nu)\, d\nu
\end{equation}

Examples:
\begin{itemize}

\item 
Sun Luminosity: $L_\odot =4 \times 10^{33}$ erg s$^{-1}$

\item 
Luminosity of a typical galaxy: $L_{\rm gal}\sim 10^{11} L_\odot$

\item Luminosity of the human body, assuming that we emit as a  black--body at
a temperature of (273+36) K  and that our skin has a surface of approximately $S=2$ m$^2$:
\begin{equation}
L_{\rm body} \, =\, S\sigma T^4 \, \sim \, 10^{10} \,\,\, {\rm erg/s} \sim \,10^3 \,\,\,  {\rm W} 
\end{equation}
This {\it is not} what we loose, since we absorb from the ambient a power $L=S\sigma T^4_{\rm amb}
\sim 8.3\times 10^9$ erg/s if the ambient temperature is 20 C (=273+20 K).

\end{itemize}

\section{Flux}
 
The flux [erg cm$^{-2}$ s$^{-1}$] is the energy passing a surface of 1 cm$^{2}$ in one second.
If a body emits a luminosity $L$ and is located at a distance $R$, the flux is
\begin{equation}
F\, =\, {L\over 4\pi R^2};
\quad
F(\nu) \, = \,  {L(\nu) \over 4\pi R^2};
\quad
F \, = \,
\int_0^\infty F(\nu) d\nu
\end{equation}

\section{Intensity}
The intensity $I$ is the energy per unit time passing through a unit surface located
perpendicularly to the arrival direction of photons, per unit of solid angle.
The solid angle appears: [erg cm$^{-2}$ s$^{-1}$ sterad$^{-1}$].
The monochromatic intensity $I(\nu)$ has units [erg cm$^{-2}$ s$^{-1}$ Hz$^{-1}$ sterad$^{-1}$].
It always obeys the Lorentz transformation:
\begin{equation}
{I(\nu)\over \nu^3} \, =\, {I^\prime(\nu^\prime)\over {(\nu^{\prime})}^3}\, =\, {\rm invariant}
\end{equation}
where primed and unprimed quantities refer to two different reference frames.
The intensity {\it does not} depend upon distance.
It is the measure of the irradiated energy along a light ray.

\section{Emissivity}

The emissivity $j$ is the quantity of energy emitted by a unit volume, in one
unit of time, for a unit solid angle 
\begin{equation}
j  \, =\, {\rm erg \over dV \, dt \, d \Omega}   
\end{equation}
If the source is transparent, there is a simple relation between $j$ and $I$:
\begin{equation}
I\, = \, j  R  \qquad ({\rm optically\,\, thin\,\, source})  
\end{equation}

\section{Radiative energy density}

We can define it as the energy per unit volume produced by a luminous source,
but we have to specify if it is per unit solid angle or not.
For simplicity consider the bolometric intensity $I$.
Along the light ray, construct the volume $dV =c dt dA$
where $dA$ (i.e. one cm$^2$) is the base of the little cylinder
of height $c dt$.
The energy contained in this cylinder is
\begin{equation}
dE\, = \, I c dt dA d\Omega 
\end{equation}
In that cylinder I find the light coming from a given direction,
I can also say that:
\begin{equation}
dE \, =\, u(\Omega) c dt dA d\Omega
\end{equation}
Therefore
\begin{equation}
u(\Omega) \, =\, {I\over c}
\end{equation}
If I want the total $u$ (i.e. summing the light coming from all directions)
I must integrate over the entire solid angle.


\section{How to go from $L$ to $u$}

There at least 3 possible ways, according if we are outside the source,
at a large distance with respect to the size of the source, or if we are inside
a source, that in turn can emit uniformly or only in a shell.

\subsection{We are outside the source}

Assume to be at a distance $D$ from the source.
Consider the shell of surface $4\pi D^2$ and height $c d t$.
The volume of this shell is $dV = 4\pi D^2 c dt$.
In the time $dt$ the source has emitted an energy $dE = L dt$.
This very same energy is contained in the spherical shell.
Therefore the energy density (integrated over the solid angle) is:
\begin{equation}
u  \, =\,  {d E \over dV} \, =\, { L\over 4\pi D^2 c} 
\label{u1}
\end{equation}

\subsection{We are inside a uniformly emitting source}

Assume that the source is optically thin and homogeneous.
In this case the energy density will depend upon the
mean escape time from the source.
For a given fixed luminosity we have that the longer the escape time,
the larger the energy density (more photons accumulate inside the source
before escaping). 
Therefore:
\begin{equation}
u  \, =\,  {L\over V} \langle t_{\rm esc} \rangle
\end{equation}
For a source of size $R$ one can think that the average escape time is 
$\langle t_{\rm esc} \rangle \sim R/c$, but we should consider that 
this is true for a photon created at the center.
If the photon is created at the border of the sphere,
it will have a probability $\sim 1/2$ to escape immediately,
and a probability {\it less} than 1/2 to pass the entire $2R$
diameter of the sphere. Therefore the mean escape time will be less
that $R/c$.
Indeed, for a sphere, it is $\langle t_{\rm esc} \rangle \sim (3/4) R/c$
Therefore:
\begin{equation}
u  \, =\,  {3 L\over 4 \pi R^3 }\, {3 R \over 4 c} = {9 \over 4} \, {L\over 4\pi R^2 c}
\label{u2}
\end{equation}
Note that this is more (by a 9/4 factor) than what could be estimated by Eq. \ref{u1}
setting $D=R$.
We stress that one can use Eq. \ref{u2} only for a thin source.

\begin{figure}[h]
\center
\includegraphics[height=9cm, width=9cm]{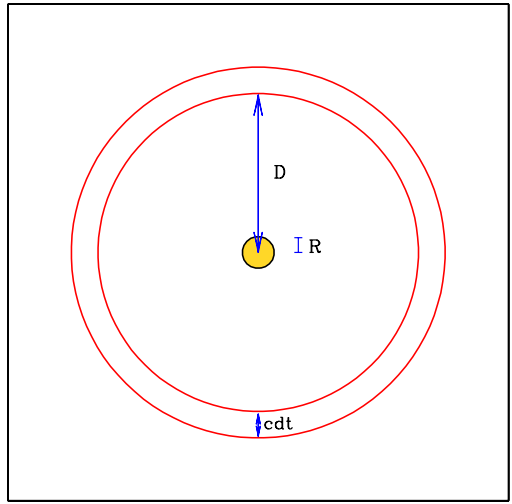}\\
\vskip -1.5 true cm
\caption[h]{
A source of radius R emits a luminosity $L$.
The way to calculate the energy density of the emitted radiation
is different if we are outside the source (at the distance $D$)
or inside the source.
}
\label{chap0:fig1}
\end{figure}{}

\begin{figure}[h]
\includegraphics[height=11cm, width=11cm]{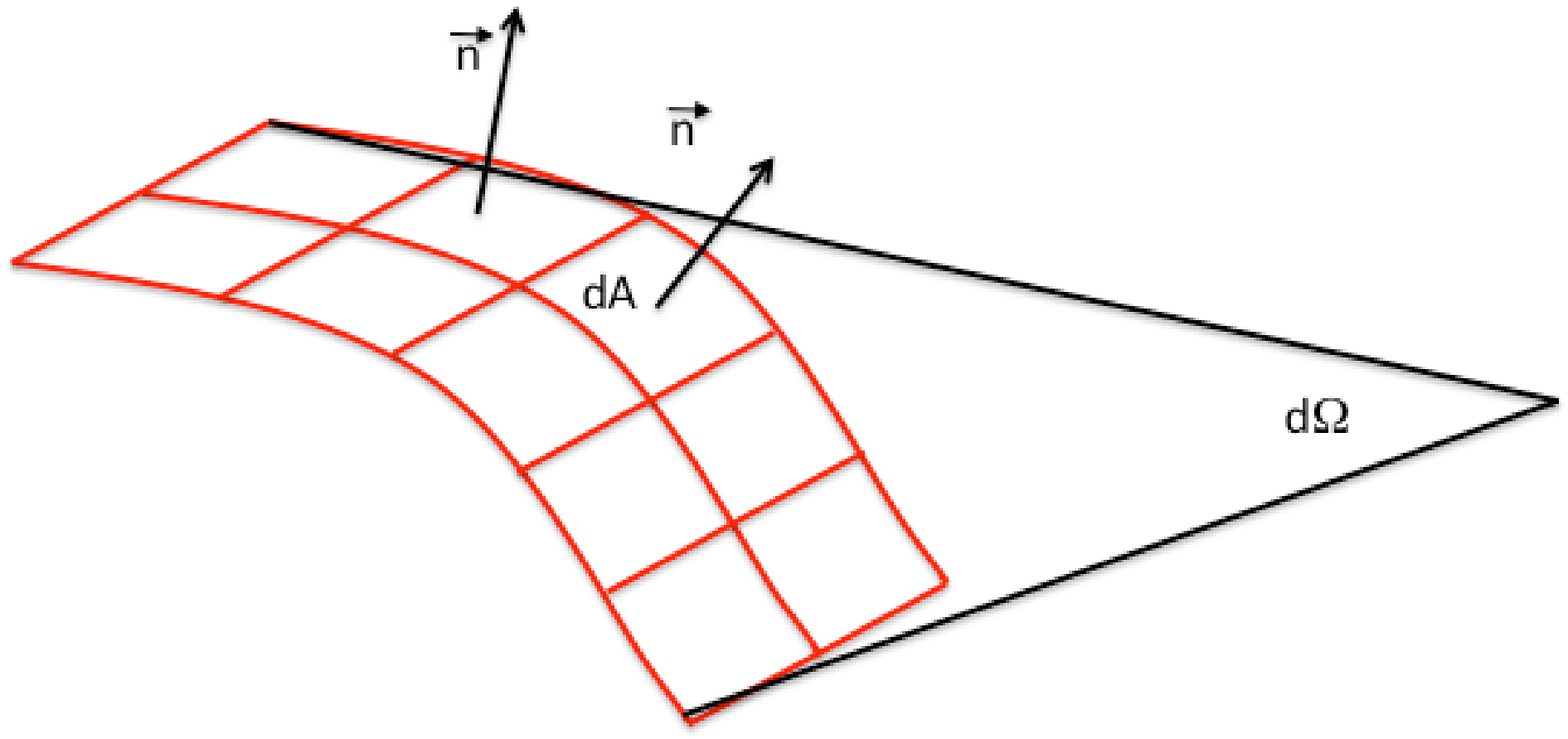}\\
\vskip -4.5 true cm
\caption[h]{This figure illustrates the fact that each element $dA$ of the surface
is seen under a different angle with respect to the normal $\vec n$.
}
\label{chap0:fig2}
\end{figure}{}
%

\subsection{We are inside a uniformly emitting shell}

Assume that a spherical shell of radius $R$ emits uniformly a luminosity $L$
In this case the radiation energy density returns to be
\begin{equation}
u  \, =\,  {L\over 4 \pi R^2 c }\, 
\label{u3}
\end{equation}
and it is equal in any location inside the shell.


\section{How to go from $I$ to $F$}

The relation between the intensity $I$ and the total flux $F$ must 
account for the fact that, in general, each element of the surface 
is seen under a different angle $\theta$. 
See Fig. \ref{chap0:fig2}.
We should then consider the projected area, and introduce a $\cos\theta$ term
in the integral:
\begin{equation}
F  \, =\,  \int I\cos\theta d\Omega
\end{equation}
%

\begin{figure}[h]
\center
\vskip -2 cm
\includegraphics[height=11cm, width=11cm]{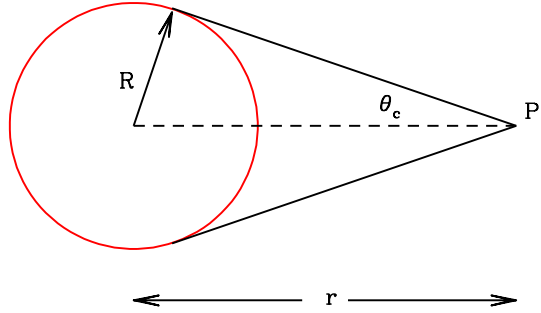}\\
\vskip -3.5 true cm
\caption[h]{The observer P sees the spherical source as a disk of angular radius
$\theta_{\rm c}$. 
Each element of the sphere has a different projection.
}
\label{chap0:fig3}
\end{figure}{}
%

\subsection{Flux from a thick spherical source}

If we see only the surface of a sphere, namely the source is
optically thick, and if the intensity is constant along the surface we have
symmetry along the $\phi$ directions and the solid angle $d\Omega = 2\pi \sin\theta d\theta$.
See Fig. \ref{chap0:fig3}. We get:
\begin{equation}
F  \, =\,  \int I\cos\theta d\Omega \, =\, 
2\pi I \int_0^{\theta_{\rm c}} \sin\theta\cos \theta d\theta
\end{equation}
Here $R=r\sin\theta_{\rm c}$.
Therefore we have
\begin{equation}
F  \, =\, 
2\pi I \left[ {\cos^2\theta\over 2}  \right]_{\cos\theta_{\rm c}}^1  \, =\, \pi I \sin\theta_{\rm c} \, 
=\, \pi I \left( {R\over r} \right)^2
\end{equation}
At the surface, $R=r$, and we have $F=\pi I$.\\
Very far away, $r \gg R$, and we have $F=\pi \theta^2_{\rm c} I = I\Omega_{\rm source}$.


\section{Radiative transport: basics}

Once the radiation is produced in a given location inside the source, 
we have to see how much of this can leave the source and reach the observer.
To calculate that, we must introduce the {\bf absorption coefficient} $\alpha_\nu$,
whose dimension is [length$^{-1}$]. 
It is defined by the following equation, describing the decrement of $I_\nu$
when passing trough an infinitesimal path of length $ds$:
\begin{equation}
dI_\nu  \, =\,  - \alpha_\nu I_\nu ds 
\end{equation}
The absorption coefficient can be thought as the product of a density of ``absorbers" times
the cross section of the absorbing process:
\begin{equation}
\alpha_\nu  \, = \, n \sigma_\nu 
\end{equation}
But inside a source, besides absorption, there is a contribution to $I_\nu$ coming from the
emitters distributed along $ds$.
The increment of $I_\nu$ is
\begin{equation}
dI_\nu  \, =\,   j_\nu ds 
\end{equation}
Therefore, combining emission and absorption, we have the basic equation of radiative transport:
\begin{equation}
{dI_\nu \over ds}  \, =\, - \alpha_\nu I_\nu +  j_\nu 
\end{equation}
We solve it in some specific cases:
\begin{enumerate}
\item
{\bf Emission only:} 
\begin{equation}
{dI_\nu \over ds}  \, =\,   j_\nu \, \to \, I_\nu =I_{\nu, 0} + \int_0^S j_\nu ds
\end{equation}
where $S$ is the total emitting path.

\item
{\bf Absorption only:}
\begin{equation}
{dI_\nu \over ds}  \, =\, - \alpha_\nu I_\nu \, \to \, {dI_\nu \over I_\nu} =  - \alpha_\nu ds
\end{equation}
Note that the form of this relation immediately implies that an exponential is involved:
\begin{equation}
I_\nu (s)  \, =\, I_\nu (s_0)\,  e^{- \int_{s_0}^s \alpha_\nu (s^\prime) ds^\prime}
\end{equation}
Before passing the layer, the intensity was $I_{\nu, 0}$.
While passing the layer of length $s$, the intensity decreases exponentially.

\item
{\bf Emission plus absorption:}
In this case it is convenient to introduce the optical depth $\tau_\nu$:
\begin{equation}
{d\tau_\nu}  \, =\,  \alpha_\nu ds \, =\, n\sigma_\nu ds
\end{equation}
The transport equation then becomes:
\begin{equation}
{dI_\nu \over \alpha_\nu ds}  \, =\, -  I_\nu +   { j_\nu\over \alpha_\nu}\,
\to \, {dI_\nu \over d\tau_\nu }  \, =\, -  I_\nu +   { j_\nu\over \alpha_\nu}\,
\label{eqrad}
\end{equation}
We can call {\bf source function} the quantity $S_\nu$ defined as
\begin{equation}
{S_\nu}  \, =\,{ j_\nu\over \alpha_\nu} \quad {\rm source \, function}
\end{equation}
Then the formal solution of Eq. \ref{eqrad} is
\begin{equation}
I_\nu (\tau_\nu) \, =\,   I_{\nu, 0}\,  e^{-\tau_\nu} +   \int_0^{\tau_\nu}
e^{-(\tau_\nu -\tau^\prime_\nu)} S_\nu(\tau^\prime_\nu) d\tau^\prime_\nu
\end{equation}
here $\tau_\nu$ is the final value of $\tau^\prime_\nu$ (i.e. when the intensity 
has travelled the entire distance $s$).
{\it If} $S_\nu$ is constant:
\begin{equation}
I_\nu (\tau_\nu) \, =\,  I_{\nu, 0} \, e^{-\tau_\nu} +   S_\nu \left( 1-e^{-\tau_\nu}\right)
\end{equation}
{\it If} $I_{\nu, 0}=0$:
\begin{equation}
I_\nu (\tau_\nu) \, =\,    {j_\nu \over \alpha_\nu} \left( 1-e^{-\tau_\nu}\right)
\end{equation}
Now, a trick: multiply and divide the RHS by $s=R$ (the dimension of the source)
to obtain:
\begin{equation}
I_\nu (\tau_\nu) \, =\,    {j_\nu R \over \alpha_\nu R } \left( 1-e^{-\tau_\nu}\right) \, =\,
j_\nu R  \left( 1-e^{-\tau_\nu} \over \tau_\nu  \right) 
\end{equation}
In this form it is immediately clear that when the source is optically thin (and $\tau_\nu \ll 1$),
we have $1-e^{-\tau_\nu} \to 1-1+\tau_\nu$ and therefore
\begin{equation}
I_\nu (\tau_\nu) \, =\, j_\nu R, \qquad (\tau_\nu \ll 1)  
\end{equation}
When instead the source becomes optically thick, and $\tau_\nu\gg 1$, then;
\begin{equation}
I_\nu (\tau_\nu) \, =\, { j_\nu R \over \tau_\nu},  \qquad (\tau_\nu \gg 1)  
\end{equation}
Usually, this happens at low frequencies.
The above equation explicitly shows that the intensity we see from a thick source comes
from a layer of width $R/\tau_\nu$, i.e. the layer that is optically thin.
In other words we always collect radiation from a layer of the source, down to the
depth at which the radiation can escape without being absorbed ($\tau_{\rm layer}=1$).

\end{enumerate}

\section{Einstein coefficients}

The Einstein coefficients concern intrinsic properties of the emitting particles.
As such, they involve more general concepts than the emission or the absorption coefficients.
They concern the probability that a particle spontaneously emits a photon, the probability to
absorb a photon, and the probability to emit a photon under the influence of another incoming photon.
The latter concept, at first sight, is weird.
In the current world we are surrounded by electronic devices using lasers, and therefore we are
used to the concept of {\it stimulated emission}.
This was however introduced by Einstein (in 1917) because, without it, he could not recover
the black--body formula.

Let us define a system in which the emitting electron
can be in one of several energy levels.
Each level has a statistical weight $g_i$.
The simple case that comes into mind is an atom.

\begin{figure}[h]
\vskip -1 cm
\center 
\includegraphics[height=11cm, width=11cm]{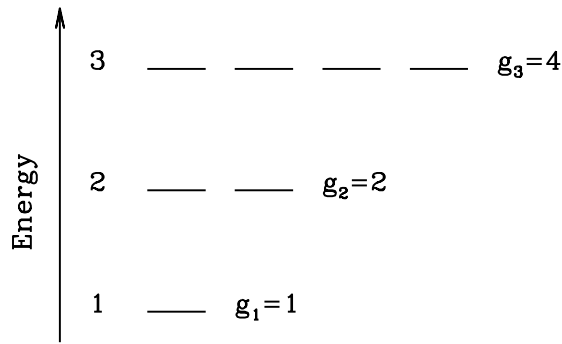}\\
\vskip -4 true cm
\caption[h]{The shown 3 energy levels have different
statistical weights $g_i$.
}
\label{levels}
\end{figure}{}

\begin{figure}[h]
\vskip -2 cm
\center 
\includegraphics[height=11cm, width=13cm]{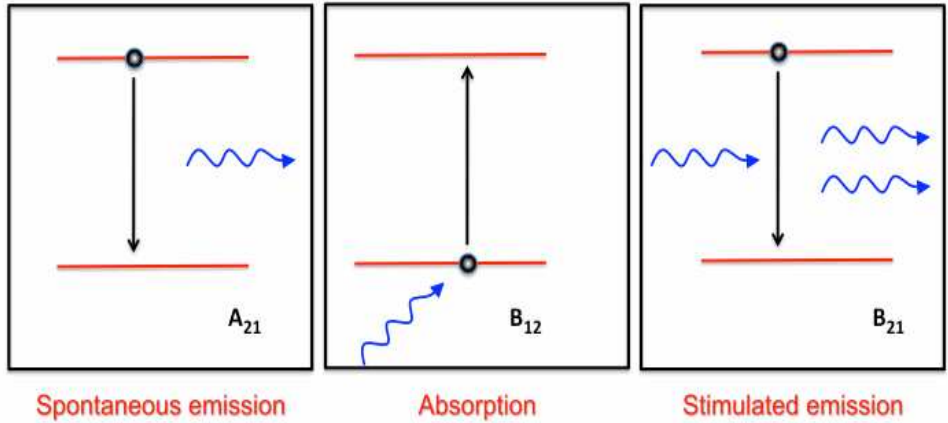}\\
\vskip -3.5 true cm
\caption[h]{Illustrative sketch for the processes of 
spontaneous emission, absorption and stimulated emission.
}
\label{levels}
\end{figure}{}

\noindent
{\bf Spontaneous emission: $A_{21}$ ---}
This process corresponds to a jump of the electrons from energy level 2 to level 1
by emitting a photon of energy $j_\nu$ corresponding to the jump in energy $E_2-E_1$.
The subscripts 1 and 2 indicate energy levels 1 and 2, but are not important, we can equivalently
talk of a jump between the energy levels 5 and 4... and so on.
The corresponding Einstein coefficient is $A_{21}$.
Therefore
\begin{equation}
A_{21}\, =\, {\rm transition\, probability\, for\, spont.\, emission\,\, [s^{-1}]} 
\end{equation}

\vskip 0.5 cm
\noindent
{\bf Absorption: $B_{12}$ ---}
This occurs when the electron is in level 1 and absorb a photon of energy $h\nu$ that corresponds
to the energy difference between levels 1 and 2. 
The probability of one electron to make this transition depends on how many photon there are,
and therefore to the intensity $J_\nu$ of the radiation field.
The latter is averaged over the solid angle: $J_\nu=(1/4\pi)\int I_\nu d\Omega$. 
\begin{equation}
B_{12}J_\nu\, =\, {\rm transition\, probability\, for\, absorption\, per\, unit\, time} 
\end{equation}

\vskip 0.5 cm
\noindent
{\bf Stimulated emission: $B_{21}$ ---}
This occurs when the electron is in level 2, and the arrival of a photon of energy
$h\nu$ corresponding to the energy difference between levels 2 and 1 makes the electron
jump to level 1 by emitting a photon.
The energy of the emitted photon is the same of the energy of the incoming one.
It can be shown that also the direction and the phase of the two photons are the same.
We then have a coherent emission, at the base of all laser and maser devices.
The probability of one electron to make the $2\to 1$ transition depends on how many photons
there are, and therefore on the radiation intensity:
\begin{equation}
B_{21}J_\nu\, =\, {\rm transition\, probability\, for\, stimulated\, emission\, per\, unit\, time} 
\end{equation}

\vskip 0.5 cm
\noindent
The Einstein coefficients are completely general, and are valid for any radiation field
(i.e. even out of equilibrium).
They are valid even when we have a free particle, (i.e. not an atom), because we can define 
even in that case the energy state of the particle.

In particular, we can find the relations between the Einstein coefficients in the particular
case of equilibrium, that is when the transitions  $2 \to 1$ are equal to the transitions $1\to 2$.
Assume then that $n_1$ and $n_2$ are the number density of electrons in the levels 1 and 2.
We must have:
\begin{equation}
n_1 B_{12}J_\nu\, =\, n_2 A_{21}+ n_2 B_{21} J_\nu
\end{equation}
Solving for $J_\nu$:
\begin{equation}
J_\nu \, =\, { A_{21}/B_{21} \over (n_1/n_2) (B_{12}/B_{21}) -1}
\label{jnu}
\end{equation}
At equilibrium the ratio $n_1/n_2$ must be:
\begin{equation}
{n_1\over n_2}\, =\, {g_1\over g_2}  \, { e^{-E/kT} \over e^{-(E+h\nu)/kT} } \, =\, {g_1\over g_2} e^{h\nu/kT}
\end{equation}
where $g_1$ and $g_2$ are the statistical weights of the levels 1 and 2.
Inserting in Eq. \ref{jnu} we have
\begin{equation}
J_\nu \, =\, { A_{21}/B_{21} \over (g_1 B_{12} /g_2 B_{21}) e^{h\nu/kT} -1}
\label{jnu}
\end{equation}
We know that at equilibrium $J_\nu$ must be equal to the black--body intensity $B_\nu$.
For this to happen it must be that:
\begin{eqnarray}
g_1 B_{12} \, &=&\, g_2 B_{21}
\nonumber \\
A_{21}\, &=& {2 h\nu^3 \over c^2} \, B_{21}
\label{einstein}
\end{eqnarray}
These are the relations among the Einstein coefficients.
Although we have found them in one particular case, these relations are always valid,
also out of equilibrium, because they concern intrinsic properties of the system
(note for instance that they do no depend on temperature).
The relations between the Einstein coefficients imply that we need to know 
only one of them to find the others.

\subsection{Emissivity, absorption and Einstein coefficients}

The emissivity [erg s$^{-1}$ cm$^{-2}$ Hz$^{-1}$ sterad$^{-1}$]
is related to the Einstein coefficient $A_{21}$.
If $n_2$ is the density of electrons in the state 2, we must have
\begin{equation}
j_\nu \, =\, {h\nu \over 4\pi} \, n_{2} A_{21}
\end{equation}
The reason of the $4\pi$ is because $n_2 A_{21}$ is the probability to emit
a photon (for spontaneous emission) {\it in any direction}, while
the emissivity $j_\nu$ is for unit of solid angle.

Also the stimulated emission produces photons. Why do we not include it 
the calculus for the emissivity?
Because, to have stimulated emission, we must have the incoming radiation,
and it is therefore more convenient to think to stimulated emission as
{\it negative} absorption.
For the {\it total} absorption, consider the quantity $\alpha_\nu I_\nu$. 
This is the amount of energy absorbed for unit volume, time, frequency and solid angle.
We also have that $n_1 B_{12} I_\nu$ is the number of the $1\to 2$ transitions per unit
time, volume and frequency.
We must multiply by $h\nu$ to have the energy (per second, Hz, cm$^3$) and divide by $4\pi$
to have the same quantity per unit solid angle. 
$I_\nu$ drops out and we have
%


\begin{eqnarray}
\alpha_\nu  &=& {h\nu \over 4\pi} \, (n_{1} B_{12} - n_2 B_{21}) 
\nonumber \\
 &=& \left({n_1 g_2\over g_1} - n_2\right) B_{21}
 \nonumber \\
 &=& \left( {n_1 g_2\over g_1} - n_2\right) {A_{21} c^2 \over 8\pi \nu^2 }
 \nonumber \\
 &=& \left( {n_1 g_2\over n_2 g_1} - 1  \right) { c^2  \over 2 h \nu^3 }\, j_\nu
\end{eqnarray}
We conclude that absorption and emission are intimately linked, and knowing
one process is enough to derive the other one, from first principles.
If a particle emits, it can also absorb.

\section{Mean free path}

The mean free path (for a photon) is the average distance $\ell$ 
travelled by a photon without interacting. 
It corresponds to a distance for which $\tau_\nu=1$:
\begin{equation}
\tau_\nu \, =\, 1 \, \to \, \sigma_\nu n \ell_\nu \, =\, 1 \, \to  \ell_\nu \, 
=\, {1\over n \sigma_\nu } 
\end{equation}
If a source has radius $R$ and total optical depth $\tau_\nu>1$ we have:
\begin{equation}
\ell_\nu \, =\, {1\over  \sigma_\nu n} \, =\, {R\over \sigma_\nu n R}\, =\, {R\over \tau_\nu}
\end{equation}

\subsection{Random walk}

Assume that a photon interacts through scattering with electrons inside a source
of radius $R$ and optical depth $\tau>1$.
It is a good approximation to assume that the scattering angle is random, so the
direction after a collision will be random. 
The photon will travel ``bouncing" from one electron to the next, each time traveling
for a mean free path.
We then ask: i) how many scatterings does it have to do before escaping?
ii) How much time does it takes?

\begin{figure}[h]
\vskip -1.5 true cm
\includegraphics[height=11cm, width=11cm]{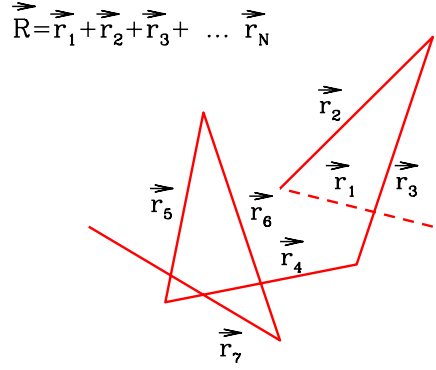}\\
\vskip -4. true cm
\caption[h]{Random walk of a photon inside a source.}
\label{chap0:fig4}
\end{figure}{}
%

Fig. \ref{chap0:fig4} illustrates the random motion of a photon inside the source.
Each segment corresponds to a free path, whose average length is $\ell$.
The total displacement after $N$ scatterings will be $\vec R$, but if we calculate $\vec R$
simply summing all the vectors $\vec r_1$, $\vec r_2$, $\vec r_3$ .... $\vec r_N$, 
we obtain 0.

To calculate the distance travelled by the photon we calculate the {\it square} of $\vec R$:
\begin{eqnarray}
\langle \vec R^2 \rangle \, &=&\, 
\langle \vec r_1^2 \rangle + 
\langle \vec r_2^2 \rangle + 
\langle \vec r_3^2 \rangle 
\nonumber\\
&+& 
2\langle \vec r_1 \cdot \vec r_2 \rangle + 
2\langle \vec r_1 \cdot \vec r_3 \rangle + ....
\end{eqnarray}
The cross products, on average, vanish, because the $\cos\theta$ between the vectors vanishes, on average.
Therefore we remain with the squares only, and each term $r_i$ is, on average, long as a mean free path.
We have $N$ of these terms.
Therefore 
\begin{equation}
\langle \vec R^2 \rangle \, =\,
N  \langle \vec r_i^2 \rangle \, =\, N \ell^2 \, \to \,  \sqrt{\langle \vec R^2 \rangle} =\sqrt{N} \ell
\end{equation}
If we want to know how many scatterings, on average, a photon undergoes before escaping a source
of radius $R$, we write:
\begin{equation}
\sqrt{N}  \, =\, {R\over \ell} \, =\, R \sigma n \, =\, \tau \to N=\tau^2
\end{equation}
Remember: this is valid only if $\tau>1$ (if $\tau<1$ then the majority of photon escape the source
without interacting...).

The other question is how long does it take to escape.
If $t_1 = \ell /c$ is the average time for one scattering, we have a total time $t_{\rm tot}$
\begin{equation}
t_{\rm tot}  \, =\, N t_1 \, =\, \tau^2 {\ell \over c} \, =\, \tau^2 \, {R \over R \sigma n c} \, =
{R\over c} \, {\tau^2 \over \tau}\, =\, \tau \, {R\over c}
\end{equation}
So, on average, the photons makes $\tau^2$ scatterings before escaping a source with 
a scattering optical depth $\tau>1$. Since each mean free path is long $R/\tau$, the
total distance travelled by the photon is $\tau R$ and of course the total time
to escape is $\tau R/c$.
Beware that this is true {\it on average}. As we shall see, the formation of 
the high energy spectrum through Comptonization involves those photons that
make more than $\tau^2$ scatterings, because in this way they can reach higher 
energies.

\section{Thermal and non thermal plasmas}

By definition, a {\it thermal plasma} is characterized by a Maxwellian distribution
of particles.
Therefore a {\it non thermal plasma} is anything else.
The non relativistic Maxwellian distribution is:
\begin{equation}
F(v) dv  \, =\, 4\pi v^2 \left(  {m \over 2\pi kT } \right)^{3/2} e^{-mv^2 /2kT} dv
\end{equation}
In this form, $F(v)$ is normalized, i.e. $\int_0^\infty F(v) dv = 1$.
This can be seen changing variable of integration, from $v$ to $x= mv^2/(2kT)$,
and remembering that $\int_0^\infty \sqrt{x} e^{-x} dx = \sqrt{\pi}/2$.

It is worth to stress that the physics is in the exponential term, while the $4\pi v^2 dv$ term
is simply equal to $dv_x dv_y dv_z$ (in three dimensions).
This then suggests the questions: it is possible to have a ``Maxwellian--like" distribution in 2 dimensions?
And in one dimension?

Instead of the {\it velocities} we may consider the {\it momenta} $p$ of the particles.
If we write
\begin{equation}
p  \, \equiv \, \gamma\beta mc 
\end{equation}
where $\gamma=1/(1-\beta^2)^{1/2}$, the above relation is valid both for non--relativistic
and for relativistic velocities.
The Maxwellian momenta distribution becomes (setting $\Theta\equiv mc^2$):
\begin{equation}
F(p) dp  \, =\, { p^2 e^{-\gamma/\Theta} \over \Theta m^3 c^3 K_2(1/\Theta) } \, dp
\end{equation}
where $K_2(1/\Theta)$ is the modified Bessel function of the second kind.

Note the following:

\begin{itemize}
\item
To define a temperature, the distribution of velocities must be {\it isotropic}.

\item
Written in terms of momenta, the Maxwellian distribution has the same form
in the relativistic limit.

\item
The Maxwellian distribution is very general, it is a result of statistical mechanics.
But to achieve this distribution, it is necessary that the particles {\it exchange energy between
themselves}.

\item
If competing processes are present (i.e. cooling) it is possible that one has a Maxwellian
distribution only in some interval of velocities/momenta (for instance for low velocities).

\item
The exponential term $e^{-E/kT}$ contains the physics, the term $p^2$ is simply due
to $dp_x dp_y dp_z = 4\pi p^2 dp$.

\end{itemize}

\subsection{Energy exchange and thermal plasmas}

There are two main ways in which particles can exchange energy among themselves:
\begin{enumerate}
\item 
Collisions

\item 
Exchange of photons (emission and absorption; scattering)

\end{enumerate}
Traditionally, one thinks to collisions as the main driver, and to Coulomb collisions as 
the main mechanism.
Of course, the more the collisions, the faster the energy exchange and the faster the
relaxation towards the thermal equilibrium.
We have
\begin{eqnarray}
{\rm \# \, of\, collisions\, of\, a\, single\, particle \over time} \, &\propto& {\rm density}\, \, n
\nonumber \\
{\rm total\, \# \,of\, collisions\, \over time} \, &\propto&  \, n^2
\end{eqnarray}
However, the density $n$ is not the only factor.
The other factor is the energy of the particle: the cross section {\it decreases} with the velocity
of the particle (and then with its energy, or the temperature of the plasma).
This means that
{\bf it is difficult to have a Maxwellian in hot and rarefied plasmas.}

Ask yourselves: what does it happen if I put particles all of the same energy in a box with
reflecting and elastic walls?
Why should the energy of a single particle change?

\subsection{Non--thermal plasmas}

In rarefied and hot plasmas, the {\it relaxation time} required to go to equilibrium,
and allow for sufficient energy exchange among particles is long, compared to the
typical timescales of other processes, as acceleration, cooling and escape.
The particle distribution responsible for the radiation we see is then shaped
by these other, more efficient, processes.

The queen of the non--thermal particle distribution is a power law:
\begin{equation}
N(E) \, =\, N_0 E^{-p}  
\end{equation}
When all particles are relativistic we can equivalently write
\begin{equation}
N(\gamma)  \, = \, K \gamma^{-p}
\end{equation}
Usually, $N$ is the density [cm$^{-3}$] of the particles,
but sometimes it can indicate the total number.
Furthermore, one can also specify if $N(\gamma	)$ is
per unit of solid angle (when one has a distribution
that is not isotropic), or not.
One must also specify in what energy range $N(\gamma)$ is valid,
so in general $N(\gamma)$ has a low  and an high energy 
cutoff ($\gamma_{\rm min}$ and $\gamma_{\rm max}$).
Within these two limits, a power--law particle distribution has no preferred
energy (there is no peak or break).

Often one deals with a {\it broken power--law} distribution, defined
by two power-laws of slopes $p_1$ and $p_2$ joining at some $\gamma_{\rm break}$.

The main acceleration process leading to a power--law distribution (but not the only one)
is shock acceleration. 
In this process a particle gains energy each time it crosses the shock,
and there is a probability, each passage, that the particle escape.
This is sufficient to yield a power--law distribution.

\section{Coulomb Collisions}

\begin{figure}[h]
\center
\includegraphics[height=12cm, width=15cm]{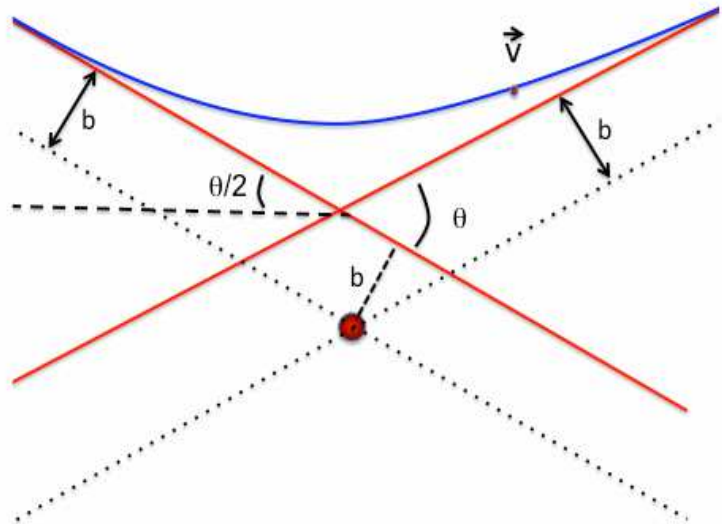}\\
\vskip -2. true cm
\caption[h]{
Sketch of the trajectory of the charge $q_1$ (of mass $m_1\ll m_2$)
that is deflected by the charge $q_2$.
}
\label{chap0:fig5}
\end{figure}

Fig. \ref{chap0:fig5} shows the trajectory of a particle of mass $m_1$ and charge $q_1$ 
when it passes near a particle of mass $m_2$ and charge $q_2$.
For the example in the figure the charges are of the same sign, and $m_2\gg m_1$.
In general, we can introduce the {\it reduced mass} $\mu$ defined as
\begin{equation}
\mu \, = \, {m_1m_2 \over m_1 +m_2}, \quad \mu  \longrightarrow m_1 \quad {\rm if\, m_2 \gg m_1}
\end{equation}
%

{}

The impact parameter $b$ corresponds to the minimum distance of the {\it unperturbed trajectory}
from the mass $m_2$.
The {\it deflection angle} $\theta$ is the angle between the incoming and outgoing velocity vectors 
(both measured at large distances from the location of $m_2$).

One also defines the quantity
\begin{equation}
\tan\theta/2 \, = \, {b_0(v) \over b}
\end{equation}
where $b_0(v)$ corresponds to the impact parameter for deflection angles of $90^\circ$
(namely, when $\tan\theta/2=1$. In this case $b_0(v)=b$).
$b_0(v)$ is related to the electrostatic force.
One makes the approximation that the interaction between the two particles is important
only when the particle 1  is close to the particle 2. 
One can then say that the interaction time is of the order of $t\sim b_0 /v$.
The acceleration can then be approximated by
$a\sim v/t \sim v^2/b_0$.
On the other hand, the force responsible for this acceleration is $F=q_1 q_2/b^2_0$,
and the acceleration is then $a=q_1q_2 /(\mu b_0^2)$.
Equating the two expression for the acceleration we have:
\begin{equation}
b_0(v) \, =\,  { q_1 q_2 \over \mu v^2}
\end{equation}
Therefore
\begin{equation}
b \, =\, {b_0(v) \over \tan\theta/2} \, =\, { q_1 q_2 \over \mu v^2 \tan(\theta/2)}
\end{equation}

\subsection{Cross section}

It is easy to understand that particles with a large impact parameter will be deflected 
much less than particles that could pass close to $q_2$.
Also, we can understand that particles with larger velocities, thus interacting
for a shorter time, will be deflected less.
We can simply write the infinitesimal cross section $d\sigma$ as
\begin{eqnarray}
d\sigma \, &=& \, 2\pi \,b \, |db|
\nonumber \\
&=& \, 2\pi \,b \, |{\partial b \over \partial \theta } \, d\theta|
\end{eqnarray}
The differential cross section (i.e. as a function of deflection angle) is
\begin{eqnarray}
{d\sigma \over d \Omega }\, &=& \, {2\pi \,b \over 2\pi \sin\theta } \, 
\left|{\partial b \over \partial \theta } \, {d\theta\over d\theta} \right|
\nonumber \\
&=& \, {b \over 2\sin(\theta/2) \cos(\theta/2)} \, \left|{\partial \over \partial\theta} 
\left[{b _0(v) \over\tan(\theta/2) }\right] \right|
\nonumber \\
&=& { (q_1 q_2)^2 \over 4 \sin^4(\theta/2) \mu^2 v^4}
\end{eqnarray}
For electron--proton interactions of non--relativistic particles:
\begin{equation}
{d\sigma \over d \Omega }\, = \, { e^4 \over 4 m_e^2 v^4}\, {1\over \sin^4(\theta/2) }, \qquad {v\ll c}
\end{equation}
In the relativistic case, accounting also for the magnetic moment of the electron:
\begin{equation}
{d\sigma \over d \Omega }\, = \, { e^4 \over 4 m_e^2 v^4}\, {1\over \gamma^2 } \, \,
{1-\beta^2\sin^2(\theta/2) \over \sin^4(\theta/2) }, \qquad {v\to c}
\end{equation}
Note the following:
\begin{enumerate}

\item 
$$ {d\sigma \over d\Omega} \propto v^{-4}$$
The number of collisions in the unit of time is proportional to 
$\langle \sigma v\rangle \propto v^{-3}$
(a faster electron encounters more protons in the unit of time).
For a Maxwellian,  $\langle v\rangle \approx T^{1/2}$.
Then, for the non--relativistic case:
\begin{equation}
{\rm \# \, of \, collisions \over time }\, \propto \, v^{-3} \, \propto \, T^{-3/2}
\end{equation}
Collisions are then less frequent in hotter plasma.
 
\item
$$ {d\sigma \over d\Omega} \propto \sin^{-4}(\theta/2)$$
Collisions with small deflections are largely favorite.
This is because it is more probable to have distant particles,
rather than close ones.

\item
As a consequence of 2), the field due to the spin (magnetic moment of the electron)
is negligible with respect to the Coulomb one.
Therefore, in general, the term $(1-\beta^2\sin\theta/2)$ is negligible.

\item 
If we have electron--proton collisions, 
$$ {d\sigma \over d\Omega} \propto {e^4 \over m_{\rm e}^2 v^{4} } = {e^4 \over m_{\rm e}^2\beta^4} =
{r_0 \over \beta^4}$$
where $r_0\equiv e^2/m_{\rm e}c^2 $ is the classical radius of the electron.
We can see that for non--relativistic particles the cross section can be much larger than
the scattering cross section $\sigma_{\rm T} = (8\pi /3)r_0^2$.

\item 
In the relativistic case:
$$ {d\sigma \over d\Omega} \propto \, { 1\over m^2 \gamma^2}$$
This simply flags the fact the relativistic mass is larger (by a factor $\gamma$):
the particle has more inertia, and deflects less, yielding a lower cross section.

\end{enumerate}


\section{The electric field of a moving charge}

The treatment follows Rybicky \& Lightman (p. 80) and Jackson (\S 14.1).
The electric and magnetic fields produced by a moving charge is:
\begin{eqnarray}
\vec E(\vec r, t) \, &=&\, \left[  {q \over k^2 R^2}\, 
{ (\vec n -\vec \beta) \over \gamma^2  }\right]_{t_{\rm ret}}
+ \, {q\over ck^3 R} 
\left\{ 
\vec n \times \left[ (  \vec n -\vec \beta ) 
\times \vec {\dot \beta} \,\, \right]  
\right\}_{t_{\rm ret}}
\nonumber \\
\vec B(\vec r, t) \, &=&\, \vec n \times \vec E
\label{eret}
\end{eqnarray}
%

\begin{figure}[h]
\center
\vskip -0.5 cm
\includegraphics[height=11cm, width=15cm]{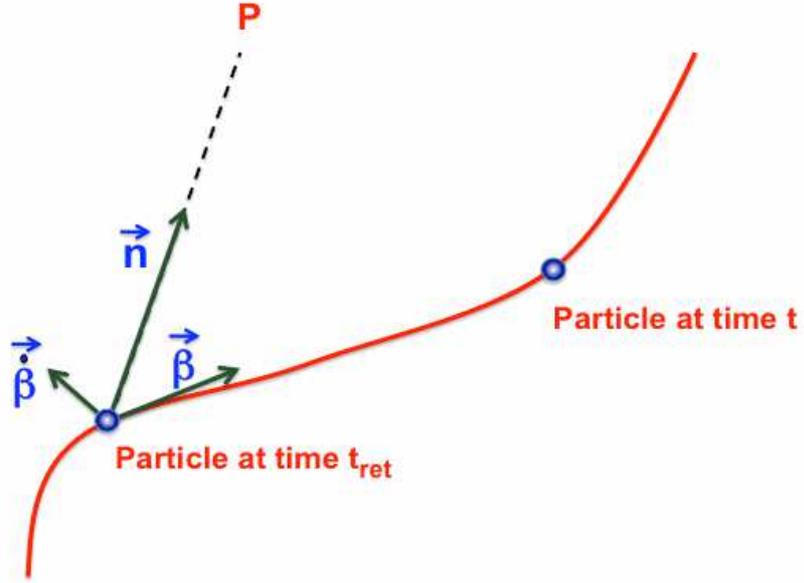}\\
\vskip -2 true cm
\caption[h]{
A charge is moving along a trajectory.
At the time $t$ we measure the electric field produced by the charge.
To find it out, we have to calculate its position, velocity and acceleration
at the retarded time $t_{\rm ret}$.
}
\label{e}
\end{figure}{}

All the quantities must be evaluated at the {\it retarded time} (it is called {\it retarded},
but is really a {\it earlier} time): at the retarded time the particle was at a distance
$R$ such that $t_{\rm ret}+R/c$ is the time $t$ (the time $t$ is the time of the observer
that measures the electric field).

We have the following definitions:
\begin{eqnarray}
t_{\rm ret}\, &=& \, t - { R(t_{\rm ret}) \over c }
\nonumber \\
k \, &=& \, 1-\vec n \cdot \vec \beta
\nonumber \\
\vec \beta \, &=& \, {\vec v \over c}
\nonumber \\
\gamma \, &=& \, (1-\beta^2)^{-1/2}
\end{eqnarray}
We see that Eq. \ref{eret} for the electric field is made by two pieces.
\begin{enumerate}
\item
The first is called the {\it velocity  field}:
acceleration does not appear, and $\vec E$ is proportional to $R^{-2}$.
\item
The second is called {\it acceleration field}: it contains $\dot \beta$, i.e.
the acceleration, and is proportional to $R^{-1}$.
It is perpendicular to $\vec n$ and to the acceleration $\vec {\dot \beta}$.
\end{enumerate}
\begin{figure}[h]
\center
\includegraphics[height=11cm, width=15cm]{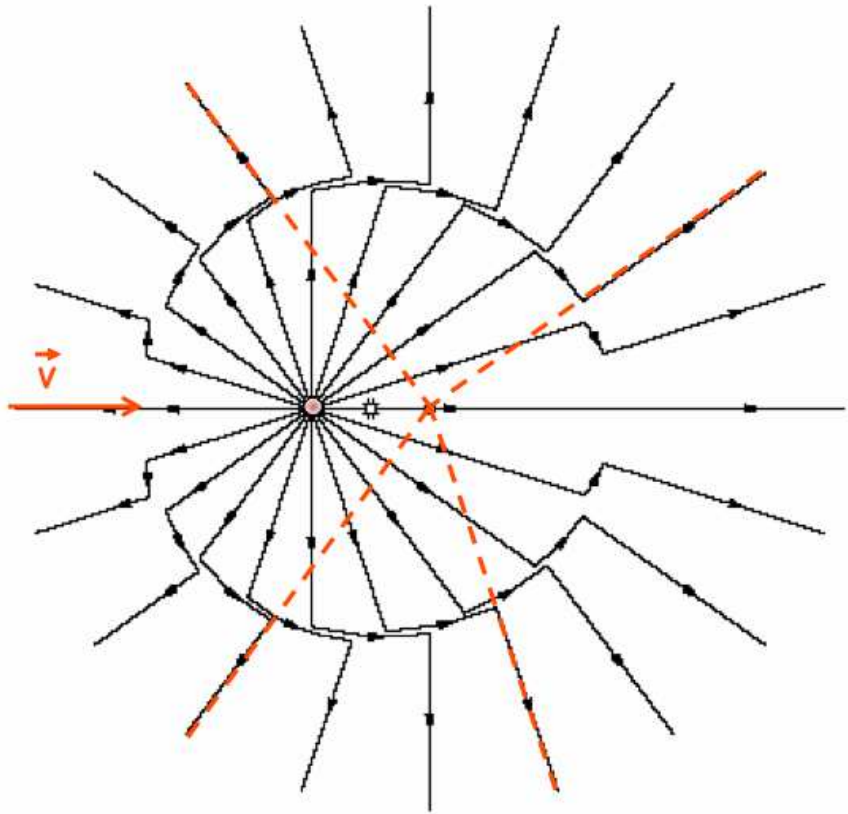} 
\caption[h]{
The electric field produced by a charge initially in uniform rectilinear motion
that is suddenly stopped.
At large distances, the electric field points to where the charge 
{\it would be if it had not been stopped}.
At closer distances, the electric field had time to ``adjust" and points to where 
the charge is.
There is then a region of space where the electric field has to change direction.
This corresponds to radiation.
The width of this region is $c\Delta t$, where $\Delta t$ is the time needed to stop
the particle, namely, the $\Delta t$ during which the charge was decelerated.
}
\label{charge}
\end{figure}{}
%
For a charge moving at a constant speed $\beta c$, we have only the velocity field.
There is something apparently weird associated to this field: it points at the location
of the particle {\it at time $t$}, not at the retarded time $t_{\rm ret}$.
This is due to the $(\vec n - \vec \beta)$ term. 
If the particle does not move, one recovers the usual $\vec E = q \vec n /R^2$ Coulomb law.
But if does move, $\vec E$ does not point in the $\vec n$ direction, even if we have to
calculate the position of the particle at the retarded time. 
The $(\vec n - \vec \beta)$ term makes the $\vec E$ vector to point somewhat along the $\vec \beta$
direction.

\vskip 0.5 cm
\noindent
Fig. \ref{charge} illustrates in a graphic way what the radiation is.
Suppose that a charge, in uniform motion along a rectilinear path,
is suddenly stopped. 
Very far way, the observer ``does not know" that the particle was stopped, and
measure an electric field assuming that the particle has instead continued its motion.
The electric vector then points to a point that would be occupied by the particle,
if it had not been stopped (dashed lines in Fig. \ref{charge}).
In a region closer to the particle, instead, the information: ``the particle is now at rest"
has been propagated, and the $\vec E$ points radially to the actual position of the particle.
In between these two regions, $\vec E$ must change direction (and value).
The change is more pronounced perpendicularly to $\vec {\dot \beta}$,
and is negligible along $\vec {\dot \beta}$ (that here coincides with the direction
of the velocity $\vec \beta$).
The region where $\vec E$ changes direction propagates in time (it goes at $c$), but its
width is always $c\Delta t $.

\subsection{Total emitted power: Larmor formula}

For simplicity, we specialize to the non relativistic case,
consider only the radiative field, and set:
\begin{eqnarray}
\vec \beta \, &\ll & 1
\nonumber \\
k \, &=& \, 1-\vec n \cdot \vec \beta \, \to\, 0
\nonumber \\
\vec n- \vec \beta \, &\to& \, \vec n
\end{eqnarray}
With these approximations we have:
\begin{eqnarray}
\vec E(\vec r, t) \, &=& \, {q\over c R} \,
\left[  \vec n \times  ( \vec n  \times \vec {\dot \beta} ) \,\,  \right]_{t_{\rm ret}}
\nonumber \\
\vec B(\vec r, t) \, &=& \, \vec n \times \vec E
\end{eqnarray}
To calculate the power per unit solid angle carried by this electromagnetic field 
let consider the Poynting vector $\vec S$ (in cgs units):
\begin{equation}
\vec S \, =\, {c \over 4\pi} \, \vec E \times \vec B \, =\, {c \over 4\pi} | \vec E |^2 \vec n
\, =\, {c \over 4\pi} \,{q^2\over c^2 R^2} \,
\left[  \vec n \times  ( \vec n  \times \vec {\dot \beta} ) \,\,  \right]^2_{t_{\rm ret}}
\vec n
\end{equation}
The power crossing a surface $dA=R^2 d\Omega$ is 
\begin{equation}
dP \, =\, S dA \, =\, S R^2 d\Omega \, \longrightarrow\, {dP \over d\Omega} \, =\, S R^2
\end{equation}
Therefore 
\begin{eqnarray}
{dP \over d\Omega} \, &=& \,  \, {q^2 \over 4\pi c} \,
\left|  \vec n \times  ( \vec n  \times \vec {\dot \beta} ) \,\,  \right|^2_{t_{\rm ret}}
\nonumber \\
&=& \,  \, {q^2 \over 4\pi c} \, \dot \beta^2 \sin^2\theta  
\nonumber \\
&=& \,  \, {q^2 \over 4\pi c^3} \, a^2 \sin^2\theta  
\label{sin2}
\end{eqnarray}
This is the Larmor formula for non relativistic particles.
The angle $\theta$ is the angle between $\vec n$ and the acceleration.
The power is null along the acceleration, it is maximum perpendicularly to it.
Integrating over the entire solid angle we have the total power:
\begin{equation}
P \, =\, \int {dP\over d\Omega} d\Omega \, =\, {2\pi \over 4\pi c^3} \,  
\int_{-1}^1 \sin^2\theta \, d(\cos\theta) \, =\, {2 q^2 \over 3 c^3} \, a^2  
\end{equation}

\subsection{Pattern}

\begin{figure}[h]
\center
\includegraphics[height=8cm, width=11cm]{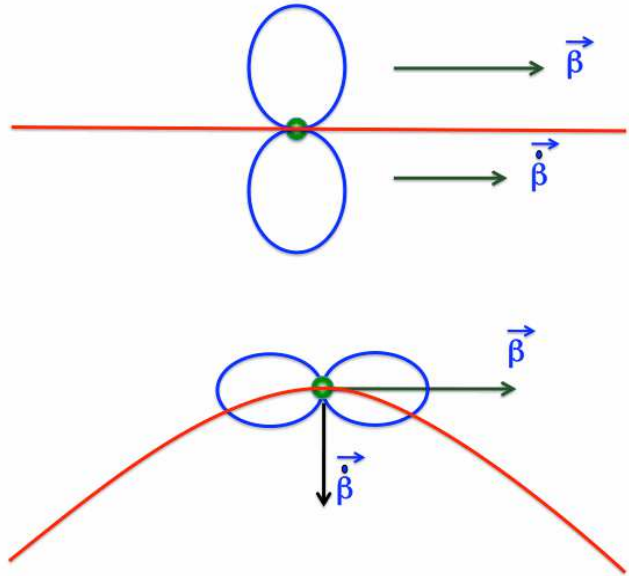} 
\caption[h]{
The pattern of the radiation emitted by a charge with its acceleration
parallel to the velocity (top) and acceleration perpendicular to the
velocity (bottom).
}
\label{pattern}
\end{figure}{}
%
Eq. \ref{sin2} shows that the {\it pattern} of the emitted radiation 
has a maximum perpendicular to the acceleration, and vanishes
in the directions parallel to it.
Fig. \ref{pattern} shows two examples: in the top part the particle has $\dot\beta \parallel \beta$
while in the bottom part $\dot\beta \perp \beta$.
The particle is non relativistic.

Just for exercise, consider an antenna consisting of a linear piece of metal.
It is called a dipole antenna. What will be the pattern of the emitted radiation?
See: 
{\tt http://en.wikipedia.org/wiki/File:Felder\_um\_Dipol.jpg}

What will be the pattern of an oscillating electron?

\vskip 0.5 cm
\noindent
We will soon see the modification occurring when the particle becomes relativistic,
both for the total emitted power and for the pattern of the emitted radiation.



\chapter{Bremsstrahlung and black body}

\section{Bremsstrahlung}

We will follow an approximate derivation. 
For a more complete treatment see Rybicki \& Lightman (1979) 
and Blumental \& Gould (1970).
We will consider an electron--proton plasma.

Definitions:
\begin{itemize}
\item
$b$: impact parameter
\item
$v$: velocity of the electron
\item
$n_{\rm e}$ number density of the electrons
\item
$n_{\rm p}$: number density of the protons
\item
$T$: temperature of the plasma: $mv^2\sim kT \to v\sim (kT/m)^{1/2}$.
\end{itemize}
We here calculate the total power and also the spectrum of bremsstrahlung
radiation.
We divide the procedure into a few steps:
\begin{enumerate}
\item
We consider the interaction between the electron and the proton only when
the electron passes close to the proton. 
The characteristic time $\tau$ is
\begin{equation}
\tau\, \approx \, {b\over v}
\end{equation}
\item
During the interaction we assume that the acceleration is constant and equal to
\begin{equation}
a \, \approx \, {e^2\over m_e b^2}
\end{equation}
\item
From the Larmor formula we get
\begin{equation}
P\, =\, {2 e^2 a^2 \over 3 c^3} \, \approx \, {e^2 \over c^3}\, {e^4\over m^2_{\rm e} c^3 b^4 }\, 
= \, {e^6 \over m^2_{\rm e} c^3 b^4}
\end{equation}
Note that we have dropped the 2/3 factor, since in this simplified treatment we neglect
all the numerical factors of order unity. 
Later we will give the exact result.

\item
Since there is a characteristic time, there is also a characteristic frequency,
namely $\tau^{-1}$:
\begin{equation}
\omega \, \approx \, {1\over \tau} \, =\, {v\over b}
\end{equation}

\item
Therefore 
\begin{equation}
P(\omega)  \, \approx \, {P\over \omega} \, =\, {e^6 \over m^2_{\rm e} c^3 v b^3}
\end{equation}

\item
We can estimate the impact factor $b$ from the density of protons:
\begin{equation}
b  \, \approx \, n^{-1/3}_{\rm p} \, \to \, b^3 = {1\over n_{\rm p}}
\end{equation}
\item
The emissivity $j(\omega)$ will be the power emitted by a single electron
multiplied by the number density of electrons.
If the emission is isotropic we have also to divide by $4\pi$, since
the emissivity is for unit solid angle:
\begin{equation}
j(\omega)  \, \approx \, {n_{\rm e} n_{\rm p} \over 4\pi}  { e^6 \over m^2_{\rm e} c^3}   
\left( {m_{\rm e} \over kT} \right)^{1/2}
\end{equation}

\item
We integrate $j(\omega)$ over frequency.
The integral will depend upon $\omega_{\rm max}$. 
What should we use for $\omega_{\rm max}$?
One possibility is to set $\hbar\omega_{\rm max}=kT$.
This would mean that an electron cannot emit a photon of
energy larger than the typical energy of the electron.
Seems reasonable, but we are forgetting all the electrons
(and the frequencies) that have energies larger than $kT$.
 In this way:
\begin{eqnarray}
j\, &=&\, \int_0^{\omega_{\rm max}}
j(\omega)  d\omega \, \sim \, {n_{\rm e} n_{\rm p} \over 4\pi}  { e^6 \over m^2_{\rm e} c^3}   
\left( {m_{\rm e} \over kT} \right)^{1/2} \, {kT\over \hbar}
\nonumber \\
&=&\, {n_{\rm e} n_{\rm p} e^6 \over 4\pi m^2_{\rm e} c^3}\,   
{( {m_{\rm e} kT} )^{1/2} \over \hbar}
\end{eqnarray}
We suspect that in the exact results there will be 
the contribution of electrons with energy larger than $kT$:
since they belong to the exponential part of the Maxwellian,
we suspect that in the exact result there will be 
an exponential...

\item
The exact result, considering also that $\nu = \omega/(2\pi)$, is 
\begin{eqnarray}
j(\nu)\, &=& \, 
{8\over 3} \left( {2\pi \over 3} \right)^{1/2}
{n_{\rm e} n_{\rm p} e^6 \over  m^2_{\rm e} c^3}   
\left( { m_{\rm e} \over kT } \right)^{1/2} e^{-h\nu/kT}\,  \bar g_{\rm ff} 
\nonumber \\
j \, &=& \, 
{4\over 3\pi} \left( {2\pi \over 3} \right)^{1/2}
{n_{\rm e} n_{\rm p} e^6 \over  m^2_{\rm e} c^3}   
{(  m_{\rm e}  kT  )^{1/2}\over \hbar  } \, \bar g_{\rm ff} 
\end{eqnarray}
\end{enumerate}

The Gaunt factor $\bar g_{\rm ff}$ depends on the minimum
impact factor which in turn determines the maximum frequency.
Details are complicated, but see Rybicki \& Lightman (p. 158--161) 
for a more detailed discussion.

We have treated the case of an electron--proton plasma.
In the more general case, the plasma will be composed by nuclei with atomic
number $Z$ and number density $n_{\rm }$. 
The emissivity will then be proportional to $Z^2$.
This is because the acceleration of the electron will be $a= Z e^2/(m_{\rm e} b^2)$
(see point 2), and we have to square the acceleration to get the power from the Larmor formula.
In cgs units we have:
\begin{eqnarray}
j(\nu) \, &=&  \, 5.4\times 10^{-39}\, Z^2 n_{\rm e} n_{\rm i} T^{-1/2} e^{-h\nu /kT} \bar g
\nonumber \\
j \,    &=&  \, 1.13 \times 10^{-28}\, Z^2 n_{\rm e} n_{\rm i} T^{1/2} \bar g
\end{eqnarray}

\subsection{Free--free absorption}

If the underlying particle distribution is a Maxwellian, we can use the
Kirchoff law to find out the absorption coefficient.
If $B_\nu$ is the intensity of black body emission, we must have
\begin{equation}
S_\nu   \, \equiv \, {j_\nu \over  \alpha_\nu}  \, =\, B_\nu \, = 
{2h\nu^3 \over c^2} \, {1\over e^{h\nu/kT} -1}
\label{kirchoff}
\end{equation}
In these cases it is very simple to find $\alpha_\nu$ once we know $j_\nu$.
Remember: this can be done only if we have a Maxwellian.
If the particle distribution is non--thermal, we cannot use the Kirchoff law and
we have to go back to a more fundamental level, namely to the Einstein coefficients.
Using Eg. \ref{kirchoff} we have:
\begin{equation}
\alpha^{\rm ff}_\nu   \, =\,{j_\nu\over  B_\nu} \, = \,  
{4\over 3} \left( {2\pi \over 3} \right)^{1/2}
{ Z^2 n_{\rm e} n_{\rm i} e^6 \over  h m^2_{\rm e} c^2}   \,
\left( { m_{\rm e} c^2 \over kT } \right)^{1/2} \,\,
{1-e^{-h\nu/kT}   \over \nu^3 } \,  \bar g_{\rm ff} 
\end{equation}
In cgs units [cm$^{-1}$] we have
\begin{equation}
\alpha^{\rm ff}_\nu   \, =\, 3.7\times 10^8 \,\,
{ Z^2 n_{\rm e} n_{\rm i}   \over  T^{1/2}} \,\,
{ 1-e^{-h\nu/kT}   \over \nu^3 } \,  \bar g_{\rm ff} 
\end{equation}
When $h\nu \ll kT$ (Raleigh--Jeans regime) this simplifies to
\begin{equation}
\alpha^{\rm ff}_\nu   \, =\, 0.018 \,\,
{ Z^2 n_{\rm e} n_{\rm i}   \over  T^{3/2} \nu^2}\, \,\bar g_{\rm ff}
\end{equation}
%
\begin{figure}[h]
\center
\vskip -1 true cm
\includegraphics[height=10cm, width=13cm]{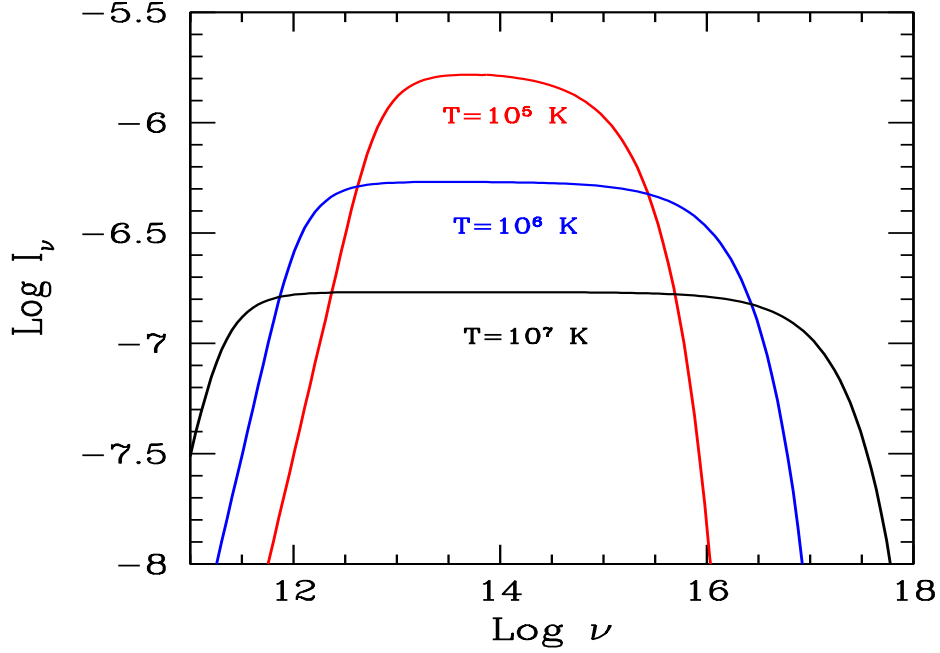} 
\vskip -0.5 true cm
\caption[h]{
The bremsstrahlung intensity from a source of radius $R=10^{15}$ cm,
density $n_{\rm e}=n_{\rm p}= 10^{10}$ cm$^{-3}$ and varying temperature.
The Gaunt factor is set to unity for simplicity.
At smaller temperatures the thin part of $I_\nu$ is larger ($\propto T^{-1/2}$), even 
if the frequency integrated $I$ is smaller ($\propto T^{1/2}$).
}
\label{brems2}
\end{figure}{}
%
Fig. \ref{brems2} shows the bremsstrahlung intensity from a source of radius $R=10^{15}$ cm
and $n_{\rm e}=n_{\rm p}=10^{10}$ cm$^{-3}$.
The three spectra correspond to different temperatures.
Note that for smaller temperatures the thin part of $I_\nu$ is larger
($I_\nu \propto T^{-1/2}$).
On the other hand, at larger $T$ the spectrum extends to larger frequencies,
making the frequency integrated intensity to be larger for larger $T$ ($I\propto T^{1/2}$).
Note also the self--absorbed part, whose slope is proportional to $\nu^2$.
This part ends when the optical depth $\tau = \alpha_\nu R \sim 1$.

\subsection{From bremsstrahlung to black body}

As any other radiation process, the bremsstrahlung emission has a self--absorbed part,
clearly visible in Fig. \ref{brems2}.
This corresponds to optical depths $\tau_\nu \gg 1$.
The term $\nu^{-3}$ in the absorption coefficient $\alpha_\nu$ ensures
that the absorption takes place preferentially at low frequencies.
By increasing the density of the emitting (and absorbing) particles,
the spectrum is self--absorbed up to larger and larger frequencies.
When {\it all} the spectrum is self absorbed (i.e. $\tau_\nu >1$ for all $\nu$),
{\it and} the particles belong to a Maxwellian, then we have a black--body.
%
\begin{figure}[h]
\center
\vskip -1 true cm
\includegraphics[height=10cm, width=13cm]{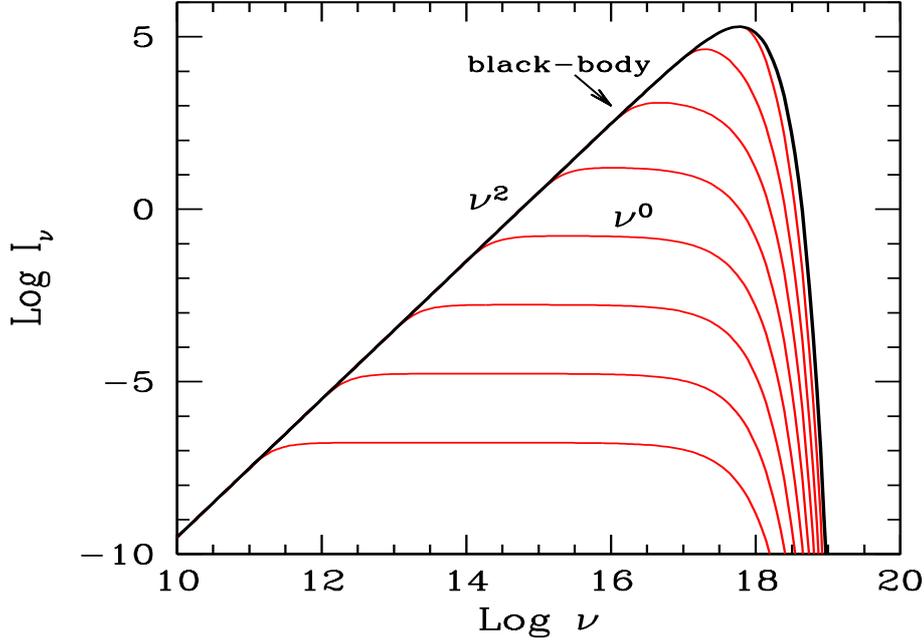} 
\vskip -0.5 true cm
\caption[h]{
The bremsstrahlung intensity from a source of radius $R=10^{15}$ cm,
temperature $T=10^7$ K. 
The Gaunt factor is set to unity for simplicity.
The density $n_{\rm e}=n_{\rm p}$ varies from $10^{10}$ cm$^{-3}$
(bottom curve) to $10^{18}$ cm$^{-3}$ (top curve),
increasing by a factor 10 for each curve.
Note the self--absorbed part ($\propto \nu^2$), the flat and the exponential
parts.
As the density increases, the optical depth also increases, and the spectrum 
approaches the black--body one.
}
\label{brems}
\end{figure}{}
%
This is illustrated in Fig. \ref{brems}: all spectra are calculated for 
the same source size ($R=10^{15}$ cm), same temperature ($T=10^7$ K),
and what varies is the density of electrons and protons (by a factor 10)
from $n_{\rm e}=n_{\rm p}=10^{10}$ cm$^{-3}$ to $10^{18}$ cm$^{-3}$.
As can be seen, the bremsstrahlung intensity becomes more and more
self--absorbed as the density increases, until it becomes a black--body.
At this point increasing the density does not increase the intensity
any longer.
This is because we receive radiation from a layer of unity optical depth.
The width of this layer decreases as we increase the densities,
but the emissivity increases, so that
\begin{equation}
I_\nu   \, =\,  {j_\nu R \over \tau_\nu } \, \propto \, 
{n_{\rm e}n_{\rm p} R \over n_{\rm e}n_{\rm p} R}
\, \to \, {\rm constant\qquad (\tau_\nu \gg 1)}
\end{equation}

\newpage
\section{Black body}

A black body occurs when ``the body is black": it is the perfect absorber.
But this means that it is also the ``perfect" emitter, since absorption and
emission are linked.
The black body intensity is given by
\begin{equation}
B_\nu (T)   \, =\, {2\over c^2} \, {h\nu^3 \over e^{h\nu/kT} -1}
\end{equation}
Expressed in terms of the wavelength $\lambda$ this is equivalent to:
\begin{equation}
B_\lambda (T)   \, =\, {2hc^2\over \lambda^5} \, {1\over e^{hc/\lambda kT} -1}
\end{equation}
Note the following:
\begin{itemize}

\item
The black body intensity has a peak.
The value of it is different if we ask for the peak of $B_\nu$ 
or the peak of $\nu B_\nu$.\\
The first is at $h\nu_{\rm peak} = 2.82 \, kT$.\\
The second is at $h\nu_{\rm peak} = 3.93 \,  kT$.

\item 
If $T_2>T_1$, then: $B_\nu (T_2)>B_\nu (T_1)$ for all frequencies.

\item
When $h\nu \ll kT$ we can expand the exponential term:
$e^{h\nu /kT} \to 1+ h\nu/kT ...$, and therefor we obtain the 
{\bf Raleigh--Jeans law}:
\begin{equation}
I_\nu^{\rm RJ} \, = \, {2\nu^2 \over c^2}\, kT
\end{equation}

\item
When $h\nu \gg kT$ we have $e^{h\nu/kT}-1 \to e^{h\nu/kT}$ and we obtain
the {\bf Wien law}:
\begin{equation}
I^{\rm W}_\nu    \, =\, {2 h\nu^3 \over c^2} e^{-h\nu/kT}
\end{equation}
%

\begin{figure}[h]
\center
\vskip -1 true cm
\includegraphics[height=10cm, width=13cm]{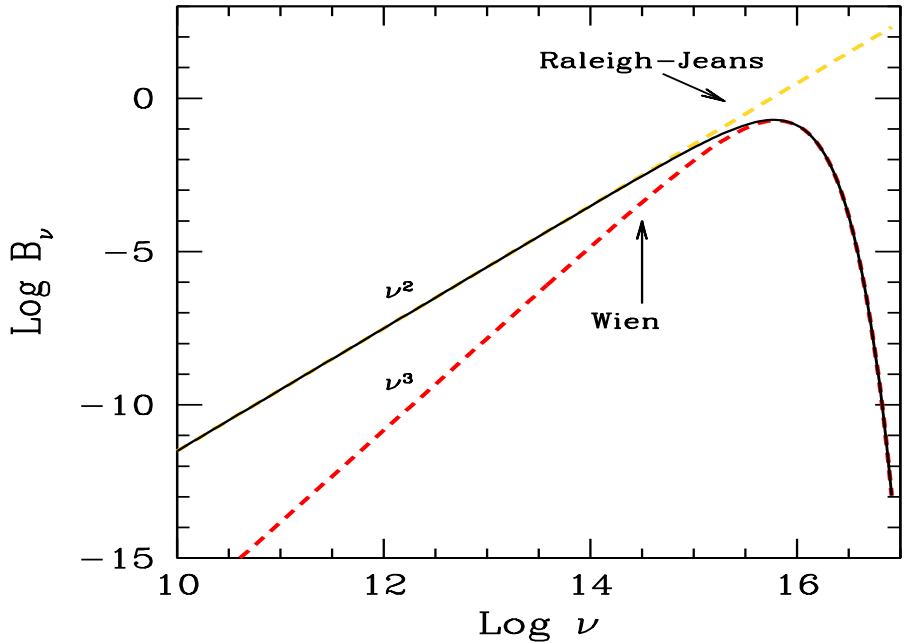} 
\vskip -0.5 true cm
\caption[h]{
The black body intensity compared with the Raleigh--Jeans and the Wien law.
}
\label{bb}
\end{figure}{}

\item
The integral over frequencies is:
\begin{equation}
\int_0^\infty B_\nu d\nu   \, =\, {\sigma_{\rm MB}\over \pi} \, T^4 , \qquad
\sigma_{\rm MB}\, =\, {2\pi^5 k^4 \over 15 c^2 h^3}
\end{equation}
The constant $\sigma_{\rm MB}$ is called Maxwell--Boltzmann constant.

\item
The energy density $u$ of black body radiation is
\begin{equation}
u \, =\, {4\pi \over c} \int_0^\infty B_\nu d\nu   \, =\, a \, T^4 , \qquad
a\, =\, {4 \sigma_{\rm MB}\over c}
\end{equation}
The two constants ($\sigma_{MB}$ and $a$) have the values:
\begin{eqnarray}
\sigma_{MB} \, &=&\, 5.67\times 10^{-5}\quad {\rm erg\, cm^{-2}\, deg^{-4}\, s^{-1}}
\nonumber \\
a\, &=& \, 7.65\times 10^{-15} \quad {\rm erg\, cm^{-3}\, deg^{-4} }
\end{eqnarray}

\item
The {\bf brightness temperature} is defined using the Raleigh--Jeans law,
since $I^{\rm RJ}_\nu =(2\nu^2/c^2) kT$ we have
\begin{equation}
T_{\rm b}\, =\, {c^2 I_\nu^{\rm RJ} \over 2 k \nu^2}
\end{equation}

\item
A black body is the most efficient radiator, for thermal plasmas 
and incoherent radiation (we can have coherent 
processes that are even more efficient).
For a given surface and temperature, it is not possible to overtake the
luminosity of the black body, at any frequency, for any emission 
process.

\item
Let us try to find the temperature of the surface of the Sun.
We know its radius (700,000 km) and luminosity ($L_\odot =4\times 10^{33}$ erg s$^{-1}$).
Therefore, from
\begin{equation}
L_\odot \, =\, \pi \, 4\pi R^2 \, \int_0^\infty B_\nu d\nu \, =\, 4\pi R^2 \sigma_{\rm MB} T^4
\end{equation}
we get:
\begin{equation}
T_\odot \, =\, \left( {L_\odot \over 4\pi R^2 \sigma_{\rm MB} } \right)^{1/4}
\, \sim 5800\,\, K
\end{equation}
\item
A spherical source emits black body radiation. We know its distance, but not its radius.
Find it.
Suppose we do not know its distance. Can we predict its angular size?
And, if we then observe it, can we then get the distance?

\end{itemize}

%



\chapter{Beaming}

\section{Rulers and clocks}

Special relativity taught us two basic notions:
comparing dimensions and flow of times in two
different reference frames, we find out that they
differ. If we measure a ruler at rest, and then
measure the same ruler when is moving, we find
that, when moving, the ruler is shorter.
If we syncronize two clocks at rest, and then 
let one move, we see that the moving clock is delaying.
Let us see how this can be derived by using the Lorentz
transformations, connecting the two reference frames
$K$ (that sees the ruler and the clock moving)
and $K^\prime$ (that sees the ruler and the clock at rest).
For semplicity, but without loss of generality, consider
a a motion along the $x$ axis, with velocity $v\equiv \beta c$ 
corresponding to the Lorentz factor $\Gamma$.
Primed quantities are measured in $K^\prime$.
We have:
\begin{eqnarray}
x^\prime \, &=& \, \Gamma(x-vt)\nonumber \\ 
y^\prime\, &=&\, y \nonumber \\
z^\prime\, &=&\, z \nonumber \\
t^\prime \, &=& \, \Gamma\left( t- \beta {x\over c}\right) 
\end{eqnarray}
with the inverse relations given by
\begin{eqnarray}
x\, &=& \, \Gamma(x^\prime + vt^\prime)  \nonumber \\
y\, &=&\, y^\prime \nonumber \\
z\, &=&\, z^\prime \nonumber \\
t\, &=& \, \Gamma\left( t^\prime + \beta {x^\prime\over c}\right). 
\end{eqnarray}
The length of a moving ruler has to be measured 
through the position of its extremes {\it at the same time $t$}.
Therefore, as $\Delta t=0$, we have 
\begin{equation}
x^\prime_2 -x^\prime_1 \, =\, 
\Gamma(x_2-x_1)-\Gamma v\Delta t \, =\, \Gamma(x_2-x_1)
\end{equation}
i.e. 
\begin{equation}
\Delta x \,=\,  {\Delta x^\prime \over \Gamma }
\, \to \, {\rm contraction}
\end{equation}
Similarly, in order to determine a time interval 
a  (lab) clock has to be compared 
with one in the comoving frame,
which has, in this frame, {\it the same position $x^\prime$}.
Then
\begin{equation}
\Delta t \, =\, \Gamma \Delta t^\prime + \Gamma \beta
\Delta {x^\prime\over c}
\, =\, 
\Gamma \Delta t^\prime \, \to \, {\rm dilation}
\end{equation}
An easy way to remember the transformations is to think to mesons
produced in collisions of cosmic rays in the high atmosphere, which 
can be detected even if their lifetime
(in the comoving frame) is much shorter than the time
needed to reach the earth's surface.
For us, on ground, relativistic mesons live longer
(for the meson's point of view, instead, it is the length of the 
travelled distance which is shorter).

All this is correct if we measure lengths by comparing
rulers (at the same time in $K$) and by comparing clocks 
(at rest in $K^\prime$)
-- the meson lifetime $is$ a clock.
In other words, {\bf if we do not use photons}
for the measurement process.

\section{Photographs and light curves}

If we have an extended moving object and 
if the information (about position and time)
are carried by photons, we {\bf must} take into account
their (different) travel paths.
When we take a picture, we detect photons arriving at the same
time to our camera: if the moving body which emitted them is 
extended, we must consider that these photons have been emitted
at different times, when the moving object occupied different
locations in space.
This may seem quite obvious. And it is. Nevertheless
these facts were pointed out in 1959 (Terrel 1959; Penrose 1959), 
more than 50 years
after the publication of the theory of special relativity.

\subsection{The moving bar}

\begin{figure}[!ht]
\includegraphics[height=10cm,width=10cm]{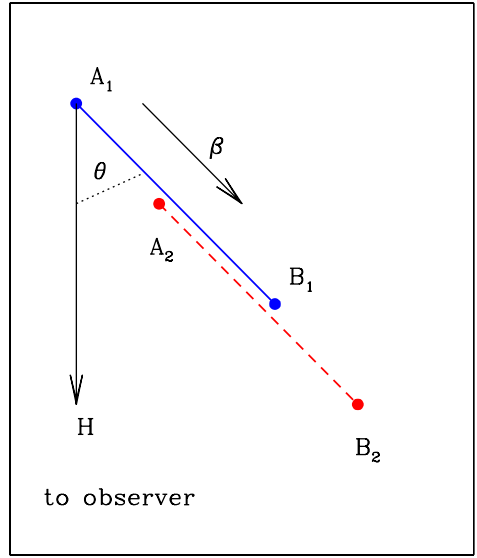}  
\vskip -1 true cm
\caption{A bar moving with velocity $\beta c$ in the direction of
its length. The path of the photons emitted by the extreme $A$ is longer
than the path of photons emitted by $B$.
When we make a picture (or a map) of the bar, we collect photons
reaching the detector simultaneously. 
Therefore the photons from $A$ have to be emitted before those
from $B$,
when the bar occupied another position.}
\label{chap1:bar}
\end{figure}{}

%

Let us consider a moving bar, of proper dimension $\ell^\prime$, moving 
in the direction of its length at velocity $\beta c$ and 
at an angle $\theta$ with respect to the line of sight 
(see Fig. \ref{chap1:bar}).
The length of the bar in the frame $K$ (according to relativity
``without photons") is $\ell =\ell^\prime/\Gamma$.
The photon emitted in $A_1$ reaches the point $H$ in the time interval 
$\Delta t_e$. 
After $\Delta t_e$ the extreme $B_1$ has reached the position $B_2$, 
and by this time, photons emitted by the other extreme of the bar can reach
the observer simultaneously with the photons emitted by $A_1$, since
the travel paths are equal. 
The length $B_1B_2=\beta c \Delta t_e$, while $A_1H=c\Delta t_e$.
Therefore  
\begin{equation}
A_1H \, =\, A_1 B_2 \cos\theta \, \to \Delta t_e \, =\, 
{\ell^\prime \cos\theta \over c \Gamma (1-\beta\cos\theta)}.
\end{equation}
Note the appearance of the term $\delta=1/[\Gamma(1-\beta\cos\theta)]$ in 
the transformation:
this accounts for both the relativistic length contraction $(1/\Gamma)$,
and the Doppler effect $[1/(1-\beta\cos\theta)]$ (see below, Eq. \ref{delta}).
The length $A_1B_2$ is then given by 
\begin{equation}
A_1B_2\, =\, {A_1H\over \cos\theta} \, =\,  
{\ell^\prime \over \Gamma(1-\beta\cos\theta)} \, =\, \delta \ell^\prime. 
\end{equation}
In a real picture, we would see the projection of $A_1B_2$, i.e.:
\begin{equation}
HB_2 \, =\,  A_1B_2\sin\theta \, =\, 
\ell^\prime\,
{\sin\theta \over \Gamma(1-\beta\cos\theta)} \, =\,  
\ell^\prime \delta \sin\theta, 
\end{equation}
The observed length depends on the viewing angle, and reaches
the maximum (equal to $\ell^\prime$) for $\cos\theta=\beta$.

\subsection{The moving square}

\begin{figure*}
\vskip -0.5 true cm
\begin{tabular}{l l}
\hskip -2.5 truecm   
\includegraphics[height=10cm, width=9cm]{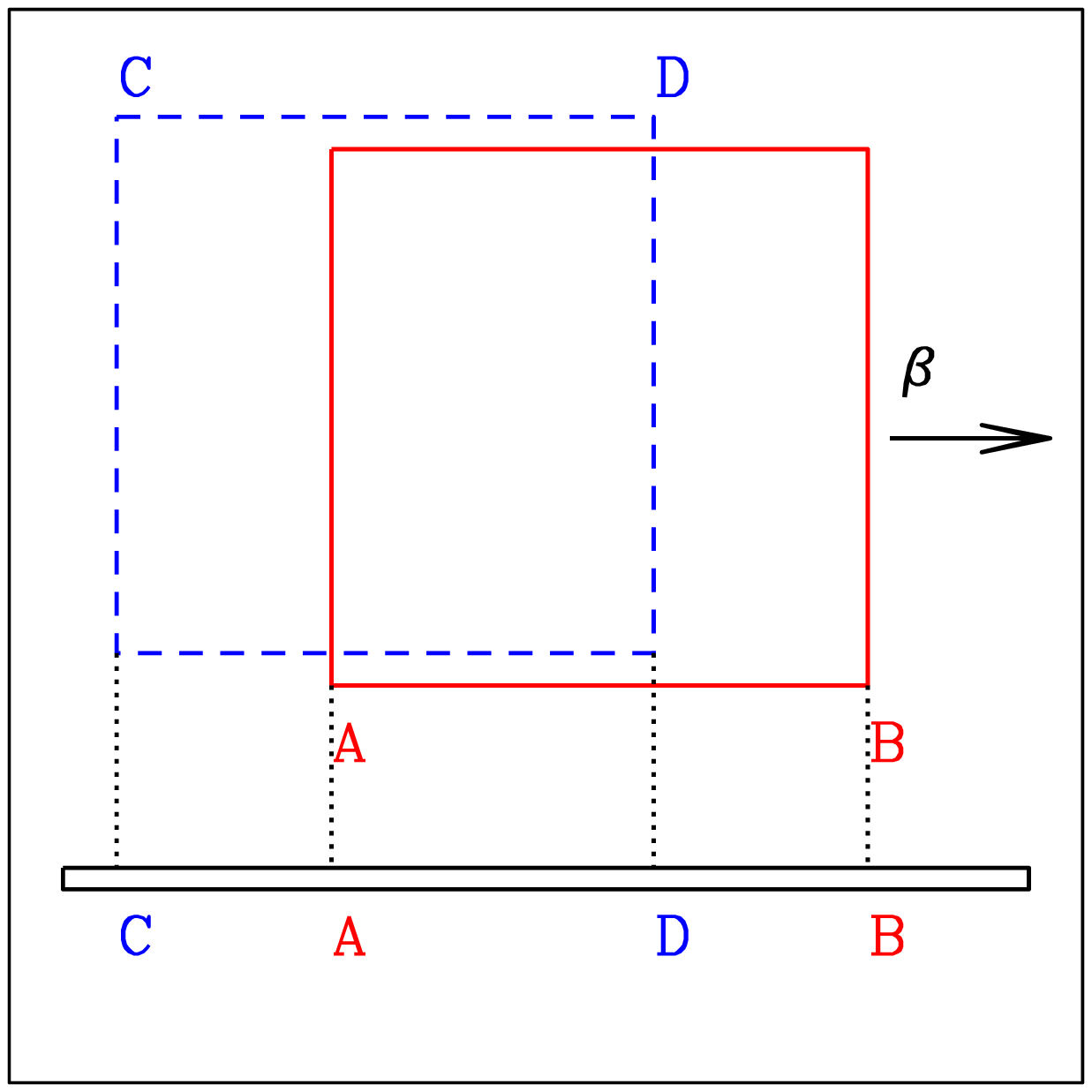} 
& \hskip -2 truecm \includegraphics[height=10cm, width=9cm]{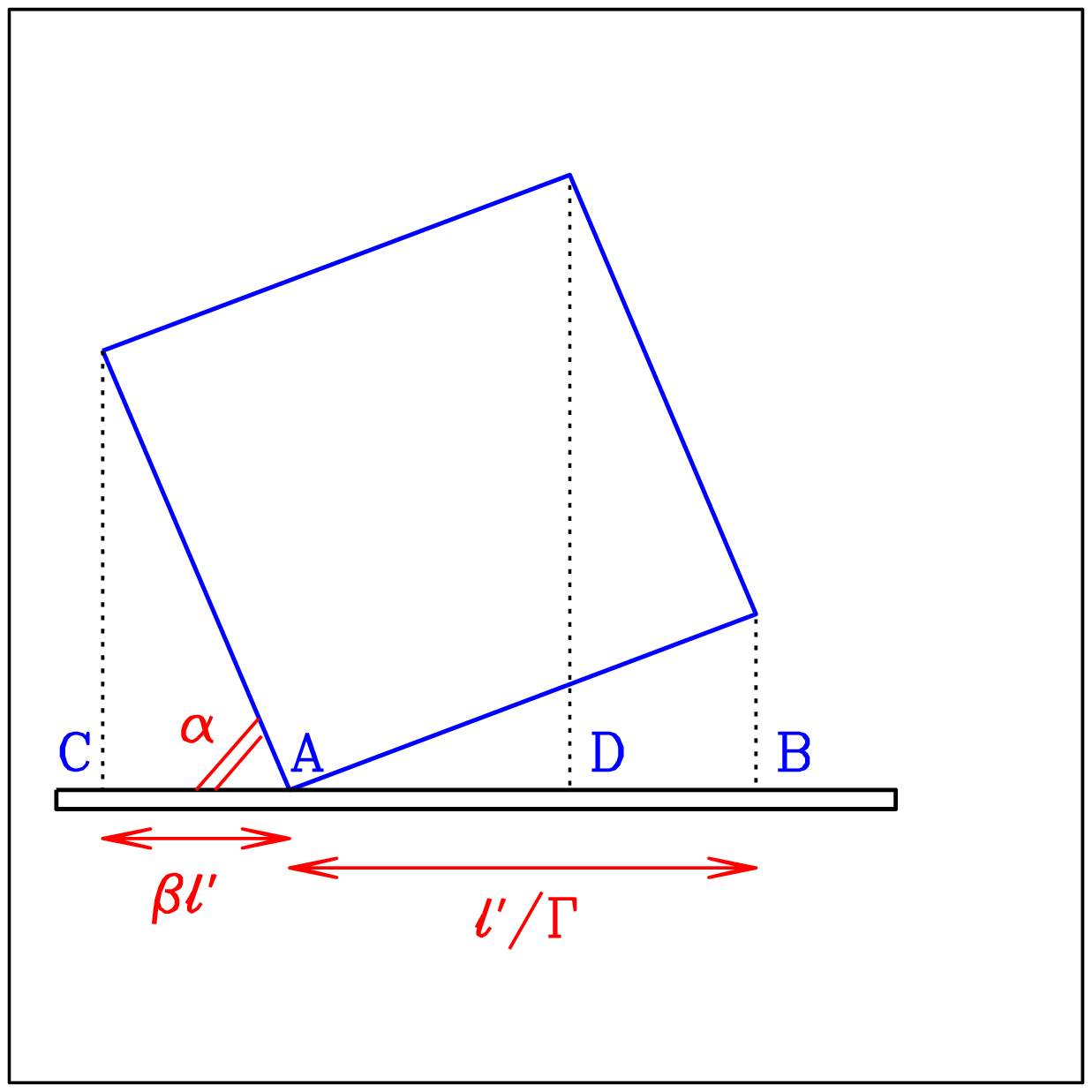} 
\end{tabular}
\vskip -2.5 true cm
\caption[h]{{\it Left:}
A square moving with velocity $\beta c$ seen at 90$^\circ$.
The observer can see the left side (segment $CA$). 
Light rays are assumed to be parallel, i.e. the square is assumed to
be at large distance from the observer.
{\it Right:} The moving square is seen as {\it rotated} 
by an angle $\alpha$ given by $\cos\alpha=\beta$.}  
\label{chap1:fig2ab}
\end{figure*}

Now consider a square of size $\ell^\prime$ in the comoving frame, 
moving at $90^\circ$ to the line of sight (Fig. \ref{chap1:fig2ab}).
Photons emitted in $A$, $B$, $C$ and $D$ have to arrive to the film plate 
at the same time. 
But the paths of photons from $C$ and $D$ are longer $\to$ they have to 
be emitted earlier than photons from $A$ and $B$: when photons from $C$ 
and $D$ were emitted, the square was in another position.

%
{}

%
The interval of time between emission from $C$ and from $A$ is  
$\ell^\prime/c$. 
During this time the square moves by $\beta \ell^\prime$, i.e.  
the length $CA$.
Photons from $A$ and $B$ are emitted and received at the same time
and therefore $AB=\ell^\prime/\Gamma$.
The total observed length is given by  
\begin{equation}
CB\, =\, CA+AB\, =\, {\ell^\prime\over \Gamma} \, (1+\Gamma\beta). 
\end{equation}
As $\beta$ increases, the observer sees the side $AB$ increasingly
shortened by the Lorentz contraction, but at the same time the length
of the side $CA$ increases.
The maximum total length is observed for $\beta=1/\sqrt{2}$,
corresponding to $\Gamma=\sqrt{2}$ and to $CB=\ell^\prime\sqrt{2}$,
i.e. equal to {\it the diagonal} of the square.
Note that we have considered the square (and the bar in the previous section)
to be at large distances from the observer, so that the emitted light rays
are all parallel.
If the object is near to the observer, we must take into account that different
points of one side of the square (e.g. the side $AB$ in Fig. 
\ref{chap1:fig2ab}) have different 
travel paths to reach the observer, producing additional distortions.
See the book by Mook and Vargish (1991) for some interesting illustrations.

\begin{figure}
\vskip -1 true cm
\hskip -1.5 true cm \includegraphics[height=15cm, width=15cm]{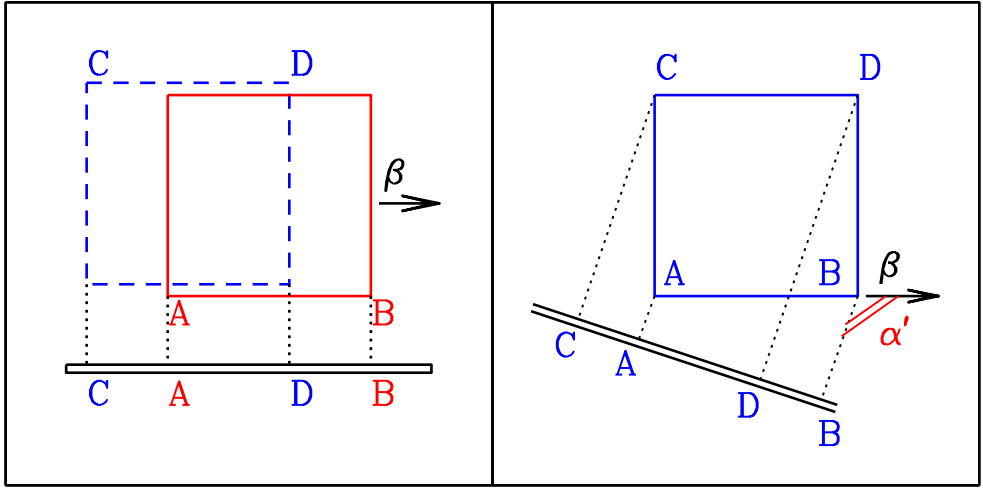}\\
\vskip -7 true cm
\caption[h]{An observer that sees the object at rest at a viewing
angle given by $\sin\alpha^\prime = \delta \sin\alpha$, will take
the same picture as the observer that sees the object moving
and making an angle $\alpha$ with his/her line of sight.
Note that $\sin\alpha^\prime=\sin(\pi-\alpha^\prime)$.}
\label{chap1:fig3}
\end{figure}{}

\subsection{Rotation, not contraction}

The net result (taking into account {\bf both} the length contraction 
{\bf and} the different paths) is an apparent {\bf rotation} of the square,
as shown in Fig. \ref{chap1:fig2ab} (right panel).
The rotation angle $\alpha$ can be simply derived (even geometrically)
and is given by
\begin{equation}
\cos\alpha \, =\, \beta
\end{equation}
A few considerations follow:
\begin{itemize}
\item If you rotate a sphere you still get a sphere:  
you {\bf do not} observe a contracted sphere.
\item The total length of the projected square,
appearing on the film, is $\ell^\prime (\beta +1/\Gamma)$.
It is maximum when the ``rotation angle" $\alpha=45^\circ \to
\beta=1/\sqrt{2} \to \Gamma=\sqrt{2}$.
This corresponds to the diagonal.
\item The appearance of the square {\it is the same as what 
seen in a comoving frame for a line of sight making an angle
$\alpha^\prime$ with respect to the velocity vector,
where $\alpha^\prime$ is the aberrated angle} given by
\begin{equation}
\sin\alpha^\prime \, =\, {\sin\alpha \over \Gamma (1-\beta\cos\alpha)}\, =\,
\delta \sin\alpha
\end{equation}
See Fig. \ref{chap1:fig3} for a schematic illustration.
\end{itemize}
The last point is particularly important, because it introduces
a great simplification in calculating not only the appearance of
bodies with a complex shape
but also the light curves of varying objects.

\begin{figure}
\vskip -0.5 true cm
\hskip -1.5 true cm \includegraphics[height=10cm, width=10cm]{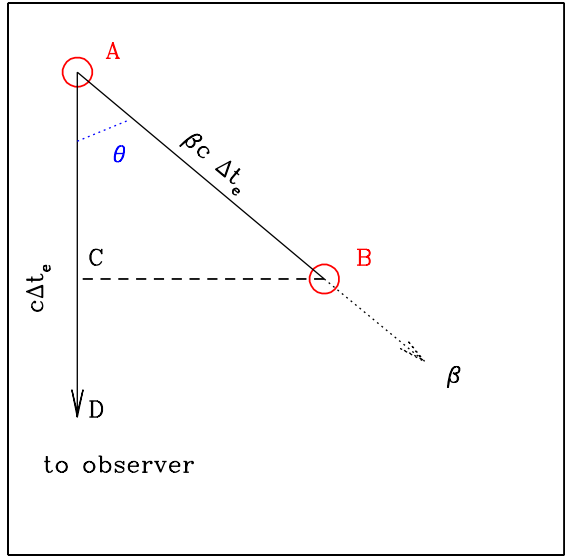}\\
\vskip -2 true cm
\caption[h]{
Difference between the proper time and the 
photons arrival time.
A lamp, moving with a velocity $\beta c$, emits photons for a time interval
$\Delta t_{\rm e}^\prime$ in its frame $K^\prime$.
The corresponding time interval measured by an observed at an
angle $\theta$, who receives the photons produced by the lamp is
$\Delta t_{\rm a} = \Delta t^\prime_{\rm e}/ \delta $.
}
\label{chap1:fig4}
\end{figure}{}

\subsection{Time}

Consider a lamp moving with velocity $v=\beta c$ at an angle 
$\theta$ from the line of sight.
In $K^\prime$, the lamp remains on for a time $\Delta t_{\rm e}^\prime$.
According to special relativity (``without photons") the measured
time in frame $K$ should be $\Delta t_{\rm e} = 
\Gamma \Delta t_{\rm e}^\prime$
(time dilation).
However, if we use photons to measure the time interval, we once again
must consider that the first and the last photons have been emitted
in different location, and their travel path lengths are different.
To find out $\Delta t_{\rm a}$, 
the time interval between the arrival of the
first and last photon, consider  Fig. \ref{chap1:fig4}. 
The first photon is emitted in $A$, the last in $B$.
If these points are measured in frame $K$, then the path $AB$ is
\begin{equation}
AB\, =\, \beta c\Delta t_{\rm e} \, =\, \Gamma \beta c\Delta t_{\rm e}^\prime                        
\end{equation}
While the lamp moved from $A$ to $B$, the photon emitted when the lamp
was in $A$ has travelled a distance $AC=c\Delta t_{\rm e}$, and is
now in point $D$. Along the direction of the line of sight, the
first and the last photons (the ones emitted in $A$ and in $B$)
are separated by $CD$. 
The corresponding time interval, $CD/c$, is the interval of time $\Delta t_{\rm a}$
between the arrival  of the first and the last photon: 
\begin{eqnarray}
\Delta t_{\rm a} \, &=&\, {CD\over c}\, =\, {AD-AC\over c} \, =\, 
\Delta t_{\rm e} -\beta \Delta t_{\rm e}\cos\theta         \nonumber \\
\, &=&\, \Delta t_{\rm e} (1-\beta\cos\theta)              \nonumber \\
\, &=&\, \Delta t^\prime_{\rm e} \Gamma (1-\beta\cos\theta) \nonumber \\
\, &=&\, {\Delta t^\prime_{\rm e} \over \delta}
\end{eqnarray}
If $\theta$ is small and the velocity is relativistic, then $\delta>1$,
and $\Delta t_{\rm a} < \Delta t_{\rm s}$, i.e. we measure a 
{\it time contraction} instead of time dilation.
Note also that we recover the usual time dilation
(i.e. $\Delta t_{\rm a} =\Gamma \Delta t_{\rm e}^\prime$)  
if $\theta=90^\circ$, because in this case all photons have to travel 
the same distance to reach us.

Since a frequency is the inverse of time, it will transform as
\begin{equation}
\nu\, =\, \nu^\prime \, \delta                      
\end{equation}
It is because of this that the factor $\delta$ is called the relativistic
Doppler factor. 
Its definition is then
\begin{equation}
\delta \, =\, {1 \over \Gamma (1-\beta\cos\theta)}
\label{delta}
\end{equation}
Note the two terms:
\begin{itemize}
\item The term $1/\Gamma$: this corresponds to the usual special relativity term.
\item The term $1/(1-\beta\cos\theta)$: this corresponds to the usual Doppler effect.
\end{itemize}
The $\delta$ factor is the result of the competition of these two terms:
for $\theta=90^\circ$ the usual Doppler term is unity, and only ``special relativity"
remains: $\delta=1/\Gamma$. For small $\theta$ the term $1/(1-\beta\cos\theta)$ becomes very large,
more than compensating for the $1/\Gamma$ factor.
For $\cos \theta=\beta$ (i.e. $\sin\theta=1/\Gamma$) we have $\delta = \Gamma$.
For $\theta=0^\circ$ we have $\delta=\Gamma (1+\beta)$.

\subsection{Aberration}

\begin{figure}[h]
\vskip -0.5 true cm
\includegraphics[height=10cm, width=10cm]{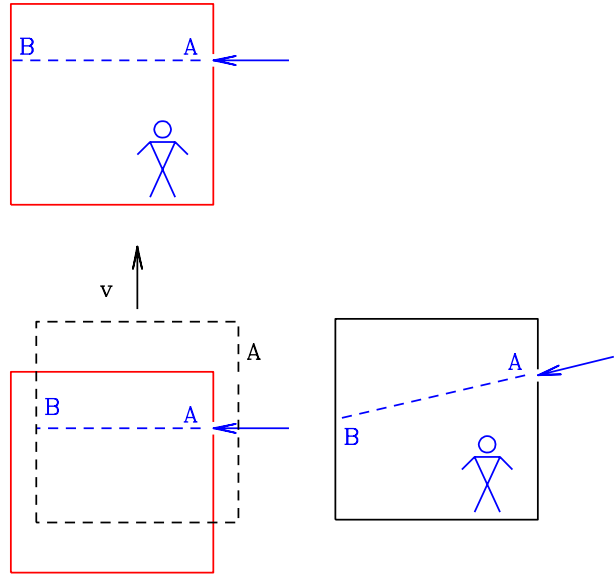}\\
\vskip -2 true cm
\caption[h]{The relativistic lift, to explain relativistic aberration
of light.
Assume first a non--moving lift, with a hole on the right wall.
A light ray, coming perpendicularly to the right wall, enter through 
the wall in $A$ and ends its travel in $B$.
If the lift is moving with a constant velocity $v$ to the top, its position 
is changed when the photon arrives to the left wall.
For the comoving observer, therefore, it appears that the light path
is tilted, since the point $B$ where the photon smashes into the left wall
is below the point $A$.
What happens if the lift, instead to move with a constant velocity,
is accelerating?
}
\label{chap1:fig5}
\end{figure}

Another very important effect happening when a source is moving is
the aberration of light.
It is rather simple to understand, if one looks at Fig. \ref{chap1:fig5}.
A source of photons is located perpendilarly to the right wall of a lift.
If the lift is not moving, and there is a hole in its right wall,
then the ligth ray enters in $A$ and ends its travel in $B$.
If the lift is not moving, $A$ and $B$ are at the same heigth.
If the lift is moving with a constant velocity $v$ to the top,
when the photon smashes the left wall it has a different location,
and the point $B$ will have, for a comoving observer, a smaller
height than $A$.
The light ray path now appears oblique, tilted.
Of course, the greater $v$, the more tilted the light ray path appears.
This immediately stimulate the question: what happens if the lift,
instead to move with a constant velocity, is accelerating?
With this example one can easily convince him/herself that the
``trajectory" of the photon would appear curved.
Since, by the equivalence principle, the accelerating lift cannot tell
if there is an engine pulling him up or if there is a planet underneath it,
we can then say that gravity bends the light rays, and make the space curved.

{}

This helps to understand why angles, between two inertial frames, change.
Calling $\theta$ the angle between the direction of the emitted
photon and the source velocity vector, we have:
\begin{eqnarray}
\sin\theta \, &=&\, {\sin\theta^\prime \over 
\Gamma (1+\beta\cos\theta^\prime )}; 
\quad\quad
\sin\theta^\prime \, =\, {\sin\theta\over \Gamma (1-\beta\cos\theta )}
\nonumber\\
\cos\theta \, &=&\, {\cos\theta^\prime +\beta\over 1+\beta\cos\theta^\prime}; 
\quad\quad\,\,\,\,\,\,\,\,
\cos\theta^\prime \, =\, {\cos\theta - \beta\over 1-\beta\cos\theta}
\end{eqnarray}
Note that, if $\theta^\prime =90^\circ$, then
$\sin\theta = 1/\Gamma$ and $\cos\theta =\beta$.
Consider a source emitting isotropically in $K^\prime$.
Half of its photons are emitted in one emisphere, namely, with
$\theta^\prime\le 90^\circ$.
Then, in $K$, the same source will appear
to emit half of its photons into a cone of semiaperture $1/\Gamma$.

Assuming symmetry around the angle $\phi$, 
the transformation of the solid angle $d\Omega$ is
\begin{equation}
d\Omega \, =\, 2\pi d\cos\theta  = 
{d\Omega^\prime \over \Gamma^2 (1+\beta\cos\theta^\prime)^2}
= d\Omega^\prime \,\Gamma^2 (1-\beta\cos\theta)^2 =
{d\Omega^\prime \over \delta^2}
\end{equation}

\subsection{Intensity}

We now have all the ingredients necessary to calculate the
transformation of the specific (i.e. monochromatic) and bolometric intensity.
The specific intensity 
has the unit of energy per unit surface, time, frequency and solid angle.
In cgs, the units are [erg cm$^{-2}$ s$^{-1}$ Hz$^{-1}$ ster$^{-1}$].
We can then write the specific intensity as 
\begin{eqnarray}
I(\nu) \, &=&\, h\nu { dN \over dt\, d\nu\, d\Omega\, dA} \nonumber \\
    \, &=&\, \delta h\nu^\prime { dN^\prime \over (dt^\prime /\delta)
\, \delta d\nu^\prime\, (d\Omega^\prime /\delta^2)\, dA^\prime}\nonumber \\
\, &=&\, \delta^3 \, I^\prime (\nu^\prime) \, =\, 
\delta^3 I^\prime (\nu/\delta)
\end{eqnarray}
Note that $dN =dN^\prime$ because it is a number, and that $dA=dA^\prime$.
If we integrate over frequencies we obtain the bolometric intensity
which transforms as
\begin{equation}
I \, =\, \delta^4 I^\prime
\label{chap1:intensity}
\end{equation}
The fourth power of $\delta$ can be understood in a simple way:
one power comes from the transformation of the frequencies,
one for the time, and two for the solid angle. They all add up.
This transformation is at the base of our understanding of relativistic
sources, namely radio--loud AGNs, gamma--ray bursts and galactic 
superluminal sources.

\subsection{Luminosity and flux}

The transformation of fluxes and luminosities
from the comoving to the observer frames is not trivial.
The most used formula is $L=\delta^4 L^\prime$, but this assumes
that we are dealing with a single, spherical blob.
It can be simply derived by noting that $L=4\pi d_{\rm L}^2 F$,
where $F$ is the observed flux, and by considering that the 
flux, for a distance source, is $F\propto \int_{\Omega_s} I d\Omega$.
Since $\Omega_s$ is the source solid angle, which is the same in the
two $K$ and $K^\prime$ frames, we have that $F$ transforms
like $I$, and so does $L$. 
But the emission from jets may come not only by a single spherical
blob, but by, for instance, many blobs, or even by a continuous 
distribution of emitting particles flowing in the jet.
If we assume that the walls of the jet are fixed, then the concept
of ``comoving" frame is somewhat misleading, because if we 
are comoving with the flowing plasma, then we see the walls of the jet
which are moving.

A further complication exists if the velocity is not uni--directional,
but radial, like in gamma--ray bursts. 
In this case, assume that the plasma is contained in a conical narrow shell
(width smaller than the distance of the shell from the apex of the cone).
The observer which is moving together with a portion of the plasma,
(the nearest case of a ``comoving observer") 
will see the plasma close to her going away from her, and more so
for more distant portions of the plasma.
Indeed, there could be a limiting distance beyond which the two portions
of the shells are causally disconnected.

Useful references are Lind \& Blandford (1985) and Sikora et al. (1997).

\subsection{Emissivity}

The (frequency integrated) emissivity $j$ is the energy emitted per unit time, 
solid angle and volume.
We generally have that the intensity, for an optically thin source,
is $I= \int_{\Delta R} j dr$, where $\Delta R$ is the length of the 
region containing the emitting particles.
The emissivity transforms like $j=j^\prime \delta^3$, namely with one
power of $\delta$ less than the intensity. 

\begin{figure}[h]
\includegraphics[height=9cm, width=10cm]{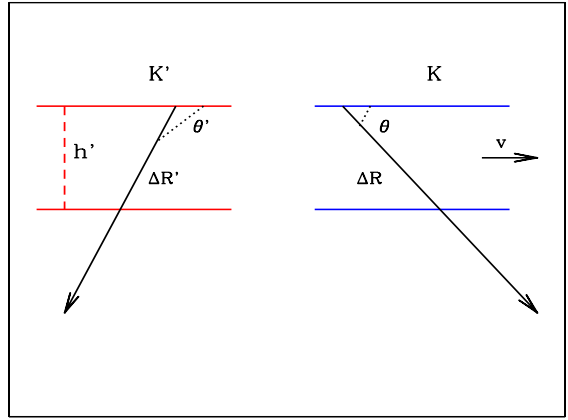} 
\vskip -3 true cm
\caption[h]{
Due to aberration of light, the travel path of the a light ray
is different in the two frames $K$ and $K^\prime$ 
}
\label{chap1:fig6}
\end{figure}{}

To understand why, consider a slab with plasma flowing with a velocity 
parallel to the walls of the slab, as in Fig. \ref{chap1:fig6}.
The observer in $K$ will measure a certain $\Delta R$ which depends on her
viewing angle.
In $K^\prime$ the same path has a different length, because of the aberration
of light.
The height of the slab $h^\prime=h$,
since it is perpendicular to the velocity.
The light ray travels a distance $\Delta R = h/\sin\theta$ in $K$,
and the same light ray travels a distance 
$\Delta R^\prime= h^\prime /\sin\theta^\prime$ in $K^\prime$.
Since $\sin\theta^\prime =\delta \sin\theta $, then 
$\Delta R^\prime = \Delta R /\delta$.
Therefore the column of plasma contributing to the emission, for $\delta>1$,
is less than what the observer in $K$ would guess by measuring $\Delta R$.
For semplicity, assume that the plasma is homogenous, allowing to simply write
$I=j \Delta R$. 
In this case:
\begin{equation}
I \, =  j \Delta R \, =\, \delta^4 I^\prime = 
\delta^4 j^\prime \Delta R^\prime
\, \to j = \delta^3 j^\prime
\end{equation}
And the corresponding transformation for the specific emissivity is
$j(\nu) = \delta^2 j^\prime(\nu^\prime)$.

\begin{figure}[h]
\includegraphics[height=10cm, width=10cm]{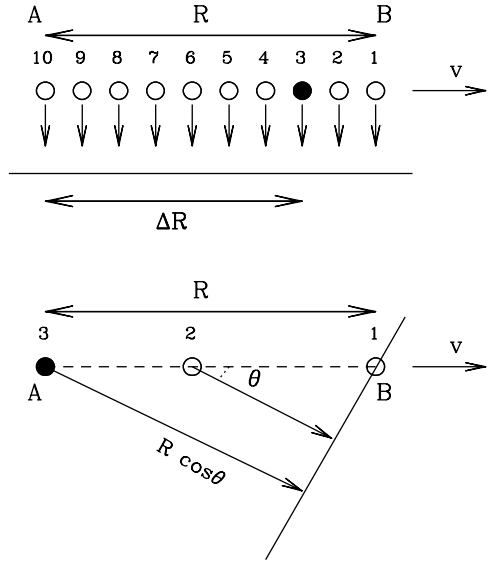} 
\vskip -2 true cm
\caption[h]{
Due to the differences in light travel time, the number
of blobs that can be observed simultaneously at any given time depends on
the viewing angle and the velocity of the blobs.
In the top panel the viewing angle is $\theta=90^\circ$ and all the 
blobs contained within a certain distance $R$ can be seen.
For smaller viewing angles, less blobs are seen.
This is because the photons emitted by the rear blobs have more distance
to travel, and therefore they have to be emitted before the photons
produced by the front blob.
Decreasing the viewing angle $\theta$ we see less blobs
(3 for the case illustrated in the bottom panel).
}
\label{chap1:fig7}
\end{figure}{}

Fig. \ref{chap1:fig7} illustrates another interesting example, 
taken from the work of Sikora et al. (1997).
Consider that within a distance $R$ from the apex of a jet
($R$ measured in $K$), at any given time there are $N$ blobs 
(10 on the specific example of Fig. \ref{chap1:fig7}),
moving with a velocity $v=\beta c$ along the jet.
To fix the ideas, let assume that beyond $R$ they switch off.
If the viewing angle is $\theta=90^\circ$, the photons emitted
by each blob travel the same distance to reach the observer,
who will see all the 10 blobs.
But if $\theta<90^\circ$, the photons produced by the rear blobs
must travel for a longer distance in order to reach the observer,
and therefore they have to be emitted before the photons produced
by the front blob.
The observer will then see {\it less} blobs.
To be more quantitative, consider a viewing angle $\theta<90^\circ$.
Photons emitted by blob numer 3 to reach blobs number 1 when it
produces its last photon (before to switch off) were emitted
when the blobs itself was just born (it was crossing point $A$).
They travelled a distance $R\cos\theta$ in a time $\Delta t$.
During the same time, the blob number 3 travelled a distance
$\Delta R= c\beta \Delta t$ in the forward direction.
The fraction $f$ of blobs that can be seen is then
\begin{equation}
f \, =  {R-\Delta R \over R} \, = \, 1- {c\beta \Delta t \over R} \, = \,
1-  \beta  \cos\theta 
\end{equation}
Where we have used the fact that $\Delta t = (R/c)\cos\theta$.
This is the usual Doppler factor.
We may multiply and divide by $\Gamma$ to obtain
\begin{equation}
f \, = \, {1 \over \Gamma \delta} 
\end{equation}
The bottom line is the following: even if the flux from a single 
blob is boosted by $\delta^4$, if the jet is made by many ($N$) equal blobs,
the total flux is not just boosted by $N\delta^4$ times the intrinsic
flux of a blob, because the observer will see less blobs if $\theta<90^\circ$.

\subsection{Brightness Temperature}

The brightness temperature is a quantity used especially in radio astronomy,
and it is defined by
\begin{equation}
T_{\rm B} \, \equiv \, { I(\nu) \over 2 k} \, {c^2 \over \nu^2} \, =\,
{ F(\nu) \over 2\pi k \, \theta_{\rm s}^2} \, {c^2 \over \nu^2} 
\end{equation}
where we have assumed that the solid angle subtended by the source
is $\Delta \Omega_{\rm s} \sim \pi \theta_{\rm s}^2$, and that the
received flux is $F(\nu)=\Delta\Omega_{\rm s} I(\nu)$.
There are 2 ways to measure $\theta_{\rm s}$:
\begin{enumerate}
\item from VLBI observations, one can often resolve the source and
hence directly measure the angular size.
In this case the relation between the brightness temperature measured
in the $K$ and $K^\prime$ frames is 
\begin{equation}
T_{\rm B} \, = { \delta^3F^\prime(\nu^\prime) 
\over 2\pi k \, \theta_{\rm s}^2} \, {c^2 \over \delta^2 ({\nu^\prime)}^2 } \, 
=\, \delta \, T^\prime_{\rm B}
\end{equation}
\item
If the source is varying, we can estimate its size by requiring
that the observed variability time--scale $\Delta t_{\rm var}$
is longer than the light travel time $R/c$, where $R$ is the typical
radius of the emission region.
In this case
\begin{equation}
T_{\rm B} \, > { \delta^3F^\prime(\nu^\prime) 
\over 2\pi k } {d_{\rm A}^2 \delta^2 \over (c{\Delta t_{\rm var}^\prime})^2} \, 
{c^2 \over \delta^2 ({\nu^\prime)}^2 } \, 
=\, \delta^3 \, T^\prime_{\rm B}
\end{equation}
where $d_{\rm A}$ is the angular distance, related to the luminosity distance
$d_{\rm L}$ by $d_{\rm A} = d_{\rm L}/(1+z)^2$.

\end{enumerate}

There is a particular class of extragalactic radio sources, 
called Intra--Day Variable (IDV) sources, showing variability time--scales 
of hours in the radio band.
For them, the corresponding observed brightess temperature can exceed $10^{16}$ K,
a value much larger than the theoretical limit for an incoherent synchrotron
source, which is between $10^{11}$ and $10^{12}$ K.
If the variability is indeed intrinsic, namely not produced by interstellar
scintillation, then one would derive a limit on the beamig factor $\delta$,
which should be larger than about 100.

\subsection{Moving in an homogeneous radiation field}

Jets in AGNs often moves in an external radiation field,
produced by, e.g. the accretion disk, or by the Broad Line Region (BLR)
which intercepts a fraction of the radiation produced by the disk
and re--emits it in the form of emission lines.
It it therefore interesting to calculate what is the energy density
seen by a an observer which is comoving with the jet plasma.

\begin{figure}[h]
\includegraphics[height=10cm, width=10cm]{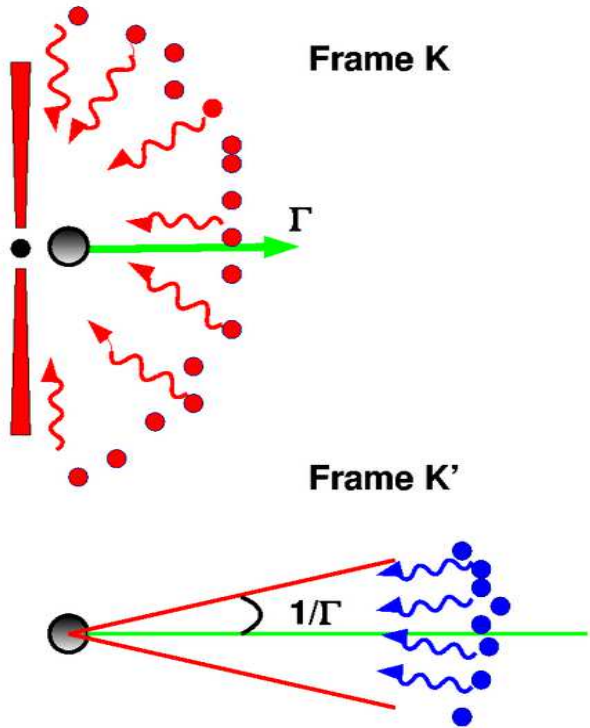} 
\caption[h]{A real case: a relativistic bob is moving within 
the Broad Line Region of a radio loud AGN, with Lorentz factor $\Gamma$. 
In the rest frame $K^\prime$ of the blob the photons coming from 
$90^\circ$ in frame $K$ are seen to come at an angle $1/\Gamma$. 
The energy density as seen by the blob is enhanced by a factor $\sim \Gamma^2$.
}
\label{blr}
\end{figure}{}

To make a specific example, as illustrated by Fig. \ref{blr},
assume that a portion of the jet
is moving with a bulk Lorentz factor $\Gamma$, velocity $\beta c$ 
and that it is surrounded by a shell of broad line clouds.
For simplicity, assume that the broad line photons are produced 
by the surface of a sphere of radius $R$ and that the jet is
within it.
Assume also that the radiation is monochromatic at some
frequency $\nu_0$ (in frame $K$).
The comoving (in frame $K^\prime$) observer will see photons
coming from a cone of semi--aperture $1/\Gamma$  
(the other half may be hidden by the accretion disk):
photons coming from the forward direction are seen blue--shifted
by a factor $(1+\beta)\Gamma$, while photons that the observer in $K$ 
sees as coming from the side (i.e. $90^\circ$ degrees) will be observed
in $K^\prime$ as coming from an angle given by
$\sin\theta^\prime=1/\Gamma$ (and $\cos\theta^\prime=\beta$)
and will be blue--shifted by a factor $\Gamma$.
As seen in $K^\prime$, each element of the BLR surface is moving
in the opposite direction of the actual jet velocity, and
the photons emitted by this element form an angle $\theta^\prime$
with respect the element velocity.
The Doppler factor used by $K^\prime$ is then
\begin{equation}
\delta' \, =\, {1 \over \Gamma(1-\beta\cos\theta^\prime)}
\end{equation}
The intensity coming from each element is see\label{chap1:intensity}n boosted as
(cfr Eq. \ref{chap1:intensity}):
\begin{equation}
I^\prime \, =\, \delta'^4 I
\end{equation}
The radiation energy density is
the integral over the solid angle of the intensity, divided by $c$:
\begin{eqnarray}
U^\prime \, &=& \, {2\pi\over c} \int_\beta^1 I^\prime 
d\cos\theta^\prime  \nonumber \\
\, &=&\,
{2\pi \over c} \int_\beta^1 {I \over 
\Gamma^4(1-\beta \cos\theta^\prime)^4} d\cos\theta^\prime  \nonumber \\
\, &=&\,
\left( 1+\beta +{\beta^2\over 3}\right) \Gamma^2 \, {2\pi I \over c}  
\nonumber \\
\, &=&\, \left( 1+\beta +{\beta^2\over 3}\right) \Gamma^2 \, U
\label{uprime}
\end{eqnarray}
Note that the limits of the integral correspond to the angles $0^\prime$
and $90^\circ$ in frame $K$.
The radiation energy density, in frame $K^\prime$, is then boosted by a factor
$(7/3) \Gamma^2$ when $\beta\sim 1$.
Doing the same calculation for a sphere, one would obtain $U^\prime=\Gamma^2U$.

Furthermore a (monochromatic) flux in $K$ is seen, in $K^\prime$, 
at different frequencies, between $\Gamma\nu_0$ and $(1+\beta)\Gamma\nu_0$,
with a slope $F^\prime(\nu^\prime) \propto \nu'^2$.
Why the slope $\nu'^2$? 
This can be derived as follows:
we already know that $I^\prime(\nu') = \delta'^3 I(\nu)=(\nu'/\nu)^3 I(\nu)$.
The flux at a specific frequency is
\begin{equation}
F^\prime(\nu')    \, =\, 2\pi 
\int_{\mu^\prime_1}^{\mu^\prime_2} 
d\mu^\prime \left({\nu'\over \nu}\right)^3 I(\nu)
\end{equation}
where $\mu^\prime \equiv \cos\theta^\prime$, and the integral
is over those $\mu^\prime$ contributing at $\nu^\prime$.
Since
\begin{equation}
{\nu^\prime \over \nu} \, =\delta^\prime \, = \,
{1\over \Gamma(1-\beta\mu^\prime)} \, \to \, \mu^\prime = {1\over \beta}\,
\left( 1-{\nu \over \Gamma \nu^\prime}\right)
\end{equation}
we have
\begin{equation}
d\mu^\prime =- {d\nu\over \beta\Gamma \nu^\prime}
\end{equation}
Therefore, if the intensity is monochromatic in
frame $K$, i.e. $I(\nu)=I_0\delta(\nu-\nu_0)$, 
the flux density in the comoving frame is
\begin{eqnarray}
F^\prime(\nu') \, &=& \, 
2\pi \int_{\nu_2}^{\nu_1} {d\nu \over \beta\Gamma \nu^\prime} 
\left({\nu'\over \nu}\right)^3 I_0 \delta(\nu-\nu_0) \nonumber \\
&=& \, {2 \pi \over \Gamma\beta} \,\, {I_0 \over \nu_0^3}\, \nu'^2;
\quad\quad \Gamma\nu_0 \le \nu^\prime \le (1+\beta)\Gamma \nu_0
\label{slopenu2}
\end{eqnarray}
where the frequency limits corresponds to photons produced 
in an emysphere in frame $K$, and between $0^\circ$ and 
$\sin\theta^\prime=1/\Gamma$ in frame $K^\prime$.
Integrating Eq. \ref{slopenu2} over frequency, one obtains
\begin{equation}
F^\prime \, = \, 2\pi I_0 \Gamma^2\left (1+\beta+{\beta^2\over 3}\right) \, =\, 
\Gamma^2\left (1+\beta+{\beta^2\over 3}\right) \, F
\end{equation}
in agreement with Eq. \ref{uprime}.

\section{Superluminal motion}

\begin{figure}[h]
\center
\includegraphics[height=13.cm, width=13.cm]{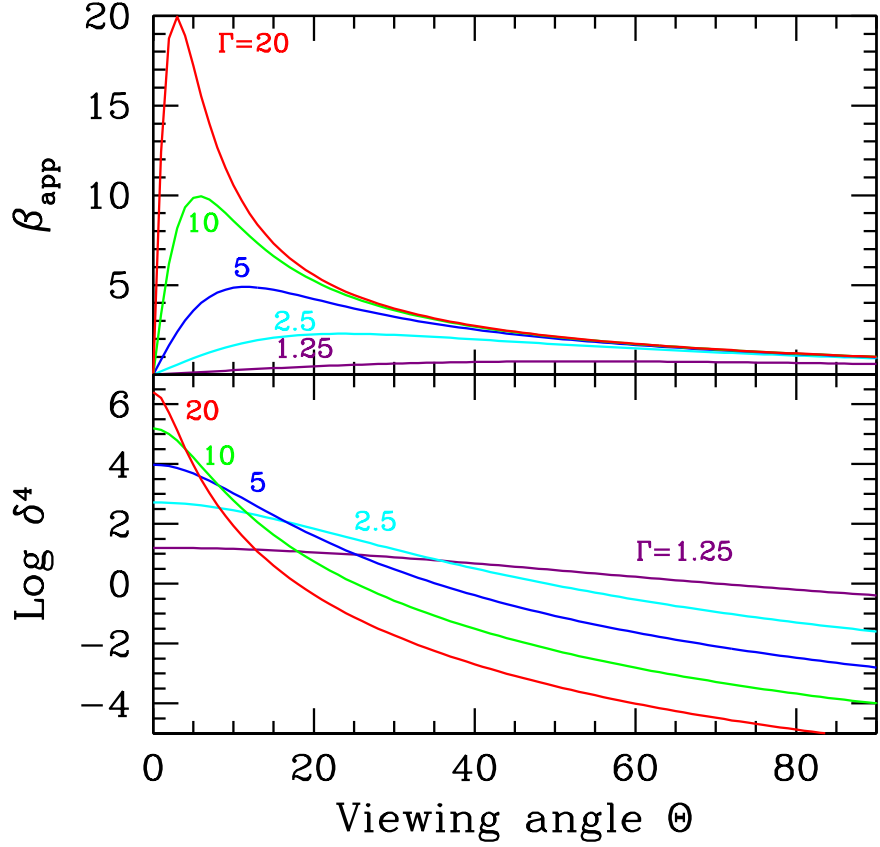} 
\vskip -0.5 cm
\caption[h]{Top: The apparent velocity $\beta_{\rm app}$ as a function
of the viewing angle $\theta$ for different values of $\Gamma$, as labelled.
Bottom: the amplification $\delta^4$ as a function of the viewing angle, for
the same $\Gamma$ as in the top panel.
}
\label{bapp}
\end{figure}{}

In 1971 the Very Long Baseline Interferometry began, linking different radio--telescopes 
that where distant even thousands of km.
The resolving power of a telescope is of the order of
\begin{equation}
\phi \, \sim \, {\lambda \over D}
\end{equation}
where $\lambda$ is the wavelength to be observed, and $D$ is the diameter of the telescope
or the distance of two connected telescopes.
Observing at 1 cm, with two telescopes separated by 1000 km (i.e.$10^8$ cm), means
that we can observe details of the source down to the milli--arcsec level (m.a.s.).
The first observations of the inner jet of radio--loud quasars revealed that the
jet structure was not continuous, but blobby, with several radio knots.
Repeating the observations allowed us to discover that the blobs were not stationary,
but were moving.
Comparing radio maps taken at different times one could measure the angular displacement
$\Delta \theta$ between the position of the blob.
Knowing the distance $d$, one could then transform $\Delta \theta$ is a linear size:
$\Delta R = d \Delta \theta$.
Dividing by the time interval $\Delta t_{\rm a}$ between the two radio maps, one obtains a velocity 
\begin{equation}
v_{\rm app} \, = \, { d \Delta \theta \over \Delta t_{\rm A}}
\end{equation}
With some surprise, in several objects this turned out to be larger than the light speed $c$.
Therefore these sources were called {\it superluminal}.
The explanation of this apparent violation of special relativity
is in Fig. \ref{chap1:fig4}: the time interval $\Delta t_{\rm a}$ can be much 
shorter than the emission time $\Delta t_{\rm e}$.
With reference to Fig. \ref{chap1:fig4}, what the observer measures in the two radio maps is the position
of the blob in point $A$ (first map) and $B$ (second map), projected in the plane of the sky.
The observed displacement is then:
\begin{equation}
CB \, = \, \beta c \Delta t_{\rm e} \sin\theta
\end{equation}
Dividing by $\Delta t_{\rm a} =\Delta t_{\rm e} (1-\beta\cos\theta)$ we have
the measured apparent velocity as
\begin{equation}
v_{\rm app} \, = \, {\beta c \Delta t_{\rm e} \sin\theta \over \Delta t_{\rm a} } \, \, \longrightarrow \,
\beta_{\rm app} \, =\, { \beta \sin\theta \over 1-\beta\cos\theta}
\end{equation}
Ask yourself: $\Gamma$ does not appear. Is it ok?
At $0^\circ$ the apparent velocity is zero. Is is ok?
At what angle $\beta_{\rm app}$ is maximized?
What is its maximum value?
Fig. \ref{bapp} shows $\beta_{\rm app}$ as a function of the viewing angle (angle between the line
of sight and the velocity) for different $\Gamma$.
Is the apparent superluminal speed given by a real motion of the emitting material?
Can it be something else?
If there are other possibilities, how to discriminate among them?

\vskip 3 cm

\begin{center}
\begin{table}[h]
\begin{tabular}{|l|l|}
\hline
\hline
                                         & \\
$\nu =  \nu^\prime   \delta  $           &frequency \\
$t = t^\prime/\delta$                    &time \\
$V =  V^\prime\delta$                    &volume  \\
$\sin\theta =\sin\theta^\prime/\delta$   &sine  \\
$\cos\theta =(\cos\theta^\prime+\beta)/(1+\beta\cos\theta^\prime)$  &cosine  \\
$I(\nu) = \delta^3 I^\prime(\nu^\prime)$  &specific intensity  \\
$I = \delta^4 I^\prime$                  &total intensity  \\
$j(\nu) = j^\prime(\nu^\prime)\delta^2 $  &specific emissivity  \\
$\kappa(\nu) = \kappa^\prime(\nu^\prime)/\delta$  &absorption coefficient  \\
$T_B =  T^\prime_B\delta$                &brightn. temp. (size directly measured)  \\
$T_B = T^\prime_B \delta^3$              &brightn. temp. (size from variability)  \\
$U^\prime=(1+\beta+\beta^2/3) \Gamma^2 U$   &Radiation energy density within an emisphere \\
                                          & \\
\hline
\hline
\end{tabular}
\caption{Useful relativistic transformations}
\end{table}
\end{center}


















\section{A question}


 
Suppose that some optically thin plasma of mass $m$ is falling onto a central 
object with a velocity $v$ and bulk Lorentz factor $\Gamma$.
The central object has mass $M$ and produces a luminosity $L$.
Assume that the interaction is through Thomson scattering and
that there are no electron--positron pairs.

a) What is the radiation force acting on the electrons?

b) What is the gravity force acting on the protons?

c) What definition of limiting (``Eddington")
luminosity would you give in this case?

d) What happens if the plasma is instead going outward?

\vskip 1 cm
\noindent
{\bf References}

\noindent
Lind K.R. \& Blandford R.D., 1985, ApJ, 295, 358

\noindent
Mook D.E. \& Vargish T., 1991, {\it Inside Relativity}, Princeton Univ. Press

\noindent
Penrose R., 1959, Proc. of the Cambridge Philos. Soc., 55, p. 137

\noindent
Sikora M., Madejski G., Moderski R. \& Poutanen J., 1997,
ApJ, 484, 108

\noindent
Terrel J., 1959, Phys. Rev., 116, 1041


\chapter{Synchrotron emission and absorption}

\section{Introduction}

We now know for sure that many astrophysical sources are magnetized
and have relativistic leptons.
Magnetic field and relativistic particles are the two ingredients
to have synchrotron radiation.
What is responsible for this kind of radiation is the Lorentz force,
making the particle to gyrate around the magnetic field lines.
Curiously enough, this force {\it does not work}, but makes
the particles to accelerate even if their velocity modulus
hardly changes.

The outline of this section is:
\begin{enumerate}

\item 
We will derive the total power emitted by the single electron.
Total means integrated over frequency and over emission angles.
This will require to generalize the Larmor formula to the
relativistic case;

\item 
We will then outline the basics of the spectrum emitted by the single 
electron. This is treated in several text--books, so we will
concentrate on the basic concepts;

\item 
Spectrum from an ensemble of electrons. Again, only the basics;

\item
Synchrotron self absorption. We will try to discuss things from
the point of view of a photon, that wants to calculate its
survival probability, and also the point of view of the electron,
that wants to calculate the probability to absorb the photon,
and then increase its energy and momentum.

\end{enumerate}

\section{Total losses}

To calculate the total (=integrated over frequencies and emission angles) 
synchrotron losses we go into the frame that is instantaneously at rest 
with the particle (in this frame $v$ is zero, but not the
acceleration!).
This is because we will use the fact that the {\it emitted} power
is Lorentz invariant:
\begin{equation}
P_{\rm e}\, =\, P_{\rm e}^\prime \, =\, {2e^2 \over 3c^3} a'^2 \,
=\, {2e^2 \over 3c^3} \left[a'^2_\parallel + a'^2_\perp\right]  
\end{equation}
where the subscript ``e" stands for ``emitted".
The fact that the power is invariant sounds natural, 
since after all,  power is energy over time, 
and both energy and time transforms the same way 
(in special relativity with rulers and clocks).
But be aware that this {\it does not mean} that the
emitted and {\it received} power are the same.
They are not!

The problem is now to find how the parallel (to the velocity vector)
and perpendicular components of the acceleration Lorentz transform.
This is done in text books, so we report the results:
\begin{eqnarray}
a'_\parallel \, &=& \, \gamma^3 a_\parallel \nonumber \\
a'_\perp \, &=& \,\gamma^2 a_\perp    
\end{eqnarray}
where $\gamma$ is the particle Lorentz factor.
One easy way to understand and remember these transformations
is to recall that the acceleration is the second derivative of
space with respect to time. The perpendicular component
of the displacement is invariant, so we have only to transform
(twice) the time (factor $\gamma^2$).
The parallel displacement instead transforms like $\gamma$, hence 
the $\gamma^3$ factor.

The generalization of the Larmor formula is then:
\begin{equation}
P_{\rm e}\, =\,P_{\rm e}^\prime \, =\, 
{2e^2 \over 3c^3} \left[a'^2_\parallel + a'^2_\perp\right]  \, =\, 
{2e^2 \over 3c^3} \gamma^4 \left[\gamma^2 a^2_\parallel + a^2_\perp\right] 
\label{larmor} 
\end{equation}
Don't be fooled by the $\gamma^2$ factor in front of $a^2_\parallel$...
this component of the power is hardly important: since the velocity,
for relativistic particles, is always close to $c$, it implies that
one can get very very small acceleration in the same direction of the
velocity. This is why linear accelerators minimize radiation losses.
For synchrotron machines, instead, the losses due to radiation can be
the limiting factor, and they are of course due to $a_\perp$: changing
the {\it direction} of the velocity means large accelerations, even
without any change in the velocity modulus.
To go further, we have to calculate the two components of the
acceleration for an electron moving in a magnetic field.
Its trajectory, in general, will have an helical shape of 
radius $r_{\rm L}$ (the Larmor radius).
The angle that the velocity vector makes with the magnetic field line
is called {\it pitch angle}. Let us denote it with $\theta$.
We can anticipate that, in the absence of electric field and 
for a homogeneous magnetic field, the modulus of the velocity
will not change: the magnetic field does not work, and so there is no
change of energy, except for the losses due to the synchrotron radiation
itself. 
So one important assumption is that {\it at least during one gyration,
the losses are not important}. This is almost always satisfied in 
astrophysical settings, but there are indeed some cases where this is 
not true. 

%
\begin{figure}
\vskip -0.7 true cm
\includegraphics[height=10cm, width=10cm]{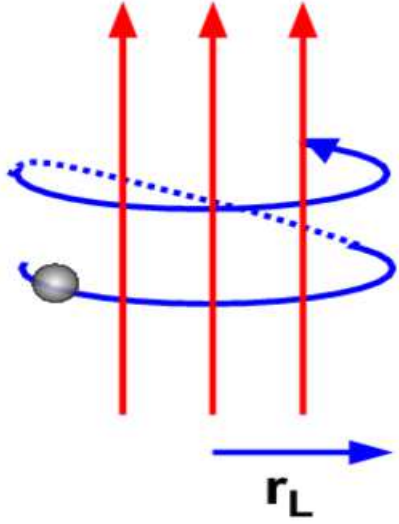} 
\vskip -0.5 true cm
\caption{A particle gyrates along the magnetic field lines.
Its trajectory has an helicoidal shape, with Larmor radius 
$r_{\rm L}$ and pitch angle $\theta$.
}
\label{pitch}
\end{figure} 
%

When there is no electric field the only 
acting force is the (relativistic) Lorentz force:
\begin{equation}
F_{\rm L}\, = {d\over dt} (\gamma m {\mathbf v})\, 
=\, {e\over c} \, {\mathbf v \times {\mathbf B} }
\end{equation}
The parallel and perpendicular components are
\begin{eqnarray}
F_{\rm L \parallel}\, = \, e \, v_\parallel  B \, =\, 0\, 
&\to&\, a_\parallel \, =\, 0 \nonumber \\
F_{\rm L \perp}\, = \gamma m {dv_\perp \over dt}\, = 
\, e \, {v_\perp \over c} B \, &\to&\, a_\perp \, =\, {e v B\sin\theta \over
\gamma m  c }
\label{aperp}
\end{eqnarray}
We can also derive the Larmor radius $r_{\rm L}$ by setting
$a_\perp = v_\perp^2/r_{\rm L}$, and so
\begin{equation}
r_{\rm L}\, = \, {v_\perp^2 \over a_\perp} \, =\,  
{\gamma m  c^2 \beta\sin\theta \over eB} 
\label{rl}
\end{equation}
The fundamental frequency is the inverse of the time occurring
to complete one orbit (gyration frequency), 
so $\nu_{\rm B} = c \beta\sin\theta/(2\pi r_{\rm L})$, giving
\begin{equation}
\nu_{\rm B}\, = \, { eB \over 2\pi \gamma m  c } \, =\ {\nu_{\rm L}\over \gamma}
\label{vl}
\end{equation}
where $\nu_{\rm L}$ is the Larmor frequency, namely the gyration frequency for
sub--relativistic particles.
Larger $B$ means smaller $r_{\rm L}$, hence greater gyration frequencies.
Vice--versa, larger $\gamma$ means larger inertia, thus larger $r_{\rm L}$, and smaller 
gyration frequencies.
Substituting $a_\perp$ given in Eq. \ref{aperp} in the generalized
Larmor formula (Eq. \ref{larmor}) we get:
\begin{equation}
P_{\rm S} \, =\, 
{2e^4 \over 3m^2 c^3} B^2 \gamma^2 \beta^2 \sin^2\theta  
\label{larmors} 
\end{equation}
We can make it nicer (for future use) by recalling that:
\begin{itemize}
\item
The magnetic energy density is $U_B \equiv B^2/(8\pi)$
\item the quantity $e^2/(m_{\rm e} c^2)$, in the case of electrons, is
the classical electron radius $r_0$
\item the square of the electron radius is proportional to the 
Thomson scattering cross section $\sigma_{\rm T}$, i.e.
$\sigma_{\rm T}= 8\pi r_0^2 /3= 6.65\times 10^{-25}$ cm$^{2}$.
\end{itemize}
Making these substitutions, we have that the synchrotron power emitted
by a single electron of given pitch angle is:
\begin{equation}
P_{\rm S}(\theta) \, =\, 
2 \sigma_{\rm T} c U_B \gamma^2 \beta^2 \sin^2\theta  
\label{pstheta} 
\end{equation}
In the case of an {\it isotropic distribution of pitch angles} we can average
the term $\sin^2\theta$ over the solid angle. The result is 2/3, giving 
\begin{equation}
\langle P_{\rm S}\rangle \, =\, 
{4\over 3} \sigma_{\rm T} c U_B \gamma^2 \beta^2 
\label{ps} 
\end{equation}
Now pause, and ask yourself:
\begin{itemize}
\item
Is $P_S$ valid only for relativistic particles, or does it describe
correctly the radiative losses also for sub--relativistic ones?
\item 
In the relativistic case the losses are proportional to the
{\it square} of the electron energy. Do you understand why?
And for sub--relativistic particles?
\item 
What happens of we have protons, instead of electrons?
\item 
What happens for $\theta\to 0$? 
Are you sure? (that losses vanishes..). Ok, but what happens
to the {\it received} power when you have the lines of the magnetic field
along the line of sight, and a beam of particles, all 
with a small pitch angles, shooting at you?
\item 
Why on earth there is the scattering cross section? Is this a
coincidence or does it hide a deeper fact?
\end{itemize}

\subsection{Synchrotron cooling time}

When you want to estimate a timescale of a quantity $A$, 
you can always write $t=A/\dot A$. 
In our case $A$ is the energy of the particle.
For electrons with an isotropic pitch angle distribution we have
\begin{equation}
t_{\rm syn} \, =\, {E\over \langle P_{\rm S}\rangle } \, =\, 
{\gamma m_{\rm e} c^2 \over  
(4/3) \sigma_{\rm T} c U_B \gamma^2 \beta^2 } 
\, \sim \, {7.75\times 10^8 \over B^2 \gamma} \,\,\, {\rm s} \, =
\, {24.57 \over  B^2 \gamma} \,\,\, {\rm yr}
\label{tcool} 
\end{equation}
In the vicinity of a supermassive AGN black hole we 
can have $B=10^3 B_3$ Gauss and $\gamma=10^3\gamma_3$, yielding
$t_{\rm syn} =0.75/(B_3^2\gamma_3)$ s.
The same electron, in the radio lobes of a radio loud quasar with 
$B=10^{-5}B_{-5}$ Gauss, cools in $t_{\rm syn} = 246$ million years.

\section{Spectrum emitted by the single electron}

\subsection{Basics}
There exists a typical frequency associated to the
synchrotron process. 
This is related to the inverse of a typical time.
If the electron is relativistic, this is {\it not} the
revolution period. 
Instead, it is the fraction of the time, for each orbit, 
during which the observer {\it receives} 
some radiation. 
To simplify, consider an electron with a pitch angle of $90^\circ$,
and look at Fig. \ref{pattern}, illustrating the typical
patterns of the produced radiation for sub--relativistic electrons
moving with a velocity parallel (top panel) or perpendicular (mid panel)
to the acceleration.
In the bottom panel we see the pattern for a relativistic electron
(with ${\mathbf v \perp {\mathbf a}}$): it is strongly beamed in the
forward direction. 
This is the direct consequence of the aberration of light, 
making half of the photons be emitted in a cone of
semi--aperture angle $1/\gamma$ (which is called {\it the beaming angle}).
Note that this {\it does not} mean that {\it half of the power}
is emitted within $1/\gamma$, because the photons inside the 
beaming cone are more energetic than those outside, and are
more tightly packed (do you remember the $\delta^4$ factor
when studying beaming?). 1

\begin{figure}[h]
\vskip -0.1  true cm
\includegraphics[height=5cm, width=6cm]{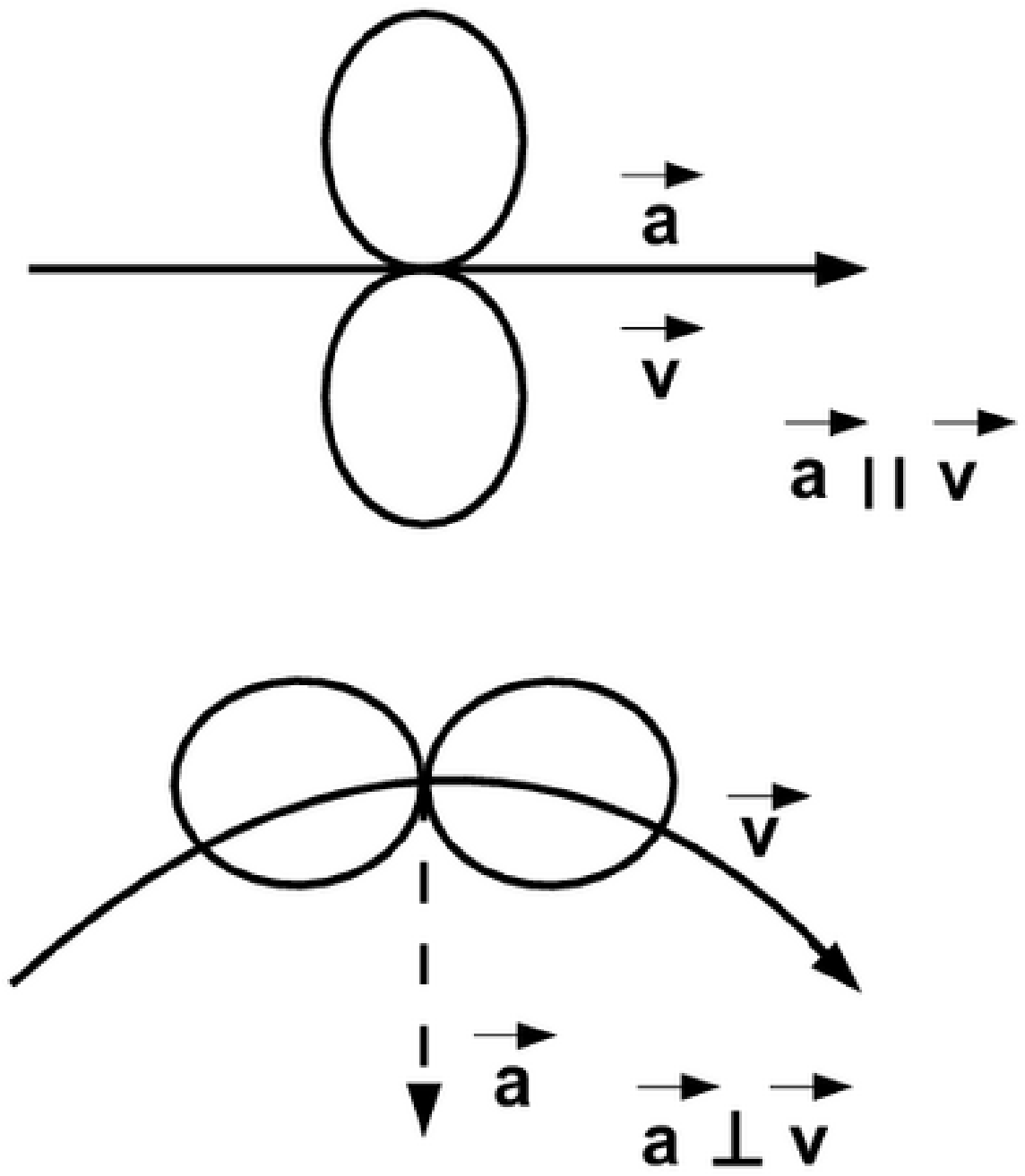} 
\vskip -0.3 true cm
\includegraphics[height=5cm, width=8cm]{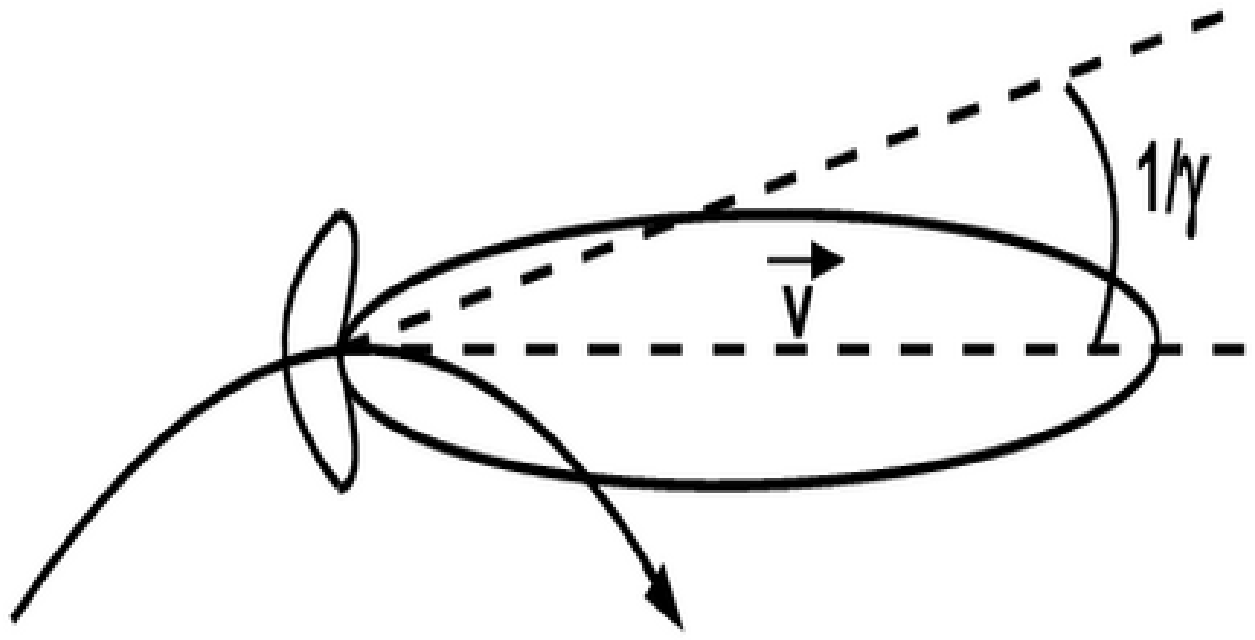} 
\vskip -0.2true cm
\caption{Radiation patterns for a non relativistic 
particle with the velocity parallel (top) or perpendicular (mid) 
to the acceleration.
When the particle is relativistic, the pattern strongly changes
due to the aberration of light, and is strongly beamed in the
forward direction.
}
\label{pattern}
\end{figure} 

\begin{figure}
\vskip -0.5 true cm
\includegraphics[height=8.5cm, width=13cm]{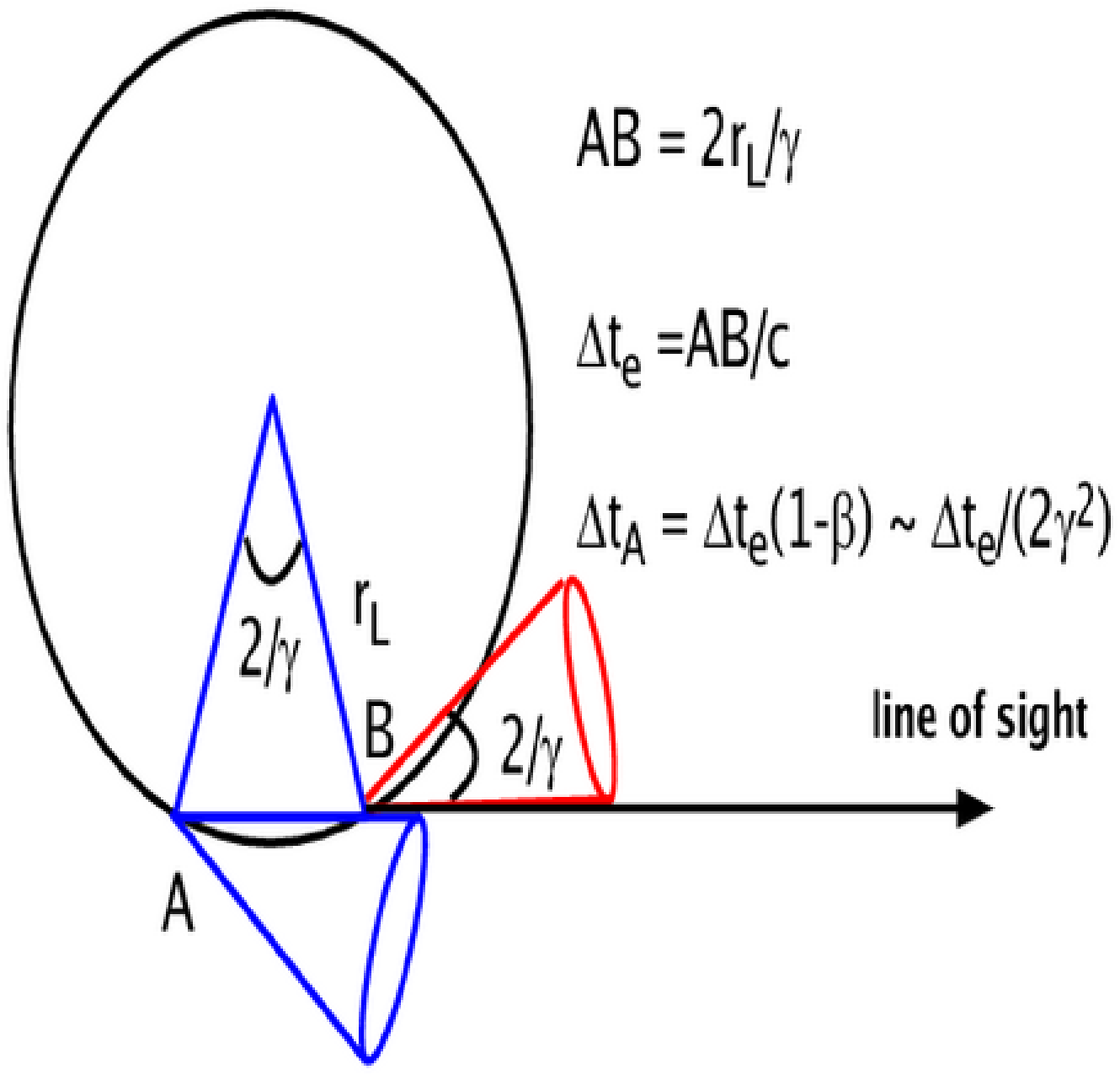} 
\vskip -0.5 true cm
\caption{A relativistic  electron is gyrating along a 
magnetic field line with pitch angle 90$^\circ$. Its trajectory is then a circle
of radius $r_{\rm L}$. Due to aberration, an observer will ``see it"
(i.e. will measure an electric field) when the beaming cone of
total aperture angle $2/\gamma$ is pointing at him.
}
\label{time}
\end{figure} 

To go further, recall what we do when we study a time series 
and we want to find the power spectrum: we Fourier transform it.
In this case we must do the same. 
Therefore if there is a typical timescale during which we {\it receive}
most of the signal, we can say that most of the power is emitted
at a frequency that is the inverse of that time.

Look at Fig. \ref{time}: the relativistic electron emits photons
all along its orbit, but it will ``shoot" in a particular direction only
for the time 
\begin{equation}
\Delta t_{\rm e} \, \sim \, {AB \over v} \, =\, {1 \over v}\, r_{\rm L} \, {\gamma \over 2} 
\, =\, {1 \over v}\, {2 mc v\over eB} \, =\, {2\over 2\pi} \, {1\over \nu_{\rm L} } \, =\, 
{2 \over 2\pi}\, {1\over  \gamma \nu_{\rm B}}
\label{te} 
\end{equation}
%
This is the {\it emitting} time during which the electron emits photons
that will reach the observer. 
We can approximate the arc $AB$ with a straight segment if
the electron is relativistic, and the observed will then measure
an {\it arrival} time $\Delta t_A$ that is shorter than $\Delta t_e$:
\begin{equation}
\Delta t_{\rm A} \, = \Delta t_{\rm e} \, (1-\beta) \, =\, 
\Delta t_{\rm e} \, {(1-\beta^2)\over 1+\beta} \, \sim \, 
{\Delta t_{\rm e} \over 2\gamma^2 } \, =\, {1 \over 2\pi \gamma^3 \nu_{\rm B}}
\label{ta} 
\end{equation}
The inverse of this time is the typical synchrotron (angular) frequency $\omega_{\rm s}=
2\pi \nu_{\rm s}$, so that:
\begin{equation}
\nu_{\rm s} \, = \, {1 \over 2\pi \Delta t_{\rm A}} \, =\,
\gamma^3 \nu_{\rm B} \, =\, \gamma^2 \nu_{\rm L}\, =\, \gamma^2 \, { eB \over 2\pi m_e c}
\label{vs} 
\end{equation}
This is a factor $\gamma^3$ greater than the fundamental frequency, and a factor
$\gamma^2$ greater than the Larmor frequency, defined as the typical frequency of
non--relativistic particles.
We expect that the particle emits most of its power at this frequency.

\begin{figure} 
\vskip -0.5 true cm
\includegraphics[height=10cm, width=12cm]{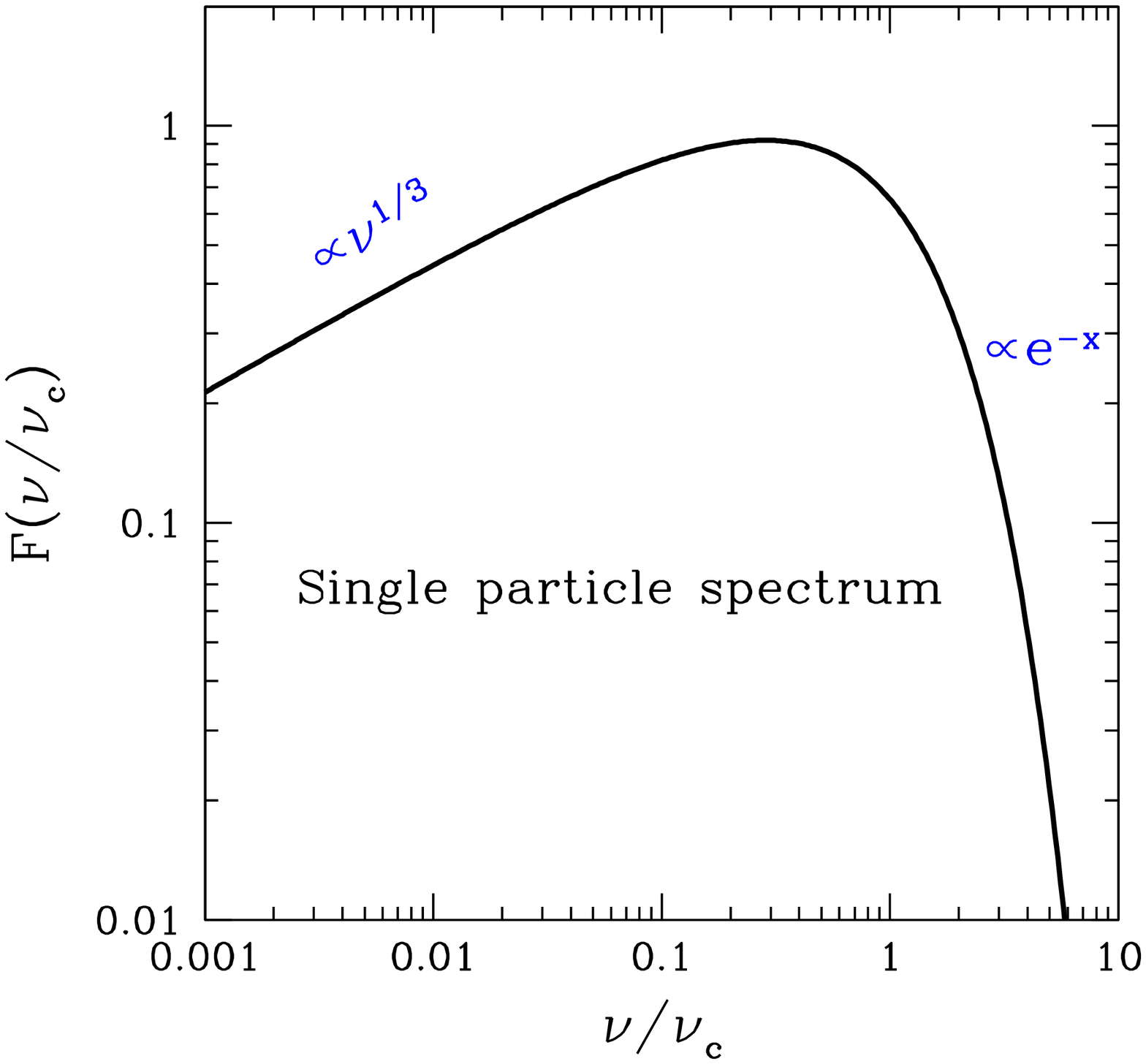}
\vskip -0.5 true cm
\includegraphics[height=10cm, width=12cm]{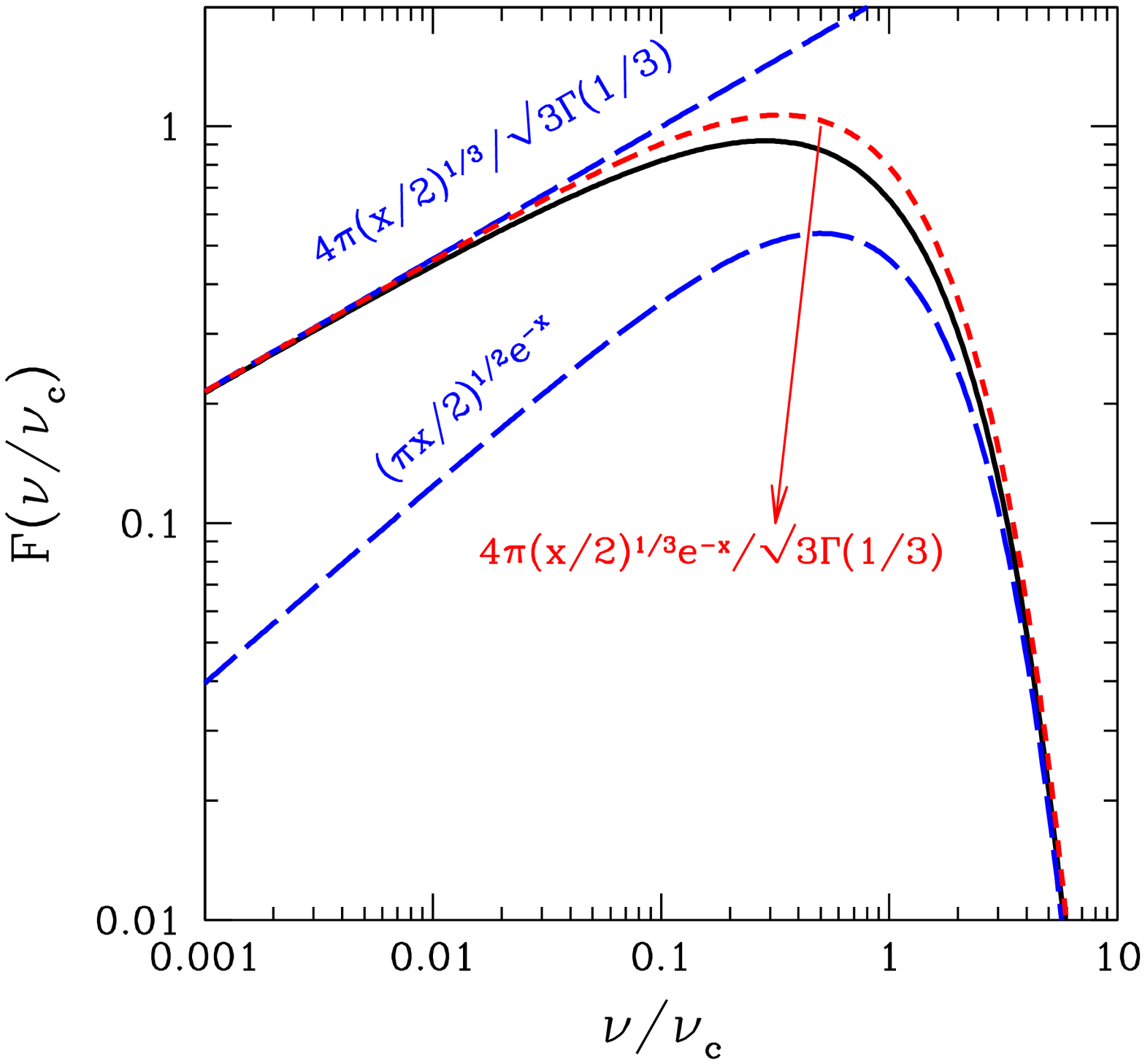}  
\vskip -0.7 true cm
\caption{Top panel: The function $F(\nu/\nu_c)$ describing the synchrotron spectrum
emitted by the single electron.
Bottom panel: $F(\nu/\nu_c)$ is compared with some approximating
formulae, as labeled. We have defined $x\equiv \nu/\nu_c$.
}
\label{singles}
\end{figure} 

\subsection{The real stuff}

One can look at any text book for a detailed discussion of the 
procedure to calculate the spectrum emitted by the single particle.
Here we report the results: the power per unit frequency emitted
by an electron of given Lorentz factor and pitch angle is:
\begin{eqnarray}
P_{\rm s}(\nu,\gamma,\theta) \, &=&\, 
{\sqrt{3} e^3 B\sin\theta  \over m_{\rm e} c^2} \, F(\nu/\nu_c) \nonumber\\
F(\nu/\nu_c)\, &\equiv & \, {\nu \over \nu_c} \, \int_{\nu/\nu_c}^\infty K_{5/3}(y)dy\nonumber  \\
\nu_c\, & \equiv & \, {3\over 2} \, \nu_{\rm s} \sin\theta
\label{psyn} 
\end{eqnarray}
This is the power integrated over the emission pattern. 
$K_{5/3}(y)$ is the modified Bessel function of order 5/3.
The dependence upon frequency is contained in $F(\nu/\nu_c)$, that is plotted
in Fig. \ref{singles}.
This function peaks at $\nu\sim 0.29\nu_c$, therefore very close to 
what we have estimated before, in our very approximate treatment.
The low frequency part is well approximated by a power law of slope 1/3:
\begin{equation}
F(\nu/\nu_c)\, \to\, {4\pi \over \sqrt{3} \Gamma(1/3)}\,  \left( {\nu \over 2 \nu_c } \right)^{1/3}
\quad (\nu\ll \nu_c)
\label{fx1} 
\end{equation}
At $\nu\gg\nu_c$ the function decays exponentially, and can be approximated by:
\begin{equation}
F(\nu/\nu_c)\, \to\, \left({\pi \over 2}\right)^{1/2} \,
\left({\nu\over \nu_c}\right)^{1/2} e^{-\nu/\nu_c}  
\quad (\nu\gg \nu_c)
\label{fx2} 
\end{equation}
Another approximation valid for all frequency, but overestimating
$F$ around the peak, is:
\begin{equation}
F(\nu/\nu_c)\,  \sim \, { 4\pi \over\sqrt{3}\Gamma(1/3)}  
\left( {\nu \over 2 \nu_c } \right)^{1/3}e^{-\nu/\nu_c}  
\label{fx3} 
\end{equation}

\subsection{Limits of validity}

One limit can be obtained by requiring that, during one orbit, 
the emitted energy is much smaller than the electron energy.
If not, the orbit is modified, and our calculations are no more valid.
For non--relativistic electrons this translates in demanding that
\begin{equation}
h\nu_{\rm B}\, <\, m_{\rm e}c^2 \, \to \, B\, <\, {2\pi m_{\rm e}^2c^3 \over h e} 
\equiv B_{\rm c}
\end{equation}
where $B_{\rm c}\sim 4.4\times 10^{13}$ Gauss is the critical magnetic field,
around and above which quantum effects appears 
(i.e. quantized orbits, Landau levels and so on).

For relativistic particles we demand that the energy emitted during one orbit
does not exceed the energy of the particle.
\begin{equation}
{P_{\rm s} \over \nu_{\rm B}} \, <\, \gamma m_{\rm e} c^2\, \to \, 
B\, <\, {e/ \sigma_{\rm T} \over \gamma^2\sin^2\theta} \, \sim \, 
{7.22\times 10^{14}\over \gamma^2\sin^2\theta} \,\,\, {\rm Gauss}
\end{equation}
Therefore for large $\gamma$ we reach the validity limit even if the magnetic 
field is sub--critical.

For very small pitch angles beware that the spectrum is not described by $F(\nu/\nu_c)$,
but consists of a blue-shifted cyclotron line.
This is because, in the gyroframe, the particle is sub--relativistic, and so it emits
only one (or very few) harmonics, that the observer sees blueshifted.

\begin{figure} 
\vskip -0.3 true cm
\includegraphics[height=16cm, width=15cm]{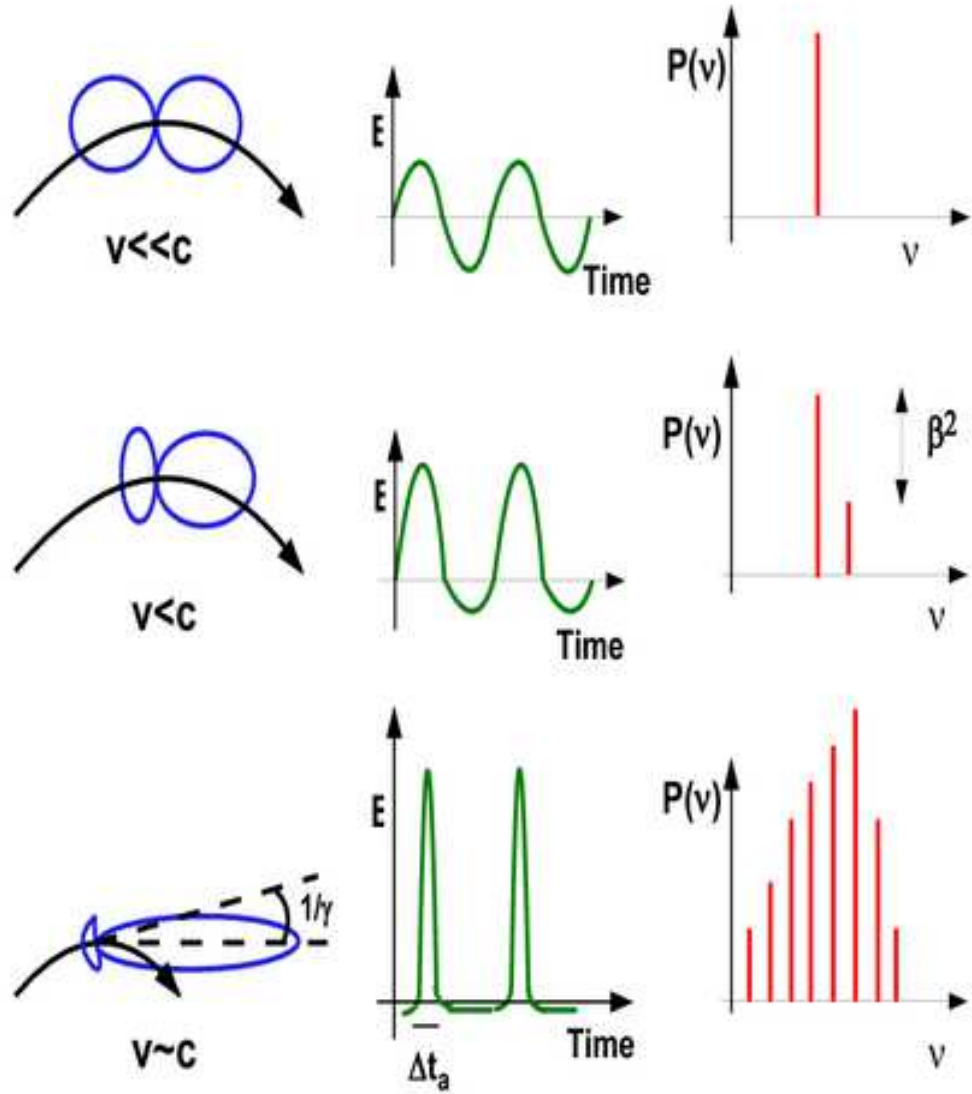} 
\caption{From cyclo to synchro: if the emitting particle has a very small velocity,
the observer sees a sinusoidal (in time) electric field $E(t)$.
Increasing the velocity the pattern becomes asymmetric, and the second harmonic appears.
For $0<\beta\ll 1$ the power in the second harmonic is a factor $\beta^2$ less than the power 
in the first.
For relativistic particles, the pattern becomes strongly beamed, the emission is concentrated
in the time $\Delta t_A$. As a consequence the Fourier transformation of $E(t)$ must contain
many harmonics, and the power is concentrated in the harmonics of frequencies
$\nu\sim 1/\Delta t_A$. 
Broadening of the harmonics due to several effects ensures that the spectrum in 
this case becomes continuous.
Note that the fundamental harmonic becomes {\it smaller} increasing $\gamma$ 
(since $\nu_B\propto 1/\gamma$).
}
\label{ctos}
\end{figure} 

\subsection{From cyclotron to synchrotron emission}

A look at Fig. \ref{ctos} helps to understand
the difference between cyclotron and synchrotron emission.
When the particle is very sub--relativistic, the observed
electric field is sinusoidal in time.
Correspondingly, the Fourier transform of $E(t)$ gives only one
frequency, the first harmonic.
Increasing somewhat the velocity (say, $\beta\sim 0.01$) 
the emission pattern starts to be asymmetric (for light aberration)
and as a consequence $E(t)$ must be described by more than just one sinusoid,
and higher order harmonics appear. 
In these cases the ratio of the power contained in successive harmonics
goes as $\beta^2$.

Finally, for relativistic (i.e. $\gamma \gg 1$) particles, the pattern is 
so asymmetric that the observers sees only spikes of electric field.
They repeat themselves with the gyration period, but all the power is
concentrated into $\Delta t_{\rm A}$.
To reproduce $E(t)$ in this case with sinusoids requires a large number of them,
with frequencies going at least up to $1/\Delta t_{\rm A}$.
In this case the harmonics are many, guaranteeing that the spectrum becomes
continuous with any reasonable line broadening effect, and the power is
concentrated at high frequencies.

\section{Emission from many electrons}

Again, this problem is treated in several text books, so we 
repeat the basic results using some approximations, tricks and shortcuts.

The queen of the particle energy distributions in high energy astrophysics
is the {\it power law} distribution:
\begin{equation}
N(\gamma) \, =\, K \, \gamma^{-p} \, =\, N(E) \, {dE \over d\gamma}; 
\qquad \gamma_{\rm min} \gamma <\gamma_{\rm max}
\label{ng}
\end{equation}
Now, assuming that the distribution of pitch angles is the same at low and high
$\gamma$, we want to obtain the synchrotron emissivity produced by these particles.
Beware that the {\it emissivity} is the power per unit {\it solid angle} produced
within 1 cm$^3$. The specific emissivity is also per unit of frequency.
So, if Eq. \ref{ng} represents a density, we should integrate over $\gamma$ 
the power produced by
the single electron (of a given $\gamma$) times $N(\gamma)$, and
divide all it by $4\pi$, if the emission is isotropic:
\begin{equation}
\epsilon_{\rm s}(\nu,\theta) \, =  \, 
{1\over 4\pi}\, \int_{\gamma_{\rm min}}^{\gamma_{\rm max}} N(\gamma) 
P(\gamma,\nu,\theta) d\gamma
\label{eps1}
\end{equation}
Doing the integral one easily finds that, in an appropriate range of frequencies:
\begin{equation}
\epsilon_{\rm s}(\nu,\theta) \, \propto  \,K B^{(p+1)/2} \nu^{-(p-1)/2}
\label{eps2}
\end{equation}
The important thing is that a power law electron distribution produces a power law
spectrum, and the two spectral indices are related.
We traditionally call $\alpha$ the spectral index of the radiation, namely
$\epsilon_{\rm s} \propto \nu^{-\alpha}$.
We then have
\begin{equation}
\alpha \, =\, { p-1 \over 2}
\label{alpha}
\end{equation}
This result is so important that it is worth to try to derive it in 
a way as simple as possible, even without doing the integral of Eq. \ref{eps1}.
We can in fact use the fact that the synchrotron spectrum emitted by 
the single particle is peaked. 
We can then say, without being badly wrong, that all the power is
emitted at the typical synchrotron frequency:
\begin{equation}
\nu_{\rm s} \, =\, \gamma^2 \nu_{\rm L}; \qquad \nu_{\rm L}\, \equiv \,{e B\over 2\pi m_{\rm e}c} 
\label{nu2}
\end{equation}
In other words, there is a tight correspondence between the energy of the electron 
and the frequency it emits.
To simplify further, let us assume that the pitch angle is $90^\circ$.
The emissivity at a given frequency, within an interval $d\nu$,
is then the result of the emission of electrons having the appropriate energy $\gamma$,
within the interval $d\gamma$
\begin{equation}
\epsilon_{\rm s}(\nu) d\nu \, =\, {1\over 4\pi} \, P_{\rm s} N(\gamma)d\gamma;
\quad \gamma = \left({\nu \over \nu_{\rm L}}\right)^{1/2};
\quad {d\gamma\over d\nu} = {\nu^{-1/2} \over 2 \nu_{\rm L}^{1/2}}
\label{eps3}
\end{equation}
we then have
\begin{eqnarray}
\epsilon_{\rm s}(\nu) \, &\propto& B^2 \gamma^2 K\gamma^{-p} {d\gamma \over d\nu}
\nonumber \\
&\propto& B^2 K \left({\nu \over \nu_{\rm L}}\right)^{(2-p)/2} {\nu^{-1/2} 
\over \nu_{\rm L}^{1/2}}
\nonumber \\
&\propto& K \, B^{(p+1)/2}\,  \nu^{-(p-1)/2}
\label{eps4}
\end{eqnarray}
where we have used $\nu_{\rm L} \propto B$.

The synchrotron flux received from a homogeneous and thin source of 
volume $V\propto R^3$, at a distance $d_{\rm L}$, is 
\begin{eqnarray}
F_{\rm s}(\nu) \, &=&\, 4\pi \epsilon_{\rm s}(\nu) {V\over 4\pi d_{\rm L}^2} 
\nonumber \\
&\propto & \, {R^3 \over d_{\rm L}^2 } \, K B^{1+\alpha} \nu^{-\alpha}
 \nonumber \\
&\propto &   \theta_s^2 RK  B^{1+\alpha} \nu^{-\alpha}
\label{flux}
\end{eqnarray}
where $\theta_s$ is the angular radius of the source (not the pitch angle!).
Observing the source at two different frequencies allows to determine $\alpha$,
hence the slope of the particle energy distribution.
Furthermore, if we know the distance and $R$, the normalization depends
on the particle density and the magnetic field: two unknowns and only one equation.
We need another relation to close the system.
As we will see in the following, this is provided by the self--absorbed flux.

\section{Synchrotron absorption: photons}

All emission processes have their absorption counterpart, 
and the synchrotron emission is no exception.
What makes synchrotron special is really the fact that it is done
by relativistic particles, and they are almost never distributed in energy
as a Maxwellian.
If they were, we could use the well known fact that the ratio between
the emissivity and the absorption coefficient is equal to the
black body (Kirchhoff law) and then we could easily find the absorption
coefficient.
But in the case of a non--thermal particle distribution we cannot do that.
Instead we are obliged to go back to more fundamental relations, the one between the $A$ and $B$
Einstein coefficients relating spontaneous and stimulated emission and ``true" absorption
(by the way, recall that the absorption coefficient is what remains subtracting
stimulated emission from ``true" absorption).
But we once again will use some tricks, in order to be as simple as possible.
These are the steps:

\begin{enumerate}
\item
The first trick is to think to our power law energy distribution
as a superposition of Maxwellians, of different temperatures.
So, we will relate the energy $\gamma m_{\rm e}c^2$ of a given electron
to the energy $kT$ of a Maxwellian.

\item We have already seen that there is a tight relation between the
emitted frequency and $\gamma$.
Since the emission and absorption processes are related, we will 
assume that a particular frequency $\nu$ is preferentially absorbed by those
electrons that can emit it.

\item As a consequence, we can associate our ``fake" temperature
to the frequency:
\begin{equation}
kT \, \sim \, \gamma m_{\rm e} c^2 \sim m_{\rm e} c^2\, 
\left({ \nu \over \nu_{\rm L}}\right)^{1/2}
\label{nuT}
\end{equation}
\item
For an absorbed source the {\it brightness temperature} $T_b$, defined by
\begin{equation}
I(\nu)  \,  \equiv \, 2 k T_b {\nu^2 \over c^2}
\label{Inu0}
\end{equation}
must be equal to the kinetic ``temperature" of the electrons, and so
\begin{eqnarray}
I(\nu)  \,  &\equiv& \, 2 k T {\nu^2 \over c^2} \, \sim \, 
2 m_{\rm e} \, \nu^2\, \left( { \nu \over \nu_{\rm L} } \right)^{1/2} \nonumber \\
&\propto& \, {\nu^{5/2} \over B^{1/2}} 
\label{Inu}
\end{eqnarray}
\end{enumerate}
These are the right dependencies.
Note that the spectrum is $\propto \nu^{5/2}$, not $\nu^2$, and this is the consequence
of having ``different temperatures".
Note also that the particle density disappeared: if you think about it is natural:
the more electrons you have, the more you emit, but the more you absorb.
Finally, even the {\it slope} of the particle distribution is not important,
it controls (up to a factor of order unity) only the normalization of $I(\nu)$
(our ultra--simple derivation cannot account for it, see the Appendix).

The above is valid as long as we can associate a specific $\gamma$ to any $\nu$.
This is not always the case. 
Think for instance to a cut--off distribution, with $\gamma_{\rm min}\gg 1$.
In this case the electrons with $\gamma_{\rm min}$ are the most efficient
emitters and absorbers of all photons with 
$\nu<\nu_{\rm min} \equiv \gamma^2_{\rm min}\nu_{\rm L}$.
So in this case we {\it should not} associate a different temperature
when dealing with different $\nu<\nu_{\rm min}$.
But if do not change $T$, we recover a self--absorbed intensity
$I(\nu)\propto \nu^2$ (i.e. Raleigh--Jeans like).

Now, going from the intensity to the flux, we must integrate $I(\nu)$
over the angular dimension of the source (i.e. $\theta_s$), 
obtaining
\begin{equation}
F(\nu)  \, 
\propto  \, \theta_s^2 {\nu^{5/2} \over B^{1/2}} 
\label{Fnu}
\end{equation}
if we could observe a self--absorbed source, of known angular size, 
we could then derive its magnetic field {\it even without knowing its
distance}.

\subsection{From thick to thin}

%
\begin{figure}
\vskip -0.5 true cm
\includegraphics[height=9cm, width=10cm]{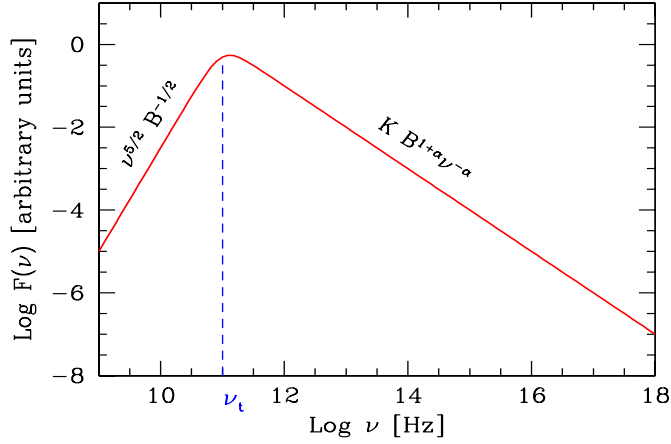} 
\vskip -2 true cm
\caption{The synchrotron spectrum from a partially self absorbed
source. 
Observations of the self absorbed part could determine $B$.
Observations of the thin part can then determine $K$ and the
electron slope $p$.
}
\label{fvsyn}
\end{figure} 
%

To describe the transition from the self absorbed to the thin regime
we have to write the radiation transfer equation.
The easiest one is for a slab.
Calling $\kappa_\nu$ the specific absorption coefficient [cm$^{-1}$]
we have
\begin{equation}
I(\nu)  \, =\, {\epsilon(\nu)\over \kappa_\nu }\, (1-e^{-\tau_\nu}); \quad
\tau_\nu \equiv R\kappa_\nu
\label{transfer}
\end{equation}
it is instructive to write Eq. \ref{transfer} in the form:
\begin{equation}
I(\nu)  \, =\, \epsilon(\nu)R \, \, { 1-e^{-\tau_\nu}  \over \tau_\nu}
\label{transfer2}
\end{equation}
because in this way it is evident that when $\tau_\nu \gg 1$ (self absorbed regime),
we simply have 
\begin{equation}
I(\nu)  \, =\, {\epsilon(\nu) R  \over \tau_\nu} \, =\, {\epsilon(\nu)\over \kappa_\nu };
\quad \tau_\nu \gg 1
\label{ithick}
\end{equation}
One can interpret it saying that the intensity is coming from electrons 
lying in a shell within $R/\tau_\nu$ from the surface.

Since we have already obtained $I(\nu)\propto \nu^{5/2}B^{-1/2}$ in the absorbed regime,
we can derive the dependencies of the absorption coefficient:
\begin{equation}
\kappa_\nu \, =\, {\epsilon(\nu) \over I(\nu)}  \, \propto {K B^{(p+1)/2} \nu^{-(p-1)/1} 
\over \nu^{5/2} B^{-1/2} } \, =\, K B^{(p+2)/2} \nu^{-(p+4)/2}
\label{kappa}
\end{equation}
Note the rather strong dependence upon frequency: at large frequencies,
absorption is small.

The obvious division between the thick and thin regime
is when $\tau_\nu=1$. We call {\it self--absorption frequency}, $\nu_t$, 
the frequency when this occurs.
We then have:
\begin{equation}
\tau_{\nu_t} \, =  \, R \kappa_{\nu_t} \, =\, 1 \, \to \,
\nu_t \, \propto \, \left[ R K \, B^{(p+2)/2} \right] ^{2/(p+4)}
\label{nut}
\end{equation}
The self--absorption frequency is a crucial quantity for studying synchrotron
sources: part of the reason is that it can be thought 
to belong to both regimes (thin and thick), the other reason is that the synchrotron
spectrum peaks very close to $\nu_t$ (see Fig. \ref{fvsyn}) even if not exactly at $\nu_t$
(see the Appendix).

%
\begin{figure}
\vskip -0.8 true cm
\includegraphics[height=11cm, width=12cm]{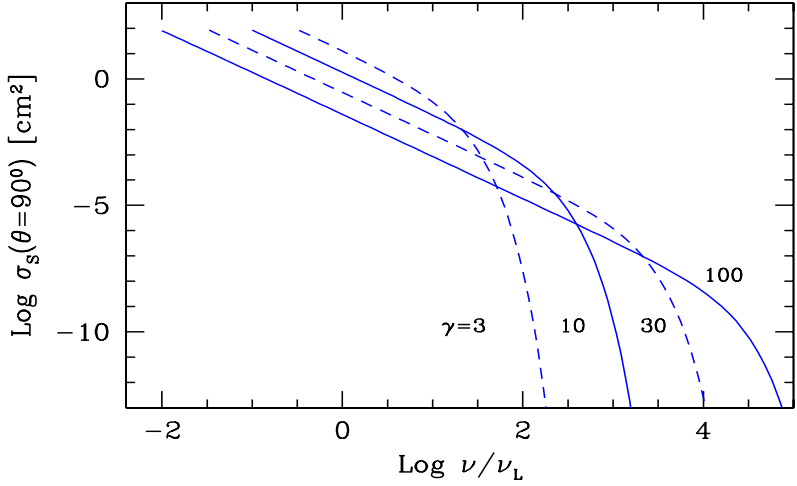} 
\vskip -3.5 true cm
\caption{The synchrotron absorption cross section
as a function of $\nu/\nu_{\rm L}$ for different
values of $\gamma$, as labeled, assuming a pitch angle of $\theta=90^\circ$
and a magnetic field of 1 Gauss.
}
\label{sigmasyn}
\end{figure}{}
%

\section{Synchrotron absorption: electrons}

In the previous section we have considered what happens to the emitted spectrum
when photons are emitted and absorbed.
This is described by the absorption coefficient.
But now imagine to be an electron, that emits and absorbs
synchrotron photons.
You would probably be interested if your budget is positive or negative,
that is, if you are loosing or gaining energy.
This is most efficiently described by a cross section, that tells you the probability
to absorb a photon.
Surprisingly, the synchrotron absorption cross section has been derived 
relatively recently (Ghisellini and Svensson 1991), and its expression is:
\begin{equation}
\sigma_{\rm s}(\nu ,\gamma, \theta) \, =\,
{16\pi^2 \over 3\sqrt{3}} \, {e\over B}\, {1\over \gamma^5 \sin\theta}
K_{5/3} \left({\nu \over \nu_{\rm c}\sin\theta}\right)
\end{equation}
For frequencies $\nu\ll\nu_{\rm c}$ this expression can be 
approximated by:
\begin{equation}
\sigma_{\rm s}(\nu ,\gamma, \theta) \, =\,
{8\pi^2 (3\sin\theta)^{2/3} \Gamma(5/3) \over 3\sqrt{3}} 
\, {e\over B}\, \left({\nu \over \nu_{\rm L}/\gamma}\right)^{-5/3};
\quad \nu\ll \nu_{\rm c}
\end{equation}
Note these features:
\begin{itemize}
\item
At the fundamental frequency $\nu_{\rm L}/\gamma$, 
the cross section does not depend on $\gamma$. 
\item
The dimensions are given by $e/B$: this factor is proportional
to the product of the classical electron radius and the Larmor 
wavelength (or radius).
Imagine an electron with $90^\circ$ pitch angle, and to see its
orbit from the side: you would see a rectangle of base $r_{\rm L}$
and height $r_0$. 
The area of this rectangle is of the order of $e/B$.
At low frequencies, $\sigma_{\rm s}$ can be
orders of magnitudes larger than the Thomson scattering cross section.

\item 
There is no explicit dependence on the particle mass.
However, protons have much smaller $\nu_{\rm L}$, and the 
dependence on mass is hidden there.
Nevertheless, electrons and protons have the same cross section (of
order $e/B$) at their respective fundamental frequencies.
\end{itemize}

Fig. \ref{sigmasyn} shows $\sigma_{\rm s}$ as a function of $\nu/\nu_{\rm L}$
for different $\gamma$. The thing it should be noticed is that this cross 
section is really large.
Can we make some useful use of it?
Well, there are at least two issues, one concerning energy, and the other concerning momentum.

First, electrons emitting and absorbing
synchrotron photons do so with a large efficiency.
{\it They can talk each other by exchanging photons}.
Therefor, even if they are distributed as a power law in energy at the beginning,
they will try to form a Maxwellian.
They will form it, as long as other competing processes are not important,
such as inverse Compton scatterings.
The formation of the Maxwellian will interest only the low energy part of the electron
distribution, where absorption is important.
Note that this thermalization process works exactly when Coulomb collisions fail:
they are inefficient at low density and high temperature, while synchrotron absorption
can work for relativistic electrons even if they are not very dense.

The second issue concerns exchange of momentum between photons and electrons.
Suppose that a magnetized region with relativistic electrons is illuminated by low
frequency radiation by another source, located aside.
The electrons will efficiently absorb this radiation, and thus its momentum.
The magnetized region will then accelerate.

\vskip 0.7 true cm
{\bf References}
\vskip 0.7 true cm

Ghisellini G. \& Svensson R., 1991, MNRAS, 252, 313 

Ghisellini G., Haardt F. \& Svensson R., 1998, MNRAS, 297, 348 

\vskip 3 cm

\section{Appendix: Useful Formulae}

In this section we collect several useful formulae
concerning the synchrotron  emission.
When possible, we give also simplified analytical
expressions.
We will often consider that the emitting electrons
have a distribution in energy which is a power law
between some limits $\gamma_1$ and $\gamma_2$.
Electrons are assumed to be isotropically distributed 
in the comoving frame of the emitting source. Their
density is
\begin{equation}
N(\gamma)\, =\, K\gamma^{-p}; \quad \gamma_1<\gamma<\gamma_2
\end{equation}
The Larmor frequency is defined as:
\begin{equation}
\nu_L\, \equiv\, {e B \over 2\pi m_{\rm e} c}
\end{equation}
%


\subsection{Emissivity}
 
The synchrotron emissivity $\epsilon_{\rm s}(\nu,\theta)$
[erg cm$^{-3}$ s$^{-1}$ sterad$^{-1}$] is
\begin{equation}
\epsilon_{\rm s}(\nu,\theta)\, \equiv \, {1\over 4\pi}\,
\int_{\gamma_1}^{\gamma_2} N(\gamma)
P_{\rm s}(\nu,\gamma,\theta) d\gamma
\end{equation}
where $P_{\rm s}(\nu, \gamma,\theta)$ is the 
power emitted at the frequency $\nu$ (integrated over all directions) 
by the single electron of energy $\gamma m_{\rm e}c^2$ and pitch
angle $\theta$. 
For electrons making the same pitch angle $\theta$ with the magnetic field,
the emissivity is
\begin{equation}
\epsilon_{\rm s}(\nu,\theta)\, =\,
{3\sigma_{\rm T} c K U_B \over 8 \pi^2 \nu_L} \, 
\left({\nu\over \nu_L} \right)^{-{p-1\over 2}} (\sin\theta)^{p+1\over 2}
\, 3^{p\over 2} \,
{ \Gamma\left( {3p-1 \over 12}\right) \Gamma\left( {3p+19\over 12}\right)  
\over p+1}
\end{equation}
between $\nu_1 \gg \gamma_1^2\nu_L$ and $\nu_2\ll \gamma_2^2\nu_L$ .
If the distribution of pitch angles is isotropic, we must
average the $(\sin\theta)^{p+1\over 2}$ term, obtaining
\begin{equation}
< (\sin\theta)^{p+1\over 2}> \, =\, \int_0^{\pi\over 2}
(\sin\theta)^{p+1\over 2} \sin\theta d\theta \, =\,
{\sqrt{\pi} \over 2}\,
{\Gamma\left( {p+5 \over 4}\right) \over
\Gamma\left( {p+7 \over 4}\right)
}
\end{equation}
Therefore the pitch angle averaged synchrotron emissivity is
\begin{equation}
\epsilon_{\rm s}(\nu)\, =\,
{3\sigma_{\rm T} c K U_B \over 16 \pi \sqrt{\pi} \nu_L} \, 
\left({\nu\over \nu_L} \right)^{-{p-1\over 2}} 
\, f_\epsilon(p)
\end{equation}
The function $f_\epsilon(p)$ includes all the products of the $\Gamma$--functions:
\begin{eqnarray}
f_\epsilon(p)\, &=&\, 
{ 3^{p\over 2}\over p+1}\, 
{ \Gamma\left( {3p-1 \over 12}\right) 
\Gamma\left( {3p+19\over 12}\right) 
\Gamma\left( {p+5 \over 4}\right) \over
\Gamma\left( {p+7 \over 4}\right)} 
\nonumber \\
&\sim& \, 3^{p\over 2}\left( {2.25 \over p^{2.2}}+0.105 \right)
\end{eqnarray}
where the simplified fitting function is accurate at the per cent level.

\subsection{Absorption coefficient}

The absorption coefficient $\kappa_\nu (\theta)$ [cm$^{-1}$]
is defined as:
\begin{equation}
\kappa_\nu (\theta) \, \equiv \,
{1\over 8\pi m_{\rm e}\nu^2}\, \int_{\gamma_1}^{\gamma_2}
{N(\gamma)\over \gamma^2} {d\over d\gamma}
\left[ \gamma^2 P(\nu, \theta)\right] d\gamma
\end{equation}
Written in this way, the above formula is valid even when 
the electron distribution is truncated.
For our power law electron distribution $\kappa_\nu (\theta)$
becomes:
\begin{equation}
\kappa_\nu (\theta) \, \equiv \,
{1\over 8\pi m_{\rm e}\nu^2}\, \int_{\gamma_1}^{\gamma_2}
{N(\gamma)\over \gamma^2} {d\over d\gamma}
\left[ \gamma^2 P(\nu, \theta)\right] d\gamma
\end{equation}
Above  $\nu = \gamma_1^2\nu_L $,  we have:
\begin{equation}
\kappa_\nu (\theta) \, = \, 
{e^2 K \over 4 m_{\rm e} c^2} \,
(\nu_L \sin\theta)^{p+2\over 2} \nu^{-{p+4 \over 2}} \, 3^{p+1\over 2} \,
\Gamma\left( {3p+22 \over 12}\right)\,
\Gamma\left( {3p+2 \over 12}\right)
\end{equation}
Averaging over the pitch angles we have:
\begin{equation}
< (\sin\theta)^{p+2\over 2}> \, =\, \int_0^{\pi\over 2}
(\sin\theta)^{p+2\over 2} \sin\theta d\theta \, =\,
{\sqrt{\pi} \over 2}\,
{\Gamma\left( {p+6 \over 4}\right) \over
\Gamma\left( {p+8 \over 4}\right)
}
\end{equation}
resulting in a pitch angle averaged absorption coefficient:
\begin{equation}
\kappa_\nu \, = \, 
{\sqrt{\pi} e^2 K \over 8 m_{\rm e} c} \,
\nu_L^{p+2\over 2} \nu^{-{p+4 \over 2}} \, f_\kappa(p)
\end{equation}
where the function $f_\kappa(p)$ is:
\begin{eqnarray}
f_\kappa(p)\, &=&\, 
3^{p+1\over 2}\, 
{ \Gamma\left( {3p+22\over 12}\right) 
\Gamma\left( {3p+2 \over 12}\right) 
\Gamma\left( {p+6 \over 4}\right)\,\over
\Gamma\left( {p+8 \over 4}\right)}
\nonumber \\
&\sim& \, 3^{p+1\over 2}\left( {1.8 \over p^{0.7}}+{p^2 \over 40} \right)
\end{eqnarray}
The simple fitting function is accurate at the per cent level.

\subsection{Specific intensity}

Simple radiative tranfer allows to calculate the 
specific intensity:
\begin{equation}
I(\nu) \, =\, {\epsilon_{\rm s}(\nu) \over \kappa_\nu }\,
\left( 1-e^{-\tau_\nu} \right)
\end{equation}
where the absorption optical depth $\tau_\nu \equiv \kappa_\nu R$
and $R$ is the size of the emitting region.
When $\tau_\nu \gg 1$, the esponential term vanishes, and the intensity 
is simply the ratio between the emissivity and the absorption coefficient.
This is the self--absorbed, ot thick, regime.
In this case, since both $\epsilon_{\rm s}(\nu)$ and $\kappa_\nu$ 
depends linearly upon $K$, the resulting self--absorbed intensity 
does not depend on the normalization of the particle density $K$:
\begin{equation}
I(\nu) \, =\, {2m_{\rm e} \over \sqrt{3}\, \nu_L^{1/2}}\,
f_I(p)\,  \left( 1-e^{-\tau_\nu} \right)
\end{equation}
we can thus see that the slope of the self--absorbed intensity
does not depend on $p$.
Its normalization, however, does (albeit weakly) depend on $p$
through the function $f_I(p)$, which in the case of averaging
over an isotropic pitch angle distribution is given by:
\begin{eqnarray}
f_I(p)\, =\, {1\over p+1} \,&=& \,
{ \Gamma\left( {3p-1 \over 12}\right) 
\Gamma\left( {3p+19\over 12}\right) 
\Gamma\left( {p+5 \over 4}\right) 
\Gamma\left( {p+8 \over 4}\right) 
\over
\Gamma\left( {3p+22\over 12}\right) 
\Gamma\left( {3p+2\over 12}\right) 
\Gamma\left( {p+7 \over 4}\right)\,
\Gamma\left( {p+6 \over 4}\right)}\, 
\nonumber \\
&\sim& \,
{5\over 4 \, p^{4/3}}
\end{eqnarray}
where again the simple fitting function is accurate at the level of 1 per cent.

\subsection{Self--absorption frequency}

The self--absorption frequency $\nu_{\rm t}$ is defined
by $\tau_{\nu_{\rm t}}=1$:
\begin{equation}
\nu_{\rm t}\, =\, \nu_L\,
\left[ {\sqrt{\pi} e^2 R K \over 8 m_{\rm e} c\nu_L} \,
f_\kappa(p)\right]^{4\over p+4} \, =\,
\nu_L\, \left[ {\pi \sqrt{\pi}\over 4}\, {e R K  \over B}\,
f_\kappa(p)\right]^{2\over p+4} 
\end{equation}
Note that the term in parenthesis is adimensional,
and since $RK$ has units of the inverse of a surface,
then $e/B$ has the dimension of a surface.
In fact we have already discussed that this is the synchrotron absorption 
cross section of a relativistic electron of energy $\gamma m_{\rm e} c^2$  
absorbing photons at the fundamental frequency $\nu_L/\gamma$.

The random Lorentz factor $\gamma_{\rm t}$ of the
electrons absorbing (and emitting) photons with
frequency $\nu_{\rm t}$ is
$\gamma_{\rm t} \sim [3\nu_{\rm t}/ (4\nu_L)]^{1/2}$.

\subsection{Synchrotron peak}

In a $F(\nu)$ plot, the synchrotron spectrum
peaks close to $\nu_{\rm t}$, at a frequency
$\nu_{s,p}$ given by solving
\begin{equation}
{d I(\nu) \over d\nu} \, =\, 0 \, \to \,{d   \over d\nu}
\left[ \nu^{5/2} \left( 1-e^{-\tau_\nu} \right)\right] \, =\, 0
\end{equation}
which is equivalent to solve the equation:
\begin{equation}
\exp \left( \tau_{\nu_{s,p}} \right) -{ p+4 \over 5} \tau_{\nu_{s,p}} -1
\, =\, 0
\end{equation}
whose solution can be approximated by
\begin{equation}
\tau_{\nu_{s,p}} \, \sim \, {2\over 5} p^{1/3}\ln{p}
\end{equation}
%


\chapter{Compton scattering}

\section{Introduction}
\noindent
The simplest interaction between photons and free electrons
is {\it scattering}.
When the energy of the incoming photons (as seen in the comoving 
frame of the electron) is small with respect to the electron rest
mass--energy, the process is called {\it Thomson scattering},
which can be described in terms of classical electro--dynamics.
As the energy of the incoming photons increases and becomes
comparable or greater than $m_ec^2$, a quantum treatment is
necessary (Klein--Nishina regime).

\section{The Thomson cross section}

Assume an electron at rest, and an electromagnetic wave of frequency
$\nu\ll m_ec^2/h$.
Assume also that the incoming wave is completely linearly polarized.
In order to neglect the magnetic force $(e/c)({\mathbf v  \times  \mathbf B} )$
we must also require that the oscillation velocity $v\ll c$.
This in turn implies that the incoming wave has a sufficiently low amplitude.
The electron starts to oscillate in response to the varying electric force $eE$,
and the average square acceleration during one cycle of duration 
$T=1/\nu$ is 
\begin{equation}
\langle a^2\rangle \,  =\, {1\over T} \int_0^T {e^2E_0^2 \over m_e^2}
\sin^2(2\pi\nu t)\, dt \, =\, 
{e^2 E_0^2 \over 2 m_e^2} 
\end{equation}
The emitted power per unit solid angle is given by the Larmor 
formula $dP/d\Omega = e^2a^2\sin^2\Theta/(4\pi c^3)$ where
$\Theta$ is the angle between the acceleration vector and the
propagation vector of the emitted radiation.
Please note that $\Theta$ is {\it not} the scattering angle,
which is instead the angle between the incoming and the scattered wave
(or photon).
We then have
\begin{equation}
{dP_e \over d\Omega}\, =\, 
{e^4 E_0^2 \over 8\pi m_e^2 c^3}\, \sin^2\Theta
\end{equation}
The scattered radiation is completely linearly polarized
in the plane defined by the incident polarization vector
and the scattering direction.
The flux of the incoming wave is $S_i= cE_0^2/(8\pi)$.
The differential cross section of the process is then
\begin{equation}
\left({ d\sigma \over d\Omega}\right)_{\rm pol}\, =\, 
{dP_e / d\Omega \over S_i}\, =\, r_0^2\, \sin^2\Theta
\end{equation}
where $r_0\equiv e^2/( m_e c^2)$ is the {\it classic electron radius},
$r_0= 2.82\times 10^{-13}$ cm.
We see that the scattered pattern of a completely polarized 
incoming wave is a torus, with axis along the acceleration direction.
The total cross section can be derived in a similar way, but considering
the Larmor formula integrated over the solid angle [$P=2e^2a^2/(3c^3)$].
In this way the total cross section is
\begin{equation}
\sigma_{\rm pol}\, =\, 
{P_e \over S_i}\, =\, {8 \pi \over 3} r_0^2
\label{spol}
\end{equation}
Note that the classical electron radius can also be derived by
equating the energy of the associated electric field
to the electron rest mass--energy:
\begin{equation}
m_{\rm e}c^2 \, =\, \int_{a_o}^\infty {E^2 \over 8 \pi} 4\pi r^2 dr
\, =\,  \int_{a_o}^\infty {e^2 \over 2 r^2}   dr
\, \to \, a_0 \, =\, {1\over 2}{e^2\over m_{\rm e} c^2}
\end{equation}
Why is $a_0$ slightly different from $r_0$?
Because there is an intrinsic uncertainty related to the 
distribution of the charge within (or throughout the surface of)
the electron. 
See the discussion in Vol. 2, chapter 28.3 of ``The Feynman Lectures on Physics'',
about the fascinating idea that the mass of the electron 
is all electromagnetic.

%
\begin{figure}
\begin{tabular}{cc}
\includegraphics[height=7cm, width=7cm]{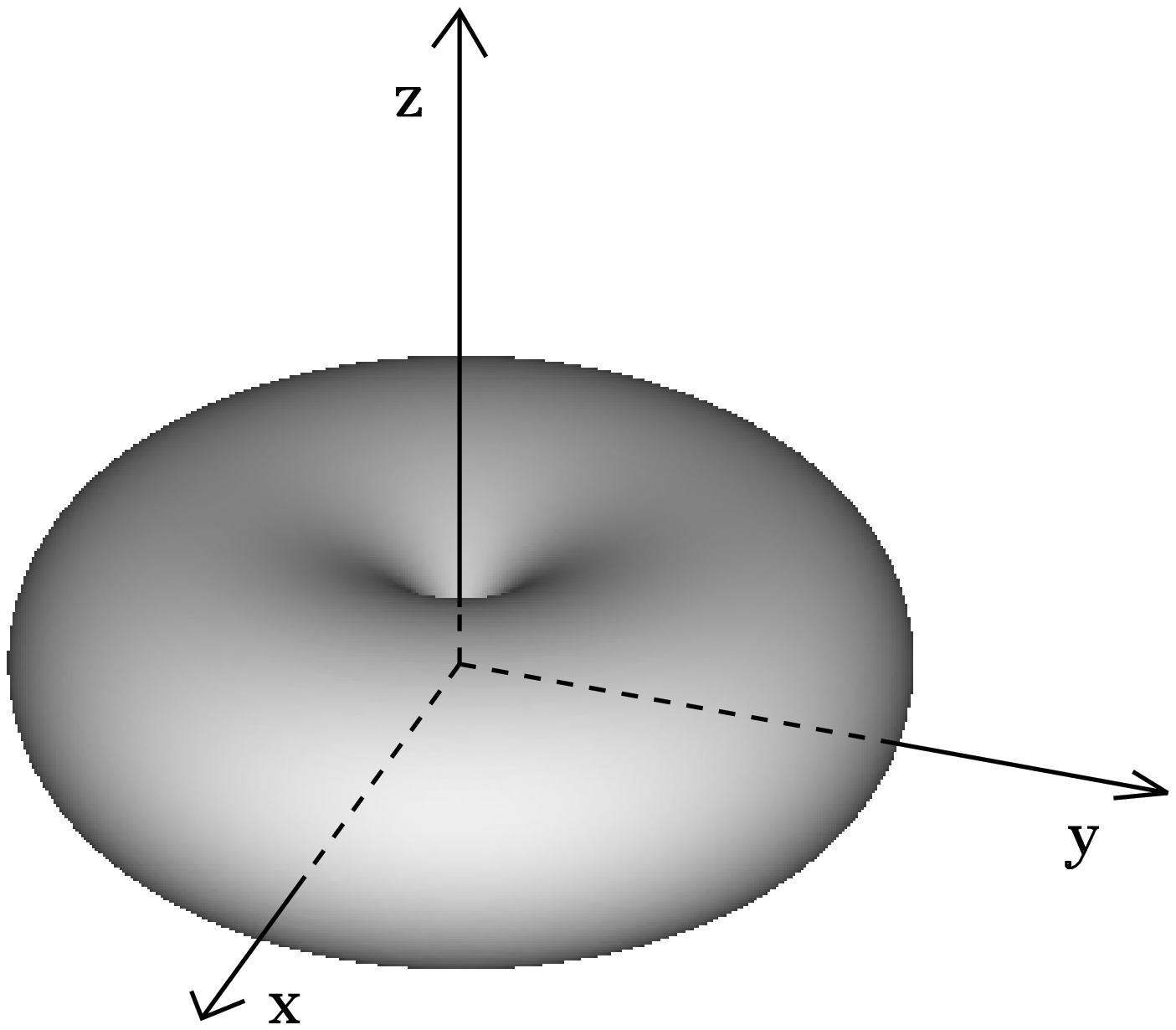}
&\includegraphics[height=7cm, width=7cm]{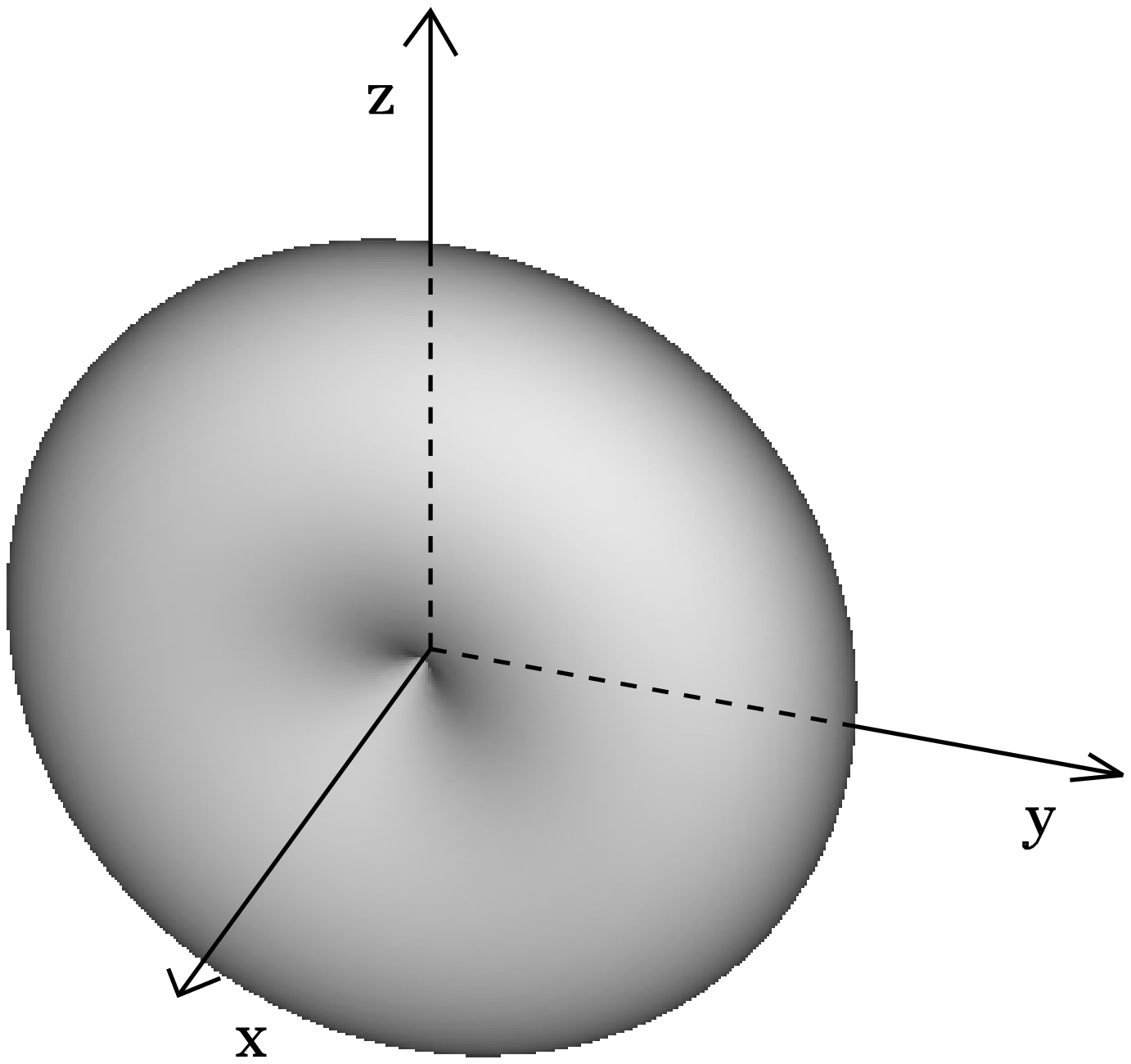} 
\end{tabular} 
\vskip -1.  true cm
\hskip 2 true cm
\includegraphics[height=11cm, width=11cm]{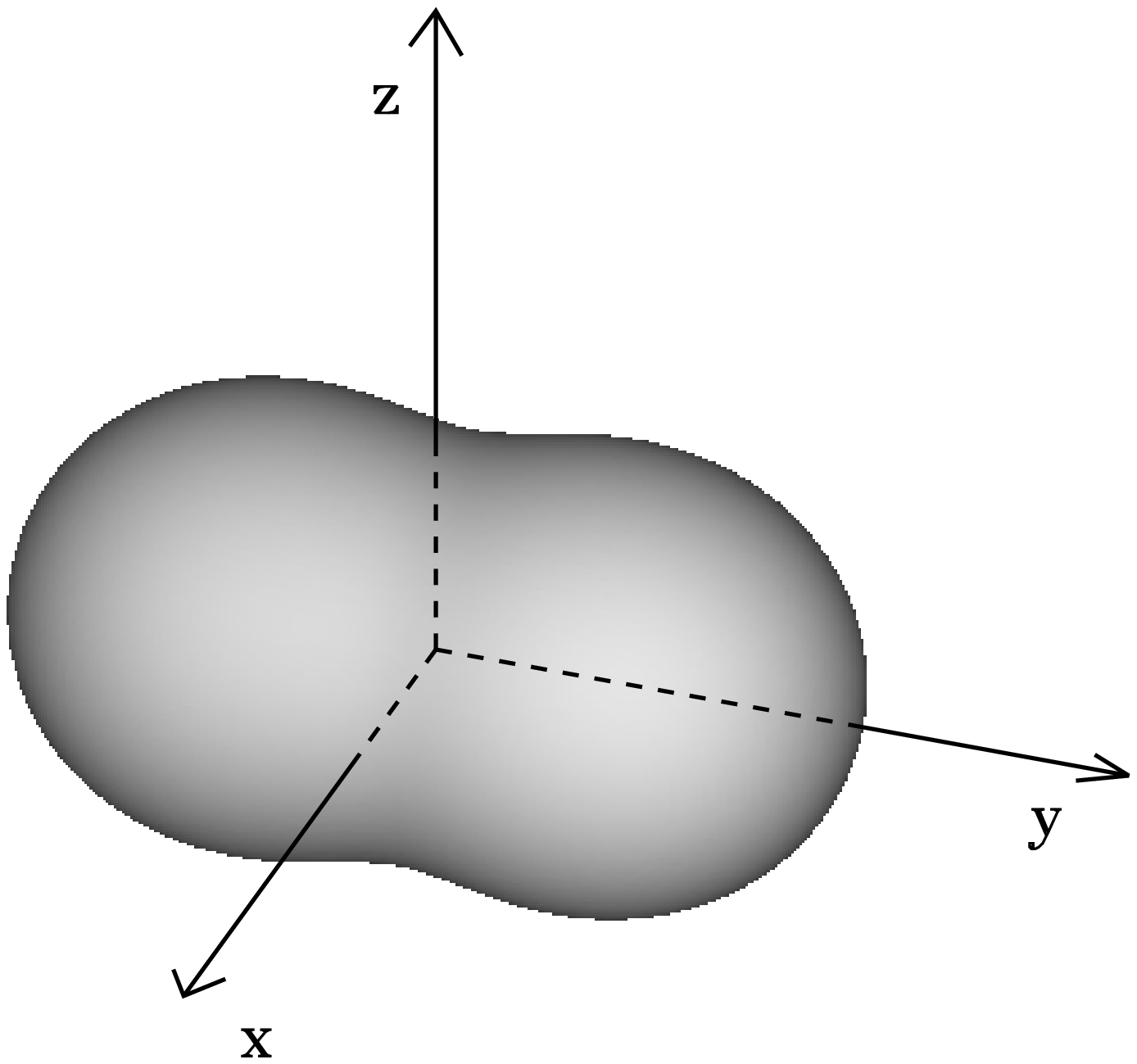}
\caption{Photons are coming along the $y$--axis.
The top panels shows the pattern of the scattered radiation for photons completely 
linearly polarized along the $z$--axis (left) and along the $x$--axis (right).
The sum of the two torii corresponds to the pattern for unpolarized
radiation (bottom panel). This explains why we have a ``peanut" shape, elongated along
the velocity vector of the incoming photons. Courtesy of Davide Lazzati.
}
\label{arachide}
\end{figure}
%

\subsection{Why the peanut shape?}

The scattering of a completely unpolarized incoming wave
can be derived by assuming that the incoming radiation 
is the sum of two orthogonal completely linearly polarized waves,
and then summing the associated scattering patterns.
Since we have the freedom to chose the orientations of the two
polarization planes, it is convenient to chose one of these planes 
as the one defined by the incident and scattered directions, 
and the other one perpendicular to this plane. 
The scattering can be then regarded as the sum of two independent scattering 
processes, one with emission angle $\Theta$, the other with $\pi/2$. 
If we note that the scattering angle (i.e. the angle between the scattered 
wave and the incident wave) is $\theta=\pi/2-\Theta$, we have
\begin{eqnarray}
\left( { d\sigma \over d\Omega}\right)_{\rm unpol}\, &=&\,
{1\over 2}\, \left[ \left( {d\sigma(\Theta) \over d\Omega}\right)_{\rm pol}
+ \left({d\sigma(\pi/2)\over d\Omega}\right)_{\rm pol}\right]\,
\nonumber \\
\, &=& {1\over 2}\, r_0^2(1+\sin^2\Theta)\, \nonumber \\
\, &=& {1\over 2}\, r_0^2(1+\cos^2\theta)
\label{pol1}
\end{eqnarray}
In this case we see that the cross section depends only
on the scattering angle $\theta$.
The pattern of the scattered radiation is then the superposition of two
orthogonal ``tori" (one for each polarization direction),
as illustrated in Fig. \ref{arachide}.
When scattering completely linearly polarized radiation, only
one ``torus" survives.
Instead, when scattering unpolarized radiation, some polarization is 
introduced, because of the difference between the two ``tori" patterns.
Both terms of the RHS of Eq. \ref{pol1} refer to completely polarized 
scattered waves (but in two perpendicular planes).
The difference between these two terms is then associated to the
introduced polarization, which is then
\begin{equation}
\Pi \, =\, {1-\cos^2\theta \over 1+\cos^2\theta}
\end{equation}
The above discussion helps to understand why the scattering process
introduces some polarization, which is maximum (100\%) if the angle
between the incoming and the scattered photons is 90$^\circ$ 
(only one torus contributes),
and zero for 0$^\circ$ and 180$^\circ$, where the two torii 
give the same contribution.

The total cross section, integrated over the solid angle,
is the same as that for polarized incident radiation (Eq. \ref{spol})
since the electron at rest has no preferred defined direction.
This is the Thomson cross section:
\begin{eqnarray}
\sigma_{\rm T} &=& \int\left( { d\sigma \over d\Omega}\right)_{\rm unpol} d\Omega 
 = {2\pi r_0^2\over 2}\, \int (1+\cos^2\theta) d\cos\theta \,
 = {8 \pi \over 3} r_0^2 \nonumber \\
 &=& 6.65\times 10^{-25}\, {\rm cm^2}
\end{eqnarray}

\section{Direct Compton scattering}

In the previous section we considered the scattering process as
an interaction between an electron and an {\it electromagnetic wave}.
This required $h\nu \ll m_ec^2$. In the general case the quantum nature 
of the radiation must be taken into account. 
We consider then the scattering process as a collision between
the electron and the {\it photon}, and apply the conservation of energy and 
momentum to derive the energy of the scattered photon.
It is convenient to measure energies in units of $m_ec^2$
and momenta in units of $m_ec$.

Consider an electron at rest and an incoming photon of energy $x_0$,
which becomes $x_1$ after scattering.
Let $\theta$ be the angle between the incoming and outgoing
photon directions.
This defines the scattering plane.
Momentum conservation dictates that also the momentum vector of the
electron, after the scattering, lies in the same plane.
Conservation of energy and conservation of momentum 
along the x and y axis gives:
%
%
\begin{equation}
x_1 \, =\, { x_0 \over 1+x_0(1-\cos\theta)}
\label{x1}
\end{equation}
Note that, for $x_0\gg 1$ and $\cos\theta \ne 1$, 
$x_1 \to (1-\cos\theta)^{-1}$.
In this case the scattered photon carries information about the scattering 
angle, rather than about the initial energy.
As an example, for $\theta=\pi$ and $x_0\gg 1$, the final energy
is $x_1=0.5$ (corresponding to $255$ keV) independently of
the exact value of the initial photon energy. 
Note that for $x_0\ll 1$ the scattered energy $x_1\simeq x_0$,
as assumed in the classical Thomson scattering. 
The energy shift implied by Eq. \ref{x1} is due to the recoil of the 
electron originally at rest, and becomes significant only when
$x_0$ becomes comparable with 1 (or more). 
When the energy of the incoming photon is comparable to the electron rest 
mass, another quantum effect  
appears, namely the energy dependence of the cross section.

\section{The Klein--Nishina cross section}

The Thomson cross section is the classical limit of the more general 
Klein--Nishina cross section (here we use $x$ as the initial photon energy,
instead of $x_0$, for simplicity):
\begin{equation}
{d\sigma_{\rm KN} \over d\Omega} \, =\, 
{3\over 16\pi}\, \sigma_{\rm T} \left({x_1\over x}\right)^2
\left( {x\over x_1}+{x_1\over x} -\sin^2\theta\right)
\end{equation}
This is a compact form, but there appears dependent quantities,
as $\sin\theta$ is related to $x$ and $x_1$.
By inserting Eq. \ref{x1}, we arrive to
\begin{equation}
{d\sigma_{\rm KN} \over d\Omega} \, =\, 
{3\over 16\pi}\, {\sigma_{\rm T}  \over [1+x(1-\cos\theta)]^2}
\left[ x(1-\cos\theta) +{1\over 1+x(1-\cos\theta)} 
+\cos^2\theta\right]
\label{dskndomega}
\end{equation}
In this form, only independent quantities appear (i.e. there is no $x_1$). 
Note that the cross section becomes smaller for increasing $x$ and that 
it coincides with $d\sigma_{\rm T} /d\Omega$ for $\theta=0$
(for this angle $x_1=x$ independently of $x$). 
This however corresponds to a vanishingly small number of
interactions, since $d\Omega\to 0$ for $\theta\to 0$).

Integrating Eq. \ref{dskndomega} over the solid angle, we obtain the 
total Klein--Nishina cross section:
\begin{equation}
\sigma_{\rm KN} \, =\, {3 \over 4}\, \sigma_{\rm T}\, 
\left\{ {1+x \over x^3} \left[ {2x(1+x) \over 1+2x} - \ln(1+2x) \right]
+{1\over 2x} \ln(1+2x) - {1+3x \over (1+2x)^2} \right\}
\end{equation}

Asymptotic limits are:
\begin{eqnarray}
\sigma_{\rm KN} \, &\simeq &\,  \sigma_{\rm T}\, 
\left( 1-2x+{26x^2 \over 5} +...\right); \quad x\ll 1 \nonumber \\
\sigma_{\rm KN} \, &\simeq &\, {3\over 8}\, { \sigma_{\rm T}\over x} \, 
\left[\ln(2x) + {1\over 2}\right]; \qquad \quad\quad x\gg 1
\label{knapprox}
\end{eqnarray}
The direct Compton process implies a transfer of energy from the photons
to the electrons.
It can then be thought as an heating mechanism.
In the next subsection we discuss the opposite process, called
{\it inverse Compton scattering}, in which hot electrons can transfer
energy to low frequency photons.

%
\begin{figure}
\includegraphics[height=13cm, width=13cm]{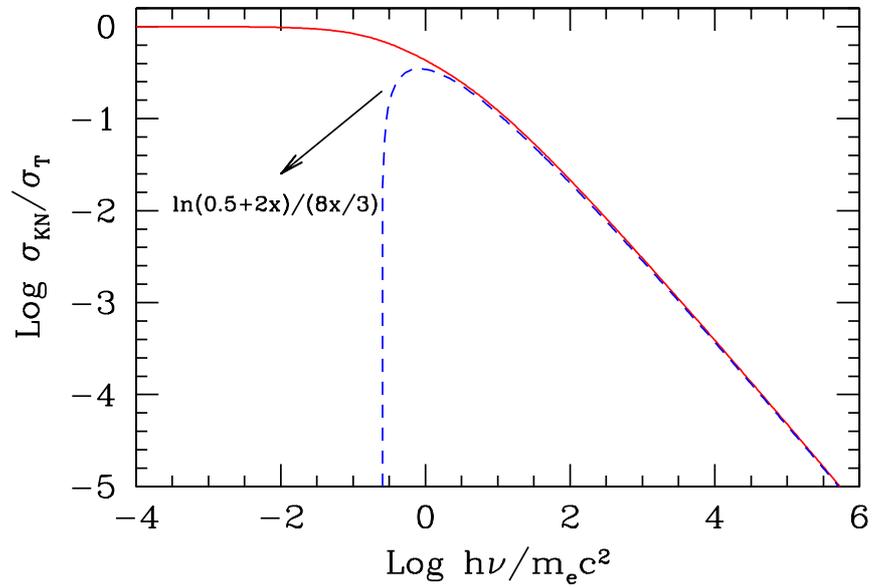} 
\vskip -4 true cm
\caption{The total Klein--Nishina cross section as
a function of energy. The dashed line is the approximation
at high energies as given in Eq. \ref{knapprox}.
}
\label{kn}
\end{figure}
%
%
\begin{figure}
\vskip -0.5 true cm
\includegraphics[height=9cm, width=11cm]{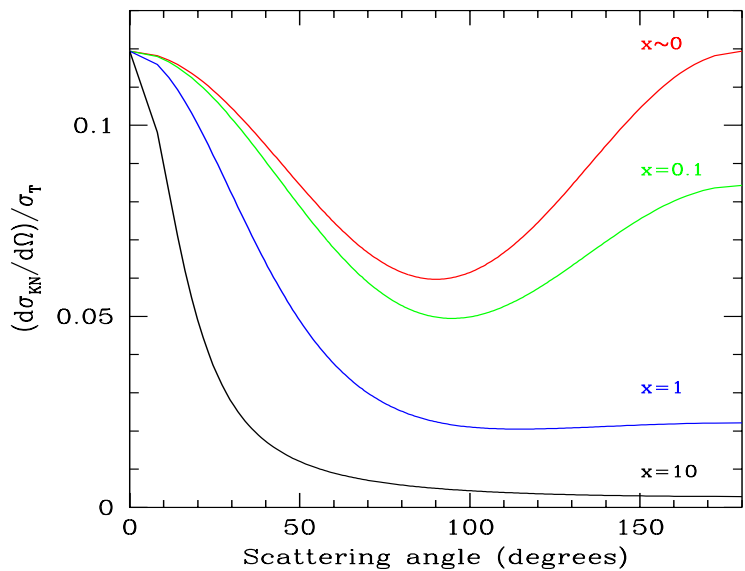}
\vskip -1 true cm
\caption{The differential Klein--Nishina cross section
(in units of $\sigma_{\rm T}$), for different
incoming photon energies.
Note how the scattering becomes preferentially forward as the energy
of the photon increases.}
\end{figure}
%
%
\begin{figure}
\vskip -0.5 true cm
\includegraphics[height=9cm, width=11cm]{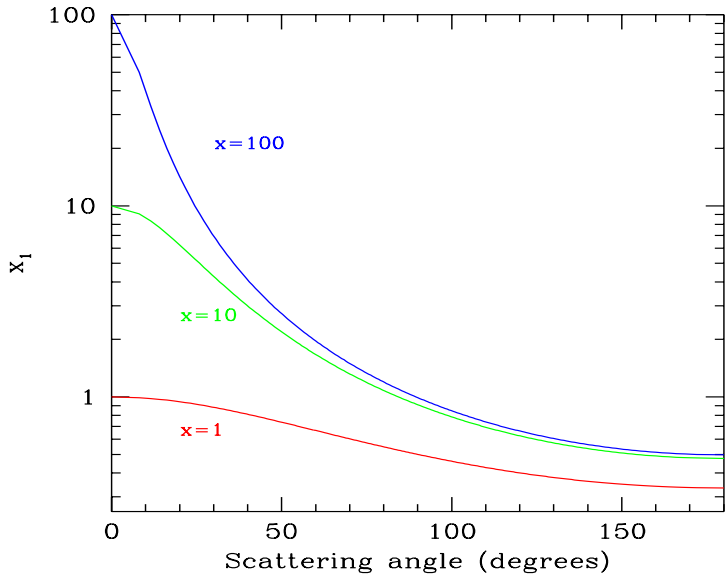}
\vskip -1 true cm
\caption{Scattered photons energies as a function of the 
scattering angle, for different incoming photon energies.
Note that, for $x\gg 1$ and for large scattering angle,
the scattered photon energies becomes $x_1\sim 1/2$, independent of
the initial photon energy $x$. }
\end{figure}
%

%
%

We have so far neglected the momentum exchange between radiation and the 
electron.
One can see, even classically, that there must be a net force acting along
the direction of the wave if one considers the action of the magnetic field
of the wave.
In fact the Lorentz force $e {\mathbf v \times {\mathbf B}}$ is {\it always} directed
along the direction of the wave (here ${\mathbf v}$ is the velocity
along the ${\mathbf E}$ field).
This explains the fact that light can exert a pressure, even classically.

\subsection{Another limit}

We have mentioned that, in order for the magnetic Lorentz force to
be negligible, the electron must have a transverse (perpendicular
to the incoming wave direction) velocity $\ll c$.
Considering a wave of frequency $\omega$ and electric field
$E=E_0 \sin(\omega t)$, this implies that: 
\begin{equation}
{v_{\perp} \over c} \, =\, \int_0^{T/2} {eE_0 \over  c m_e} \sin(\omega t) dt 
\, =\, {2 eE_0 \over m_e c \omega}
\, \ll\, 1
\end{equation}
This means that the scattering process can be described by the Thomson
cross section if the wave have a sufficiently low amplitude and
a not too small frequency (i.e. for very small frequencies the electric
field of the wave accelerates the electron for a long time, 
and then to large velocities).

\subsection{Pause}

Now pause, and ask if there are some ways to apply what we have done
up to now to real astrophysical objects.
\begin{itemize}
\item
The Eddington luminosity is derived with the Thomson cross section,
with the thought that it describes the smallest probability of
interaction between matter and radiation.
But the Klein--Nishina cross section can be even smaller,
as long as the source of radiation emits at high energy.
What are the consequences? If you have forgotten the definition
of the Eddington luminosity, here it is:
\begin{equation}
L_{\rm Edd} \, =\, 
{4 \pi G M m_{\rm p}c \over \sigma_{\rm T} } \, =\, 1.3\times 
10^{38} \, {M \over M_\odot}\, \, \, 
{\rm erg~s^{-1}}
\end{equation}

\item In Nova Muscae, some years ago a (transient) annihilation line was detected,
together with another feature (line--like) at 200 keV.
What can this feature be?

\item
It seems that high energy radiation can suffer less scattering and therefore
can propagate more freely through the universe.
Is that true?
Can you think to other processes that can kill high energy photons
in space?

\item 
Suppose to have an astrophysical source
of radiation very powerful above say -- 100 MeV.
Assume that at some distance there is a very efficient ``reflector'' 
(i.e. free electrons) and that you can see the scattered radiation.
Can you guess the spectrum you receive?
Does it contain some sort of ``pile--up'' or not?
Will this depend upon the scattering angle?

\end{itemize}

\vskip 0.5 true cm

\section{Inverse Compton scattering}

When the electron is not at rest, but has an energy
greater that the typical photon energy, there can be a transfer
of energy from the electron to the photon.
This process is called {\it inverse} Compton to distinguish 
it from the {\it direct} Compton scattering, in which
the electron is at rest, and it is the photon to give part
of its energy to the electron.

We have two regimes, that are called the {\it Thomson} and
the {\it Klein--Nishina} regimes.
The difference between them is the following:
we go in the frame where the electron is at rest, and in that
frame we calculate the energy of the incoming photon.
If the latter is smaller than $m_e c^2$ we are in the
Thomson regime. In this case the recoil of the electron,
even if it always exists, is small, and can be neglected.
In the opposite case (photon energies larger than $m_e c^2$),
we are in the Klein--Nishina one, and we cannot neglect the recoil.
As we shall see, in both regimes the typical photon {\it gain} energy,
even if there will always be some arrangements of angles for which
the scattered photon looses part of its energy.

\subsection{Thomson regime}

Perhaps, a better name should be ``inverse Thomson" scattering, as
will appear clear shortly.
\subsection{Typical frequencies}

\begin{figure}[h]
\vskip -0.3 true cm
\includegraphics[height=9cm, width=13.3cm]{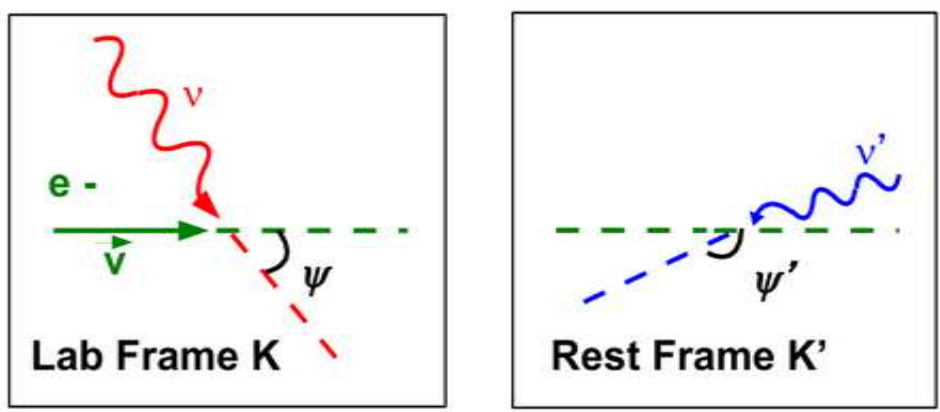}
\vskip -0.5 true cm
\caption{In the lab frame an electron is moving with velocity ${\mathbf v}$.
Its velocity makes an angle $\psi$ with an incoming photon
of frequency $\nu$.
In the frame where the electron is at rest, the photon is coming
from the front, with frequency $\nu^\prime$, 
making an angle $\psi^\prime$ with the direction of the velocity.
}
\label{scat1}
\end{figure}

In the frame $K^\prime$ comoving with the electron, the incoming photon 
energy is
\begin{equation}
x^\prime \, =\, x\gamma (1-\beta\cos\psi)
\label{xxprime}
\end{equation}
where $\psi$ is the angle between the electron velocity and the
photon direction (see Fig \ref{scat1}). 

At first sight this is different from $x^\prime= x\delta$ derived in Chapter 3.
But notice that i) in this case the angle $\psi$ is measured in the lab frame;
ii) it is not the same angle going into the definition of $\delta$ (i.e. in $\delta$
we use the angle between the line of sight and the velocity of the emitter, 
i.e. $\theta^\prime =\pi -\psi^\prime$).
Going to the rest frame of the electrons we should use
(recalling Eq. 3.16 for the transformation of angles): 
\begin{equation}
\cos\psi\, = \, { \beta + \cos\psi^\prime \over 1+\beta\cos\psi^\prime}
\end{equation}
Substituting this into equation \ref{xxprime} we have
\begin{equation}
x^\prime \, =\, {x \over \gamma (1+\beta\cos\psi^\prime)}
\label{xxprime2}
\end{equation}
Finally, consider that $\cos\theta^\prime= \cos(\pi -\psi^\prime)=-\cos\psi^\prime$,
validating $x^\prime = x\delta$.

If $x^\prime \ll 1$, we are in the Thomson regime.
In the rest frame of the electron the scattered photon will have the same
energy $x^\prime_1$ as before the scattering, independent of the scattering angle.  
Then 
\begin{equation}
x_1^\prime=x^\prime
\end{equation}

This photon will be scattered at an angle $\psi^\prime_1$ with respect to 
the electron velocity. The pattern of the scattered radiation will follow the pattern of 
the cross section (i.e. a peanut).
{\it Think to the scattering in the comoving frame as a re-isotropization process:
even if the incoming photons are all coming from the same direction, after
the scattering they are distributed quasi--isotropically}.
Going back to $K$ the observer sees
\begin{equation}
x_1 \, =\, x^\prime_1 \gamma (1+\beta\cos\psi^\prime_1)
\end{equation}
Recalling again Eq. 3.16, for the transformation of angles: 
\begin{equation}
\cos\psi^\prime_1\, = \, { \cos\psi_1 -\beta \over 1-\beta\cos\psi_1}
\end{equation}
we arrive to the final formula:
\begin{equation}
x_1 \, =\, x \, { 1-\beta\cos\psi \over 1-\beta\cos\psi_1}
\label{x1x}
\end{equation}
Now all quantities are calculated in the lab--frame.

%
\begin{figure}[h]
\vskip 0.3  true cm
\includegraphics[height=7.7cm, width=13.2cm]{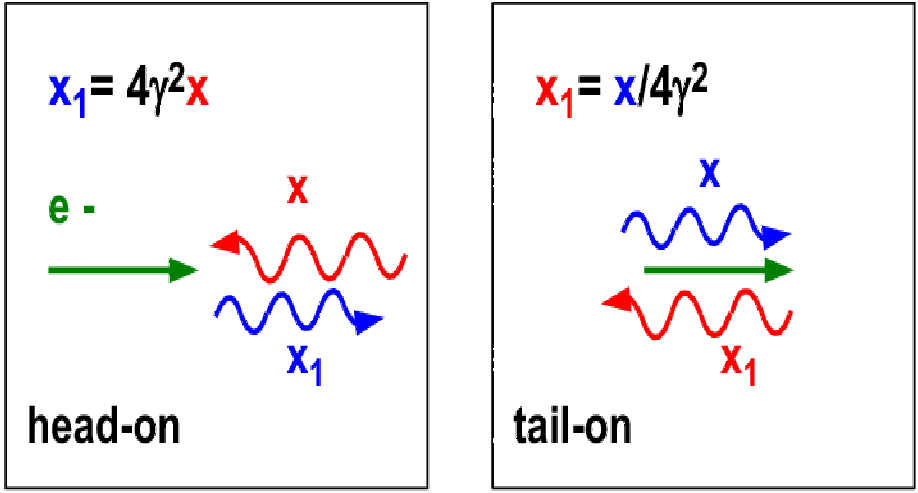}
\vskip -0.5 true cm
\caption{Maximum and minimum scattered frequencies.
The maximum occurs for head-on collisions, the minimum
for tail--on ones. These two frequencies are one the inverse 
of the other.
}
\label{x1x}
\end{figure}
%

Let us see the minimum and maximum energies.
The maximum is when $\psi=\pi$ (head on collision), and when 
$\psi_1=0$ (the photon is scattered along the electron velocity vector).
In these head--on collisions:
\begin{equation}
x_1 \, =\, x \, { 1+\beta \over 1-\beta} \, =\, \gamma^2 (1+\beta)^2 x
\, \to \, 4\gamma^2x; \quad {\rm head-on}
\end{equation}
where the last step is valid if $\gamma\gg 1$.
The other extreme is for $\psi_1=\pi$ and $\psi=0$.
In this case the incoming photon ``comes from behind'' and
bounces back. In these ``tail--on'' collisions:
\begin{equation}
x_1 \, =\, x \, { 1-\beta \over 1+\beta} \, =\, {x \over \gamma^2 (1+\beta)^2 }
\, \to \, { x\over 4\gamma^2}; \quad {\rm tail-on}
\end{equation}
where again the last step is valid if $\gamma\gg 1$.
Another typical angle is $\sin\psi_1=1/\gamma$, corresponding to $\cos\psi_1=\beta$.
This corresponds to the aperture angle of the beaming cone.
For this angle:
\begin{equation}
x_1 \, =\,  \, {1-\beta\cos\psi \over 1-\beta^2} x \, =\, 
\gamma^2(1-\beta\cos\psi) x ;
\,  \quad {\rm beaming~cone}
\end{equation}
which becomes $x_1=x/(1+\beta)$ for $\psi=0$, $x_1=\gamma^2 x$ for $\psi=\pi/2$ and
 $x_1=\gamma^2 (1+\beta)x$ for $\psi=\pi$.

For an isotropic distribution of incident photons and for $\gamma\gg 1$
the average photon energy after scattering is (see Eq. \ref{averx1}):
\begin{equation}
\langle x_1 \rangle \, =\,  {4\over 3}\gamma^2 x 
\end{equation}

\subsubsection{Total loss rate}

We can simply calculate the rate of scatterings per electron considering
all  quantities in the lab--frame.
Let $n(\epsilon)$ be the density of photons of energy $\epsilon=h\nu$,
$v$ the electron velocity and $\psi$ the angle between the electron
velocity and the incoming photon.
For mono--directional photon distributions, we have:
\begin{equation}
{dN \over dt} \, =\, \int \sigma_{\rm T} v_{\rm rel} n(\epsilon)d\epsilon
\end{equation}
$v_{\rm rel}=c-v\cos\psi$ is the relative velocity between 
the electron and the incoming photons.
We then have
\begin{equation}
{dN \over dt} \, =\, \int  \sigma_{\rm T} c  (1-\beta\cos\psi)
n(\epsilon)d\epsilon
\end{equation}
Note that the rate of scatterings in the lab frame, when the electron and/or
photon are anisotropically distributed, 
can be described by an effective cross section 
$\sigma_{\rm eff}\equiv 
\int \sigma_{\rm T} (1-\beta\cos\psi) d\Omega /4\pi$.
For photons and electrons moving in the same direction the scattering 
rate (hence, the effective optical depth) can be greatly reduced.

The power contained in the scattered radiation is then
\begin{equation}
{dE_\gamma \over dt}\, =\,  {\epsilon_1 dN \over dt}\, =\, 
\sigma_{\rm T} c\int   {(1-\beta\cos\psi)^2 \over
1-\beta\cos\psi_1} \epsilon n(\epsilon)d\epsilon
\end{equation}
Independently of the incoming photon angular distribution,
the average value of $1-\beta\cos\psi_1$ can be calculated
recalling that, in the rest frame of the electron, the scattering
has a backward--forward symmetry, and therefore 
$\langle\cos\psi_1^\prime\rangle=\pi/2$.
The average value of $\cos\psi_1$ is then $\beta$, leading to
$\langle 1-\beta\cos\psi_1\rangle=1/\gamma^2$.
We therefore obtain
\begin{equation}
{dE_\gamma \over dt}\, =\,  
\sigma_{\rm T} c \gamma^2 \int (1-\beta\cos\psi)^2 
\epsilon n(\epsilon)d\epsilon
\label{aniso}
\end{equation}
If the incoming photons are isotropically distributed,
we can average out $(1-\beta\cos\psi)^2$ over the solid angle, 
obtaining $1+\beta^2/3$.
The power produced is then
\begin{equation}
{dE_\gamma \over dt}\, =\,  
\sigma_{\rm T} c \gamma^2 \left( 1+{\beta^2 \over 3}\right) U_{\rm r}
\end{equation}
where 
\begin{equation}
U_{\rm r} \, =\,  \int \epsilon n(\epsilon) d\epsilon
\end{equation}
is the energy density of the radiation before scattering.
This is the power contained in the scattered radiation. 
To calculate the energy loss rate of the electron, we have to subtract 
the initial power of the radiation eventually scattered
\begin{equation}
P_{\rm c}(\gamma) \, \equiv \, 
{dE_e \over dt}\, =\,  {dE_\gamma \over dt}-\sigma_{\rm T} c U_{\rm r} \, =\, 
{4\over 3} \sigma_{\rm T} c \gamma^2 \beta^2  U_{\rm r}
\label{pic}
\end{equation}
A simple way to remember Eq. \ref{pic} is:
\begin{eqnarray}
P_{\rm c}(\gamma)\, &=&\,  \left( { {\rm \#~of~collisions \over sec}}\right) \,
\left( {\rm average~ phot.~ energy~ after~ scatt.} \right) \nonumber \\
&=& \, \left( \sigma_{\rm T} c \, { U_{\rm r} \over \langle h\nu \rangle } \right) \, 
\left( {4\over 3} \langle h\nu \rangle  \gamma^2 \right)
\label{pic2}
\end{eqnarray}
Note the similarity with the synchrotron energy loss. 
The two energy loss rates are identical, once the radiation energy density
is replaced by the magnetic energy density $U_{\rm B}$.
Therefore, if relativistic electrons are in a region with
some radiation and magnetic energy densities, they will emit by both
the synchrotron and the Inverse Compton scattering processes.
The ratio of the two luminosities will be
\begin{equation}
{L_{\rm syn} \over L_{\rm IC}}\, =\,  
{ P_{\rm syn}  \over P_{\rm c}}\, =\,
{U_{\rm B} \over U_{\rm r}}
\label{ll}
\end{equation}
where we have set $dE_{\rm IC}/dt=dE_e/dt$.
This is true unless one of the two processes is inhibited for some reason.
For instance:
\begin{itemize}
\item At (relatively) low energies, electrons could emit {\it and absorb} synchrotron
radiation, so the synchrotron cooling is compensated by the heating due
to the absorption process.

\item At high energies, electrons could scatter in the Klein--Nishina regime:
in this case, since the cross section is smaller, they will do less scatterings,
and cool less.

\end{itemize}

%
\begin{figure}[h]
\vskip -0.5  true cm
\includegraphics[height=8cm, width=13.cm]{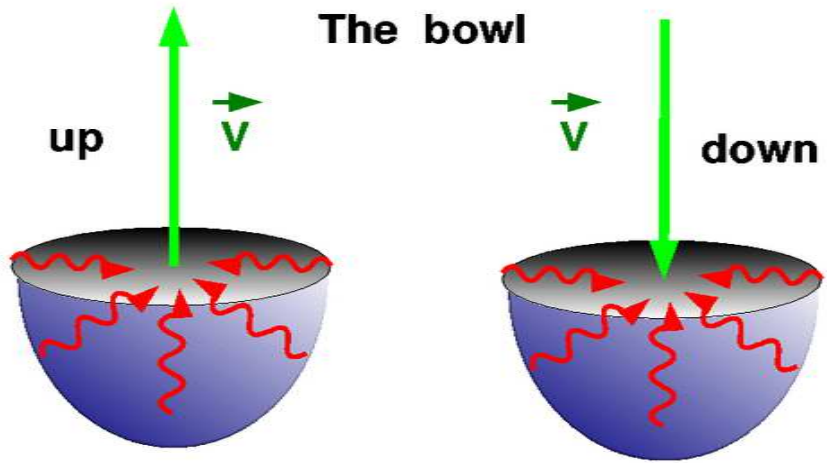}
\vskip -0.5 true cm
\caption{
In the center of a semi--sphere (the ``bowl'') we have relativistic
electrons going down and going up, all with the same $\gamma$.
Since the seed photon distribution is anisotropic, so is the 
scattered radiation and power. The losses of the electron going down are
7 times larger than those of the electron going up (if $\gamma\gg 1$).
Since almost all the radiation is produced along the velocity vector
of the electrons, also the downward radiation is 7 times more powerful
than the upward radiation. 
}
\label{bowl}
\end{figure}
%

But let us go back to Eq. \ref{aniso}, that is the starting point when dealing
with {\it anisotropic} seed photon distributions.
Think for instance to an accretion disk as the producer of the seed 
photons for scattering, and some cloud of relativistic electrons
above the disk. 
If the cloud is not that distant, and it is small with respect
to the disk size, then this case is completely equal to the
case of having a little cloud of relativistic electrons located
at the center of a semi--sphere.
That is, we have the ``bowl'' case illustrated in Fig \ref{bowl}.
Just for fun, let us calculate  the total power emitted by an electron
going ``up'' and by its brother (i.e. it has the same $\gamma$) going down.
Using Eq. \ref{aniso} we have:
\begin{equation}
{P_{\rm down} \over P_{\rm up}} \, =\, { \int_{-1}^0 (1-\beta \mu)^2 d\mu \over
\int_0^{1} (1-\beta \mu)^2 d\mu } \, =\, 
{ 1+\beta +\beta^2/3 \over 1-\beta + \beta^2/3}\, \to \, 7
\label{updown}
\end{equation}
where $\mu\equiv \cos\psi$ and the last step assumes $\beta\to 1$.
Since almost all the radiation is produced along the velocity vector
of the electrons, also the downward radiation is  more powerful
than the upward radiation (i.e. 7 times more powerful for $\gamma\gg 1$). 
What happens if the cloud of electrons is located at some height
above the bowl? Will the $P_{\rm down}/P_{\rm up}$ be more or less?

\subsection{Cooling time  and compactness}

The cooling time due to the inverse Compton process is
\begin{equation}
t_{\rm IC} \, =\, {E \over dE_e /dt} \, =\,
{3 \gamma m_e c^2 \over 4 \sigma_{\rm T} c \gamma^2 \beta^2  U_{\rm r}} 
\, \sim \, {3 m_e c^2 \over 4\sigma_{\rm T} c \gamma  U_{\rm r}};
\qquad \gamma \epsilon \ll m_{\rm e} c^2
\label{tccool}
\end{equation}
This equation offers the opportunity 
to introduce an important quantity, namely {\it the compactness}
of an astrophysical source, that is essentially the luminosity $L$ over
the size $R$ ratio.
Consider in fact how $U_{\rm r}$ and $L$ are related:
\begin{equation}
U_{\rm r} \, =\, { L\over 4\pi R^2 c}
\label{urad1}
\end{equation}
Although this relation is almost universally used, there are subtleties.
It is surely valid if we measured $U_{\rm r}$ {\it outside} the source,
at a distance $R$ from its center.
In this case $4\pi R^2 c$ is simply the volume of the shell crossed by the
source radiation in one second.
But if we are {\it inside} an homogeneous, spherical transparent source, 
a better way to calculate $U_{\rm r}$ is to think to the average 
time needed to the typical photon to exit the source. This is $t_{\rm esc}=3R/(4c)$.
It is less than $R/c$ because the typical photon is not born at the center
(there is more volume close to the surface).
If $V=(4\pi/3) R^3$ is the volume, we can write:
\begin{equation}
U_{\rm r} \, =\, { L\over V }\, t_{\rm esc} \, =\, {3 L \over 4\pi R^3 } {3 R \over 4c}
\, =\, { 9L \over 16 \pi R^2 c}
\end{equation}
This is greater than  Eq. \ref{urad1} by a factor 9/4.
Anyway, let us be conventional and insert Eq. \ref{urad1} in Eq. \ref{tccool}:
\begin{equation}
t_{\rm IC}\, =\, {3 \pi m_e c^2  R^2   \over  \sigma_{\rm T}  \gamma  L} \,\, \to\,\,
{t_{\rm IC} \over R/c} \, =\, 
{3 \pi\over \gamma}\, { m_e c^3  R \over \sigma_{\rm T} L}
\, \equiv \, {3 \pi\over \gamma}\, {1 \over \ell}
\label{elle1}
\end{equation}
where the dimensionless compactness $\ell$ is defined as
\begin{equation}
\ell \, =\, {\sigma_{\rm T} L \over m_{\rm e} c^3 R }
\label{elle}
\end{equation}
For $\ell$ close or larger than unity, we have that even low energy electrons cool
by the Inverse Compton process in less than a light crossing time $R/c$.

There is another reason why $\ell$ is important, related to the fact that
it directly measures the optical depth (hence the probability to occur)
of the photon--photon collisions that lead to 
the creation of electron--positron pairs.
The compactness is one of the most important physical parameters when 
studying high energy compact sources (X--ray binaries,
AGNs and Gamma Ray Bursts).

\subsection{Single particle spectrum}

As we did for the synchrotron spectrum, we will not repeat the 
exact derivation of the single particle spectrum, but we 
try to explain why the typical frequency of the scattered
photon is a factor $\gamma^2$ larger than the frequency
of the incoming photon.
Here are the steps to consider:
\begin{enumerate}
\item Assume that the relativistic electron travels in a region
where there is a radiation energy density $U_{\rm r}$ made
by photons which we will take, for simplicity, monochromatic,
therefore all having a dimensionless frequency $x=h\nu/m_{\rm e} c^2$.

\item 
In the frame where the electron is at rest, half of the photons
appear to come from the front, within an angle $1/\gamma$.

\item The typical frequency of these photons is $x^\prime \sim \gamma x$
(it is twice that for photons coming exactly head on).

\item 
Assuming that we are in the Thomson regime means that 
i) $x^\prime<1$;
ii) the cross section is the Thomson one; 
iii) the frequency of the scattered photon is the same of the incoming one,
i.e. $x_1^\prime = x^\prime \sim \gamma x$, and  iv) the pattern of the scattered 
photons follows the angular dependence of the cross section, therefore the ``peanut''.

\item
Independently of the initial photon direction, and therefore independently of the 
frequencies seen by the electrons, all photons after scatterings are isotropized.
This means that all observers (at any angle $\psi_1^\prime$ in this frame
see the same spectrum, and the same typical frequency.
Half of the photons are in the semi-sphere with $\psi_1^\prime\le \pi/2$.

\item 
Now we go back to the lab--frame.
Those photons that had $\psi_1^\prime\le \pi/2$ now have $\psi_1\le 1/\gamma$.
Their typical frequency if another factor $\gamma$ greater than what they had
in the rest frame, therefore
\begin{equation}
x_1 \, \sim \, \gamma^2 x
\label{g2}
\end{equation}
This is the typical Inverse Compton frequency.  
\end{enumerate}
The exact derivation can be found e.g. in Rybicki \& Lightman (1979) and
in Blumenthal \& Gould (1970).
We report here the final result, valid for a monochromatic and isotropic
seed photons distribution, characterized by a specific intensity
\begin{equation}
{ I(x) \over x}  \, =\, {I_0 \over x} \delta(x-x_0)
\label{i0}
\end{equation}
Note that $I(x)/x$ is the analog of the normal intensity, but
it is associated with the number of photons. If we have $n$ electrons
per cubic centimeter we have:
\begin{equation}
\epsilon_{\rm IC}(x_1) \, =\, 
{\sigma_{\rm T} n I_0 (1+\beta) \over 4\gamma^2\beta^2 x_0}\, 
F_{\rm IC}(x_1) 
\label{epsc}
\end{equation}
%
%
\begin{figure}[h]
\vskip -0.5  true cm
\includegraphics[height=11.5cm, width=13cm]{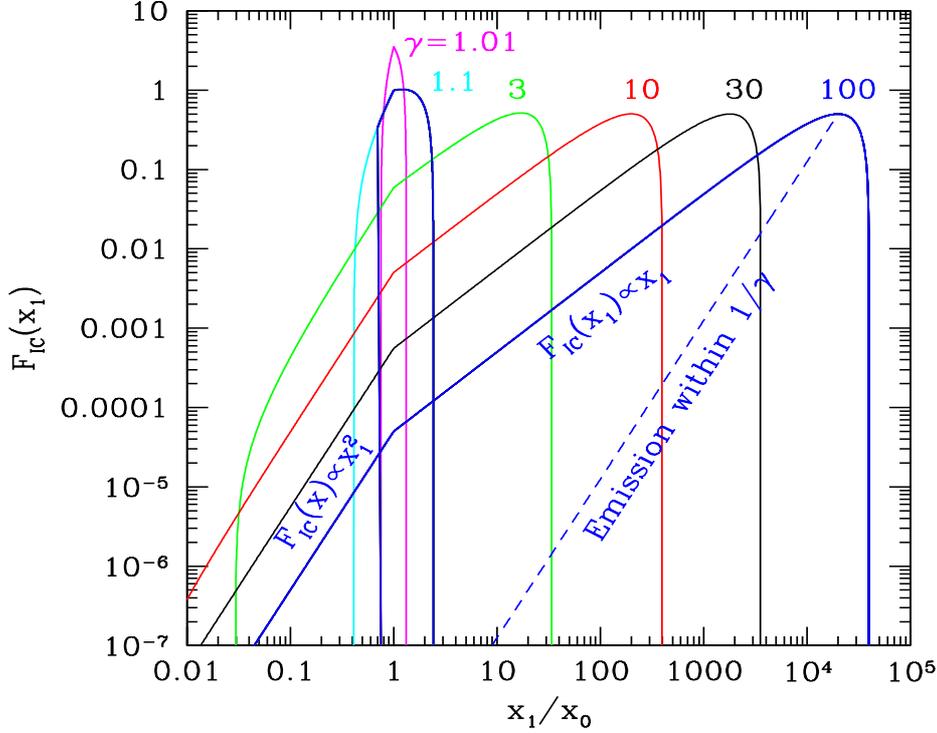}
\vskip -0.9 true cm
\caption{Spectrum emitted by the Inverse Compton process by electrons
of different $\gamma$ (as labeled) scattering an isotropic monochromatic
radiation field of dimensionless frequency $x_0$.
The dashed line corresponds to the spectrum emitted within the $1/\gamma$
beming cone: it always contains the 75\% of the total power, for any $\gamma$.
For $x_1<x_0$ we have {\it downscattering}, i.e. the photons
loose  energy in the process.
Note also the power law segments arising when $\gamma\gg 1$: 
$F_{\rm IC}(x_1) \propto x_1^2$ for the downscattering
tail, and $F_{IC}(x_1)\propto x_1$ for the upscattering segment.
}
\label{icsingle}
\end{figure}
%
The function $F_{\rm IC}$ contains all the frequency dependence:
\begin{eqnarray}
F_{\rm IC}(x_1)  &=& 
{x_1 \over x_0} \, \, \left[{x_1\over x_0} - {1\over (1+\beta)^2\gamma^2}\right];
\quad {1\over (1+\beta)^2\gamma^2} < {x_1\over x_0} < 1
\nonumber \\
F_{\rm IC}(x_1) &=&  
{x_1 \over x_0} \, \, \left[ 1 - {x_1 \over x_0} \,{1\over (1+\beta)^2\gamma^2}\right]; 
\quad 1<{x_1\over x_0} < (1+\beta)^2\gamma^2 
\end{eqnarray}
The first line corresponds to {\it downscattering}: the scattered photon has {\it less}
energy than the incoming one.
Note that in this case $F_{\rm IC}(x_1) \propto x_1^2$.
The second line corresponds to {\it upscattering}: in this case 
$F_{\rm IC}(x_1) \propto x_1$ except for frequencies close to the maximum ones.
The function $F_{\rm IC}(x_1)$ is shown in Fig. \ref{icsingle} for different
values of $\gamma$.
The figure shows also the spectrum of the photons contained in the beaming cone $1/\gamma$:
the corresponding power is always 75\% of the total.

The average frequency of $F_{\rm IC}(x_1)$ is
\begin{equation}
\langle x_1 \rangle \, =\, 2\gamma^2 x_0; \quad {\rm energy~ spectrum}
\label{averenergy}
\end{equation}
This is the average frequency {\it of the energy spectrum}.
We sometimes want to know the average energy {\it of the photons}, i.e.
we have to calculate the average frequency of {\it the photon spectrum}
$F_{\rm IC}(x_1)/x_1$. This is:
\begin{equation}
\langle x_1 \rangle \, =\, {4\over 3} \gamma^2 x_0; \quad {\rm photon~ spectrum}
\label{averx1}
\end{equation}

\section{Emission from many electrons}

We have seen that the emission spectrum from a single particle is 
peaked, and the typical frequency is boosted by a factor $\gamma^2$.
This is equal to the synchrotron case.
Therefore we can derive the Inverse Compton emissivity as we did 
for the synchrotron one.
Again, assume a power--law energy distribution for the relativistic electrons:
\begin{equation}
N(\gamma) \,=\, K \gamma^{-p}\, =\, N(E)\, {dE \over d\gamma}; 
\quad \gamma_{\rm min} < \gamma < \gamma_{\rm max}
\label{ngamma}
\end{equation}
and assume that it describes an isotropic distribution of electrons.
For simplicity, let us assume that the seed photons are isotropic and monochromatic,
with frequency $\nu_0$ (we now pass to real frequencies, since we are getting closer
to the real world..). 
Since there is a strong link between the scattered frequency 
$\nu_{\rm c}$ and the electron
energy that produced it, we can set:
\begin{equation}
\nu_{\rm c} \, = {4\over 3} \gamma^2 \nu_0  \, \to \, \gamma\, =\, 
\left( { 3\nu_{\rm c}  \over 4\nu_0} \right)^{1/2} \, \to \, 
\left|{d\gamma \over d\nu}\right| \, = 
\, {\nu_{\rm c}^{-1/2}\over 2} \left({ 3\over 4\nu_o}\right)^{1/2}
\label{nucnu0}
\end{equation}
Now, repeating the argument we used for synchrotron emission, we can 
state that the power lost by the electron of energy $\gamma m_{\rm e} c^2$ 
within $m_{\rm e} c^2 d\gamma$
goes into the radiation of frequency $\nu$ within $d\nu$.
Since we will derive an emissivity (i.e. erg cm$^{-3}$ s$^{-1}$ sterad$^{-1}$ Hz$^{-1}$)
we must remember the $4\pi$ term (if the emission is isotropic).
We can set:
\begin{equation}
\epsilon_{\rm c}(\nu_{\rm c}) d\nu_{\rm c} \, =\, 
{1\over 4\pi}\, m_{\rm e} c^2 P_{\rm c}(\gamma) N(\gamma)  d\gamma
\label{jic1}
\end{equation}
This leads to:
\begin{equation}
\epsilon_{\rm c}(\nu_{\rm c}) \, =\, 
{1\over 4\pi}\, {(4/3)^\alpha \over 2} \, \sigma_{\rm T} c K {U_{\rm r} \over \nu_0}
\left( {\nu_{\rm c} \over \nu_0}\right)^{-\alpha}
\label{jic2}
\end{equation}
Again, a power law, as in the case of synchrotron emission by a power law 
energy distribution. Again the same link between $\alpha$ and $p$:
\begin{equation}
\alpha \, = \, { p-1 \over 2}
\label{alpha}
\end{equation}
Of course, this is not a coincidence: it is because both the Inverse Compton and 
the synchrotron single electron spectra are peaked at a typical frequency that
is a factor $\gamma^2$ greater than the starting one.

Eq. \ref{jic2} becomes a little more clear if 
\begin{itemize}
\item
we express $\epsilon_{\rm c}(\nu_{\rm c})$
as a function of the photon energy $h\nu_{\rm c}$.
Therefore $\epsilon_{\rm c}(h \nu_{\rm c}) = \epsilon_{\rm c}(\nu_{\rm c})/h$;
\item
we multiply and divide by the source radius $R$;
\item 
we consider a proxy for the scattering optical depth of the 
relativistic electrons setting
$\tau_{\rm c} \equiv \sigma_{\rm T} K R$.
\end{itemize}
Then we obtain:
\begin{equation}
\epsilon_{\rm c}(h \nu_{\rm c}) \, =\, 
{1\over 4\pi}\, {(4/3)^\alpha \over 2} \,
{ \tau_{\rm c} \over R/c} \, {U_{\rm r} \over h\nu_0}
\left( {\nu_{\rm c} \over \nu_0}\right)^{-\alpha}
\label{jic3}
\end{equation}
In this way: $\tau_{\rm c}$ (for $\tau_{\rm c}<1$) is the fraction of the
seed photons $U_{\rm r} /h \nu_0$ undergoing scattering in a time $R/c$, 
and $\nu_{\rm c}/\nu_0 \sim \gamma^2$ is the average gain in energy
of the scattered photons.

\subsection{Non monochromatic seed photons}

It is time to consider the more realistic case in which the seed
photons are not monochromatic, but are distributed in frequency.
This means that we have to integrate Eq. \ref{jic2} over the
incoming photon frequencies.
For clarity, let us drop the subscript 0 in $\nu_0$. We have
\begin{equation}
\epsilon_{\rm c}(\nu_{\rm c}) \, =\, 
{1\over 4\pi}\, {(4/3)^\alpha \over 2} \,
{ \tau_{\rm c} \over R/c} \, \nu_{\rm c}^{-\alpha} 
\, \int_{\nu_{\rm min}}^{\nu_{\rm max}}
{U_{\rm r}(\nu) \over \nu} \nu^\alpha d\nu
\label{jic3}
\end{equation}
where $U_{\rm r} (\nu)$ [erg cm$^{-3}$ Hz$^{-1}$]
is the specific radiation energy density at the frequency $\nu$.
The only difficulty of this integral is to find the correct limit
of the integration, that, in general, {\it depend on $\nu_{\rm c}$}.
Note also another interesting thing.
We have just derived that if the same electron population produces
Inverse Compton and synchrotron emission, than the slopes of the two
spectra are the same.
Therefore, when $U_{\rm r}(\nu)$ is made by synchrotron photons, then
 $U_{\rm r}(\nu) \propto \nu^{-\alpha}$. 
The result of the integral, in this case, will be $\ln(\nu_{max}/\nu_{min})$.
%
\begin{figure}[h]
\vskip -0.7  true cm
\includegraphics[height=11.cm, width=14cm]{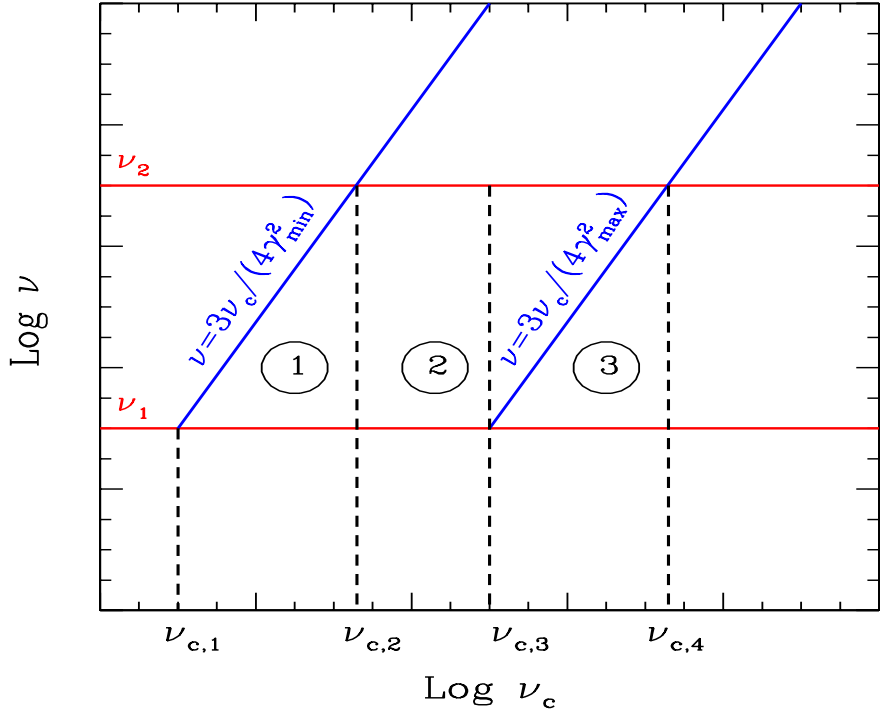}
\vskip -0.9 true cm
\caption{The $\nu$--$\nu_{\rm c}$ plane.
The two diagonal lines delimit the regions of the
seed photons that can be used to give a given frequency $\nu_{\rm c}$.
 }
\label{iclimits}
\end{figure}
%

Fig. \ref{iclimits} helps to understand what are the right
$\nu_{max}$ and $\nu_{min}$ to use.
On the y--axis we have the frequencies of the seed photon
distribution, which extend between 
$\nu_1$ and $\nu_2$.
On the x--axis we have the scattered frequencies,
which extend  between $\nu_{\rm c,1} = (4/3)\gamma_{\rm min}^2 \nu_1$
and $\nu_{\rm c,4} = (4/3)\gamma_{\rm max}^2 \nu_1$.
The diagonal lines are the functions
\begin{equation}
\nu \, =\, {3\nu_{\rm c} \over 4 \gamma_{\rm min}^2} 
\qquad 
\nu \,=\, {3\nu_{\rm c} \over 4 \gamma_{\rm max}^2} 
\end{equation}
that tell us what are the appropriate $\nu$ that can give $\nu_{\rm c}$
once we change $\gamma$.

There are three zones:
\begin{enumerate}
\item In zone (1), between $\nu_{\rm c,1}$ and $\nu_{\rm c,2}=
(4/3)\gamma_{\rm min}^2 \nu_2$
the appropriate limits of integration are:
\begin{equation}
\nu_{\rm min} \, =\, \nu_1
\qquad 
\nu_{\rm max} \, =\, {3\nu_{\rm c} \over 4 \gamma_{\rm min}^2} 
\end{equation}

\item In zone (2), between $\nu_{\rm c,2}$ and 
$\nu_{\rm c,3}=(4/3)\gamma_{\rm max}^2 \nu_1$
the limits are:
\begin{equation}
\nu_{\rm min} \, =\, \nu_1
\qquad 
\nu_{\rm max} \, =\, \nu_2
\end{equation}

\item In zone (3), between $\nu_{\rm c,3}$ and $\nu_{\rm c,4}=
(4/3)\gamma_{\rm max}^2 \nu_2$
the  limits  are:
\begin{equation}
\nu_{\rm min} \, =\,  {3\nu_{\rm c} \over 4 \gamma_{\rm max}^2} 
\qquad 
\nu_{\rm max} \, =\, \nu_2
\end{equation}
\end{enumerate}
We see that only in zone (2) the limits of integration coincide with the
extension in frequency of the seed photon distribution, and are therefore constant.
Therefore $\epsilon_{\rm c}(\nu_{\rm c})$ will be a power law of slope $\alpha$ only
in the corresponding frequency limits.
Note also that for a broad range in $[\nu_1; \nu_2]$ or a narrow range in 
$[\gamma_{\rm min}; \gamma_{\rm max}]$ we do not have a power law, since there is
no $\nu_{\rm c}$ for which the limits of integrations are both constants.

\section{Thermal Comptonization}

With this term we mean the process of {\it multiple} scattering
of a photon due to a {\it thermal} or {\it quasi--thermal}
distribution of electrons.
By {\it quasi--thermal} we mean a particle distribution that 
is peaked, even if it is not a perfect Maxwellian.
Since the resulting spectrum, by definition, is due
to the superposition of many spectra, each corresponding
to a single scattering, the details of the particle distribution 
will be lost in the final spectrum, as long as the  distribution 
is peaked. 
The ``bible'' for an extensive discussion about this process
is Pozdnyakov, Sobol \& Sunyaev (1983).

There is one fundamental parameter measuring the importance
of the Inverse Compton process in general, and of multiple scatterings
in particular: {\it the Comptonization parameter}, usually
denoted with the letter $y$.
Its definition is:
\begin{equation}
y \, = \, \left[{\rm average~ \# ~of~ scatt. }\right] \times
\left[ {\rm average~ fractional~ energy~ gain~ for ~scatt.}\right]
\end{equation}
If $y>1$ the Comptonization process is important, because the
Comptonized spectrum has more energy than the spectrum of the
seed photons.

\subsection{Average number of scatterings}

This can be calculated thinking that the photon, before 
leaving the source, experience a sort of random walk 
inside the source.
Let us call 
\begin{equation}
\tau_{\rm T}\, =\, \sigma_{\rm T} n R
\end{equation}
the Thomson scattering optical depth, where $n$ is the electron density
and $R$ the size of the source.
When $\tau_{\rm T} < 1$ most of the photons leave the source
directly, without any scattering.
When $\tau_{\rm T}>1$ then the mean free path is $d=R/\tau_{\rm T}$
and the photon will experience, on average, $\tau_{\rm T}^2$ scatterings
before leaving the source.
Therefore the total path travel by the photon, from the time of its birth
to the time it leaves the source is:
the photon is born, is
\begin{equation}
c \Delta t\, =\, \tau_{\rm T}^2 \, {R \over \tau_{\rm T}}\, =\, \tau_{\rm T}R
\end{equation}
and $\Delta t$ is the corresponding elapsed time.

\subsection{Average gain per scattering}

\subsubsection{Relativistic case}

If the scattering electrons are relativistic, 
we have already seen that the photon energy is amplified 
by the factor $(4/3)\gamma^2$ (on average).
Therefore the problem is to find what is $\langle\gamma^2\rangle$ in
the case of a relativistic Maxwellian, that has the form
\begin{equation}
N(\gamma) \, \propto \gamma^2 e^{-\gamma/\Theta}; 
\quad \Theta\equiv {kT \over m_{\rm e} c^2}
\end{equation}
Setting $x_0 = h\nu_0/(m_{\rm e} c^2)$ we have that the average energy 
of the photon of initial frequency $x_0$ after a single scattering with
electron belonging to this Maxwellian is:
\begin{eqnarray}
\langle x_1\rangle \, &= &\, {4\over 3}\langle \gamma^2\rangle \, =\,
{4\over 3}\, x_0 \,{ \int_1^\infty \gamma^2 \gamma^2 e^{-\gamma/\Theta}d\gamma \over
 \int_1^\infty   \gamma^2 e^{-\gamma/\Theta} d\gamma}
\nonumber \\
&= &\,{4\over 3}\, x_0  \Theta^2 {\Gamma(5)\over \Gamma(3) } 
\nonumber \\
&= &\,{4\over 3}\, x_0 {4! \over 2!} \, =\, 16 \Theta^2 x_0
\end{eqnarray}

\subsubsection{Non relativistic case}
In this case the average gain is proportional to the electron energy,
not to its square.
The derivation is not immediate, but we must use a trick.
Also, we have to account that in any Maxwellian, but especially
when the temperature is not large, there will be electrons
that have {\it less} energy than the incoming photons.
In this case it is the photon to give energy to the electron:
correspondingly, the scattered photon will have less energy
than the incoming one.
Averaging out over a Maxwellian distribution, we will have:
\begin{equation}
{\Delta x \over x}  \, =\, {x_1-x_0 \over x_0} \, =\, \alpha\Theta - x
\end{equation}
Where $\alpha \Theta $ is what the photon gains and the $-x$ term corresponds
to the downscattering of the photon (i.e. direct Compton).
We do not know yet the value for the constant $\alpha$.
To determine it we use the following argument.
We know (from general and robust arguments) what happens when photons
and electrons are in equilibrium under the only process of scattering,
and neglecting absorption (i.e. when the {\it number of photon is conserved}).
What happens is that the photons follow the so--called {\it Wien distribution}
given by:
\begin{equation}
F_{\rm W}(x)  \, \propto \, x^3 e^{-x/\Theta} \to N_{\rm W}(x) \, = \,
{ F_{\rm W}(x) \over x} \, \propto x^2 e^{-x/\Theta} 
\end{equation}
where $F$ correspond to the radiation spectrum,  $N$ to the 
photon spectrum, and $\Theta$ is the dimensionless electron temperature.
When a Wien distribution is established 
we must have $\langle \Delta x\rangle =0$, since we are at equilibrium,
So we require that, on average, gains equal losses:
\begin{equation}
\langle \Delta x \rangle  \, = 0 \, \to \, 
\alpha\Theta \langle x\rangle  -\langle x^2\rangle \, =\, 0
\end{equation}
Calculating $\langle x\rangle$ and $\langle x^2\rangle$ for a photon Wien
distribution, we have:
\begin{eqnarray}
\langle x\rangle \, &= &\, { \int_0^\infty x^3 e^{-x/\Theta}dx \over
 \int_0^\infty  x^2 e^{-x/\Theta} dx}
\, =\, {\Gamma(4) \over \Gamma(3)}\, \Theta \, = {3! \over 2!}\, \Theta \,
= \, 3\, \Theta 
\nonumber \\
\langle x^2\rangle \, &= &\, { \int_0^\infty x^4 e^{-x/\Theta}dx \over
 \int_0^\infty  x^2 e^{-x/\Theta} dx}
\, =\, {\Gamma(5) \over \Gamma(3)}\, \Theta^2 \, =\, {4! \over 2!} \, \Theta^2
\, = \, 12\Theta^2 
\end{eqnarray}
This implies that $\alpha=4$ not only at equilibrium, but always, and we finally have
\begin{equation}
{\Delta x \over x} \, =\, 4 \Theta - x
\end{equation}
Combining the relativistic and the non relativistic cases, we have an
expression valid for all temperatures:
\begin{equation}
{\Delta x \over x} \, =\, 16\Theta^2 + 4 \Theta - x
\end{equation}
going back to the $y$ parameter we can write:
\begin{equation}
y \, =\, \max(\tau_{\rm T}, \tau_{\rm T}^2) \times [ 16\Theta^2 + 4 \Theta - x]
\end{equation}
Going to the differential form, and neglecting downscattering, we have\:
\begin{equation}
{dx\over x} \, =\, [ 16\Theta^2 + 4 \Theta]\, dK \, \to 
x_{\rm f}  \, =\, x_0 \, e^{(16\Theta^2 + 4 \Theta)K}\, \to \, 
x_{\rm f}  \, =\, x_0\,  e^y
\end{equation}
where now $K$ is the number of scatterings.
If we subtract the initial photon energy, and consider that the above equation
is valid for all the $x_0$ of the initial seed photon distribution,
of luminosity $L_0$, we have
\begin{equation}
{L_{\rm f} \over L_0}\, =\, e^y -1
\end{equation}
Then the importance of $y$ is self evident, and also the fact that
it marks the importance of the Comptonization process when it is larger than 1.

\subsection{Comptonization spectra: basics}

We will illustrate why even a thermal (Maxwellian) distribution of
electrons can produce a power law spectrum.
The basic reason is that the total produced spectrum is the superposition
of many orders of Compton scattering spectra: when they are
not too much separated in frequency (i.e. for not too large temperatures)
the sum is a smooth power law.
We can distinguish 4 regimes, according to the values of $\tau_{\rm T}$
and $y$. 
As usual, we set $x\equiv h\nu /(m_{\rm e}c^2)$ and 
$\Theta\equiv kT/(m_{\rm e}c^2)$.

\subsubsection{The case $\tau_{\rm T}<1$}

Neglect downscattering for simplicity.
The fractional energy gain is $\Delta x/x = 16\Theta^2 +4\Theta$,
so the amplification $A$ of the photon frequency at each scattering is
\begin{equation}
A\, \equiv \, {x_1 \over x} \, =\, 16\Theta^2 +4\Theta+1 \, 
\sim \, {y \over \tau_{\rm T}}
\end{equation}
We can then construct Table \ref{table1}.
\begin{table}
\begin{tabular}{lll}
\hline
\# scatt.  &Fraction of escaping                    &$\langle x \rangle$ \\
           &photons                                 & \\
\hline 
0          &$e^{-\tau_{\rm T}} \to 1-\tau_{\rm T}$  &$x_0$  \\
1          &$\sim\tau_{\rm T}$                      &$x_0A$ \\
2          &$\sim\tau_{\rm T}^2$                    &$x_0A^2$ \\
3          &$\sim\tau_{\rm T}^3$                    &$x_0A^3$ \\
4          &$\sim\tau_{\rm T}^4$                    &$x_0A^4$ \\
.......    &......                                  &.......\\
n          &$\sim\tau_{\rm T}^n$                    &$x_0A^n$ \\
\hline
\end{tabular}
\caption{
When $\tau_{\rm T}<1$, a fraction $e^{-\tau_{\rm T}}$ of the 
seed photons escape without doing any scattering, and a fraction
$1-e^{-\tau_{\rm T}}\to \tau$ undergoes at least one scattering.
We can then repeat these fractions for all scattering orders.
Even if a tiny fraction of photons does several scatterings,
they can carry a lot of energy. 
}
\label{table1}
\end{table}
%

%
\begin{figure}[h]
\vskip -0.5  true cm
\hskip -1  true cm
\includegraphics[height=12.cm, width=14.7cm]{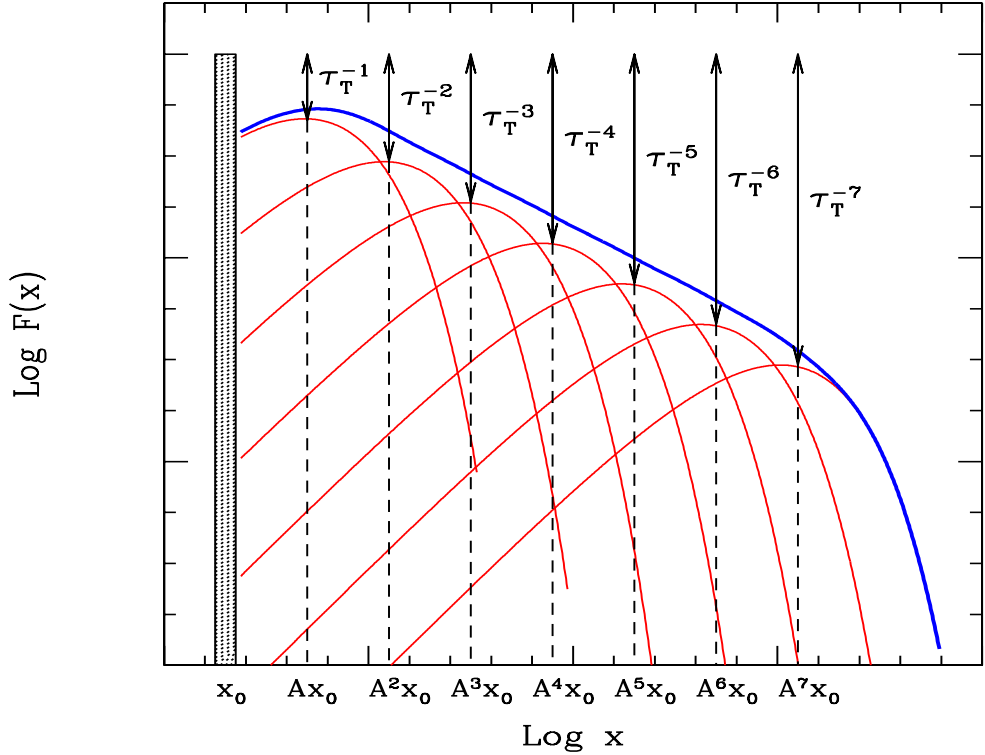}
\vskip -0.9 true cm
\caption{Multiple Compton scatterings when $\tau_{\rm T}<1$.
A fraction $\tau_{\rm T}$ of the photons of the previous scattering
order undergoes another scattering, and amplifying the frequency
by the gain factor $A$, until the typical photon frequency equals
the electron temperature $\Theta$.
Then further scatterings leave the photon frequency unchanged.
}
\label{alpha}
\end{figure}
%
A look to Fig. \ref{alpha} should convince you that the sum of all the
scattering orders gives a power law, and should also make clear how
to find the spectral slope.
Remember that we are in a log--log plot, so the spectral index
is simply $\Delta y/\Delta x$. We can find it
considering two successive scattering orders:
the typical (logarithm of) frequency is enhanced by $\log A$,
and the fraction of photons doing the scattering is $-\log\tau_{\rm T}$.
Remember also that we use $F(x)\propto x^{-\alpha}$ as the definition
of energy spectral index.

Therefore
\begin{equation}
\alpha \, = \, - {\log \tau_{\rm T} \over \log A} 
\, \sim \,  - {\log \tau_{\rm T} \over \log y- \log \tau_{\rm T}} 
\end{equation}
When $y\sim 1$, its logarithm is close to zero, and we have
$\alpha\sim 1$.
When $y>1$, then $\alpha<1$ (i.e. {\it flat, or hard}), 
and vice--versa,  when $y<1$, then 
its logarithm will be negative, as the logarithm of $\tau_{\rm T}$,
and  $\alpha>1$ (i.e. {\it steep, or soft}).


Attention! when $\tau_{\rm T} \ll 1$ and  $A$ is large (i.e. big 
frequency jumps between one scattering and the next),
then the superposition of all scattering orders 
(by the way, there are fewer, in this case) will not guarantee 
a perfect power--law.
In the total spectrum we can see the ``bumps'' corresponding
to individual scattering orders.

\subsubsection{The case $\tau_{\rm T} \gsim 1$}

This is the most difficult case, as we should solve a famous equation, 
the equation of Kompaneet. The result is still a power law,
whose spectral index is approximately given by
\begin{equation}
\alpha \, = \, - {3\over 2} +\sqrt{ {9\over 4} + {4 \over y}}
\end{equation}

\subsubsection{The case $\tau_{\rm T} \gg 1$: saturation}

In this case the interaction between photons and matters is so intense
that they go to equilibrium, and they will have the same temperature.
But instead of a black--body, the resulting photon spectrum has a
Wien shape.
This is because the photons are conserved (if other
scattering processes such as induced Compton or two--photon scattering
are important, then one recovers a black--body, because these
processes {\it do not} conserve photons).
The Wien spectrum has the slope:  
\begin{equation}
I(x) \, \propto \, x^3 \, e^{-x/\Theta}
\end{equation}
At low frequencies this is {\it harder} than a black--body.

\subsubsection{The case $\tau_{\rm T} > 1$, $y> 1$: quasi--saturation}

\begin{figure}[h]
\vskip -0.7   true cm
\hskip -1  true cm
\includegraphics[height=11.cm, width=14.7cm]{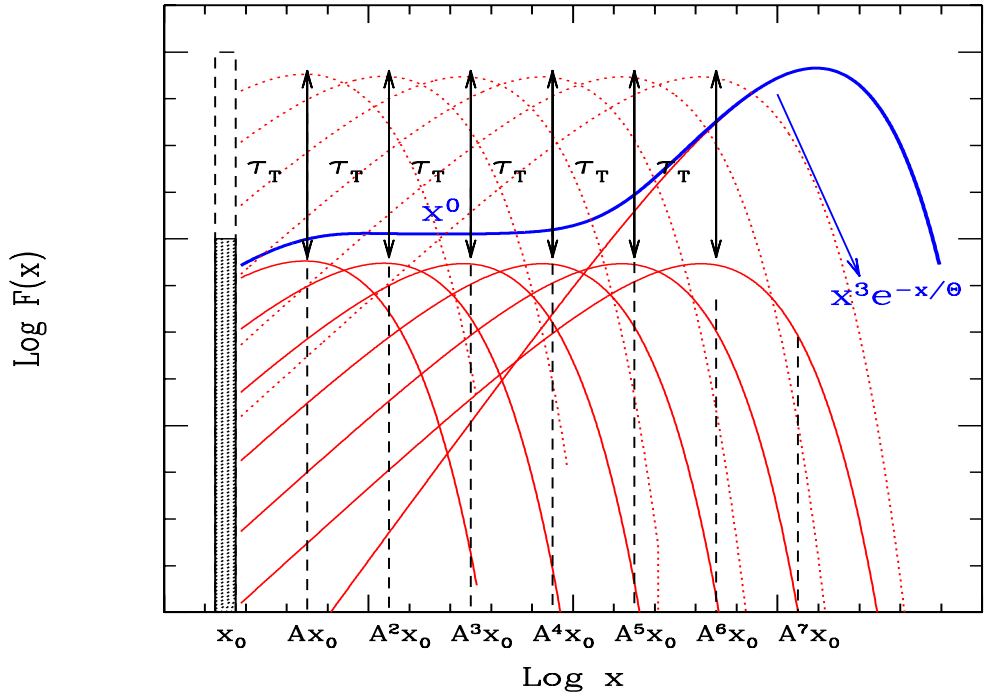}
\vskip -1 true cm
\caption{Multiple Compton scatterings when $\tau_{\rm T}> 1$
and $y\gg 1$.
For the first scattering orders, {\it nearly  all} photons are
scattered: only a fraction $1/\tau_{\rm T}$ can escape.
Therefore the number of photons escaping at each scattering order is the same.
This is the reason of the flat part, where $F(x)\propto x^0$.
When the photon frequency is of the order of $\Theta$, photons
and electrons are in equilibrium, and even if only a small fraction 
of photons can escape at each scattering order, they do not change
frequency any longer, and therefore they form the {\it Wien bump},
with the slope  $F(x)\propto x^3e^{-x/\Theta}$.
If we increase $\tau_{\rm T}$, the flux with slope $x^0$ decreases,
while the Wien bump stays the same.
}
\label{alpha0}
\end{figure}

Suppose that in a source characterized by a large $\tau_{\rm T}$ 
the source of soft photons is spread throughout the source.
In this case the photons produced close to the surface,
in a skin of optical depth $\tau_{\rm T}=1$, leave the source
without doing any scattering (note that having the source of
seed photons concentrated at the center is a different case).
The remaining fraction, $1-1/\tau_{\rm T}$, i.e. {\it almost} all photons, 
remains inside.
This can be said for each scattering order.
This is illustrated in Fig. \ref{alpha0}, where 
$\tau_{\rm T}$ corresponds to the ratio between the  
flux of photon inside the source at a given frequency and
the flux of photons that escape.
If I start with 100 photons, only 1 -- say -- escape,
and the other 99 remain inside, and do the first scattering.
After it, only one escape, and the other 98 remain inside, and
so on, until the typical photon and electron energies are equal,
and the photon therefore stays around with the same final frequency
until it is its turn to escape.
This ``accumulation'' of photons at $x\sim 3\Theta$ gives the Wien
bump.
Note that since at any scattering order only a fixed number of photons
escape, always the same, then the spectrum in this region 
will always have $\alpha=0$.
This is a ``saturated'' index, i.e. one obtains always zero
even when changing $\tau_{\rm T}$ or $\Theta$.
What indeed changes, by increasing  $\tau_{\rm T}$, 
is that i) the flux characterized by $x^0$ decreases,
ii) the Wien peak will start to dominate earlier (at lower frequencies),
while nothing happens to the flux of the Wien peak (it stays there).
Increasing still $\tau_{\rm T}$ we fall in the previous case
(equilibrium, meaning only the Wien spectrum without
the $x^0$ part).
%





\vskip 4 true cm
\noindent
{\bf References}
\vskip 0.5 true cm
\noindent
Blumenthal G.R. \& Gould R.J., 1970, Rev. of Modern Physics 42, 237

\noindent
Rybicki G.B \& Lightman A.P., 1979, 
{\it Radiative processes in Astrophysics} (Wiley \& Sons)

\noindent
Pozdnyakov L.A., Sobol I.M. \& Sunyaev R.A., 1983,
Astroph. Space Phys. Rev. Vol. 2 p. 189--331


\chapter{Synchrotron Self--Compton}

Consider a population of relativistic electrons in a 
magnetized region.
They will produce synchrotron radiation, and therefore
they will fill the region with photons.
These synchrotron photons will have some probability
to interact again with the electrons, by the Inverse
Compton process. 
Since the electron ``work twice'' (first making synchrotron
radiation, then scattering it at higher energies) this
particular kind of process is called synchrotron self--Compton,
or SSC for short.

\section{SSC emissivity}

The importance of the scattering will of course be high
if the densities of electrons and photons are large.
If the electron distribution is a power law [$N(\gamma) =K\gamma^{-p}$],
then we expect that the SSC flux will be $\propto K^2$,
i.e. {\it quadratic} in the electron density.

\begin{figure}[h]
\includegraphics[height=13cm, width=13cm]{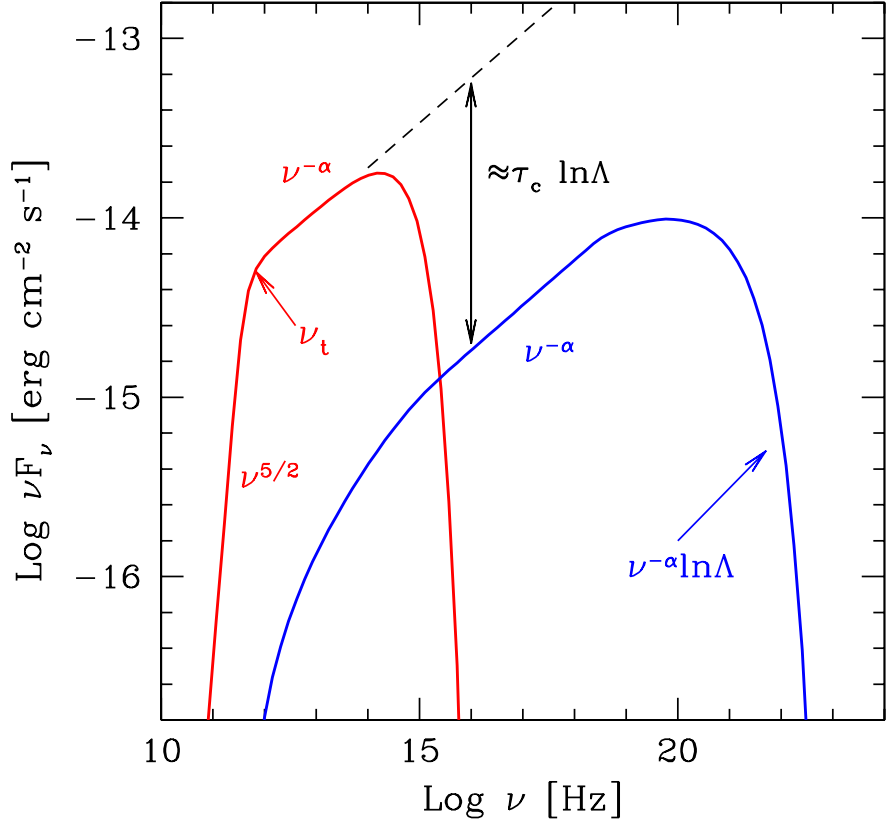} 
\vskip -0.5 true cm
\caption{Typical example of SSC spectrum, shown in the
$\nu F_\nu$ vs $\nu$ representation.
The spectral indices instead correspond to the $F_\nu \propto \nu^{-\alpha}$
convention.
}
\label{ssc}
\end{figure}

We should remember Eq. \ref{jic3}, and, instead of a generic $U_{\rm r}(\nu)$,
we should substitute the appropriate expression for the specific
synchrotron radiation energy density.
We will then set:

\begin{equation}
U_{\rm s}(\nu) \, =\, {3R \over 4c}\, {L_{\rm s}(\nu) \over V} \, =\,
4\pi \,{3R \over 4c}\,  \epsilon_{\rm s}(\nu)
\end{equation}
where $3R/(4c)$ is the average photon source--crossing time, and
$V$ is the volume of the source. 
Now a simple trick: we write the specific synchrotron emissivity as
\begin{equation}
\epsilon_{\rm s}(\nu)\, =\, \epsilon_{\rm s,0} \, \nu^{-\alpha}
\end{equation}
Remember: the $\alpha$ appearing here {\it is the same} index
in Eq. \ref{jic3}.
Substituting the above equations into Eq. \ref{jic3} we have
\begin{equation}
\epsilon_{\rm ssc}(\nu_{\rm c}) \, =\, 
 { (4/3)^{\alpha-1} \over 2} \,  \tau_{\rm c}  
\, \epsilon_{\rm s,0} \nu_{\rm c}^{-\alpha}
\, \int_{\nu_{\rm min}}^{\nu_{\rm max}} {d\nu \over \nu} 
\label{jssc0}
\end{equation}
As you can see, $ \epsilon_{\rm s,0} \nu_{\rm c}^{-\alpha} =\epsilon_{\rm s}(\nu_{\rm c})$ 
is nothing else than
the specific synchrotron emissivity calculated at the 
(Compton) frequency $\nu_{\rm c}$. 
Furthermore, the integral gives a logarithmic term,
that we will call $\ln \Lambda$.
We finally have:
\begin{equation}
\epsilon_{\rm ssc}(\nu_{\rm c}) \, =\, 
 { (4/3)^{\alpha-1} \over 2} \,  \tau_{\rm c}  
\, \epsilon_{\rm s}(\nu_{\rm c}) \ln \Lambda
\label{jssc}
\end{equation}
In this form the ratio between the synchrotron and the SSC flux
is clear, it is $[(4/3)^{\alpha-1}/2] \tau_{\rm c} \ln \Lambda
\sim \tau_{\rm c}  \ln \Lambda$.
It is also clear that since $\tau_{\rm c}\equiv \sigma_{\rm T} R K$ and
$\epsilon_{\rm s}(\nu_{\rm c})  \propto KB^{1+\alpha}$,
then, as we have guessed, the SSC emissivity  
$\epsilon_{\rm ssc}(\nu_{\rm c}) \propto K^2$ (i.e. electrons work twice).
Fig. \ref{ssc} summarizes the main results.

%
%

\section{Diagnostic}

If we are confident that the spectrum of a particular source is indeed
given by the SSC process, then we can use our theory to estimate a number
of physical parameters.
We have already stated (see Eq. \ref{Fnu}) that observations of the synchrotron
spectrum in its self--absorbed part can yield the value of 
the magnetic field if we also know
the angular radius of the source (if it is resolved).
Observations in the thin part can then give us the 
product $RK \equiv \tau_{\rm c}/\sigma_{\rm T}$ (see Eq. \ref{flux}).
But $\tau_{\rm c}$ is exactly what we need to predict the high energy
flux produced by the SSC process.
Note that if the source is resolved (i.e. we know $\theta_s$)  
we can get these information even without knowing the distance
of the source.
To summarize:
\begin{eqnarray}
F_{\rm thick}^{\rm syn}(\nu) \, & \propto & \theta_s^2 {\nu^{5/2} \over B^{1/2}} 
\,\qquad \quad \,\, \,\to \, \,\, {\rm get}\,\, B
\nonumber \\
F_{\rm thin}^{\rm syn}(\nu)\, & \propto & \theta_s^2 RK  B^{1+\alpha} \nu^{-\alpha}
\, \to \, {\rm get}\,\,\, \tau_{\rm c} = R K /\sigma_{\rm T}
\end{eqnarray}
There is an even simpler case, which for reasons outlined below,
is the most common case employed when studying radio--loud AGNs.
In fact, if you imagine to observe the source at the self absorption
frequency $\nu_{\rm t}$, then you are both observing the thick and
the thin flux at the same time.
Then, let us call the flux at $\nu_{\rm t}$ simply $F_{\rm t}$.
We can then re-write the equation above:
\begin{eqnarray}
B \, & \propto &  {\theta_s^4 \nu_{\rm t}^5 \over F^2_{\rm t} }
\nonumber \\
\tau_{\rm c} \, & \propto &\,  {F_{\rm t}\nu_{\rm t}^\alpha \over 
\theta_s^2 B^{1+\alpha} }
\nonumber \\
F_{\rm ssc}(\nu_{\rm c}) \, & \propto & \tau_{\rm c} F_{syn}(\nu_{\rm c}) \,
\propto \tau_{\rm c}^2 B^{1+\alpha} \nu_{\rm c}^{-\alpha}
\nonumber \\
\, & \propto & \,
F_{\rm t}^{2(2+\alpha)} 
\nu_{\rm t}^{-(5+3\alpha)}
\theta_s^{-2(3+2\alpha)}
\nu_{\rm c}^{-\alpha}
\end{eqnarray}
Once again: on the basis of a few observations of only the synchrotron
flux, we can calculate what should be the SSC flux at the frequency
$\nu_{\rm c}$.
Note the rather strong dependencies, particularly for $\theta_s$,
in the sense that the more compact the source is, the larger the SSC flux.

If it happens that we do observe the source at high frequencies,
where we expect that the SSC flux dominates, then we can check if
our model works.
Does it? For the strongest radio--loud sources, almost never.
The disagreement between the predicted and the observed flux
is really severe, we are talking of several orders of magnitude.
Then either we are completely wrong about the model,
or we miss some fundamental ingredient.
We go for the second option, since, after all, we do not find any
mistake in our theory.

The missing ingredient is relativistic bulk motion.
If the source is moving towards us at relativistic 
velocities, we observe an enhanced flux and blueshifted 
frequencies. Not accounting for it, our estimates of the 
magnetic field and particles densities are wrong, in the sense
that the $B$ field is smaller than the real one, and the
particle densities are much greater (for smaller $B$ we need more particle
to produce the same synchrotron flux).
So we repeat the entire procedure, but this time assuming that 
$F(\nu) =\delta^{3+\alpha} F'(\nu)$, where 
$\delta=1/[\Gamma(1-\beta\cos\theta)]$
is the Doppler factor and $F'(\nu)$ is the flux received by a comoving
observer at the same frequency $\nu$.
Then
\begin{eqnarray}
F_{\rm thick}^{\rm syn}(\nu) \, & \propto & \theta_s^2 {\nu^{5/2}
\over B^{1/2}} \,  \delta^{1/2}
\nonumber \\
F_{\rm thin}^{\rm syn}(\nu)\, & \propto & \theta_s^2 RK  B^{1+\alpha} \nu^{-\alpha}
\delta^{3+\alpha} 
\end{eqnarray}
The predicted SSC flux then becomes
\begin{equation}
F_{\rm ssc}(\nu_{\rm c}) \,  \propto \,
F_{\rm t}^{2(2+\alpha)} 
\nu_{\rm t}^{-(5+3\alpha)}
\theta_s^{-2(3+2\alpha)}
\nu_{\rm c}^{-\alpha} \delta^{-2(2+\alpha)}
\end{equation}
If we now compare the predicted with the observe SSC flux,
we can estimate $\delta$.
And indeed this is one of the most powerful $\delta$--estimators,
even if it is not the only one.

\section{Why it works}

\begin{figure}[h]
\vskip -0.8 true cm
\includegraphics[height=9.5cm, width=13cm]{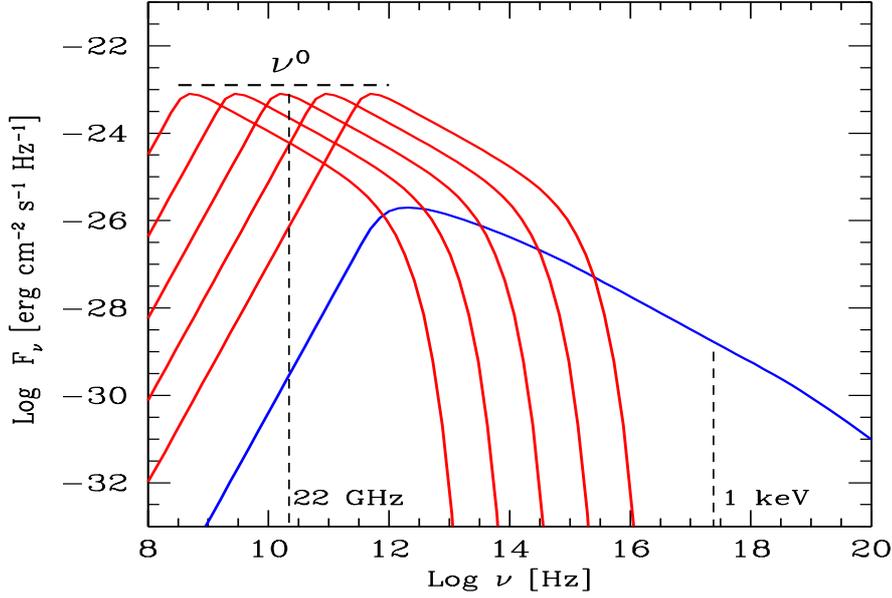} 
\vskip -1 true cm
\caption{Typical example of the composite spectrum of a flat
spectrum quasars (FSRQ) shown in the
$F_\nu$ vs $\nu$ representation, to better see the flat spectrum
in the radio.
The reason of the flat spectrum is that different parts
of the jet contributes at different frequencies, but in a coherent way.
The blue line is the SSC spectrum.
Suppose to observe, with the VLBI, at 22 GHz: in this framework
we will always observe the jet component peaking at this frequency.
So you automatically observe at the self--absorption frequency
of that component (for which you measure the angular size).
}
\label{flat}
\end{figure}

We have insisted on the importance of observing the 
synchrotron flux both in the self--absorbed and in the
thin regime, to get $B$ and $\tau_{\rm c}$.
But the self--absorbed part of the synchrotron spectrum,
the one $\propto \nu^{5/2}$ is very rarely observed in
general, and never in radio--loud AGNs.
So, where is the trick?
It is the following.
In radio--loud AGN the synchrotron emission, at radio frequencies,
comes partly from the radio lobes (extended structures, hundreds of
kpc in size, very relaxed, unbeamed, and usually self--absorbing 
at very small frequencies) and from the jet.
The emission from the latter is beamed, and it is the superposition
of the fluxes produced in several regions:
the most compact ones (closer to the central engine)
self--absorb at high radio frequencies (say, at 100 GHz),
and the bigger they are, the smaller their self--absorbed 
frequency.
But what is extraordinary about these jets is that the peak
flux of each component (i.e. the flux at the self--absorption frequency)
is approximately constant (in the past, this phenomenon was called
{\it cosmic conspiracy}).
Therefore, when we sum up all the components, we have a flat radio
spectrum, as illustrated by Fig. \ref{flat}.

%
%

Of course the emission components of the jet, to behave in
such a coherent way, must have an electron density
and a magnetic field that decrease with the distance from
the central engine in an appropriate way.
There is a family of solutions, but the most appealing
is certainly $B(R)\propto R^{-1}$ and $K(R)\propto R^{-2}$.
It is appealing because it corresponds to conservation of the 
total number of particles, conservation of the bulk power
carried by them (if $\Gamma$ does not change) and conservation
of the Poynting flux (i.e. the power carried by the magnetic field).

To our aims, the fact that the jet has many radio emission sites
self--absorbing at different frequency is of great help.
In fact suppose to observe a jet with the VLBI, at one
frequency, say 22 GHz.
There is a great chance to observe the emission zone which is
contributing the most to that frequency, i.e. the one which 
is self--absorbing at 22 GHz. At the same time you measure the size.
Then, suppose to know the X--ray flux of the source.
It will be the X--ray flux not only of that component 
you see with the VLBI, but an integrated flux from all the 
inner jet (with X--ray instruments the maximum angular resolution
is about 1 arcsec, as in optical).
But nevertheless you know that your radio blob cannot exceed the
measured, total, X--ray flux.
Therefore you can put a limit on $\delta$ (including constants): 
\begin{eqnarray}
\delta \, &>&\, (0.08\alpha+0.14)\, (1+z)\, 
\left( {F_{\rm t} \over {\rm Jy} } \right) 
\left( {F_x \over {\rm Jy} } \right)^{-{1 \over 2(2+\alpha)}}
\left( { \nu_x\over {\rm 1~ keV} } \right)^{-{\alpha \over 2(2+\alpha)}}
\nonumber \\
& \times & 
\left( { \nu_{\rm t}\over {\rm 5~ GHz} } \right)^{-{5+3\alpha\over 2(2+\alpha)}}
\left( { 2\theta_s \over {\rm m.a.s.} } \right)^{-{3+2\alpha \over 2+\alpha}}
\left[ \ln\left({\nu_{\rm s,max} \over \nu_{\rm t} } \right) 
\right]^{1 \over 2(2+\alpha)}
\end{eqnarray}
For some sources you would find $\delta> 10$ or 20, i.e. rather
large values.

\def\gsim{ \lower .75ex \hbox{$\sim$} \llap{\raise .27ex \hbox{$>$}} }
\def\lsim{ \lower .75ex\hbox{$\sim$} \llap{\raise .27ex \hbox{$<$}} }
\def\crexp{{\rm\thinspace km^{2} \thinspace sr \thinspace yr}}
\def\ergcms{{\rm\thinspace erg \thinspace cm^{-2} \thinspace s^{-1}}}
\def\kmps{{\rm\thinspace km \thinspace s^{-1}}}
\def\mpc{{\rm\thinspace Mpc}}
\def\ev{{\rm\thinspace eV}}
\def\kev{{\rm\thinspace keV}}
\def\hi{{\rm H}\,{\small\rm I}}
\def\sc{Schwarzschild}
\def\beq{\begin{equation}}
\def\eeq{\end{equation}}
\def\msun{$M_\odot$}
\def\sc{Schwarzschild}
\def\Omm{{\Omega_m}}
\def\Ommz{{\Omega_m^{\,z}}}
\def\Omr{{\Omega_r}}
\def\Omk{{\Omega_k}}
\def\Oml{{\Omega_{\Lambda}}}

\chapter{Pairs}

In the presence of energetic particles and photons one has to 
wonder about the possibility that there are collisions between them. 
One result of these collisions is the production 
of e$^\pm$.
If we have a photon--photon collision, then the original photons might
disappear, so that this process becomes an important absorption process.

The importance of this process came initially from the realization that
the virial temperature of protons, in the vicinity of black holes, can be very large.
As an estimate, the kinetic energy of a proton at 3 \sc\ radii is
\begin{equation}
E_{\rm kin}\, =\, {GM m_{\rm p} \over R} \, \simeq
150\, \left( {3R_{\rm S}\over R}\right) \,\, \, {\rm MeV}
\end{equation}
therefore particles can be very energetic in accreting systems.
Be aware, on the other hand, that protons can efficiently give their energy to electrons,
that will emit this energy.
This would keep the protons much cooler than the above value.

The other key ingredient, realized in the `60s, is that Active Galactic Nuclei (AGNs) 
vary quickly (with a variability timescale $t_{\rm var}$).
By the causality argument they cannot be larger than $R\sim c t_{\rm var}$.
Therefore their emitting regions must be small, and yet produce a huge
luminosities. 
Densities are thus very large, and the collisions between particles,
between particles and photons, and between photons must be probable.
The discovery that blazars (AGNs with jets pointing at us) are very efficient 
$\gamma$--ray producers is now a proof that high energy particles indeed exist
in these objects, that can in turn produce pairs.
And finally, also Gamma Ray Bursts (GRBs) are emitting $\gamma$--rays. 

The importance of electron--positron pair production in the nuclei of AGNs was realized quite
early (Jelley 1966) and at the beginning of the `70s' two seminal papers
appeared (one by Bysnovaty--Kogan, Zeldovich \& Sunyaev 1971, the other by Bonometto \& Rees 1971):
the first concerned thermal plasmas, the second was about pairs in non--thermal plasmas.
Thereafter the field was sleepy for about one decade, when the paper by Lightman (1982) and Svensson (1982)
resurrected the interests of the scientific community.

At those time it was believed that pair processes were fundamental for the formation of the spectrum
of {\it all} AGNs, galactic binaries and GRBs. 
Now we know that they do not play such a key role for AGNs and binaries, but they still could for GRBs. 
Nevertheless the entire subject is worth studying, because, at the very least, it allows to pose
robust and important limits on the physics of all compact objects.

We will divide the study into thermal and non--thermal plasmas.
We will see two fundamental results of pairs in thermal plasmas,
and will see in more detail what happens when the particles
have a relativistic (and not--Maxwellian) energy distribution.

\section{Thermal pairs}

Consider particles characterized by a Maxwellian distribution
and a temperature $T$.
Since we are dealing with the production of e$^\pm$, it is convenient to 
measure all energies in units of $m_{\rm e} c^2$ (=511 keV).
Let us define:
\begin{equation}
\Theta   =  {kT\over m_{\rm e} c^2}, \qquad\quad x  = {h\nu \over m_{\rm e} c^2}
\end{equation}
Furthermore, define:
\begin{eqnarray}
n_\pm  &=& {\rm number \, density \, of \, positrons\, (+)\, and \, electrons\, (-)}
\nonumber \\
n_\gamma  &=& {\rm number \, density \, of \, photons}
\nonumber \\
\dot n_\pm  &=& {\rm production\, rate\, of \,  pairs}
\nonumber \\
\dot n_{\rm A}  &=& {\rm rate\, of \, annihilation\, of\, pairs}
\end{eqnarray}
To these processes we should add the possibility that the leptons escape
the source, with a rate 
\begin{equation}
\dot n_{\rm esc}  = {\rm rate\, of \, escape\, of\, leptons}
\end{equation}

The process that can produce e$^\pm$ are:

\begin{eqnarray}
\gamma\gamma\to e^+e^-: \qquad \dot n_+ &=& r_{\rm e}^2 c \, n_{\gamma\gamma}^2 \, F_{\gamma\gamma}
\nonumber \\
\gamma p \to e^+e^- p:  \qquad \dot n_+ &=& \alpha_{\rm F} \, r_{\rm e}^2 c \, n_{\gamma}n_{\rm p} \, F_{\gamma\rm p}
\nonumber \\
\gamma e \to e^+e^- e:  \qquad \dot n_+ &=& \alpha_{\rm F} \, r_{\rm e}^2 c \, n_{\gamma}n_{\rm e} \, F_{\gamma\rm e}
\nonumber \\
e e \to e^+e^- ee:      \qquad \dot n_+ &=& \alpha^2_{\rm F} \, r_{\rm e}^2 c \, (n_+ + n_{-})^2 \, F_{\rm ee }
\nonumber \\
e p \to e^+e^- ep:      \qquad \dot n_+ &=& \alpha^2_{\rm F} \, r_{\rm e}^2 c \, (n_+ + n_{-}) n_{\rm p} \, F_{\rm ep}
\end{eqnarray}
while the processes that annihilate pairs or that correspond to the escape of pairs and electrons are:
\begin{eqnarray}
e^+e^- \to \gamma\gamma:  \qquad \dot n_{\rm A} &=& 2 r_{\rm e}^2 c \, n_+n_{-}   \,F_{\rm A}
\nonumber \\
{\rm Escape}:            \qquad \dot n_{\rm esc} &=& r_{\rm e}^2 c \, (n_+ + n_{-})^2 \, F_{\rm esc}
\end{eqnarray}
The $F$--factors are averages of the energy dependent part of the cross sections over the Maxwellian distribution.
They are dimensionless.
The geometrical part of the cross section is always of the form of $\alpha_{\rm F}^a r^2_{\rm e}$:
$\alpha_{\rm F}=1/137$ is the fine structure constant, and $r_{\rm e}$ is the classical electron radius.
For photon--photon interaction, $a=0$, for particle--photon $a=1$, for particle--particle $a=2$.
Consider that pair production processes {\it have an energy threshold}: 
there must be enough energy to produce a pair.
But, when dealing with a Maxwellian distribution, one has particles (even if a few) at all energies,
so the process can occur even if $\Theta\ll 1$.
Here is a brief summary of the $F$--factors:

\begin{table}[h]
\begin{center}
\begin{tabular}{|l|l|l|l|}
\hline
\hline
 & & & \\
 &$\Theta \ll 1$ &$\Theta\gg 1$ &Comment \\
 & & & \\
\hline
 & & & \\
$F_{\gamma\gamma}$ &${\pi^2 e^{-2/\Theta} \over 8\Theta^3}$   &${\pi \, {\rm ln}\Theta\over 2\Theta^2}$  &Wien\\
 & & & \\
$F_{\gamma\gamma}$ &...                     &${2\over 3} \left( {\Theta \over 2 }\right)^{2\alpha} $ &power law ($\alpha>-1)$\\
 & & & \\
$F_{ee}$           &...           &${112\over 27\pi }\, {\rm ln}^3(2\Theta)$ & \\
 & & & \\
$F_{ep}$           &...           &$ {{\rm ln}^3 (2\Theta) \over \pi} $ & \\
 & & & \\
$F_{A}$            & $\pi$          &$ {\pi \, {\rm ln}\Theta \over 2\Theta^2 }$ & \\
 & & & \\
$F_{\rm esc}$      &$ {8\pi \over 3\tau_{\rm T}}\beta_{\rm esc}$           &$ {8\pi \over 3\tau_{\rm T}}\beta_{\rm esc}$    & \\
 & & & \\
\hline
\hline
\end{tabular}
\caption{Summary of the $F$--factors for the main rates of 
pair production mechanism. 
Only photon--photon and particle--particle processes are considered.}
\label{f}
\end{center}
\end{table}

When dealing with processes involving photons, we have to know their spectral
distribution. 
This is why in Tab. \ref{f} there are the rates for a Wien distribution 
[$ L(x) \propto x^3\exp (-x/\Theta)$], and for a power law [$L(x)\propto x^{-\alpha}$].

\subsection{First important result}

We impose that the source is stationary. 
This implies a detailed balance between the created and the destroyed pairs:
\begin{equation}
\dot n_+ \, =\, \dot n_{\rm A}
\end{equation}
For simplicity, we neglected escape as a way to get rid of pairs (but in reality 
there will always be some escape).
To solve the above equation, we have to know the $F$--terms, that are functions of $\Theta$.
Now the important point:
{\it for particle--particle pair production processes, the $F$--terms are 
increasing functions of the temperature.}
On the contrary, the pair {\it annihilation rate decreases} with the temperature.
Compare this behavior with the ``Klein--Nishina" decline of the scattering cross section.

\begin{figure}[h]
\begin{center}
\includegraphics[height=9cm, width=10cm]{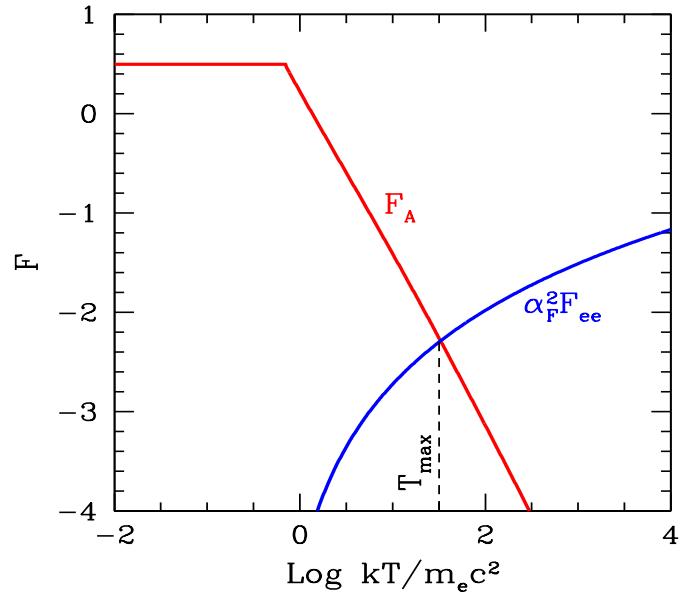} 
\vskip -0.5 true cm
\caption[h]{To illustrate pair balance: the annihilation rate can balance pair production 
at low temperatures, but as $\Theta\equiv kT/m_{\rm e}c^2$ becomes large, the
particle--particle processes increase their rate. 
Above $T_{\rm max}$, annihilation cannot keep up with pair production, pair
balance is impossible, and the number of pairs increases. 
Having more particles to share the available energy it is very likely that
the temperature then decreases, making the system to return to pair balance.
}
\label{chap6:fig1}
\end{center}
\end{figure}

Therefore, if the temperature is high enough, there is the possibility that we produce more pairs
than what we can destroy. 
The source then becomes non--steady.
As a result, if we want a steady source, we require that the temperature cannot be 
greater than a critical value.
This value is:
\begin{equation}
\Theta=24  \quad \to \quad kT_{\rm max} \, =\, 12 \,\, {\rm MeV}
\end{equation}
To see this, consider a plasma that is pair--dominated (i.e neglect the original
electrons associated with protons).  Let us set $n_{\rm e}=n_{+}+n_{-}$.
Neglect also all the particle--photon processes.
Stationarity demands:
\begin{eqnarray}
\alpha_{\rm F}^2 n_{\rm e}^2 F_{\rm ee} + n_\gamma^2 F_{\gamma\gamma} &=& 2 n_{\rm e}^2 F_{\rm A}
\nonumber \\
\rightarrow\,\,  \left( {n_\gamma \over n_{\rm e}} \right)^2  &=& {2 F_{\rm A} -\alpha^2_{\rm F} F_{\rm ee} 
\over F_{\gamma\gamma} }
\, \ge\,  0
\end{eqnarray}
Fig. \ref{chap6:fig1} shows how the two functions $F_{\rm A}$ and $\alpha^2_{\rm F}F_{\rm ee}$ behave.
We require that $F_{\rm A}>\alpha^2_{\rm F}F_{\rm ee}$, since otherwise the annihilation rate cannot keep the
pace with the production rate, the number of particles increases, and if they are at the same temperature,
then they continue to produce even more pairs... (and this would quickly make an explosion).

But in reality what happens is that the energy stored in the system is limited: if the number of particles
increases, then {\it they will share the available energy among themselves}.
There will be less energy per particle $\to$ the temperature will decrease.
This is the second important result, that we will see now in more detail.

{}

\subsection{Second important result}

Usually, when we increase the energy of a system, the temperature increases.
But when $kT$ approaches $m_{\rm e} c^2$, the pair production processes start to
be important. 
{\it New particles are created:} the energy per particle decreases $\to kT$ decreases.

\vskip 0.5 cm
\noindent
{\bf Example: Wien plasmas}
\vskip 0.3 cm
\noindent
The name ``Wien plasma" indicates a plasma in which the photons and particles
are in equilibrium, {\it reached through scattering}.
We must have $\tau_{\rm T}\equiv \sigma_{\rm T}n_{\rm e} R > 1$.
The spectrum of the radiation {\it is not} a black body, because we do not have 
absorption and re--emission, but only scattering.
In these conditions, whatever the process of photon production (i.e. bremsstrahlung)
the spectrum of the radiation is 
\begin{equation}
\dot n_\gamma(\epsilon) \propto \epsilon^2 \exp(-\epsilon/kT) \quad \to \quad
\dot n_\gamma(x) \propto x^2 \exp(-x/\Theta) 
\end{equation}
where $\epsilon=h\nu$.
$\dot n_\gamma(\epsilon)$ is the photon production rate.
To convert it in a density of photons, we use a very simplified equation of
radiative transport:
\begin{equation}
n_\gamma(x) =\dot n_\gamma(x) \, {R\over c} \, (1+\tau_{\rm T})
\end{equation}
where $R$ is the size of the region.

Pause and observe the above equation: when $\tau_{\rm T}\ll 1$, the
difference between $n_\gamma$ and $\dot n_\gamma$ is the light crossing
of the photon inside the source. 
It is the time needed for a photon created in the center to escape.
When $\tau_{\rm T}\gg 1$, we have that the ``transit time" is longer
by a factor $\tau_{\rm T}$.
This corresponds to have done $\tau^2_{\rm T}$ scatterings (remember the random walk)
before escaping. The distance travelled between two consecutive scatterings
is the mean free path, namely $R/\tau_{\rm T}$.
Hence the time needed to escape is $\tau^2_{\rm T}  R/(\tau_{\rm T}c) = \tau_{\rm T} R/c$.
With this link between the production rate and the density of photons, we
can write the luminosity
\begin{eqnarray}
L \, &=& \, {4\over 3} \pi R^3 \int \epsilon \dot n_\gamma (\epsilon) d\epsilon
\nonumber \\
&=& \, {4 \pi R^3 \over 3 R/c} \int \epsilon  {n_\gamma (\epsilon) \over 1+\tau_{\rm T}} d\epsilon
\end{eqnarray}
When $\tau_{\rm T}\gg 1$, we can set $\tau_{\rm T}+1 \sim \tau_{\rm T}= \sigma_{\rm T}R n_{\rm e}$.
In this case
\begin{equation}
L\, = \, {4 \pi R c \over 3 \sigma_{\rm T} } \, \langle\epsilon\rangle \, {n_\gamma \over n_{\rm e}}
\end{equation}
where $\langle\epsilon\rangle$ is the mean energy of the photons.
Note these remarkable facts:
\begin{enumerate}
\item 
$L$ depends on the ratio $n_\gamma/n_{\rm e}$.
For pair plasmas (i.e. when pairs outnumber original electrons), this ratio
is fixed by the $F$ terms.

\item
For a Wien spectrum, $\langle \epsilon \rangle= 3 kT$ (and then $\langle x \rangle= 3 \Theta$).

\item
For large luminosities, the dominating pair production process  
is photon--photon.

\item
Therefore $n_\gamma /n_{\rm e}$ is fixed by $(F_{\rm A}/F_{\gamma\gamma})^{1/2}$. From 
Tab. \ref{f}, assuming $\Theta\ll 1$, we have 
\begin{eqnarray}
{n_\gamma \over n_{\rm e}} \, &=& \, \left( { F_{\rm A} \over F_{\gamma\gamma} }\right)^{1/2} 
= \left( { 8\pi \Theta^3 \over \pi^2 e^{-2/\Theta} } \right)^{1/2}
\nonumber \\
{n_\gamma \over n_{\rm e}} \, &=& \, 2 \sqrt{2/\pi } \,\, \Theta^{3/2}\,\, e^{1/\Theta}
\end{eqnarray}
\end{enumerate}
%

\begin{figure}[h]
\includegraphics[height=9cm, width=10cm]{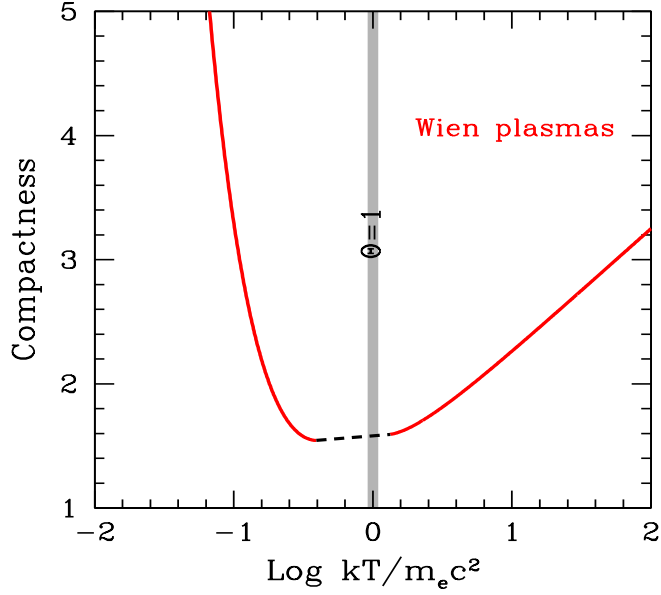} 
\vskip -0.5 true cm
\caption[h]{
Wien plasmas: assuming that the pairs outnumber original electrons,
there is a fixed relation between the compactness and the temperature.
Large compactnesses (and then luminosities, for a given size),
are possible only if the temperature is small.
}
\label{chap6:fig2}
\end{figure}{}

We can see how the luminosity of a pair dominated, steady Wien source depends on the temperature:
\begin{eqnarray}
L\, &=&\,  {4\over 3} \pi {Rc\over \sigma_{\rm T}}\, 
3\, \Theta \, m_{e}c^2  \, 2 \sqrt{2/\pi } \,\, \Theta^{3/2}\,\, e^{1/\Theta}
\nonumber \\
\Rightarrow 
L \, &\propto & \,  \Theta^{5/2}\,\, e^{1/\Theta}
\end{eqnarray}
For a given $R$, {\it high luminosities are possible only for low temperatures!}
Introducing the compactness
\begin{equation}
\ell \, \equiv \, {\sigma_{\rm T} L \over R m_{\rm e} c^3}
\end{equation}
the luminosity--temperature relation has a simple form:
\begin{equation}
\ell_{\rm Wien} \, =\, 16 \pi \sqrt{2/\pi }\, \,  \Theta^{5/2}\,\, e^{1/\Theta}\, ,
\quad\qquad \Theta\ll 1
\end{equation}
Fig. \ref{chap6:fig2} shows $\ell_{\rm Wien}$ as a function of $\Theta$.
Note the minimum for $\theta\sim 1$. 
The dashed line  corresponds to the region where the asymptotic approximations are invalid.

\section{Non--thermal pairs}

\begin{figure}[h]
\includegraphics[height=8cm, width=11cm]{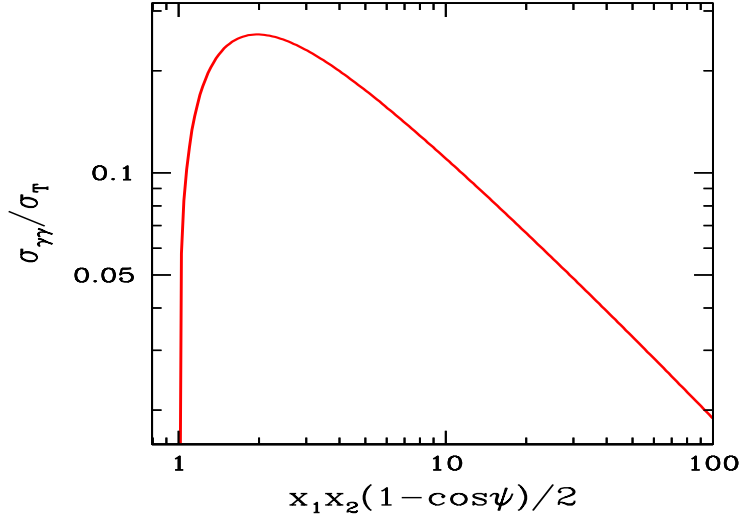} 
\vskip -0.5 true cm
\caption[h]{The photon--photon cross section, as a function
of $x_1x_2(1-\cos\psi)$, where $\psi$ is the angle between the direction of the
two photons (i.e. $\psi=180^\circ$ corresponds to head on collisions).
The cross section is in units of the Thomson cross section $\sigma_{\rm T}$.
}
\label{sigmagg}
\end{figure}{}

In non--thermal plasmas the density of the particles is small, and the particle
energies can be large.
If so, the production rate is usually large, corresponding to a large
density of very energetic photons even if the plasma is rarefied.
In these conditions the main pair production process is photon--photon.

The threshold energy for pair production is ($x\equiv h\nu /m_{\rm e}c^2$):
\begin{equation}
x_1x_2 \ge {2 \over 1-\cos\theta}
\end{equation}
where $\theta$ is the angle between the two photons: $\theta=2\pi$ means head on collisions.
In this case the threshold is $x_1x_2\ge 1$.
For tail on encounters ($\theta=0$) the process cannot occur.

Naively, one would have thought that the threshold condition were $x_1+x_2>2$, namely that the
{\it sum} of the photon energies is larger than $2m_{\rm e}c^2$.
But the threshold condition {\it must be calculated} in a reference frame where the energies
of the two photons are equal.
To illustrate this point, consider the simple case of an head on collision,
with $x_1 \ne x_2$.
Let us go in a frame where $x_1^\prime = x_2^\prime$.
To do that, let us go in a reference frame having a Lorentz factor $\Gamma$ such that
the two energies become equal.
In this frame the angle would still be $2\pi$ (head on collisions), and the two energies will be
$x_1^\prime= x_1/(2\Gamma)$ and $x_2^\prime= 2\Gamma x_2$.
In this frame we require that
\begin{eqnarray}
x_1^\prime &+& x_2^\prime \, \ge \, 2 \qquad {\rm and}\qquad  x_1^\prime = x_2^\prime \qquad  
\Rightarrow  2\Gamma = \left( {x_1 \over x_2}\right)^{1/2} 
\nonumber \\
{x_1\over 2\Gamma} &+& 2\Gamma x_2  \, \ge\,  2 \quad \Rightarrow \quad x_1x_2 \, \ge \, 1
\end{eqnarray}
The cross section resembles the Klein--Nishina cross section, and it has a peak
for $x_1x_2 \sim 2$ where its value is
\begin{equation}
\sigma_{\gamma\gamma} \sim {\sigma_{\rm T} \over 5}, \qquad  (x_1x_2\sim 2)
\end{equation}
The cross section is of course zero below threshold, and decreases after the peak.
This means that a $\gamma$--ray photon with energy $x$ can interact with all photons 
with energy $x_{\rm T}>1/x$, but it will interact preferentially with those photons
near its threshold, i.e. $x_{\rm T} = 1/x$  (here the subscript ``T" stands for ``target").

\subsection{Optical depth and compactness}

We want to calculate the optical depth $\tau_{\gamma\gamma}$ for a source of a given luminosity.
Assume that this luminosity is at high frequencies.
A photon with energy $\sim$0.5--1 MeV will interact with photons of the same energy (i.e. 0.5--1 MeV)
to produce pairs (photons of 5 MeV will have $x=10$ and will interact mainly with photons of
energy $x_{\rm T}\sim 1/10$, i.e. 50 keV. 

Like all the optical depths, $\tau_{\gamma\gamma}$ depends on the cross section, the density of the 
targets, and the size for which the process can occur.
The density of targets is
\begin{eqnarray}
&~& n_\gamma({\rm 1\, MeV}) \simeq { U_{\rm rad}({\rm 1\, MeV}) \times {\rm 1\, MeV} \over {\rm 1\, MeV} }\,
\simeq {L_X \over 4\pi R^2 c m_{\rm e} c^2 }
\nonumber \\
&\Rightarrow & \tau_{\gamma\gamma}({\rm 1\, MeV}) 
\simeq {\sigma_{\rm T}\over 5} \, R\, { L_X \over 4\pi R^2 m_{\rm e}c^3 }
\, =\, {\sigma_{\rm T}\over 20\pi } \, {L_X \over R m_{\rm e}c^3 }
\end{eqnarray}
Therefore $\tau_{\gamma\gamma}$ depends (apart from the numerical factor) upon the so--called
{\it compactness} parameter $\ell$, defined as:
\begin{equation}
\ell \, \equiv \, {L \over R m_{\rm e}c^3 }
\end{equation}
and we have:
\begin{equation}
\tau_{\gamma\gamma}({\rm 1\, MeV}) \, \simeq \, {\ell \over 60} 
\end{equation}
If we measure $L$ in Eddington units, and $R$ in \sc\ radii, we have
\begin{eqnarray} 
&~& \ell \, = \, {2\pi \over 3} \, { m_{\rm p} \over m_{\rm e} }  \, { L\over L_{\rm Edd} } \, 
{R_{\rm S}\over R}
\nonumber \\
&\Rightarrow& \, \ell_{\rm Edd} \, = \, \ell_{\rm max} \, \simeq \, 10^4
\end{eqnarray}

\subsection{Photon--photon absorption}

Let us assume that a compact source produces a spectrum with a power law shape:
$F(x)\propto x^{-\alpha}$, where $\alpha$ is the energy spectral index.
Assume also that this spectrum extends above the threshold for pair production.
We want to calculate the optical depth for the $\gamma$--$\gamma \to e^\pm$ process.

\begin{figure}[h]
\includegraphics[height=9cm, width=10cm]{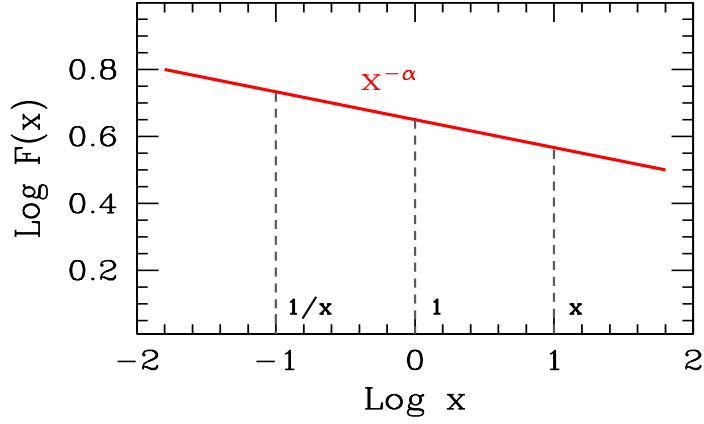} 
\vskip -2.2 true cm
\caption[h]{Photons of energy $x>1$ can interact with photons of energy $x_{\rm T}>1/x$
to produce pairs. Since the cross section is decreasing when increasing the energy of
the targets, and since for a power law spectrum with $\alpha>0$ the number of photons
is also decreasing when increasing their energy, we can approximate the number of
target photons as $(1/x)\cdot n(1/x)$.
Note the ``$x$--$1/x$" symmetry when we plot $F(x)$ in a logarithmic scale.
}
\label{chap6:fig3}
\end{figure}{}

Consider a $\gamma$--ray photon with $x>1$.
First we need to know the number density of the targets. 
These are all photons of energy $x_{\rm T}> 1/x$. 
On the other hand, we know that the cross section decreases when
increasing the energy of the targets.
In most cases, furthermore, we deal with spectral indices $\alpha>0$, 
meaning that also the number of photons decreases with energy
(but be aware that this ``rule" has an important exception, since several Gamma Ray Bursts
have very flat spectra).

We then adopt the following important simplification:
\begin{eqnarray} 
\tau_{\gamma\gamma} \, &=& \, \sigma_{\gamma\gamma} R \int_{1/x}^\infty n(x)dx
\nonumber \\
&\sim& \, {\sigma_{\rm T} \over 5} R \, {1\over x} \, n\left( {1\over x} \right)
\nonumber \\
&\propto& \,  {1\over x} \, { F(1/x)\over x}\, \propto \,  {1\over x} \, \left( {1\over x}\right)^{-\alpha-1}\,
\nonumber \\
&\propto& \, x^\alpha
\end{eqnarray}
Therefore $\tau_{\gamma\gamma}$ increases increasing $x$: there are more targets (if $\alpha>0$).
Consider also that the spectral index $\alpha$ {\it is the spectral index of the targets}: the same
calculation would hold if we had a broken power law, with the break at --say-- $x=1$.
Let us calculate the resulting spectrum, recalling the simplest equation of radiative
transfer through a slab (a slab, instead of a sphere, is here used for simplicity):
\begin{equation}
I(x) \, =\, {\epsilon(x) R \over \kappa(x) R } \left[ 1-e^{-\tau_{\gamma\gamma}(x)} \right]
\end{equation}
where $I(x)$ is the monochromatic intensity; $\epsilon(x)$ is the emissivity and $\kappa(x)$
is the absorption coefficient.
The little trick that we have used here is to multiply both the numerator and the denominator by $R$,
the size of the source, because in this way $\kappa(x) R=\tau_{\gamma\gamma}(x)$, and this allows to write:
\begin{equation}
I(x) \, =\, \epsilon(x) R 
\left[ { 1-e^{-\tau_{\gamma\gamma}(x)} \over \tau_{\gamma\gamma}(x)}\right]
\end{equation}
Written in this way, it is easy to see the behavior when $\tau_{\gamma\gamma}(x)$ is much smaller
or much larger than unity:
\begin{eqnarray} 
\tau_{\gamma\gamma}(x) \, &\ll 1& \quad \Rightarrow \quad I(x) \, =\, \epsilon(x) R
\,  \propto \, x^{-\alpha}
\nonumber \\
\tau_{\gamma\gamma}(x) \, &\gg 1& \quad \Rightarrow \quad I(x) \, =\, {\epsilon(x) R \over \tau_{\gamma\gamma}(x) }
\,  \propto \, x^{-2\alpha}
\end{eqnarray}
There is a well defined break in the spectrum, occurring at the energy
$\tilde x$ for which  $\tau_{\gamma\gamma}(\tilde x) \sim 1$...

\centerline{\bf \Large BUT}
\noindent
this is not the end of the story.

\begin{figure}[h]
\includegraphics[height=9cm, width=10cm]{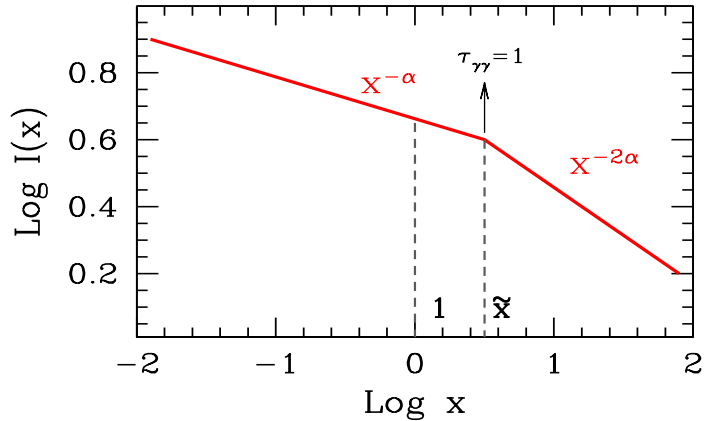} 
\vskip -2.2 true cm
\caption[h]{The expected spectrum after photon--photon absorption,
but without reprocessing.
The initial spectrum is a single power law of index $\alpha$.
Absorption produces a break at the energy $\tilde x$, where the
optical depth $\tau_{\gamma\gamma}$ is unity.
After the break, the spectrum becomes $\propto x^{-2\alpha}$.
}
\label{chap6:fig4}
\end{figure}{}

In fact the produced pairs emit, and contribute to the spectrum (if they remain into the source).
In general, each pair of electron and positron will share the energy of the photon creating them:
(i.e. the electron and positron will have $\gamma m_{e}c^2\sim h\nu/2$, i.e. $\gamma=x/2$).
Therefore the radiation they can produce will be at energies smaller than $x$.
Depending on the energies, the secondary photons can themselves pair--produce.
Alternatively, the secondary photons can be used as targets, enhancing the 
rate of the process, and the number of pairs that the source can produce.
Therefore the pair production process is

\centerline{\bf \large highly non linear}
\noindent
photons that produce pairs that produce photons that produce pairs...

\subsection{An illustrative case: saturated pair production}

If the source is compact, meaning that the compactness parameter $\ell>1$,
the photon density is huge. 
This means that the cooling timescale for inverse Compton scattering is very short, 
shorter than the light crossing time [remember that $\ell$ measures not only
the importance of the $\gamma$--$\gamma \to e^\pm$ process, but also the ratio
$t_{\rm cool}/(R/c)$, see Eq. 4.38].
Therefore, since all leptons cool in a short time,
to have a steady source we have to replenish the ``dead leptons" with new ones.
In general, we have two ways to do that.

The first is to have ``heating". 
With this word we mean that, besides cooling, leptons are also heated 
by some mechanism.
Then there is a competition between energy losses and energy gains.
Since the cooling rate for inverse Compton is $\propto \gamma^2$, it is very
likely that cooling ``wins" at high energies, and heating at low energies.
There will be a specific energy where heating and cooling balance.
All leptons will quickly move (in the energy space) towards this energy.
In the absence of diffusion (in the energy space) the leptons will form
a monoenergetic distribution.
Diffusion will broaden this, and we may have the formation of a Maxwellian particle 
distribution even if the particles do not directly exchange energy between them.

The second way to ``refill" the source with fresh energy is to continuously inject
new energetic particles throughout the source.
A shock does exactly that: at the shock front the particles are accelerated
in a time that must be shorter than the cooling time.
Particles are therefore continuously injected downstream, where they cool.
Proton--proton collisions is another example (but also photon--photon collisions
producing pairs can do).
There can be differences among these processes: for instance, we expect a stratification 
in the downstream region of a shock, with the most energetic particles closer to the shock front,
while we might expect a more homogeneous distribution of particles following
the injection through proton--proton or photon--photon.
We will neglect these differences in the following, and assume, for simplicity,
that the injection and cooling mechanisms are homogeneous.

Let us assume that we inject $Q(\gamma)$ leptons per second and per cm$^3$:
\begin{equation}
Q(\gamma)  \, =\, { \rm \left[ { number \, of\, particles \over cm^3\,\, s } \right] \quad  \to\quad  injection}
\end{equation}
This {\it is not} the quantity we need to calculate the spectrum. We want the {\it density}:
\begin{equation}
N(\gamma)  \, =\, { \rm \left[ { number \, of\, particles \over cm^3} \right] \quad  \to\quad  density}
\end{equation}
The injection rate and the density are linked by the continuity equation.
In its simplest form it reads:
\begin{equation}
{\partial N(\gamma) \over \partial t} \, =\, {\partial \over \partial \gamma}
\left[ \dot\gamma N(\gamma) \right] \, +\, Q(\gamma) 
\end{equation}
It describes how $N(\gamma)$ evolves in time. 
If we want a steady source, we require $\partial N(\gamma)/\partial t=0$.
This allows to obtain the solution:
\begin{equation}
N(\gamma) \, =\, {\int_\gamma^{\gamma_{\rm max}} Q(\gamma^\prime) d\gamma^\prime 
\over \dot \gamma } 
\end{equation}
To go further let us assume that:

\noindent 
1) the injection rate is a power law in energy: $Q(\gamma)=Q_0\gamma^{-s}$; 

\noindent
2) the cooling rate is quadratic with energy: $\dot\gamma\propto \gamma^2$;

\noindent
3) all particles cool down to $\gamma=1$, and then ``disappear" (i.e. we neglect the
accumulation of ``dead leptons").

We have  2 cases:
\begin{eqnarray} 
1)\,\,\, &s<1& \quad \Rightarrow \qquad N(\gamma)\propto \gamma^{-2} 
\nonumber \\
2)\,\,\, &s>1& \quad \Rightarrow \qquad N(\gamma)\propto \gamma^{-(s+1)} 
\end{eqnarray}
If we inject a monoenergetic $Q(\gamma)=Q_0\, \delta(\gamma-\gamma_{\rm max})$
we have $N(\gamma)\propto \gamma^{-2}$.

We have learnt that if $N(\gamma)$ is a power law, the emitted spectrum will also be
a power law, with a well defined link between the slope $p$ of $N(\gamma)$ and the 
radiation spectrum $\alpha$.
We have $\alpha =(p-1)/2$.

Now we have all the ingredients to see the effects of the {\it reprocessing of pairs}
on the emitted spectrum.
To fix the ideas, consider a region of radius $R$ close to an accretion disk emitting UV radiation.
Assume to continuously inject, throughout this region, relativistic leptons.
Just to make it simpler, consider a monoenergetic injection at some $\gamma=\gamma_{\rm max}$.
Assume that the inverse Compton process is the dominant one. 
Let us see what happens in a few steps:
\vskip 0.3 cm
\noindent
1) After cooling, the particle density $N(\gamma)\propto \gamma^{-2}$, and 
the inverse Compton spectrum will have $L(x)\propto x^{-0.5}$.
Assume that the maximum photon energy ($x_{\rm max}$) is well above 
the threshold for pair production, and that the compactness is very large
for all photons with $x>1$.
In these condition, all high energy photons get absorbed and produce pairs
({\it saturated pair cascade}).

\vskip 0.3 cm
\noindent
2) The generation of pairs means that, besides $Q(\gamma)$, there is another
injection term. Call it $P_1(\gamma)$.
We know the slope of $P_1(\gamma)$.
In fact it will be the same of the photon spectrum, whose {\it energy} index is $\alpha=0.5$, and
therefore its {\it photon} index is $\alpha+1=1.5$.
\begin{eqnarray} 
P_1(\gamma)\, &=& \, 2\dot n_\gamma(2\gamma)\, \propto {L(x)\over x},  \quad x>1
\nonumber \\
P_1(\gamma)\, &\propto& \, \gamma^{-\alpha-1} \, =\, \gamma^{-1.5}
\end{eqnarray}

\vskip 0.3 cm
\noindent
3) This new injection term originates a corresponding density of the first generation of pairs:
\begin{equation}
N_1(\gamma) \, \propto \, {\int_\gamma^{\gamma_{\rm max}} P_1(\gamma^\prime) d\gamma^\prime 
\over  \gamma^2 } \, \propto \gamma^{-2.5}
\end{equation}
And consequently the radiation spectrum due to this first generation of pairs will have a
spectral index
\begin{equation}
\alpha_1 \, =\, {2.5-1\over 2}\, =\, 0.75
\end{equation}
%

\begin{figure}[h]
\vskip -0.5 true cm
\includegraphics[height=10cm, width=11cm]{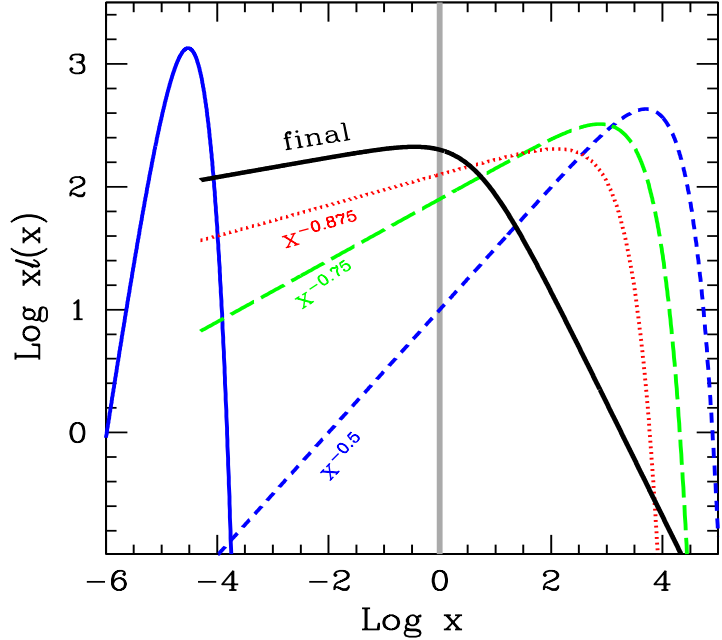} 
\vskip -0.8 true cm
\caption[h]{Reprocessing due to pairs. 
A source produces a soft photon distribution (here it is a black body, blue solid line).
The relativistic primary leptons injected throughout the source
produce, by inverse Compton, a primary high energy spectrum $\propto x^{-0.5}$ (blue short--dashed line).
All photons above threshold get absorbed and transformed in pairs.
They emit, and the spectrum of this first generation of pairs 
is $\propto x^{-0.75}$ (green long--dashed line).
Again, the photons above thresold are absorbed and create the second generation of
pairs, whose spectrum $\propto x^{-0.875}$ (dotted red line).
The process can continue, and the limiting spectrum has a spectral index close
to unity.
The thick black solid line shows how the final spectrum would look like: it is a broken power
law with indices $\sim$1 and $\sim$2 below and above $x\sim 1$.
}
\label{chap6:fig4}
\end{figure}{}

\vskip 0.3 cm
\noindent
4) The cycle repeats. If pairs have enough energy to emit above threshold, their
photons will get absorbed, and generate the second generation of pairs.
\begin{eqnarray} 
P_2(\gamma)\, &\propto& \, \gamma^{-\alpha_1-1} \, =\, \gamma^{-1.75}
\nonumber \\
N_2(\gamma) \, &\propto& \, {\int_\gamma^{\gamma_{\rm max}} P_2(\gamma^\prime) d\gamma^\prime 
\over  \gamma^2 } \, \propto \gamma^{-2.75}
\nonumber \\
\alpha_2 \, &=& \, {2.75-1\over 2}\, =\, 0.875
\end{eqnarray}

\vskip 0.3 cm
\noindent
5) If these second--generation pairs have enough energy to produce photons above threshold,
we will have a third generation of pairs, and so on.
A very interesting result of this exercise is that the final spectral index
tends towards unity.
\begin{equation}
\alpha_{i+1} \, =\, {\alpha_i\over 2} +{1\over 2}
\end{equation}
where $i$ stands for the $i$--th generation of pairs.
The equation above is valid even if we start with a steep spectrum ($\alpha>1$).
But in this case the reprocessed energy will be a minor fraction of the total:
most of the power is emitted at low frequencies.
So, even if we correctly derive the slope of the reprocessed spectrum,
this will be energetically unimportant.

If the initial spectrum is flat (i.e. $\alpha<1$), instead, the reprocessing due
to pairs can be very important, and will have the effect to soften the
overall spectrum, redistributing photons at lower frequencies.

\subsection{Importance in astrophysics}

The pair production processes were intensively studied in the `80s, especially
in the field of Active Galactic Nuclei (AGNs).
Researchers, at that time, hoped that the reprocessing due to pairs could
explain the ``universality" of the X--ray spectrum of AGNs, that has $\alpha_x\sim 0.7$.
This would have implied a non--thermal origin of the spectrum, required to extend
much above threshold.
In other words, it was though that the typical AGN spectrum would resemble the one shown in 
Fig. \ref{chap6:fig4}. 
It was then predicted that all AGNs, i.e. also the radio--quiet ones, emit in the $\gamma$--rays,
with a relatively steep power law shape.
The first results obtained at high energies, coming from the {\it Compton Gamma Ray Observatory, CGRO} 
satellite showed instead an exponential cutoff at a few hundreds keV, and no signs of $\gamma$--rays
in radio--quiet AGNs.
It was then concluded that 1) the X--ray spectra of radio--quiet AGNs have a thermal origin
(i.e. Comptonization) and 2) that pairs have no or little role in the formation of this spectrum.

\vskip 0.3 cm

On the other hand, the main surprise of {\it CGRO} was the discovery that radio--loud
AGNs are strong $\gamma$--ray emitters, as a class (i.e. we believe that {\it all} of them do).
The properties of the observed spectra tell us that, again, photon--photon absorption
inside the source rarely occurs (we do see a lot of $\gamma$--rays...).
The absence of this effect poses a problem, because the high energy flux of these
sources is varying quickly, requiring very compact sources, and the observed
large luminosities imply very large values of the compactness parameter.
The process {\it should} indeed occur, and yet it does not.
The solution of this puzzle is relativistic motion, shortening variability timescales,
blue--shifting the observed frequencies, and enhancing the apparent luminosities.
Indeed, requiring that the sources are optically thin for photon--photon absorption,
we can derive a limit on the bulk Lorentz factor of the emitting sources.

\vskip 0.3 cm

Another interesting field where pairs can be important is Gamma Ray Bursts (GRBs).
Again, these sources emit high energy radiation that varies quickly.
Since we now know that they are cosmological, the implied powers
are huge, as well as the estimated photon densities.
As in radio--loud AGNs, we can estimate a limit to their bulk Lorentz
factor requiring that they are optically thin to the photon--photon absorption
process.

\vskip 0.3 cm

At this point you may ask if pairs are indeed produced in some class of sources...
The answer is yes, even if the reprocessing features outlined above
do not have (yet) a clear application to specific sources.
At the beginning of the `90s, it was possible to detect TeV photons from the ground,
through optical Cherenkov telescopes.
Soon, low power blazars (i.e. BL Lac objects) were found to strongly emit at
these frequencies.
Some of them are distant enough that the interaction of TeV photons with 
the cosmic IR background photons is not negligible.
We do see signs of the corresponding absorption.
The importance of the latter depends, of course, of the density
of the target photons, i.e. the level of the cosmic IR background.
Since the latter is not precisely known, one can use the absorption features of the
observed spectrum to estimate the background itself.

\vskip 1 cm
\noindent
{\bf References}

\noindent
Bisnovatyi--Kogan G.S., Zeldovich Ya.B. \& Syunyaev R.A., 1971, Soviet Astr., 15, 17

\noindent
Bonometto S. \& Rees M.J., 1971, MNRAS, 152, 21	
	
\noindent
Jelley J.V., 1966, Nature, 211 472

\noindent
Lightman A.P., 1982, ApJ, 253, 842 

\noindent
Svensson R., 1982, ApJ, 258, 321

\noindent
Svensson R., 1982, ApJ, 258, 335

%
%
%



\chapter{Active Galactic Nuclei}

\section{Introduction}

Up to the seventeenth century, the unaided eye was the only receiver that 
humanity could use to observe the Universe.
Evolution was able to apt this ``instrument" to be sensitive to
the light of the star we happen to be orbiting, the Sun.
The invention of the telescope amplified the sensitivity of the eye
and its angular resolution, letting humanity discover, less than
a century ago, that other galaxies exist, far beyond out Milky way, 
and that these galaxies are moving apart: the Universe expands.
However, all we could discover using the eye and its extension, the
optical telescope, concerned a tiny, very tiny, part of the entire
electromagnetic spectrum. 
As soon as technology enabled humanity to open new windows, we discovered
other phenomena, other objects, and could push our knowledge farther
out in space and in time.

The opening of the radio window made the sixties the golden decade
for astronomy, with the discovery of the microwave background and of pulsars.
The third great discovery made in that decade was that of quasars.
The term ``quasar" originally stood for ``quasi--stellar radio source", 
and refers
to the fact that when an optical telescope is pointed towards the direction
of some radio source, which can be as extended as minutes of arc in radio maps,
the resulting optical plate shows a source which looks like a star, i.e. a 
not extended, a ``pointlike" object.
This apparently innocuous point is instead 
a gigantic energy plant, able to produce much more power that an entire
galaxy like our own, in a volume which is ridiculously small, if compared with
the dimension of the Galaxy, and comparable with our Solar System.
{\it The process that powers the stars, thermonuclear reactions, is not enough
to power quasars.
}
For this reason we believe that at the core of these sources there is a massive
black hole, with a mass between a million and a billion the mass
of the Sun.
Matter around the hole is attracted by the black hole gravity, it is compressed,
heated, and then radiates.
This was realized quite soon (Salpeter 1964; Zeldovich 1964; Lynden--Bell 1969;
Shakura \& Sunjaev 1973)\footnote{
Lynden--Bell D., 1969, Nature, 223, 690;\\
Salpeter E.E., 1964, ApJ, 140, 796;\\
Shakura N.I. \& Sunjaev R.A., 1973, A\&A, 24, 337;\\
Zeldovich Ya. B., 1964, Dock. Akad. Nauk SSSR, 157, 67
}.

Another major advance came with the opening of the X--ray window, first 
(in the sixties) with rocket experiments pioneered by Riccardo Giacconi, 
Bruno Rossi and others, and then with the first X--ray satellites, in the early 
seventies.
The Uhuru, Ariel 5, HEAO--1 and then {\it Einstein} satellites made clear
that all kinds of quasars were strong X--ray emitters:
at the same time, people started to believe that quasars were 
the major contributors to the cosmic diffuse X--ray background, 
already discovered in 1962 (Giacconi et al. 1962)\footnote{
Giacconi R., Gursky H., Paolini F., Rossi B., 1962 
      Phys. Rev. Lett., 9, 439. Giacconi was awarded the Nobel Prize in Physics in 2002.
}.

A third qualitative ``jump" was the improvements of the interferometric
technique in the radio band, in the early seventies.
Radio telescopes in different continents, looking at the same source,
can resolve details as close as a few tenths of a millisecond of arc.
This enabled us to discover that some radio emitting quasars
have spots of radio emission which are observed to move.
Given the huge distances, this motion corresponds to velocities
that exceed the speed of light.
Far from challenging special relativity, this ``superluminal" motion,
as it is now called, was even predicted before it was observed,
by Martin Rees in 1966\footnote{
Rees M.J., 1966, Nature, 211, 468; Martin Rees was born in 1942. 
He was 24 in 1966.
}, 
and corresponds to real fast motion (but
slower than the velocity of light!) at an angle close to our 
line of sight.

\section{The discovery}
\subsection{Marteen Schmidt discovers the redshift of 3C 273}

From Maarteen Schmidt (Schmidt 1990)\footnote{
Schmidt M., 1990, in Modern Cosmology in Retrospect; (Cambridge University Press)
}:

``The puzzle was suddenly resolved in the afternoon of February 5, 1963,
while I was writing a brief article about the optical spectrum of 3C 273.
Cyril Hazard had written up the occultation results for publication
in Nature and suggested that the optical observations be published in 
an adjacent article.
While writing the manuscript, I took another look at the spectra.
I noticed that four of the six lines in the photographic spectra
showed a pattern of decreasing strength and decreasing spacing from 
red to blue.
For some reason, I decided to construct an energy--level diagram
based on these lines. 
I must have made an error in the process which seemed
to contradict the regular spacing pattern.
Slightly irritated by that, I decided to check the regular
spacing of the lines by taking the ratio of their wavelengths
to that of the nearest line of the Balmer series.
The first ratio, that of the 5630 line to H--$\beta$, was 1.16.
The second ratio was also 1.16. 
When the third ratio was 1.16 again, it was clear that I was looking at
a Balmer spectrum redshifted by 0.16.

.....

I was stunned by this development: stars of magnitude 13 are not
supposed to show large redshift!
When I saw Jesse Greenstein minutes later in the hallway and told him 
what had happened, he produced a list of wavelengths of emission lines
from a just completed manuscript about the spectrum of 3C 48.
Being prepared to look for large redshift, it took us
only minutes to derive a redshift of 0.37.

.....

The interpretation of such large redshifts was an extraordinary challenge.
Greenstein and I soon found that an explanation in terms of gravitational 
redshift was essentially impossible on the basis of spectroscopic arguments.
We recognized that the alternative explanation in terms of
cosmological redshifts, large distances, and enormous luminosities and energies
was very speculative but could find no strong arguments against it.
The results for 3C 273 and 3C 48 were published six weeks later
in four consecutive articles in {\it Nature} 
(Hazard, et al. 1963; Schmidt 1963; Oke 1963;
Greenstein and Matthews 1963)\footnote{
Greenstein J.L. \& Matthews T.A.,1963, Nature, 197, 1041,\\
Hazard C., Mackey M.B. \& Shimmins A.J., 1963, Nature, 197, 1037;\\
Oke J.B., 1963, Nature, 197, 1040;\\
Schmidt M., 1963, Nature, 197, 1040.
}".

\section{Basic components}

The first quasars that were discovered were radio--loud,
but we now know that these radio--loud AGNs are only roughly
the 10\% of all AGNs.
The majority are {\it radio--quiet}.

The basic structure of AGNs includes:
\begin{itemize}

\item A black hole, with $10^6 M_\odot <M < 10^{10} M_\odot$.
It is probably spinning at some level, even if we do not have
secure measurements of the spin value.

\item An accretion disk. Matter with even a small amount of angular
momentum, attracted by the black hole gravity, spirals in and forms
a disk.
This is a major source of power.

\item An X--ray corona, sandwiching the accretion disk. 
It is supposed to be a hot layer, or an ensemble of clumpy
regions particularly active in the inner parts of the disk.

\item 
An obscuring torus located at several parsec from the black hole,
intercepting some fraction of the radiation produced by the disk
and re--emitting it in the infrared.

\item A region of many small clouds at a distance of $\sim 10^{17}$--$10^{18}$ cm
from the hole (i.e. less than a parsec) moving rapidly ($\sim 3000$ km s$^{-1}$).
They intercept $\sim$10\% of the ionizing radiation of the disk, and re--emit
it in the form of lines. Doppler shifts broaden the observed lines.
This is why this region is called {\it Broad Line Region (BLR)}.
\item At larger distance ($\sim$100 pc)
there is another region where less dense clouds are moving, less rapidly.
This is called {\it Narrow Line Region (NLR)}.

\item 
About 10\% of AGNs, besides accreting matter, are able to expel it in two
oppositely directed jets. 
Their direction likely traces the rotational axis of a spinning black hole.
The material inside these jets is moving at relativistic speeds.
Therefore the jet emission is highly beamed, and their appearance depends
on the viewing angle. 
AGNs whose jets are pointing at us are called {\it blazars}.
AGNs whose jets are pointing elsewhere are called {\it radio--galaxies}.

\end{itemize}

\section{The supermassive black hole}

\begin{figure}[h]
\center
\includegraphics[height=7 cm, width=12cm]{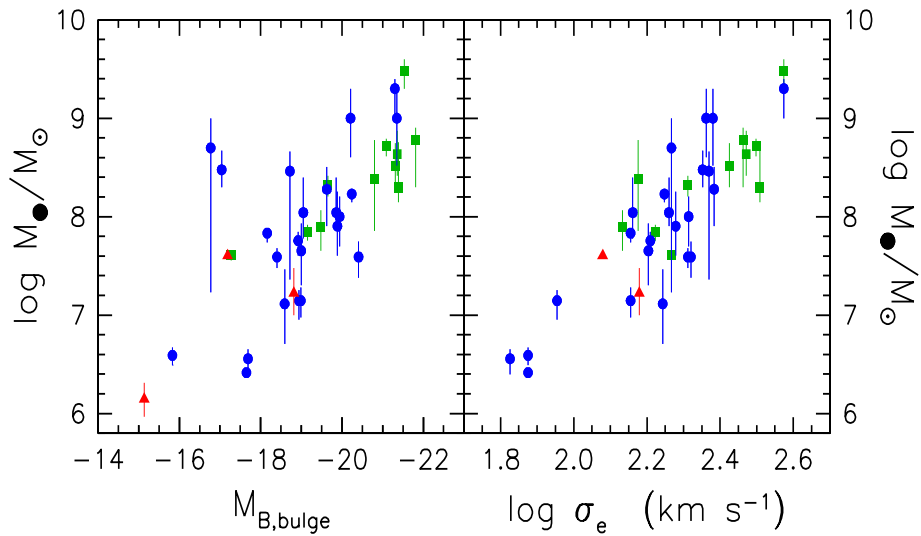} 
\caption[h]{The mass of the supermassive black hole
correlates with the luminosity of the bulge of the host galaxy and with the
velocity dispersion of the stars of the central regions of the 
host galaxy. From http://chandra.as.utexas.edu/$\sim$kormendy/bhsearch.html.
}
\label{bh}
\end{figure}{}

\noindent
The black hole in the center of AGNs has a mass that ``knows" of the
mass of the host galaxy.
In fact there is a correlation between the black hole mass and the
luminosity of the bulge, due to the stars
(see Fig. \ref{bh}, left panel).
The black hole mass correlates even more with the velocity dispersion of
these stars
(see Fig. \ref{bh}, right panel).
The correlation with the bulge luminosity is linear:
$M_{\rm BH}\sim 2\times 10^{-3} M_{\rm bulge}$, while the correlation
with the velocity dispersion $\sigma$ is between
$M_{\rm BH}\propto \sigma^4$ and $M_{\rm BH}\propto \sigma^5$.

\section{Accretion disk: Luminosities and spectra}

\begin{figure}[h]
\center
\includegraphics[height=10cm, width=12cm]{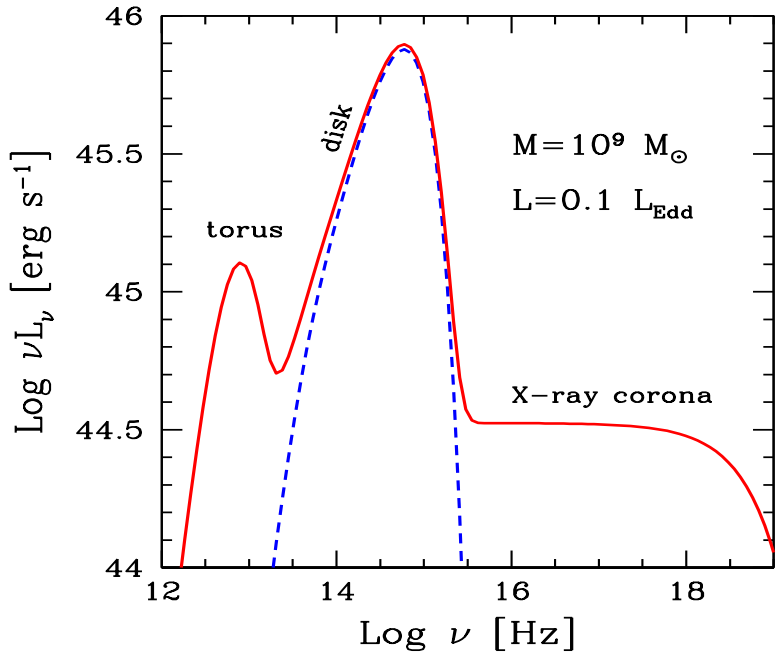} 
\vskip -1 true cm
\caption[h]{Spectrum from a standard accretion disk, the infrared torus
and the X--ray emitting corona.
}
\label{disk}
\end{figure}{}

The luminosity of the accretion disk depends on the amount of the mass
accretion rate $\dot M$. 
It is usually measured in solar masses per year.
The process has a very large efficiency $\eta$, defined as the fraction of rest
mass energy that is converted into radiation, before being swallowed by the black hole.
We have 
\begin{equation}
L_{\rm disk}\, =\, \eta \dot M c^2
\end{equation}
The efficiency $\eta$ depends on the details of the accretion mode,
but it is of the order of $\eta\sim 0.1$.
To produce a disk luminosity of $L_{\rm disk}= 10^{46} L_{46}$ erg s$^{-1}$ we then require
\begin{equation}
\dot M \, = \, {L_{\rm disk}\over \eta c^2 } \, \sim \, 1.75\, L_{46} \,\, M_\odot \, {\rm yr^{-1}}
\end{equation}
There is a limit on the disk luminosity.
This is called {\it Eddington luminosity}.
It corresponds to radiation pressure balancing gravity. 
Radiation pressure acts on electrons, gravity acts both on electrons and
their companions protons, that are heavier.
\begin{equation}
\sigma_{\rm T}\, {L_{\rm Edd}\over 4\pi R^2 c} \,=\, { GMm_{\rm p} \over R^2}
\, \longrightarrow\, L_{\rm Edd} \, \equiv \, {4\pi c GM m_{\rm p} \over \sigma_{\rm T}}
\, =\, 1.3 \times 10^{38} {M\over M_\odot} \,\, {\rm erg\, \, s^{-1}}
\end{equation}
The Eddington luminosity depends only on the black hole mass.
If the accretion rate for some reason produces a luminosity $L_{\rm disk}>L_{\rm Edd}$,
then the accreting mass feels a radiation pressure that contrasts gravity,
and the accretion stops, making $L_{\rm disk}$ to decrease below $L_{\rm Edd}$.
We have a {\it self--regulating} process.

Ask yourself:
\begin{itemize}
\item
What {\it implicit} assumptions have we made to derive $L_{\rm Edd}$?
\item 
Did we neglect some other forces?
\item
Why is there $\sigma_{\rm T}$?
\item
What happens if we have electron--positron pairs?
\item
Can you envisage cases in which the Eddington limit can be avoided?
\end{itemize}

\subsection{Spectrum}

Accreting matter, by spiraling inside, has to dissipate angular momentum
and potential energy.
The potential energy of a proton at distance $R$ (much larger than the
Schwarzschild radius $R_{\rm S} \equiv 2 GM/c^2$) is
\begin{equation}
E_{\rm g} \, = \, {GM m_{\rm p} \over R}
\end{equation}
namely, it increases (or decreases, if you take it negative) as $R$ decreases.
More energy has to be dissipated close to the black hole than far from it.
Besides, the surface $2\times 2\pi r dR$ (there are two faces of the disk...)
of the ring at a distance $R$ and width $dR$ 
becomes smaller and smaller as you move close to the black hole.
More energy has to be dissipated from a smaller surface.
Assuming that the produced disk radiation is a blackbody, this inevitably implies that
the temperature must increase as $R$ decreases.
If all rings of the disk emit black--body radiation, the temperature profile is:
\begin{eqnarray}
T  \, &=&\, \left[ {  3 R_{\rm S}  L_{\rm disk }  \over 16 \pi\eta\sigma_{\rm MB} R^3 }  \right]^{1/4}
\left[ 1- \left( {3 R_{\rm S} \over  R}\right)^{1/2} \right]^{1/4}
\nonumber \\
& \propto & \, R^{-3/4},  \qquad {\rm for}\,\, R \gg 3 R_{\rm S}
\label{temp}
\end{eqnarray}
where $\sigma_{\rm MB}$ is the Maxwell--Boltzmann constant. 
The term in the second square brackets allows for the fact that not all the orbits are stable
around a black hole, but only those beyond the marginally stable orbit: for a Schwarzschild
black hole this orbit is at $3 R_{\rm S}$.
Matter at $R<3R_{\rm S}$ falls directly onto the hole in a short time.
If we measure $R$ in units of the Schwarzschild radius $R_{\rm S}$, $x\equiv R/R_{\rm s}$ 
we have:
\begin{eqnarray}
T  \, &=&\, \left[ {  3 \dot M c^6 \over 64 \pi \sigma_{\rm MB} x^3 G^2 M^2 }  \right]^{1/4}
\left[ 1- \left( {3  \over  x}\right)^{1/2} \right]^{1/4}
\nonumber \\
\, &\propto &  { {\dot M}^{1/4}\over M^{1/2}} \, x^{-3/4}, \qquad {\rm for} \,\, x \gg 3 
\end{eqnarray}
%

\begin{figure}[h]
\center
\includegraphics[height=10cm, width=12cm]{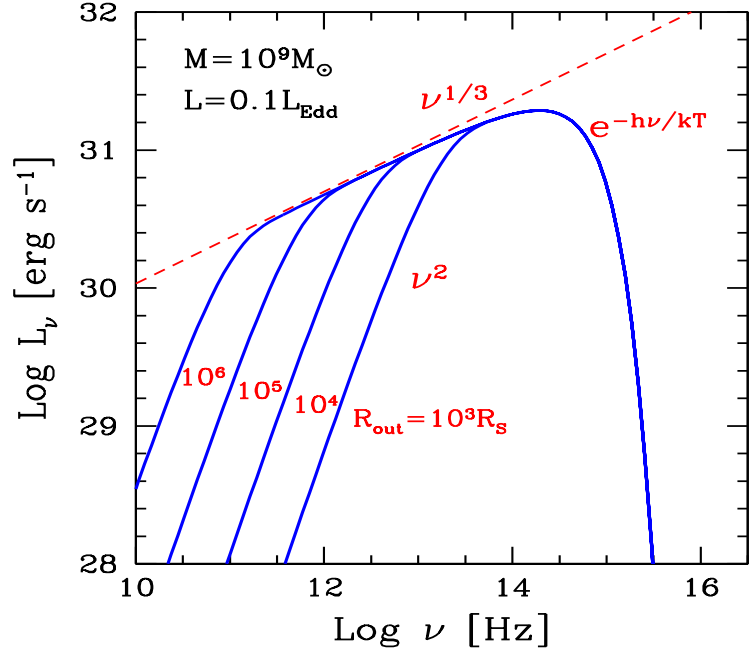} 
\vskip -0.7 true cm
\caption[h]{The spectrum from a standard accretion disk, emitting 
black--body radiation from all its surface, with a temperature profile given
by Eq. \ref{temp}.
The disk extends from $R_{\rm in}=3R_{\rm S}$ to different values
of $R_{\rm out}$, as labelled. Note that the $\nu^{1/3}$ slope 
is present only for very large values of $R_{\rm out}$.
}
\label{disk2}
\end{figure}{}

We can now have an idea of the emitted spectrum.
It will be the sum of black--bodies peaking at different temperatures:
each ring of width $dR$ emits a luminosity $dL$:
\begin{equation}
dL \, = \, 2\times 2\pi \sigma_{MB} R T^4 dR \, \longrightarrow \, {dL \over dR } \, =
\, 2\times 2\pi\sigma_{MB} T^{-4/3} T^4
\end{equation}
We can then set:
\begin{equation}
{dL \over dT} = {dL \over dR } {dR \over dT} \, \propto \,  \sigma_{MB} T^{-4/3} 
T^4 {dR \over dT} \, \propto T^{1/3}
\end{equation}
where we have used $R\propto T^{-4/3}$ and then $(dR/dT)\propto T^{-7/3}$.
This is valid as long as $R\gg 3R_{S}$.
Since there is a one to one correspondence between the temperature and the peak
frequency of the black--body, we have that
\begin{equation}
L_{\rm disk}(\nu) \, \propto \, \nu^{1/3}; \qquad h\nu \ll kT_{\rm max}
\end{equation}
up to a frequency corresponding to the maximum temperature.
Beyond that the emission is not a superposition of black--bodies anymore, and only
the region at the maximum temperature contributes.
Therefore, at large frequencies, we will see an exponential drop.
At the opposite side (small frequencies) we see only the black--body produced
by the outer radius of the disk $R_{\rm out}$, and correspondingly, 
we see the Raleigh--Jeans emission of that radius.
Therefore we can approximate the disk spectrum as:
\begin{eqnarray}
L_{\rm disk} \, & \propto & \, \nu^2, \qquad \qquad\qquad \quad h\nu < kT(R_{\rm out}) 
\nonumber \\
 &\propto &   \nu^{1/3} e^{-h\nu/kT_{\rm max}}, \qquad  h\nu > kT(R_{\rm out}) 
 \end{eqnarray}
The peak is determined by $T_{\rm max}$.
We can estimate it by assuming that most of the luminosity is produced within --say--
$10R_{\rm S}$:
\begin{eqnarray}
L_{\rm disk}  \, \approx \, 2\times 2\pi 10^2  R^2_{\rm S} \sigma_{MB} T^4_{\rm max} \, 
& \longrightarrow & T_{\rm max} \sim  \left({ L_{\rm disk} 
\over 400 \pi R^2_{\rm S} \sigma_{MB}}\right)^{1/4}
\nonumber \\
&\sim & 3.5\times 10^4  \, \left( {L_{46} \over M_9^2}\right)^{1/4}\, {\rm K} 
\end{eqnarray}
where $M_9$ is the black hole mass in units of $10^9 M_\odot$. 
This temperature corresponds to a frequency
\begin{equation}
\nu_{\rm peak}  \, \sim {kT_{\rm max} \over h}\, \sim\, 7.4\times 10^{14} 
\,\left({L_{46} \over M_9^2}\right)^{1/4}\,\, {\rm Hz}
\end{equation}
i.e. in the near ultraviolet. For disks emitting at the same Eddington ratio (i.e.
$L_{\rm disk}/L_{\rm Edd}=$const), we have $\nu_{\rm peak}\propto M^{-1/4}$.
Fig. \ref{disk2} shows the spectrum calculated numerically according to the assumed temperature
distribution of Eq. \ref{temp}, for different values of the outer radius of the disk
$R_{\rm out}$. 
Note that the $\nu^{1/3}$ behaviour can be seen only if $R_{\rm out}$ is very large.

\section{Emission lines}

A distinctive characteristic of AGNs is the presence, in their optical spectra,
of emission lines. They comes in two flavors: {\it the broad and the narrow.}

\subsection{Broad emission lines}

Fig. \ref{disk} shows an atlas (i.e. the average spectrum of many sources) 
of the optical--UV spectrum of typical AGNs.
Typical lines are labelled. The most prominent are the Hydrogen Ly$\alpha$
(transitions from n=2 to n=1), the lines from partially ionized Carbon,
Magnesium, Oxygen and then the Hydrogen H$\alpha$ (transitions from n=3 to n=2).
If you look carefully, you should notice that most of these lines are broad, while 
the [OIII] line is narrow.
The square brackets indicate that this is a ``forbidden" line. 
So why do we see it?
The reason to call this line ``forbidden" is that when the density is larger than 
some critical value, the de--excitation of the excited state is made by collisions, 
occurring faster than the radiative ones. 
In our labs the densities are
always larger than the critical ones, so in the lab we do not see the line.
This gives some hints of the density where the narrow lines are made.

\begin{figure}[h]
\center
\vskip 0.5 cm
\includegraphics[height=9cm, width=12cm]{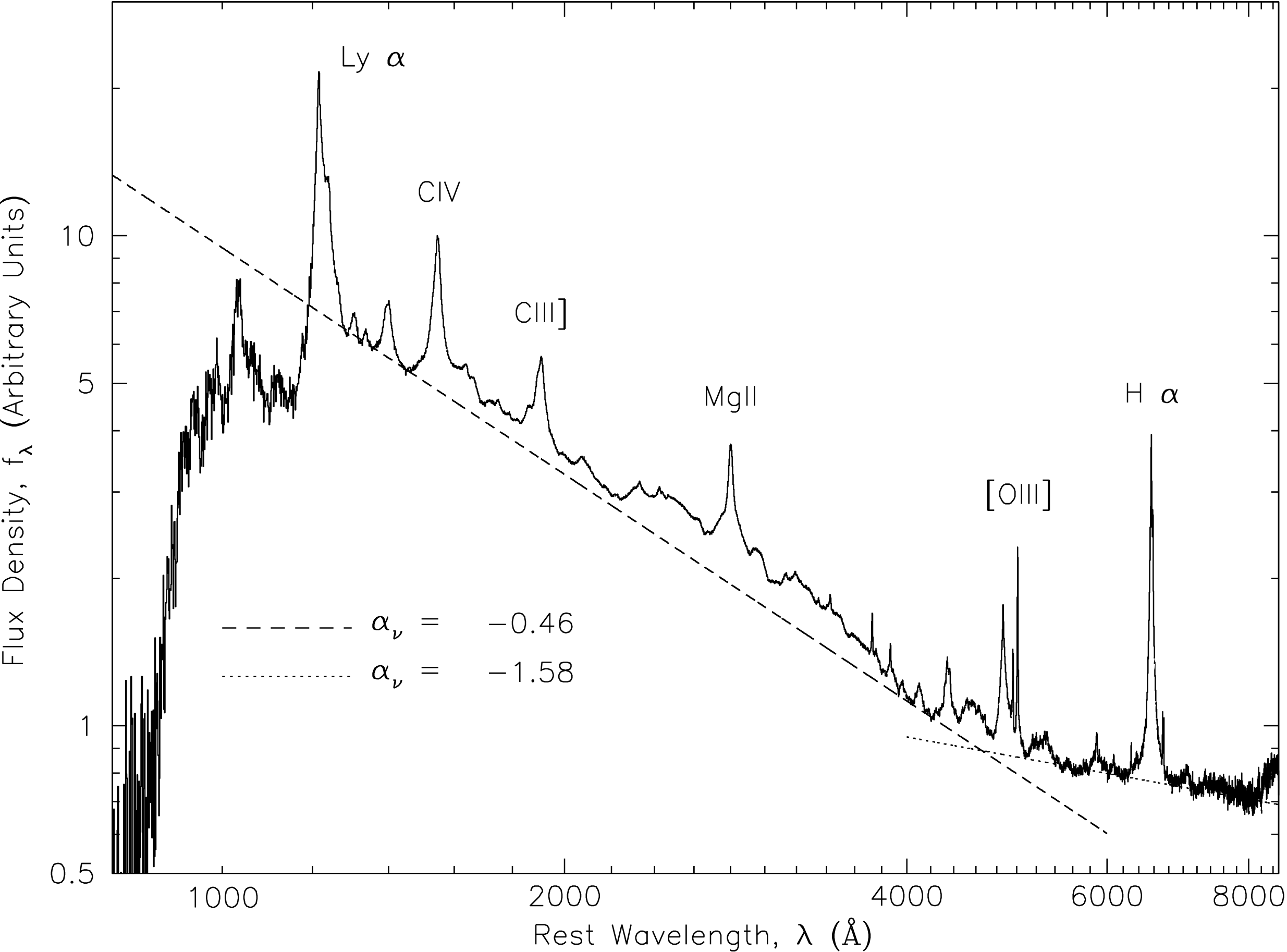}\\
\vskip -0.3 true cm
\caption[h]{Atlas optical/UV spectrum for optically--selected quasars in the Sloan Digital Sky Survey, 
from the work of Vanden Berk et al. (2001, AJ, 122, 549). 
Notice the presence of {\it broad} lines and also of the forbidden [OIII] {\it narrow} line.
}
\label{disk}
\end{figure}{}

The most used way to quantify the width of a line is to measure it
at a flux level which is half of the peak flux.
In this case we speak of Full Width at Half Maximum (FWHM).
Full width means that it is the entire width, not only one half as in the case
of a $\sigma$ of a Gaussian.
The units can be \AA\ or km s$^{-1}$.
In fact the width can be related immediately with a Doppler shift, and thus to a velocity.

The typical values for the broad emission lines are:
\begin{itemize}

\item {\bf FWHM:} from 1000 to 10,000 km s$^{-1}$.  
They are too large to be due to thermal motions. 
In fact, with a temperature of $T\sim10^4$ K, we have that thermal broadening implies:
\begin{equation}
v  \, \sim \left( {kT \over m_{\rm p}}\right)^{1/2} \, \sim\, 10^6 \,\,\, {\rm cm\,\, s^{-1}} 
\, =\, 10  \,\,\, {\rm km\,\, s^{-1}}
\end{equation}
Therefore we need bulk motions (in different directions with respect to our line of sight) 
of the material emitting the lines.

\item {\bf Density:} between $10^9$ and  $10^{11}$ cm$^{-3}$. This comes from the presence or
absence of specific lines.

\item {\bf Temperature:} around $10^4$ K. Again, this comes from the presence or
absence of specific lines.

\item {\bf Covering factor:} around 0.1. A simple estimate comes from the luminosity of all 
the broad lines with respect to the continuum.

\item {\bf Distance:} depends on the luminosity of the ionizing continuum. 
Recent studies use the technique of ``reverberation mapping": taking spectra of a source
regularly, one can see the time delay between a variation of the continuum and the
variations of the lines. From these studies one gets a size:
\begin{equation}
R_{\rm BLR} \sim 10^{17} L^{1/2}_{\rm ion,45}\sim 10^{17} L^{1/2}_{\rm disk,45}\,\, {\rm  cm}
\label{rblr}
\end{equation}
Where we have assumed that most of the disk radiation can ionize the atoms in the BLR. 
Please note that this is only approximately true.

One useful quantity when studying lines is the ``ionization parameter $\xi$". 
$\xi$ can be defined as the ratio of the densities of the ionizing photons and the 
electrons:
\begin{equation}
\xi \, =\,   {1\over n_{\rm e}}\, 
\int_{\nu_{\rm i}}^\infty  {L(\nu)\over 4\pi R^2 c\,  h\nu} d\nu \, =\, 
{n_{\rm \gamma,i} \over n_{\rm e} } \, \propto {\rm Ionization\,\, rate \over Recombination \,\, rate}
\end{equation}
%

\end{itemize}

So, what is then the Broad Line Region? We do not know exactly yet.
It can be an ensemble of clouds in Keplerian motions
around the black hole, or material that is infalling, or outflowing,
with a range of velocities. 
Broad lines could also be produced within the accretion disk itself, but
in this case we should see ``double horned" lines, because for a typical inclination of the disk,
the opposite sides of the disk produce redshited and blueshifted lines.
They are seen sometimes, but not often.
It has been proposed also that broad line clouds are stars with bloated envelopes, 
coming close to the black hole. But a large number of them is required, possibly too large.
If broad lines are due to individual clouds, we must have many of them, because each one
contributes for only 10 km s$^{-1}$ to the broadening, and if they were a few, we should see
a not completely smooth profile of the line.
Estimates range from $10^5$ to billions.
But the total required mass is tens of solar masses, a very small quantity.

\subsection{Narrow emission lines}

The main properties of the narrow emission lines are:

\begin{itemize}

\item {\bf FWHM:} around 300--500 km s$^{-1}$. 
These lines are ``narrow", but in any case much larger than thermal broadening.

\item {\bf Density:} around $10^3$ cm$^{-3}$. This comes from the presence of forbidden lines,
requiring small densities. 

\item {\bf Temperature:} around $10^4$ K. This comes from the presence/absence of
specific lines.

\item {\bf Covering factor:} around 0.01. This comes from the entire NLR luminosity
divided by the continuum luminosity.

\item {\bf Distance:} around 100 parsec. They can be resolved in nearby objects.
Sometimes they appear contained in a double cone.

\end{itemize}

\noindent
Since the NLR is located at greater distances than the BLR, they are unaffected
by the possible presence, in a region close to the accretion disk, of
absorbing material. 
This is a crucial fact, as we will soon see.

\section{The X--ray corona}

\begin{figure}
\begin{tabular}{cc}
\includegraphics[height=6.7cm, width=6.5cm]{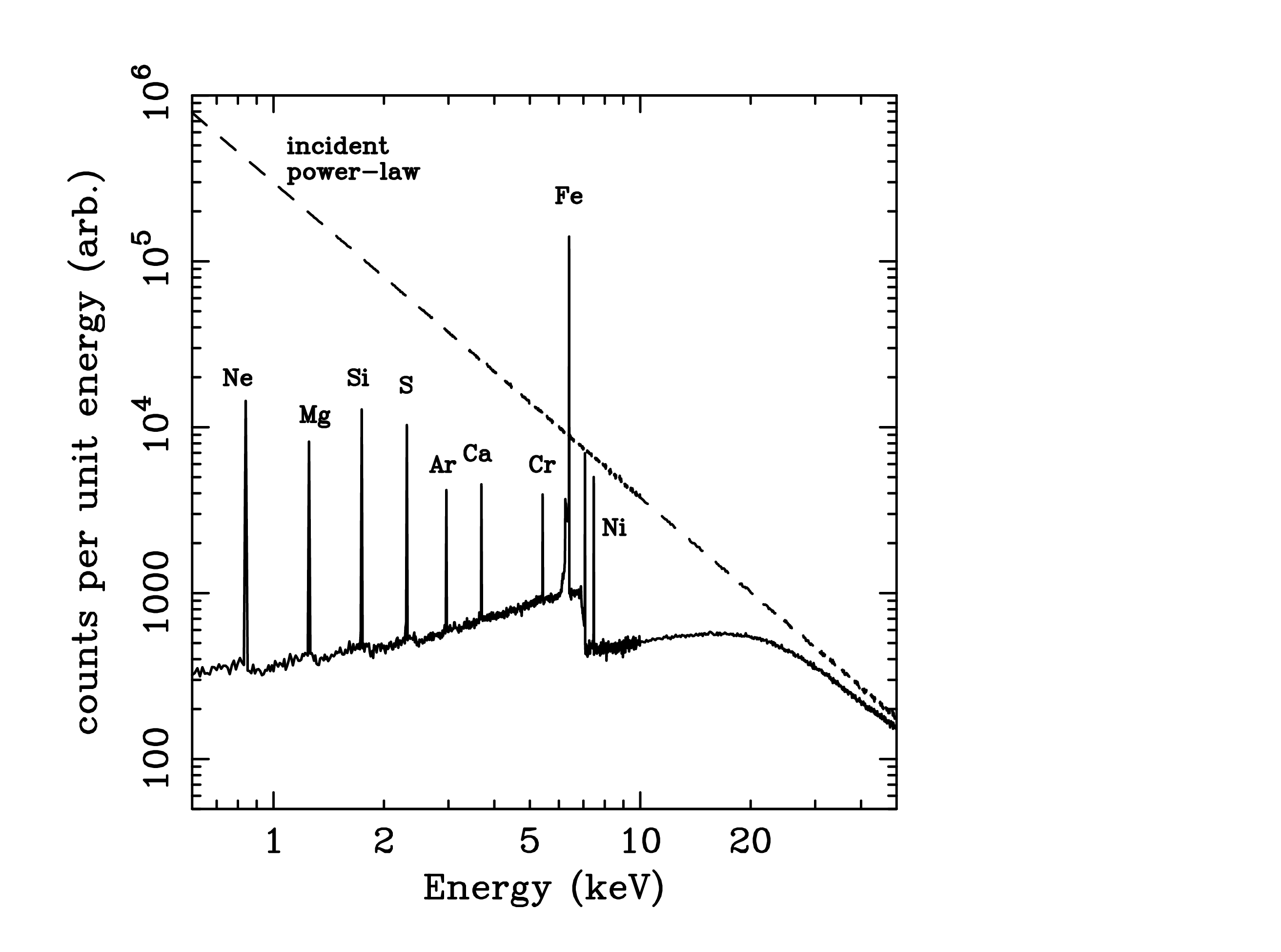}
&\includegraphics[height=6.5cm, width=6.5cm]{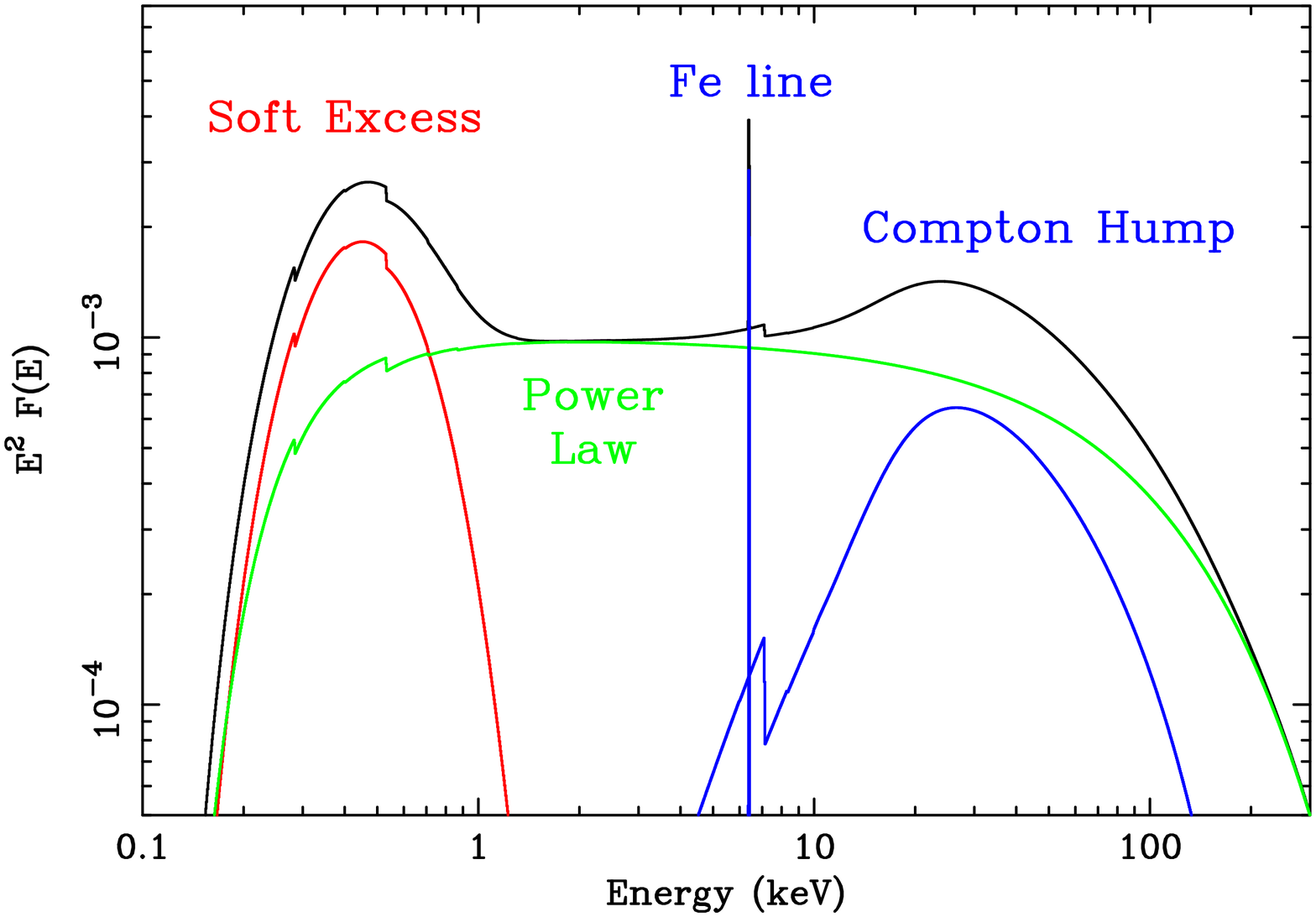} 
\end{tabular} 
\hskip 2 true cm
\caption{{\it Left:} The ``Compton reflection hump", together with the produced emission lines.
{\it Right:} X--ray spectrum and its components: a power law continuum, a soft excess,
and a ``Compton reflection", together with the Iron K$\alpha$ line. From 
Fabian \& Miniutti 2005 (astro--ph/0507409).
}
\label{refle}
\end{figure}

From Eq. \ref{temp} it is clear that the standard accretion disk is relatively 
``cold", in the sense that even the inner regions cannot reach $\sim$keV temperatures
(1 keV $\sim 10^7$ K).
On the other hand, AGNs are powerful X--ray emitters, and their X--ray spectrum extends out to 
hundreds of keV.
We must assume that there is an additional component, besides the accretion disk,
responsible for this emission.
Observed in the energy range 0.1--10 keV, with low spectral resolution, 
the X--ray spectrum can be approximated by a power law: 
$F_X(\nu) \sim \nu^{-\alpha_X} $, with $\alpha_X \sim$ 0.7--0.9.
At first sight, it seems a {\it non--thermal} spectrum, because
it is characterized by a power law.
But then more accurate observations were made, that revealed three
important features:
\begin{itemize}

\item 
When high energy observations in the 10--200 keV were available,
a cut--off was observed, of an exponential type. The spectrum was thus
a power law ending with an exponential cut: 
\begin{equation}
F_X(\nu) \propto \nu^{-\alpha_X}\, e^{-\nu/\nu_{\rm c}} 
\end{equation}
where $\nu_{\rm c}$ is somewhat different for different objects, and
it is in the range 40--300 keV.

\item 
When spectral resolution was improved in the soft X--rays (soft means 0.1--10 keV),
an emission line appeared at $\sim 6.4$ keV (see Fig. \ref{refle}).
This is the K$\alpha$ line from cold Iron.
It is a {\it fluorescent} line.
It means that a photon of enough energy interacts with one electron
of the inner shell of the Iron atom, and ionizes it.
There is then a ``vacant place", triggering a $n=2\to n=1$ transition.
The energy of this transition depends on the ionization state of the Iron.
If it is partially ionized, the energy of the jump is more.
So the energy of the line tells us about the ionization state of
the Iron.
For almost completely ionized Iron, we should observe a line at 6.7 keV
(and sometimes we do). 
A line at 6.4 keV means that it is not largely ionized, and therefore
it is relatively cold (i.e. less than $10^6$ K).

\item
At the same time, the improved spectral resolution revealed that the
slope of the spectrum was more complex than a simple power law.
There is the presence of a hump, peaking at roughly 30 keV,
superposed to the power law.
So the previously determined slope of the power law was not correct,
because it included the hump.
Depurating from this extra component, the spectral index of the power law
becomes $\alpha_X\sim$ 0.9--1.
The right panel of Fig. \ref{refle} illustrates this point.
The nature of this 30 keV ``hump" was soon interpreted as the emission
coming from an {\it irradiated disk}, and was baptized ``Compton hump"
or ``Compton reflection".
\end{itemize}
We will now discuss in somewhat more details these three components:
i) the power law;
ii) the Compton reflection, and
iii) the iron line.

\subsection{The power law: Thermal Comptonization}

We have seen in the previous chapters that a power law does not 
necessarily imply a non--thermal emission.
Bremsstrahlung and thermal Comptonization are made by thermal electrons,
but their shape is a power law (ending in an exponential cut).
Following this idea we will now postulate that there is a
region above and below the disk, where the electrons are hot,
much hotter than the disk.
It is something similar to the corona of our Sun, and therefore
this region is called X--ray corona.
It cannot be located far from the black hole for several reasons:
\begin{enumerate}
\item
Its luminosity is comparable (even if smaller) than the luminosity
of the accretion disk. If its power depends on the release of the gravitational
energy, it must be close to the black hole (i.e. remember $E_{\rm g}\propto 1/R$).
\item The X--ray flux varies rapidly. 
\item It must illuminate the disk and produce the ionization flux for the production of the
fluorescent Iron line. As we shall see, the Iron line is sometimes very broad,
as a result of Doppler broadening due to the high Keplerian velocities 
in the inner regions of the disk. So also the ``illuminator" must be close.
\end{enumerate}

\noindent
Can we derive some more information from the spectral shape?
We can exclude bremsstrahlung as the main radiation process because 
it is too inefficient, meaning that we would require too many electrons
to produce what we see. 
Variability tells us that the emitting volume is small, so the electron density
would be large, and the Thomson optical depth would be much larger than unity.
In these conditions, Compton scatterings would be more important.
Therefore let us assume that the spectrum is due to thermal Comptonization.
The temperature can be inferred from the exponential cut:
\begin{equation}
\Theta \,  \equiv \, {kT \over m_{\rm e} c^2}\, \sim\, {h\nu_{\rm c} \over m_{\rm e} c^2}
\, =\, 0.2 \, {h\nu_{\rm c} \over 100 \,\, {\rm keV}}
\end{equation}
For each scattering, the energy gain $A=x_1/x$ of the photon  is

\begin{equation}
A \, =\, 16\Theta^2+4\Theta+1 \, \sim 1.9 \qquad {\rm (for \,\, \Theta=0.2)} 
\end{equation}
If $\tau_{\rm T}<1$, to be checked a posteriori, we have a relation between
the spectral shape, the amplification factor $A$ and the optical depth:
\begin{equation}
\alpha_X \, =\, -{\log \tau_{\rm T} \over \log A}\, \longrightarrow
\log \tau_{\rm T} \, =\, -\alpha_X \log A \, 
\longrightarrow \, \tau_{\rm T} \,=\, A^{-\alpha_X} = 0.5
\end{equation}
where the last equality assumes $\alpha_X=1$ and $A\sim 2$.
The Comptonization parameter $y$ must be of the order of unity,
because $A\sim y/\tau_{\rm T}$.
What derived here are typical values.
There are sources where we need $\tau_{\rm T}>2$: for these sources
the simple equation we have used to relate $\alpha_X$, $\tau_{\rm T}$ and $A$
is not valid, and another relation must be used.

Why the corona is so hot, even if the disk is cold?
What are the mechanisms able to extract energy from the accreting matter and
to deposit it in the corona?
We do not know exactly, even if we suspect that  
the magnetic field plays a crucial role.

\subsection{Compton Reflection}

If the corona emits isotropically, half of its flux will be intercepted by the disk.
We have seen that the disk in AGNs should be relatively cold, in the sense that
most of its material is not ionized.
X--ray photons will interact with the material in the disk in the following ways:
\begin{enumerate}
\item
at low energies ($h\nu<$10 keV) they will be photoelectrically absorbed by the metals
present in the disk;
\item at intermediate energies ($10<h\nu<40$ keV) they will be Thomson scattered
and part of them will be scattered in the upward direction;
\item at large energies ($h\nu> 40$ keV) the scattering will not be in the Thomson
regime, but Klein--Nishina effects will start to be important: the scattering
is preferentially forward directed (and therefore the scattered photons will penetrate more
deeply into the disk) and the photon energy will be reduced.
The decrease of the energy means that 
photoelectric absorption, initially negligible, is again important,
and the photon will be absorbed.
\end{enumerate}
The result is that some of the incoming radiation will be scattered back,
with a modified shape. 
This {\it Compton reflection} component will be peaked at $\sim$30 keV,
where photoelectric absorption and Klein--Nishina effects are not 
important.
See Fig. \ref{refle} to see the shape of this component, and
how it modifies the original power law spectrum.

If the disk is hot, and partially ionized, the amount of free electrons will
increase the importance of Thomson scattering, and the ``left shoulder" of the
Compton reflection increases.
Increasing the ionization state of the disk implies that the Compton
reflection is not peaked any longer, but it retains, at low energies, the
slope of the original power law.

\subsection{The Iron Line}

Among the metals in the disk irradiated by the hot corona, Iron is the
one producing the most prominent emission line. 
This is due to a combination of two factors: it is one of the most
abundant elements (but not the most abundant), and it is the one ``suffering"
less from the Auger effect. This is, in very simple terms, a non--radiative
transition: the energy of the $n=2\to n=1$ transition is
 used to expel an electron from the atom instead of producing a photon.
See the left panel of Fig. \ref{refle} to see the relative strength
of the emission lines.
The energy, luminosity, width, and profile of the iron line are a powerful diagnostic
of the condition of the inner parts of the accretion disk.

\vskip 0.2 cm
\noindent
{\bf The energy} of the K$\alpha$ Fe line (from 6.4 to 6.7) tells about the
ionization state of the iron, and thus the temperature of the disk.

\vskip 0.2 cm
\noindent
{\bf The luminosity} of the line tells about the amount of iron, and thus about
the abundance of metals of the disk.
The ratio about the line to continuum luminosity tells about the geometry
of the emitting region and possible anisotropy of the continuum.

\vskip 0.2 cm
\noindent
{\bf The width} of the line tells about the velocities of the irradiated material
forming the line.

\vskip 0.2 cm
\noindent
{\bf The profile} (symmetric, double horned, skewed) tells about Doppler boosting
and gravitational redshift.

\begin{figure}[h]
\center
\includegraphics[height=10cm, width=16cm]{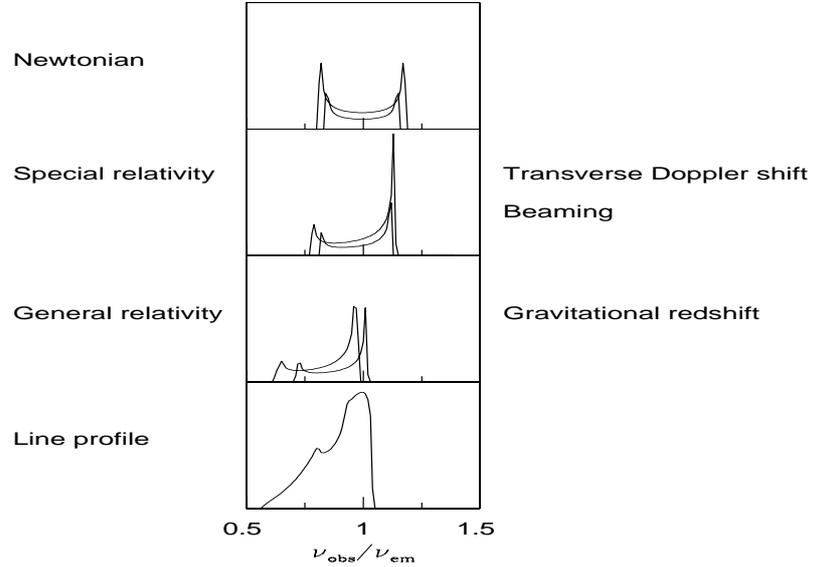} 
\vskip -2 true cm
\caption[h]{The profile of an intrinsically narrow emission line is modified by 
the interplay of Doppler/gravitational energy shifts, relativistic beaming, and gravitational 
light bending occurring in the accretion disc. 
{\it Upper panel:} the symmetric doubleÐ-peaked profile from two annuli 
of a nonÐrelativistic Newtonian disc. 
{\it Second panel:} the transverse 
Doppler shifts make  the profiles redder;  the relativistic beaming 
enhances the blue peak with respect to the red. 
{\it Third panel:} 
the gravitational redshift shifts the overall profile to the red side and 
reduces the blue peak strength. 
{\it Bottom panel:} The combination of all the effects gives
rise to a broad, skewed line profile.  From  Fabian \& Miniutti (2005) (astro--ph/0507409).
}
\label{profiles}
\end{figure}{}

The studies of iron lines in AGNs received a lot of attention in the recent past.
The most intriguing observation concerns the finding of {\it relativistically 
broadened lines}.
Fig. \ref{profiles} shows in a schematic way what happens.
Suppose to see an accretion disk from an intermediate angle
(i.e. neither face on or edge on). 
If the emitting material is moving with a non relativistic velocity (Newtonian
case, upper panel) we see a symmetric double peaked line, because the parts of the disk that are approaching us
emit a line that we observe blueshifted, while the parts of the disk moving away from us 
emit a line that we observe redshifted.
The symmetry refers to the fact that the blue and red lines have equal fluxes.
But if the material is orbiting very close to the black hole, the Doppler
enhancement of the flux becomes important, and  the blue peak will have more
flux than the red one (second panel from the top).
If the emission takes place {\it very} close to the black hole (within a few
Schwarzschild radii) also the gravitational redshift is important:
then all frequencies are redshifted by an amount that is larger as the emitting
regions are closer to the black hole (third panel).
Furthermore, we have to account for gravitational light bending, changing the emission pattern
and thus the received flux.

All these effects have opened the way to {\it the possibility to test 
general relativity in the strong field case}.

\begin{figure}[h]
\center
\vskip 0.5 cm
\includegraphics[height=9cm, width=11cm]{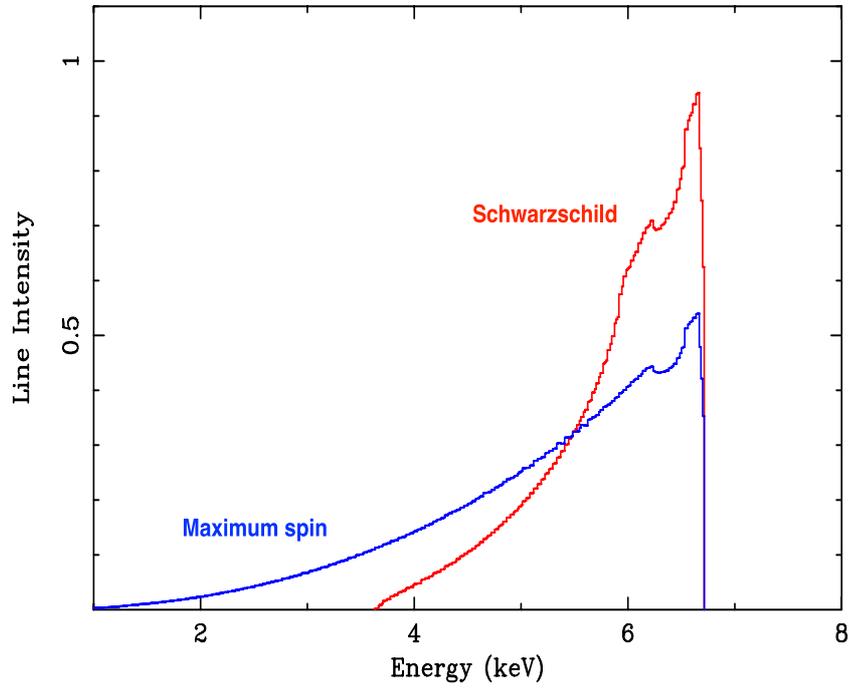}\\
\vskip -0.3 true cm
\caption[h]{The K$\alpha$ Fe line from a Schwarzschild (no spin) and a maximally rotating
Kerr black hole. From Fabian \& Miniutti (2005) (astro--ph/0507409).
}
\label{diskline}
\end{figure}{}

Besides this, there is another, crucial effect, related to {\it the spin of the black hole.}
This is conceptually simple: for a Kerr black hole fastly rotating the innermost stable
orbit is not $3 R_{\rm S}$ (i.e. 6 gravitation radii $= 6 R_{\rm g}$), but it moves closer
to the hole, becoming equal to one gravitational radius for a maximally rotating hole.
The closer to the hole the emitting material is, the larger the velocities, and the stronger the
gravitational effects.
So the broadening of the line becomes larger, as it is illustrated in Fig. \ref{diskline}.
{\it We can tell the spin of the black hole} by observing the profile
of the iron line.

\newpage

\section{The torus and the Seyfert unification scheme}

\begin{figure}[h]
\center
\includegraphics[height=10cm, width=12cm]{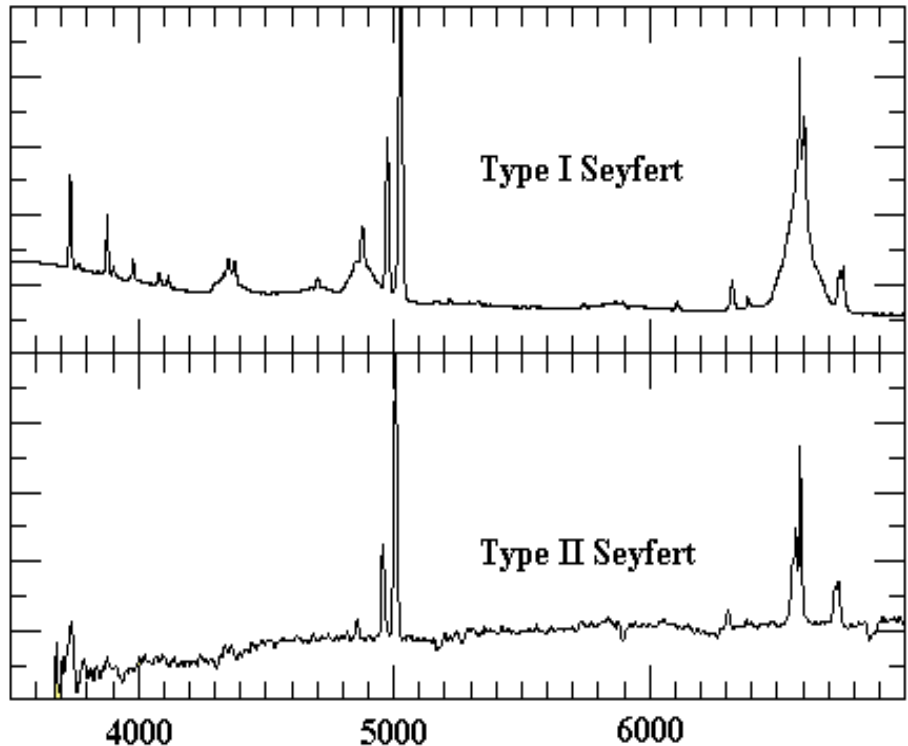}
\caption[h]{The optical spectrum of a Seyfert Type 1 and Type 2.
Note that Type 1 Seyferts have both broad and narrow lines,
while Type 2 Seyferts have only narrow lines.
}
\label{seyfert}
\end{figure}

Seyfert galaxies are spiral galaxies with a bright core, that host 
an Active Galactic Nucleus.
They come in two flavors: the Type 1 and the Type 2.
Type 1 Seyferts have both broad and narrow emission lines,
while Type 2 have only narrow lines. 
See Fig. \ref{seyfert}.

{}
%

\begin{figure}[h]
\center
\includegraphics[height=11cm, width=14cm]{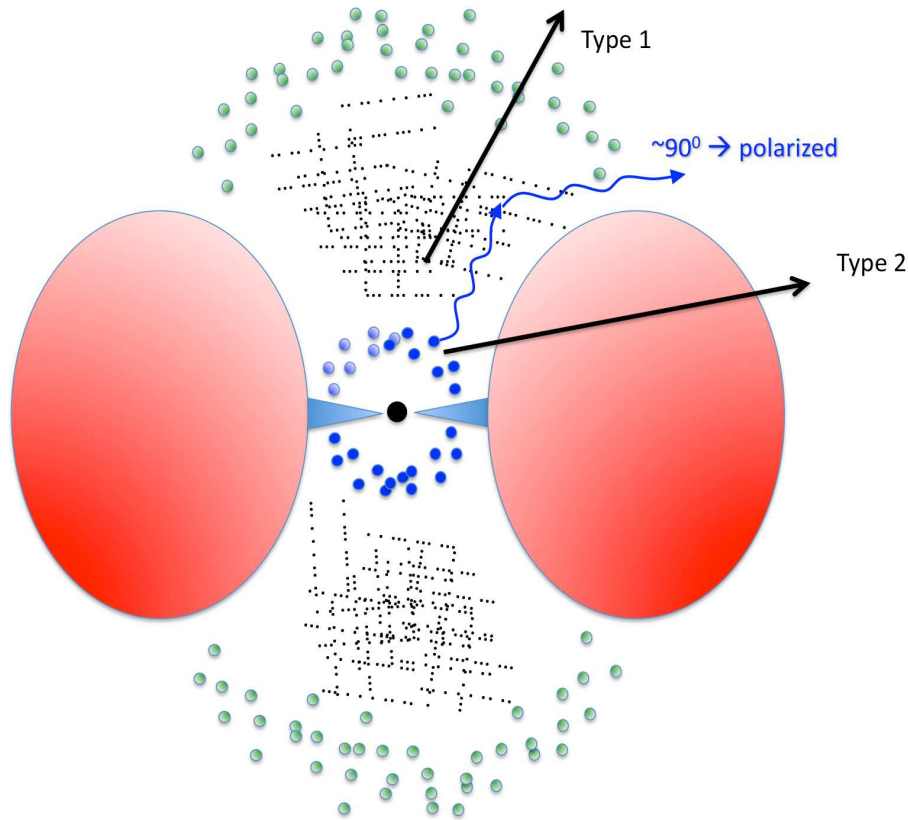}
\caption[h]{Schematic illustration of the disk+torus system,
with the broad line clouds that are close to the disk, 
and the narrow line clouds external to the torus.
Narrow lines are then observable at all viewing angles, while side
observers cannot see the disk and the BLR.
However, if a populations of free
electrons fill the region (black little dots), 
they can scatter the emission line photons
into the line of sight. Since the scattering angle is close to $90^\circ$,
the scattered photons are polarized.
Observing in polarized light the broad emission lines emerge.
}
\label{torus}
\end{figure}

Antonucci \& Miller (1985)\footnote{Antonucci R.R.J. 
\& Miller J.S., 1985, ApJ, 297, 621} 
made spectropolarimetry observations of NGC 1068,
a nearby and bright Seyfert 2 galaxy.
In total light the broad lines were not seen, but in polarized light they emerged.
Therefore the broad line clouds are there, but for some reason their emission
is overwhelmed by the unpolarized continuum, and in total light 
their emission is not seen.
Antonucci \& Miller then argued that this is due to obscuration.
Fig. \ref{torus} illustrates the point.
If there is a big and dusty torus surrounding the accretion disk,
and the observer is looking from the side, the emission from the
disk and the BLR is absorbed by the dust in the torus.
Instead the narrow lines, coming from an extended region $\sim$100 pc in size,
are not intercepted by the torus, and can reach observers located at any
viewing angle.

Furthermore, if the ``funnel" of the torus is filled with free electrons,
they can scatter the emission of the disk and the broad lines also into
the direction of side observers. 
The amount of this scattered radiation can be tiny, but it is polarized
(remember: for scattering angles close to $90^\circ$ the scattering photons
are maximally polarized).
Observing the polarized light we select these photons and discard the
much more intense unpolarized light: as a result the broad lines
now appear.
{\it The free electrons act as a periscope.}

{}

This simple idea implies that all Seyferts are intrinsically equal,
but their appearance depends on the viewing angle: observers looking face on 
can see all the components (and thus also the broad lines): they see a Type 1 object.
Observers looking from the side see only the narrow lines and the emission of the 
torus.
Being at some distance from the accretion disk, the torus in fact intercepts
part of the disk emission and re--emits it in the infrared.
We can even predict approximately at which frequencies 
the torus emission will start to be important.
This is because the dust cannot exist at temperatures hotter than $\sim$2000 K.
Taking this temperature as reference, we have $\nu\sim 3 kT/h\sim 10^{14}$ Hz,
which corresponds to a wavelength of 3 $\mu$.
If we are correct, then we predict that all the radiation that the torus
intercepts from the disk comes out at frequencies smaller than $10^{14}$ Hz.

\subsection{The X--ray background}

Another very  important consequence of the presence of the torus concerns
the X--ray emission, because also X--rays are absorbed by the torus,
especially in the soft energy range (soft X--rays are much more absorbed than
hard X--rays).
In addition, the walls of the torus will also ``reflect" X--rays,
producing a ``Compton reflection hump", like the disk.
The main difference is that the fluorescent iron line will in this case
be extremely narrow (since the typical velocities of the torus material,
located far from the black hole, are very small).

\begin{figure}[h]
\center
\vskip 0.5 cm
\includegraphics[height=8cm, width=12cm]{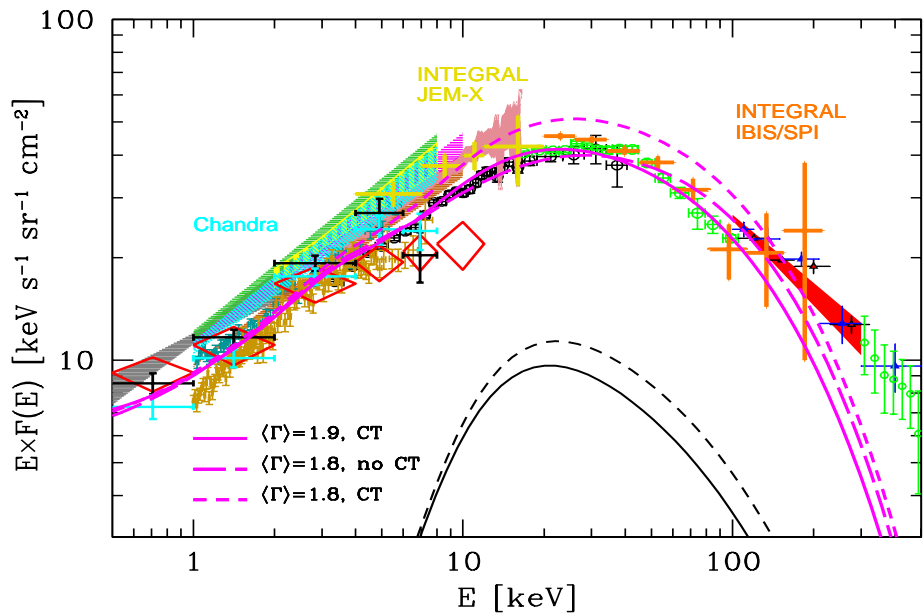}\\
\vskip -0.3 true cm
\caption[h]{The X--ray background. The lines are synthesis models
assuming some prescriptions about the intrinsic slope of the X--ray spectrum and
the ratio of ``thick to thin" AGNs.
From Gilli et al. (2007) (astro--ph/0704.1376).
}
\label{xrb}
\end{figure}{}

The importance of the absorption depends on the metals present along the line
of sight.
For solar abundances, there is a well defined relation between the
amount of metals and the amount of hydrogen.
So it is customary to express the amount of absorption in terms
of the ``column of hydrogen" along the line of sight. 
It is $N_{\rm H} = n_{\rm H} R$, where R is the size of the absorbing region, and
$n_{\rm H}$ is the hydrogen number density.
The units of $N_{\rm H}$ are thus cm$^{-2}$.
When $N_{\rm H}>10^{24}$ cm$^{-2}$ we have that $\tau_{\rm T} = \sigma_{\rm T} N_{\rm H}$
becomes close or larger than unity, and we speak of {\it Compton thick} sources.

With these properties, a mixture of Type 1 and Type 2 objects can
account for the {\it X--ray background}, namely, the integrated X--ray emission
from all sources.
It was a mystery for decades, because its spectrum in the 0.5--10 keV energy range
is very hard ($\alpha_X\sim 0.4)$, unlike most of the known extragalactic 
X--ray sources (that have $\alpha_X\sim 1$).
Fig. \ref{xrb} shows a modern collection of data defining the X--ray background
from 0.5 to 300 keV.

To explain it, we need a population of very hard X--ray sources and the Type 2 ones
have indeed the right properties: since soft X--rays are more absorbed than hard
ones, their spectrum is very hard.
Since the X--ray continuum is depressed, but the reflection hump (from the torus)
is not, they can contribute significantly at $\sim$30 keV, just where the X--ray 
background peaks.

\newpage

\section{The jet}

About 10\% of AGNs, besides accreting matter, are able to expel
part of it at relativistic speeds in two oppositely directed jets.
These {\it radio--loud} AGNs were the first to be discovered, due to their 
relevant radio emission.
Radio jets can be spectacular, since they can reach a few Mpc in size,
tens of times the radius of their host galaxies.
AGNs with relativistic jets were believed to always have 
giant ellipticals as host galaxies, but now this paradigm is
challenged by the observations of Narrow Line Seyfert 1 Galaxies
(beware that the term ``narrow" here indicates lines that are
in any case broad, but ``less" broad than usual, namely with 
FWHM$<2000$ km s$^{-1}$).

The radio emission produced by jets is only a small fraction
of the entire electro--magnetic power they emit.
We now know that most of it is produced in the range mm--optical and in the
$\gamma$--ray band.
In turn, the electro--magnetic output is only a small fraction of the total
power carried by the jet.
Most of it is carried in the form of bulk kinetic energy of the matter
flowing relativistically, and by the moving magnetic field (i.e. Poynting flux);
in powerful sources it reaches the large radio structure: the
hot spots and the radio lobes.
In less powerful sources, however, these structures are absent.
We call the powerful sources {\it FR II} radio--galaxies, and the
weaker sources {\it FR I} radio--galaxies.
The name FR comes from Fanaroff \& Riley, that classified radio--galaxies
in this way.

\subsection{Flat and steep radio AGNs}

Since the emitting material is moving with a bulk Lorentz factor $\Gamma$,
its radiation is beamed.
What we see is therefore amplified if the jet points at us, and
depressed if the jet points elsewhere.
The large structures (hot spots and radio--haloes), instead, are
not moving, and their emission is isotropic.
The ratio between the jet emission and the radio--lobe emission 
is thus a strong function of the viewing angle.

\begin{figure}[h]
\vskip -0.3 true cm
\center
\includegraphics[height=10cm, width=12cm]{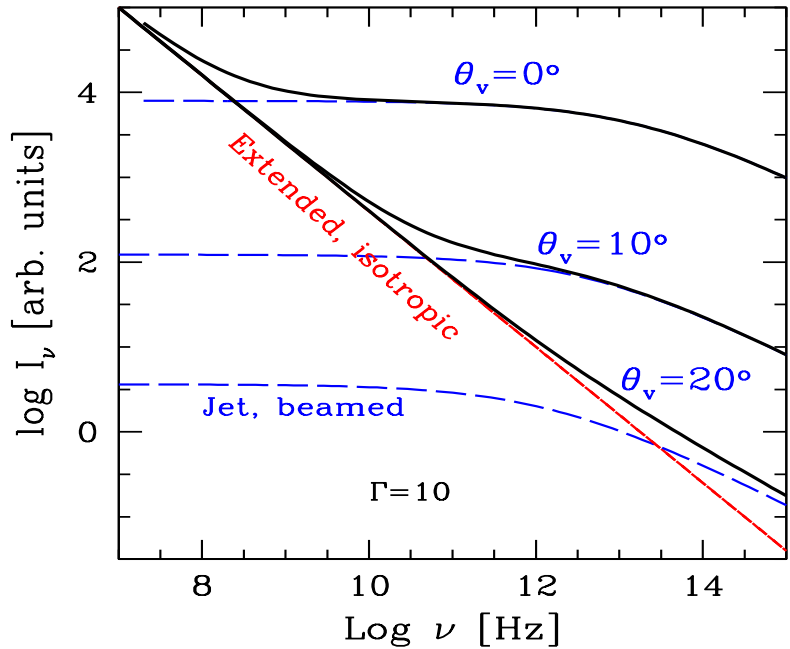}
\vskip -0.5 true cm
\caption[h]{The emission of the jet is beamed, and depends on the viewing angle. 
Since it is a superposition of emission from several blobs
of different sizes, the spectrum is flat up to the self--absorption frequency of the
most compact component (hundreds of GHz).
The emission from the radio--lobe is instead isotropic, and is optically thin.
Since it is not beamed, it {\it does not} depend on the viewing angle.
The overall appearance of the radio spectrum depends therefore on the viewing angle.
Steep radio spectra correspond to large $\theta_{\rm v}$ (and small $\delta$), 
flat radio spectra to small $\theta_{\rm v}$ (and large $\delta$).
}
\label{flatsteep}
\end{figure}{}

Fig. \ref{flatsteep} illustrates this point: the jet is producing a 
very flat spectrum [i.e. $F(\nu)\propto \nu^0$] up to $10^{12}$ Hz.
The observed intensity is $I_{\rm jet}(\nu)\propto  \delta^3 I_{\rm jet}^\prime(\nu^\prime)$,
and changes a lot changing the viewing angle.
The emission coming from the extended, unbeamed, region is characterized
by a steeper slope [for instance: $I_{\rm lobe}(\nu)\propto \nu^{-1}$].
Therefore:
\begin{itemize}
\item At very low frequencies we expect that the emission is always
dominated by the lobe, for all viewing angles.
\item Lobe and jet emission are equal at some frequency that is larger
as the viewing angle increases.
\item For misaligned jets the depressing factor due to beaming is so large
that we do not see the jet, we see only the lobe emission.
\item Aligned sources show a {\it flat radio spectrum}.
Misaligned sources show a {\it steep radio spectrum}.
The slope of the radio spectrum is thus an indication of the viewing angle.
\item If we perform a radio survey with some limiting flux, the number
of aligned and misaligned sources changes according to the observing frequency:
if we use 178 MHz, we preferentially look at the lobe emission, and we can pick up
aligned and misaligned sources; if we observe at 30 GHz we pick up preferentially
the flat, aligned sources.  
\end{itemize}
To summarize, and simplifying a bit:\\
{\bf flat radio spectrum $\,\,\,\,\, \longrightarrow$ aligned, beamed}\\
{\bf steep radio spectrum $\longrightarrow$ misaligned, de--beamed}.

\subsection{Blazars}

Sources whose jet is pointing at us are called {\it blazars}.
The sub--classification of blazars is rather complex, reflecting
either the way they were discovered (radio or X--rays) or
the location of their synchrotron peak (in $\nu F_\nu$), or
the presence or absence of broad emission lines.
Particularly relevant in this respect 
is the concept of {\it equivalent width of a line} (EW).
It is defined as
\begin{equation}
EW \, =\, \int {F_0 - F_\lambda \over F_0 }d\lambda 
\end{equation}
where $F_\lambda$ is the total flux (line + continuum) and $F_0$ is the flux
of the continuum.
This describes the EW of both absorption and emission lines, but in the case
of emission lines the result is negative, so sometimes the absolute value is given.
The units of the EW are \AA.

The classical subdivision of blazars is \\
\noindent
{\bf BL Lac objects (BL Lacs) ---} They have weak or absent lines. Their equivalent width 
is EW$<5$ \AA.  \\
\noindent
{\bf Flat Spectrum Radio Quasars (FSRQs) ---} They have strong emission lines, 
whose EW$>5$ \AA.  

\noindent
Since we are considering very variable objects, this subdivision depends
on the state of the source when it is observed: this is because the
EW is a ratio between the line and (beamed) optical flux,
and the latter varies.
Now we believe that the absence or weakness of the lines in BL Lacs
is not due to a particularly amplified continuum, but it is an intrinsic 
property. In other words, the luminosity of the lines in BL Lacs
is less than in FSRQs.

\subsubsection{The overall spectrum of blazars}

\begin{figure}[h]
\center
\vskip -0.5 cm
\includegraphics[height=12cm, width=14cm]{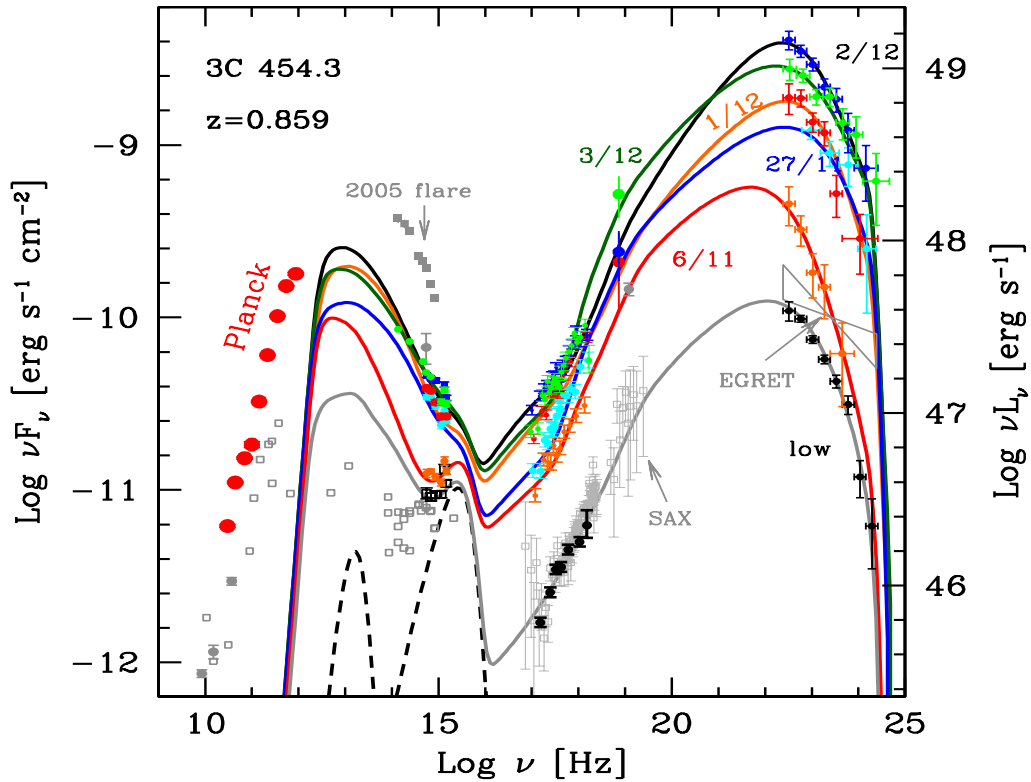}
\vskip -1 true cm
\caption[h]{The overall electromagnetic spectrum of 3C 454.3, 
the most luminous $\gamma$--ray source up to now.
Note the large amplitude variability, even day--by--day.
Dates refer to the year 2009. Lines correspond to fitting models.
See Bonnoli et al. (2011, MNRAS, 410, 368). 
}
\label{454}
\end{figure}

Fig. \ref{454} shows the source of 3C 454.3, a powerful FSRQ, that
was the brightest blazar detected in the $\gamma$--ray band
for a few years.
One can see the extraordinary variability, encompassing 
2 orders of magnitude in flux, both in the $\gamma$--ray band and
in the optical.
When the optical flux is in the low state, one can also see
the contribution of the accretion disk (see the little upturn
of the optical--UV data at $10^{15}$ Hz).
The dashed black line illustrates a model for the disk emission
($10^{15}$ Hz)and the torus emission (at $10^{13}$ Hz).

{}

Radio sources in general, and blazars in particular, emit over 
the entire electromagnetic spectrum, from the radio (down to $10^7$ Hz)
to the TeV band (up to $10^{27}$ Hz).
The main characteristics are:
\begin{itemize}
\item The overall spectral energy distribution (SED), once plotted
in $\nu F_\nu$, shows two broad peaks.
The location of the peak frequencies varies from object to object,
but in general the first peak is between the mm and the soft--X--rays,
while the high energy peak is in the MeV--GeV band.

\item They are variable, at all frequencies, but especially at high energies.
Minimum variability timescales range between weeks and tens of minutes.

\item In restricted frequency ranges, their spectrum is a power law.

\item The variability is often (even if not always) coordinated 
and simultaneous in different energy bands (excluding the radio).

\item They are often polarized, in the radio and in the optical.

\item The high energy hump often dominates the power output. 

\end{itemize}
%
\begin{figure}[h]
\center
\vskip -1 cm
\includegraphics[height=12cm, width=14cm]{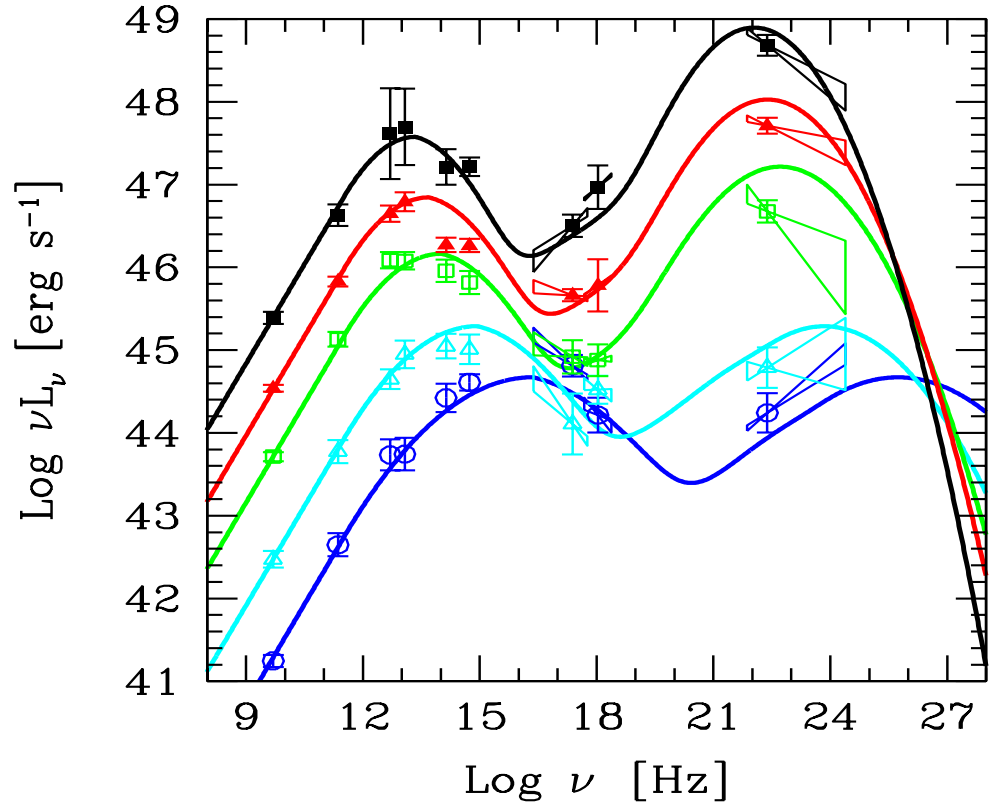}
\vskip -0.8 true cm
\caption[h]{The blazar sequence.
As the bolometric luminosity increases, the peak of the two
humps shifts to smaller frequencies, and the high energy
hump becomes more important. From Fossati et al. (1998, MNRAS, 299, 433);
Donato et al. (2001  A\&A, 375, 739).
}
\label{gfos}
\end{figure}{}

Besides these common features, it seems that blazars form a sequence
of SEDs, according to their observed bolometric luminosity.
Fig. \ref{gfos} illustrates this point.
It was constructed taking the average luminosity in selected bands,
and considering $\sim$100 blazars coming from radio and X--ray complete samples
(complete here means: it contains all blazar whose flux is greater than a limiting
one).
Note that:
\begin{itemize}
\item Low power blazars, that are BL Lacs (i.e. no lines),
are {\it bluer} than powerful blazars, that are FSRQs (with lines).
Bluer means that the peak frequencies of both peaks
are larger.

\item The high energy hump increases its relevance as we increase the
bolometric luminosity.
At low luminosities both humps have the same power,
while the most powerful FSRQ have an high energy hump that is $\sim 10\times$ the
low energy one.

\end{itemize}

\subsubsection{Interpretation of the SED of blazars}

{\it First hump ---} 
It is rather straightforward to interpret the first hump of the SED
as synchrotron emission.
Since it sometimes varies rapidly and simultaneously (at least in
optical--UV--soft--X--rays bands) we believe that it comes
from a single region of the jet.

\noindent
{\it High energy hump ---} It is believed (but with no unanimity)
that it is produced by the Inverse Compton process, by the same
electrons producing the synchrotron.
The seed photons could be the synchrotron photons themselves 
(especially in low power blazars) and/or the emission line photons
(when the lines are present, so only in powerful objects...).
Remember that in the comoving frame of the jet, the lines
are seen blueshifted (factor $\Gamma$) and the arrival time is contracted
(another factor $\Gamma$) so that in the comoving frame
the radiation energy density of the lines is seen enhanced by $\Gamma^2$.

\subsubsection{Simple estimates}

Suppose to know the luminosity of both peaks of a blazar,
call them $L_{\rm S}$ and $L_{\rm C}$, and
the location of their peak frequency $\nu_{\rm S}$ and $\nu_{\rm C}$.
You also have information about the minimum variability timescale
$t_{\rm var}$.
Assume that the emitting region is only one, and approximate it with a
sphere of radius $R$. 

Let us consider two cases. 
For the first we will assume that the high energy emission is due
to the synchrotron self--Compton process (SSC), in the second
we will assume that it is due to inverse Compton on the photons of the 
broad line region. For both cases the size $R$ is given by
\begin{equation}
R \, \sim \, c t_{\rm var}\, {\delta \over 1+z}
\label{tvar}
\end{equation}
%

\begin{figure}[h]
\center
\vskip -0.5 cm
\includegraphics[height=12cm, width=13cm]{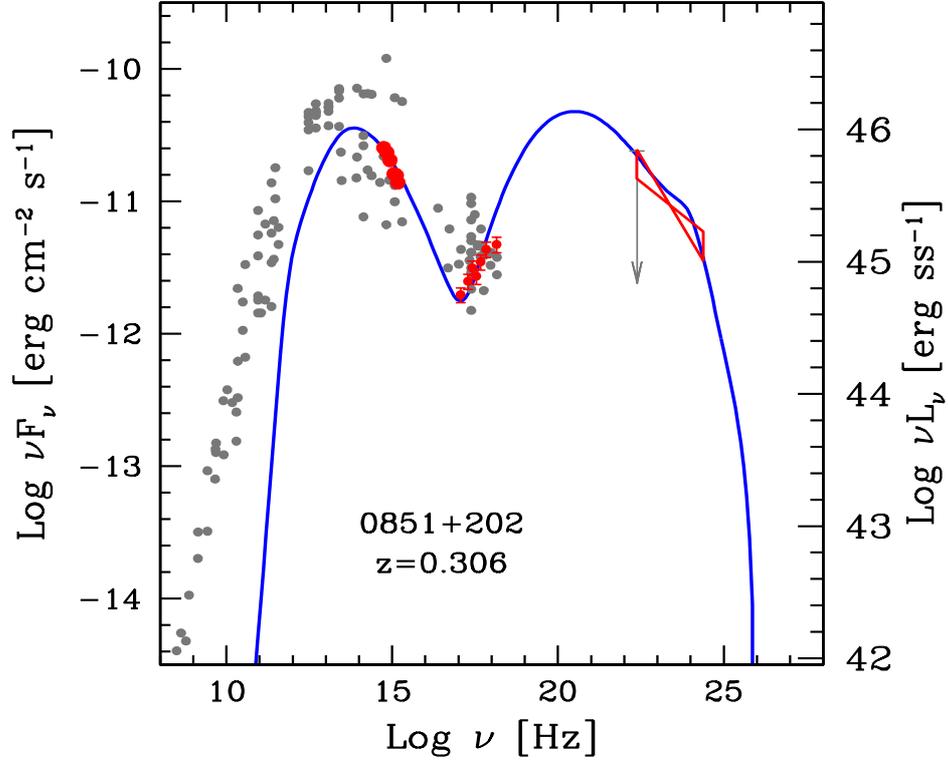}
\vskip -0.8 true cm
\caption[h]{The SED of the BL Lac 0851+202, together with a
fitting model.
Red points are simultaneous, grey points are archival data.
}
\label{0851}
\end{figure}{}

{\it SSC ---} In this case the ratio between $L_{\rm C}$ and $L_{\rm S}$ 
is given by:
\begin{equation}
{L_{\rm C}\over L_{\rm S} } \, =\, {U^\prime_{\rm rad} \over U^\prime_B} \,
= \, { L_{\rm S} \over 4\pi R^2 c \, \delta^4} \, {1\over U^\prime_B}
\label{lcls}
\end{equation}
where primed quantities refer to the comoving frame.
Note the $\delta^4$ factor to pass from the observed to the comoving
luminosity.
We have assumed that $U^\prime_{\rm rad}$ is entirely due to the
synchrotron luminosity, with no contribution from the SSC one.
Is that correct?
It is, if most of the SSC radiation is at frequencies so large that the 
corresponding scattering process occurs in the Klein--Nishina regime, and 
thus can be neglected.
From Eq. \ref{lcls} we can derive ($U^\prime_B=B^{\prime 2}/8\pi$):
\begin{equation}
B^\prime \delta^2 \, =\, {L_{\rm S} \over R } \, 
\left( {2 \over c\,  L_{\rm C} } \right)^{1/2} \, \longrightarrow \,
B^\prime \delta^3 \, =\,
(1+z)\, {L_{\rm S}  \over c \, t_{\rm var} } \, 
\left( {2 \over c\,  L_{\rm C} } \right)^{1/2}
\label{bd3}
\end{equation}
Now let us use the peak frequencies: the ratio between them is
\begin{equation}
{\nu_{\rm C} \over \nu_{\rm S} }\, =\, \gamma^2_{\rm peak}
\label{gpeak2}
\end{equation}
Note that this is a {\it ratio}, therefore there is no $\delta$, no $(1+z)$.
The observed synchrotron peak frequency is given by
\begin{equation}
\nu_{\rm S} \, \sim {4\over 3} \nu^\prime_{\rm B} \gamma^2_{\rm peak} \, {\delta \over 1+z}
\, = \, {4\over 3}\, {eB^\prime \over 2\pi m_{\rm e} c}  
\gamma^2_{\rm peak} \, {\delta \over 1+z}
\label{vsin}
\end{equation}
Inserting Eq. \ref{gpeak2} we arrive to
\begin{equation}
B^\prime \delta \, =\, {3\pi m_{\rm e} c \over 2 e} \, {\nu^2_{\rm S} \over \nu_{\rm C}}
\, (1+z)\, =\, {1\over 3.7\times 10^6}
{\nu^2_{\rm S} \over \nu_{\rm C}} \, (1+z)
\label{bd}
\end{equation}
We have two equations (Eq. \ref{bd3} and Eq. \ref{bd}) for the two unknowns $B^\prime$
and $\delta$.

\vskip 0.3 true cm
\noindent
{\it Example SSC ---} Fig. \ref{0851} shows the SED of the BL Lac 0851+201 (=0J 287), 
at $z=0.306$. For this source we have:
$L_{\rm C} \sim L_{\rm S}\sim 10^{46}$ erg s$^{-1}$.
We can guess $\nu_{\rm C}\sim 10^{20}$ Hz
extrapolating the X--ray and the $\gamma$--ray spectra.
Let us take $\nu_{\rm S}\sim 5\times 10^{13}$ Hz, and assume that
$t_{\rm var} = 10^4$ s.

From Eq. \ref{bd3} we have $B^\prime\delta^3 = 3.5\times 10^3$ G
and from Eq. \ref{bd} we derive $B^\prime \delta= 8.8 $ G.
Therefore $\delta=20$ and $B^\prime=0.44$ G.
Notice that if $\nu_{\rm C}$ were $10^{21}$ Hz instead of $10^{20}$ Hz,
than we would have found $\delta=63$ and $B^\prime=0.014$ G.
These estimates are very sensitive to the $\nu_{\rm C}/\nu_{\rm S}$ ratio.
The larger the separation between the two peaks, the larger $\delta$ and 
the smaller $B^\prime$.

\vskip 0.5 true cm
\noindent
{\it Inverse Compton with broad line photons ---}
In this case let us assume that most of the radiation energy density
in the comoving frame comes from the BLR.
We have

\begin{equation}
U^\prime_{\rm rad} \, \sim\, \Gamma^2 \, { L_{\rm BLR}\over 4\pi R^2_{\rm BLR} c}
\label{uext1}
\end{equation}
We may remember that $L_{\rm BLR}\sim 0.1 L_{\rm disk}$, and
that $R_{\rm BLR} \sim 10^{17} L_{\rm ion, 45}$ cm.
This assumptions lead to (assuming that $L_{\rm ion} \sim L_{\rm disk}$):
\begin{equation}
U^\prime_{\rm rad} \, \sim\, {\Gamma^2 \over 12 \pi}
\label{uext2}
\end{equation}
a remarkable result, due to the $R_{\rm BLR}\propto L_{\rm disk}^{1/2}$
dependence. 
As long as the emitting region of the jet is within the BLR,
then $U^\prime_{\rm rad}$ depends only from $\Gamma$, but not on
the power of the disk or the actual value of the size of the BLR.
By making the ratio with $U^\prime_B$ we have
\begin{equation}
{L_{\rm C}\over L_{\rm S} } \, = \, 
{U^\prime_{\rm rad} \over U^\prime_B}\, = \, {2 \Gamma^2 \over 3 B^{\prime 2}}
\, \longrightarrow\,
{ B^\prime \over \Gamma } \, =\, \left( {2 L_{\rm S}  \over 3 L_{\rm C}  } \right)^{1/2}
\label{bgamma}
\end{equation}
The synchrotron peak frequency is still given by Eq. \ref{vsin},
{\it if $\nu_{\rm S}$ is larger than the self absorption frequency $\nu_{\rm t}$}.
The peak frequency of the Compton hump is now given by:

\begin{equation}
\nu_{\rm C} \, = {4 \over 3} \, \gamma^2_{\rm peak} \nu_{\rm Ly\alpha} \, 
{\delta \Gamma \over 1+z} 
\label{vcext}
\end{equation}
where we have taken the frequency of the Ly$\alpha$ line as the characteristic frequency
of the seed photons (since the Ly$\alpha$ line is the most intense).
The factor $\Gamma\delta$ comes from considering the scattering in the comoving frame,
where the typical frequency is $\nu^\prime_{\rm Ly\alpha}  \sim \Gamma \nu_{\rm Ly\alpha}$,
and then blueshifting the frequency after the scattering in the comoving frame by
the factor $\delta$.
Making the ratio $\nu_{\rm C}/\nu_{\rm S}$ we have
\begin{equation}
{\nu_{\rm C} \over \nu_{\rm S} } \, 
= { (4/3) \, \gamma^2_{\rm peak} \nu_{\rm Ly\alpha}\delta \Gamma/(1+z)  \over
(4/3) \, \gamma^2_{\rm peak} \nu^\prime_{\rm B} \delta / (1+z)} \, 
=\, {\Gamma \nu_{\rm Ly\alpha} \over \nu^\prime_{\rm B}}
\label{vcvs}
\end{equation}
From this equation we have:
\begin{equation}
{ B^\prime \over \Gamma } \, \sim \, 8.8\times 10^8\, 
\, {\nu_{\rm S} \over \nu_{\rm C}  }  
\label{bgamma2}
\end{equation}
Unfortunately, Eq. \ref{bgamma2} and Eq. \ref{bgamma} give two relations
for the same quantity: $B^\prime / \Gamma$.
We cannot derive $B^\prime$ and $\Gamma$ separately.
But we can use them to derive a  relation between the peak frequencies
and the synchrotron and inverse Compton luminosities:
\begin{equation}
 {\nu_{\rm C} \over \nu_{\rm S}  }  \, \sim \, 10^9 
 \, \left( {L_{\rm C} \over L_{\rm S}} \right)^{1/2}; \qquad \nu_{\rm S}>\nu_{\rm t}
\label{vsvclslc}
\end{equation}
This can be thought as a {\it consistency check}, to see if our assumption of
inverse Compton on the BLR photons is reasonable.
Instead, to the aim to derive the parameters, we have to use additional information.
One important information comes from the {\it superluminal motion} of the sources.
Usually, it is observed at 5--22 GHz. 
At these frequencies the flux is due to a region of the jet self--absorbing there.
This region {\it is not} the same as the one producing most of the emission
(we know it because of variability: most of the emission is produced
in a much more compact region, self--absorbing at $\nu_{\rm t}\sim$100--1000 GHz).
Therefore we must make the assumption that both the viewing angle and
the bulk Lorentz factor do not change from the two regions.
If this is true, and assuming that the viewing angle is $\theta_{\rm v} \sim 1/\Gamma$,
we can set
\begin{equation}
\Gamma  \, \sim \, \beta_{\rm app} 
\label{gbetapp}
\end{equation}
and thus solve for $B^\prime$.
Additional information may come from the details of the SED, if we see 
some signs of the SSC emission: 
even if not dominating the bolometric luminosity, it may contribute,
especially in the X--rays.
\begin{figure}[h]
\center
\vskip -0.5 cm
\includegraphics[height=12cm, width=13cm]{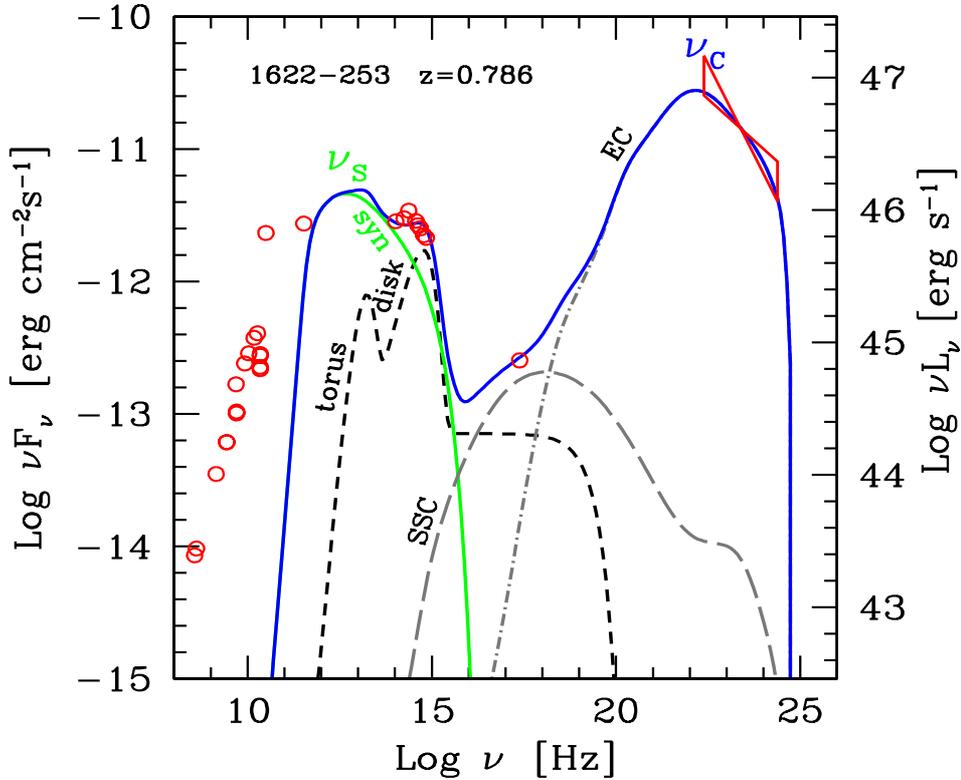}
\vskip -0.8 true cm
\caption[h]{The SED of the blazar 1622--253, together with a
fitting model.
The peak of the Compton emission $\nu_{\rm C}=10^{22}$ Hz,
and the ratio $L_{\rm C}/L_{\rm S}=10$. 
The figure shows also the contribution from the accretion disk, its
X--ray corona and the IR torus (short dashed line) and SSC (long dashed).
}
\label{1622}
\end{figure}{}

\vskip 0.3 true cm
\noindent
{\it Example EC ---} By EC we mean ``External Compton", namely that 
the relevant seed photons for scattering are produced {\it externally}
to the jet, as in the case of the BLR photons.
Suppose that we observe a source as the one shown in Fig. \ref{1622}.
Assume $\beta_{\rm app} =10$. From the figure $L_{\rm C}/ L_{\rm S} =10$.
Assume $\nu_{\rm C}\sim 10^{22}$ Hz. 
We do not observe directly the synchrotron peak, but the radio spectrum
is flat ($\propto \nu^0$) and the optical emission is steep 
[i.e. $F(\nu) \propto \nu^{-1.5}$ in the optical, once the contribution from
the accretion is subtracted].

From Eq. \ref{bgamma} we have $B^\prime/\Gamma = 0.26$. 
Assuming $\Gamma=\beta_{\rm app}$, we derive $B^\prime=2.6$ G.
From Eq. \ref{bgamma2} we have 
$\nu_{\rm S} = \nu_{\rm C}\, (B^\prime/\Gamma)/8.8\times 10^8 = 3\times 10^{12}$ Hz.
Is that consistent with what we observe?
Yes: the radio and optical spectrum indicate a peak (in $\nu F_\nu$)
between the last observable radio frequencies (usually 100 GHz =$10^{11}$ Hz)
and the optical spectrum.
The value of $\nu_{\rm S}=3\times 10^{12}$ Hz is 
larger than the self--absorption frequency of our emitting region.

\subsection{The power of the jet}

It is not easy to estimate the total power of the jet.
There are a few ways to do it, including studying the
energy content of the radio lobes, thought to be the
calorimeter of the source, and, more recently, the energy
required for the jet to produce an ``X-ray cavity" in cluster of galaxies, around
the central giant galaxies that are also radio--loud sources.

The jet power has different components:
\begin{itemize}
\item The jet emits, and the produced radiation must be at the expenses
of the jet power. Calculating the power (in all directions) in the form
of radiation sets a lower limit to the jet power.

\item The jet transports matter. If anything, we must have the electrons
responsible for the radiation we see. 
They are relativistic in the comoving frame (lower case $\gamma$), 
and have also a relativistic bulk  motion (capital $\Gamma$).

\item The jet transports also protons. How many? This is controversial,
because if the emission is totally produced by electron--positrons,
there are no protons.
If instead the emitting plasma a is a normal electron--proton one, than there is one
proton per emitting electron.

\item The jet transports a magnetic field. There is a Poynting flux.

\end{itemize}
All these components can be formally accounted for by writing
\begin{equation}
P_{\rm i} \, =\, \pi R^2 \Gamma^2  \beta\,  c \, U^\prime_{\rm i}
\end{equation}
where $U^\prime_{\rm i}$ is the comoving energy density of the i{\it th} component:
when dealing with the emitting electrons we will have
\begin{equation}
U^\prime_{\rm e} \, =\, m_{\rm e} c^2 \int \gamma N(\gamma) d \gamma \,
=\, \langle \gamma\rangle n^\prime_{\rm e}\, m_{\rm e} c^2 
\end{equation}
When dealing with the magnetic field we have 
\begin{equation}
U^\prime_{\rm B} \, =\, {B^{\prime 2} \over 8\pi}
\end{equation}
and when dealing with the radiation we have
\begin{equation}
U^\prime_{\rm rad} \, =\, 
{L^\prime  \over 4\pi R^2 c} \, = {L  \over 4\pi R^2 c \, \delta^4 }
\end{equation}
For the protons, assuming them cold, 
\begin{equation}
U^\prime_{\rm p} \, = U^\prime_{\rm e} \, {m_{\rm p} \over m_{\rm e} }\, 
{n^\prime_{\rm p} \over \langle \gamma\rangle n^\prime_{\rm e}}
\end{equation}
The logic is the following:
we sit at some place along the jet, cut a cross sectional surface of it,
and count how many photons, protons, electrons and magnetic field are passing by 
in one second. One factor $\Gamma$ accounts for the different mass or energy 
as seen in the lab frame (i.e. a proton will have a mass $\Gamma m_{\rm p}$)
and the other factor $\Gamma$ comes because the density we see is more than
the comoving density (lengths are smaller in the direction of motion).
This is a flux of energy, i.e. a power.

It is found that the jet is powerful, often as powerful as the accretion disk, and
sometimes more.
Furthermore, the jet of powerful blazars {\it cannot be} dominated by
the Poynting flux.
This is to be expected, since the synchrotron component in these sources
is only a minor fraction of the bolometric luminosity, dominated by the 
inverse Compton component.
This limits the possible values of the magnetic field.

\subsection{Lobes as energy reservoirs}

It is also found that the energy dissipated into radiation is
only a minor fraction of the jet power, at least in powerful FSRQs.
Most of the power goes to energize the lobes.
There, we have a relaxed region, in the sense that the dimensions are
huge, and the magnetic field must be small (i.e. $B\sim 10^{-5}$ G).
Large dimensions also means that the radiation energy density is small.
Therefore the radiative cooling timescales are very long.
Particles have not enough time to emit their energy, that instead goes
to increase the dimension of the region.
Extended radio lobes are therefore a reservoir of the energy that the
jet has provided through the years.
If we could measure the energy content of the lobe and its lifetime, than
we could measure ($E/t$): the {\it average} power of the jet.

Lobes emit in the radio, through synchrotron radiation.
They are so big that we can easily resolve them. So we know
their size.
To produce synchrotron radiation, there must be some magnetic field and
relativistic particles.
For a given observed synchrotron luminosity, we could have relatively large
magnetic fields and fewer particles, or, on the contrary,
less magnetic field and more particles.
These solutions are not energetically equivalent, as we shall see.
When the energy content in the magnetic field and particles balance,
then we have the minimum total energy. 
To see this, consider a lobe of size $R_{\rm lobe}$ 
emitting $L_{\rm lobe}$ by synchrotron:
\begin{eqnarray}
L_{\rm lobe} \, &= &\, {4\pi \over 3} R^3_{\rm lobe} m_{\rm e} c^2 \int N(\gamma) 
\, \dot\gamma_{\rm S} \, d\gamma 
\nonumber \\
&=&\,  {4\pi \over 3} R^3_{\rm lobe} 
{4\over 3} \sigma_{\rm T}\, c\,{B^2 \over 8\pi} 
\langle \gamma^2 \rangle \, n_{\rm e} 
\end{eqnarray}
This gives $n_{\rm e}$:
\begin{equation}
n_{\rm e} \, =\,  
{ 9 L_{\rm lobe}  \over 
2 R^3_{\rm lobe}  \sigma_{\rm T} c \, B^2  \langle \gamma^2 \rangle  }
\end{equation}
The energy contained in the magnetic field is
\begin{equation}
E_{\rm B} \, =\,  
{ 4\pi \over 3} R^3_{\rm lobe} { B^2 \over 8\pi}  
\end{equation}
The energy contained  in the particles  depends if we have cold or hot protons.
We do not know. 
We can parametrize our ignorance saying that the proton energy is $k$
times the energy contained in the relativistic electrons.
Therefore the particle energy in the lobe is
\begin{eqnarray}
E_{\rm e,p} \, &=&\,  
{ 4\pi \over 3} R^3_{\rm lobe} \langle \gamma\rangle \,  n_{\rm e} m_{\rm e} c^2 \, (1+k)
\nonumber\\
& =& \, 
 6\pi   \,  { \langle \gamma\rangle m_{\rm e} c^2 \over \langle \gamma^2 \rangle} \,  
{L_{\rm lobe}  \over \sigma_{\rm T} c \, B^2  } \, (1+k)
\end{eqnarray}
Note the dependence on $B^2$: $E_{\rm B} \propto B^2$ , while $E_{\rm e,p} \propto 1/B^2$.

\begin{figure}[h]
\center
\vskip -0.5 cm
\includegraphics[height=12cm, width=13cm]{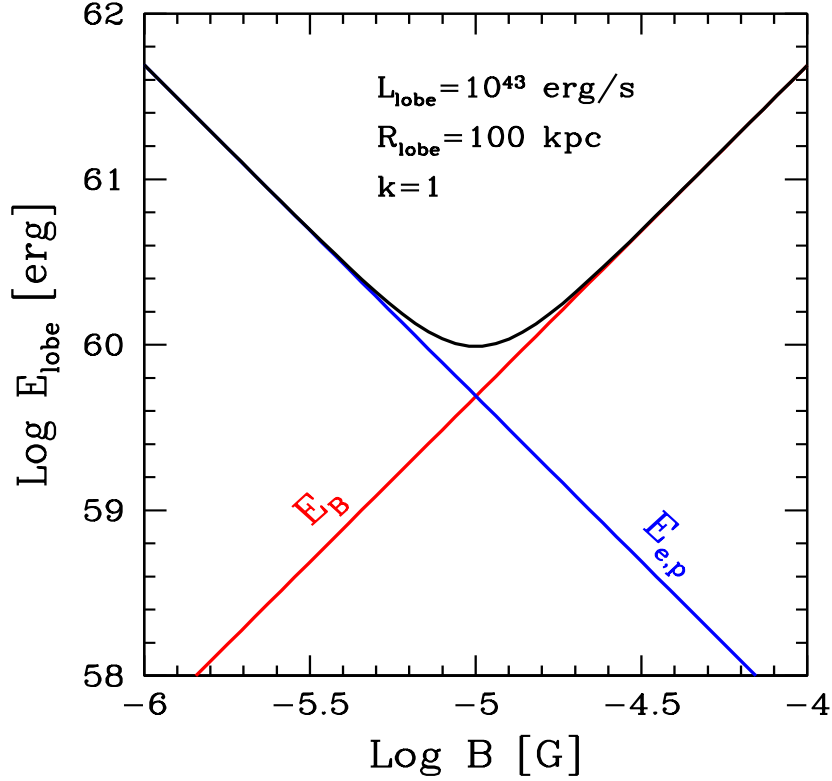}
\vskip -0.8 true cm
\caption[h]{The energy contained in the lobe in magnetic field 
($E_{\rm B}$) and in particles ($E_{\rm e,p}$).
The black line is the sum.
We have assumed that the lobe of 100 kpc in size emits
synchrotron luminosity $L_{\rm lobe}=10^{43}$ erg s$^{-1}$ and that
protons have the same energy of the emitting electrons (i.e. $k=1$).
For the particle distribution, we have assumed $\gamma_{\rm min}=1$, 
$\gamma_{\max}=10^5$ and $p=2.5$ [i.e. $N(\gamma)\propto \gamma^{-2.5}$].
}
\label{equip}
\end{figure}{}

Due to these scalings, the total energy $E_{\rm tot}=E_{B}+E_{e,p}$
{\it will have a minimum} for $E_{B}=E_{e,p}$.
This condition of minimum energy also corresponds to {\it equipartition}.
Fig. \ref{equip} shows an example, for which the {\it equipartition magnetic field}
is $B_{\rm eq}\sim 10^{-5}$ G: with this field we minimize the total energy requirement.
In the specific case illustrated in Fig. \ref{equip}, the minimum energy
contained in the lobe is $E_{\rm tot}\sim 10^{60}$ erg.
If we knew its lifetime, we could calculate the average jet power.
Assuming 10 million years $\sim 3\times 10^{14}$ s, we have
$\langle P_{\rm jet} \rangle \sim 3\times 10^{45}$ erg s$^{-1}$.
In this case the lobe synchrotron luminosity is only a tiny fraction
(i.e. $\sim 3\times 10^{-3}$) of the received power.
This estimates neglects adiabatic losses, likely to be important.
Suppose to include them: how they modify the estimate on the average 
jet power?

\section{Open issues}

Research in the AGN field is still active, because the open issues are many.
The previous notes tried to give a glimpse of the ``common wisdom" picture
of the mainframe about AGNs, but we should avoid to take it as absolutely
not controversial. 
The following is an incomplete list of open problems.

\begin{itemize}

\item{\bf Black hole masses.} We have several ways to estimate the black hole masses,
but the uncertainties are still large. 
Knowing the mass more precisely, we could know all the quantities in
Eddington units, which is more physical.

\item{\bf Black hole growth.} When did the supermassive black hole form?
It takes time to build them up. 
Can we live with black hole seeds of stellar size or do we need
something more exotic?

\item {\bf Accretions modes.}
When  the density is small, protons cannot interact efficiently with electrons
(responsible for the emitted radiation and the cooling).
The energy remain in the protons, that do not cool. 
The pressure increases, and the disk puffs up, and the density becomes smaller still.
At what accretion rate is there the transition? 
Are Coulomb collisions the only way for protons and electrons to interact?

\item{\bf What is the broad line region?} We still do not know if it is an ensemble
of clouds in Keplerian motion, or some form of inflow or outflow.

\item{\bf The infrared ``torus".} It should be unstable.
Is it formed by a clumpy medium? Can it be, instead, a tilted disk?
Where does it begin: is it a separate structure or just a continuation of the disk?

\item {\bf The acceleration mechanisms in the jet.} Shocks are the best bet,
but shocks accelerate protons more than electrons.
Accounting for hot (and relativistic) protons would increase the estimates for the jet power.
What is the role of magnetic reconnection?

\item {\bf The power of jets.} We just started to estimate them in a reliable way,
but we are still far from robust conclusions.
Main uncertainties: the amount of electron--positron pairs, and the contribution
of hot protons.

\item{\bf The spin of the black hole.} The rotational energy of the black hole is a huge
energy reservoir that can be extracted.
Does the black hole spin rapidly? Is it the spin energy that is used to accelerate
relativistic jets? 
Is the spin at the root of the radio--loud/radio--quiet dicotomy?

\end{itemize}




\end{document}